\documentclass[12pt]{article}
\usepackage[]{graphicx}
\usepackage[]{color}
\usepackage[a4paper, margin=1in]{geometry}
\usepackage{amsmath}
\usepackage{pdfpages}
\usepackage{multirow}
\usepackage{ulem}

\usepackage[utf8]{inputenc}
\usepackage[LAE, T1]{fontenc}
\usepackage{amsmath}
\usepackage{amsfonts}
\usepackage{amssymb}
\usepackage{rotating}

\def\data{\hbox{Data}}
\def\ln{\hbox{ln}}

\def\min{\hbox{min}}
\def\max{\hbox{max}}

\makeatletter
\def\maxwidth{
  \ifdim\Gin@nat@width>\linewidth
    \linewidth
  \else
    \Gin@nat@width
  \fi
}
\makeatother
\usepackage[caption=false]{subfig}
\usepackage{changepage}

\definecolor{fgcolor}{rgb}{0.345, 0.345, 0.345}

\usepackage{framed}
\makeatletter
 {\par\unskip\endMakeFramed%
 \at@end@of@kframe}
\makeatother

\definecolor{shadecolor}{rgb}{.97, .97, .97}
\definecolor{messagecolor}{rgb}{0, 0, 0}
\definecolor{warningcolor}{rgb}{1, 0, 1}
\definecolor{errorcolor}{rgb}{1, 0, 0}
\usepackage {placeins}
\usepackage{titlesec}
\usepackage{amsfonts}
\usepackage{amsmath}
\usepackage{mathabx}
\usepackage{amssymb}
\usepackage{rotating}

\usepackage{alltt}
\IfFileExists{upquote.sty}{\usepackage{upquote}}{}

\begin{document}
\title{A simulation study of the estimation quality in the double-Cox model with shared frailty for  non-proportional hazards survival analysis. }

\date{\today}
\author{Alexander Begun and  Elena Kulinskaya\\  (University of East Anglia,
Norwich,United Kingdom)\\}
\date{\today}
\maketitle
\begin{abstract}
The Cox regression, a semi-parametric method of survival analysis, is  extremely popular in biomedical applications. The proportional hazards assumption is a key requirement in the Cox model.  To accommodate non-proportional hazards, we propose to parameterise  the shape parameter of the baseline hazard function using the additional, separate Cox-regression term which  depends on the vector of the covariates. We call this model the double-Cox model. The R programs for fitting the double-Cox model are available on Github. We formally introduce the double-Cox model with shared frailty and investigate, by simulation, the estimation bias and the coverage of the proposed point and interval estimation methods for the Gompertz and the Weibull baseline hazards.  In applications with low frailty variance and a large number of clusters, the marginal likelihood estimation  is almost unbiased and
the profile likelihood-based confidence intervals provide good coverage for all model parameters.  We also compare the results from the over-fitted double-Cox model to those from the standard Cox model with frailty  in the case of the scale-only proportional hazards.  Results of our simulations  on the  bias and coverage of the model parameters are provided in 12 Tables and  in 145 A4 Figures, 178 pages in total.
\end{abstract}

\newpage
\section{Introduction}
 The  Cox regression \cite{Cox1972}, a semi-parametric method of survival analysis, is  very popular  in biomedical applications.
The semi-parametric approach is based on the maximum likelihood Breslow estimator of the cumulative baseline hazard function where this function is regarded as infinite-dimensional parameter and the maximum partial likelihood estimators for the Cox-regression parameters \cite{Lin}.  The proportional hazards assumption is a key requirement in the Cox model.   In the parametric approach, the baseline hazard function has a parametric form corresponding to the exponential, Weibull, Gompertz or other survival distribution. Although the parametric approach guarantees more power than the  semi-parametric one if the form of the baseline hazard function is known, the semi-parametric method is preferable if it is not the case.

 Proportional hazard model under frailty setting is a natural extension of the simple proportional hazards model. A non-negative random variable $Z$ (frailty) is added to the model, so that   the  instantaneous conditional risk of failure at the moment $t$, $h(t|\mathbf u, Z)=Z\chi (\mathbf u)h_0(t)$ is proportional to the baseline hazard function $h_0(t)$, the non-negative function $\chi (\mathbf u)$ for the vector of observed covariates $\mathbf u$, and a value of $Z$. The function $\chi (\mathbf u)$ is usually defined as $\exp (\beta \mathbf u)$ (the Cox-regression term) and specifies the fixed effects. The frailty $Z$ is included to take into account the influence of unobserved factors and to avoid specification of the joint distributions for dependent  failures (e.g. deaths of individuals belonging to the same family or to the same cluster).

However, for real-world data, the proportional hazards assumption may not be satisfied. 
  Stratification is a  useful simple option  for non-proportional hazards analysis, allowing to take into account different forms of the baseline hazard functions in different strata. This option is widely available in  statistical software relating to survival analysis. In the parametrical approach, the baseline hazard functions in different strata need to be specified parametrically. 
  
In this paper, we consider a new method of specifying the hazard functions in different strata. Namely, we assume that the shape parameter of the baseline hazard function can be specified  using the additional, separate Cox-regression term. That is, the shape parameter $b(\mathbf u)$ of the hazard function is written as  $b\exp (\beta _b\mathbf u)$ and depends on the vector of the covariates. This parametrization retains the general form of the hazard function over the strata  and is similar to one in \cite{devarajan} in the case of the Weibull distribution, but differs for other hazard functions. We call this model a double-Cox model.


 The double-Cox  model with shared frailty was  successfully applied  to analysis of time-to-failure of hip replacements \cite{Begun2019} and to analysis of effects of HRT \cite{HRT2022} and of a diagnosis of type-2 diabetes mellitus \cite{Ncube2022} on longevity. In present paper, we formally introduce the model and investigate, by simulation,  the estimation bias and the coverage of the proposed estimation methods. In Section 2, we discuss point and interval parameter estimation. Simulation design is described in Section 3.  The results of the simulations are described in Section 4.   We also compare the results from the overfitted double-Cox model to those from the standard Cox model and from the {\it parfm} R package for parametric frailty models in the case of the scale-only proportional hazards, \cite{parfm}.  Summary and discussion are in Section 6. The R programs for fitting the double-Cox model are available on Github, see https://rdrr.io/github/AB5103/doubleCoxr/ for details.

\section{Model and ML estimation}
\subsection{The double-Cox model with shared frailty}


 Consider a non-proportional hazard regression model with frailty $Z$ and a specified two-parameter (scale and shape) baseline hazard function, where both parameters log-linearly depend on the observed covariates. In this study we consider  two baseline distributions, the Gompertz and the Weibull. The Gompertz hazards are typically used in human longevity studies and the Weibull hazards in modelling time-to-failure  of technical devices.  The cumulative conditional hazard function is defined as
\begin{eqnarray}
\begin{aligned}
\tilde H(t|\mathbf u,Z)=&ZH(t|\mathbf u)=&Ze^{\beta _{scale}\mathbf u} \left (\frac {t}{a}\right )^{b\exp(\beta _{shape}\mathbf u)} \label{HWeib}
\end{aligned}
\end{eqnarray}
in the case of the Weibull model and as
\begin{eqnarray}
\begin{aligned}
\tilde H(t|\mathbf u,Z)=&ZH(t|\mathbf u)=&{Zae^{\beta _{scale}\mathbf u}}\frac{(e^{b\exp(\beta _{shape}\mathbf u)t}-1)}{b\exp(\beta _{shape}\mathbf u)}\label{HGomp}
\end{aligned}
\end{eqnarray}
in the case of the Gompertz model. Here $a>0$ and {$b>0$ are  the scale and the shape parameters of the baseline survival distribution, respectively, $\mathbf u$ is the vector-column of the covariates, $\beta _{scale}$ and $\beta _{shape}$ are the vector-rows of the Cox-regression parameters.

We consider the gamma-distributed frailty $Z$  with mean 1, variance $\sigma ^2$ and probability density function $f(z|\sigma ^2)$ and define the \lq\lq random effect" by $\omega=\ln Z$. We also assume that the population  is divided into $N_{cl}$ clusters and all individuals from the cluster $i$, $i=1,...,N_{cl}$, share the same frailty $Z_i$.

The conditional and the marginal survival functions are given by
\begin{eqnarray}
\begin{aligned}
S(t|\mathbf u, Z)=&\exp(-\tilde H (t|Z,\mathbf {u}))\nonumber, \\
S(t|\mathbf u)=&\mathbb {E}S(t|\mathbf u, Z)=(1+\sigma ^2H(t|\mathbf u))^{-1/\sigma ^2}.\nonumber
\end{aligned}
\end{eqnarray}
The conditional likelihood function is defined by
\begin{eqnarray}
\begin{aligned}
\mathcal L_c(\data|\zeta,Z_1,...,Z_{N_{cl}})=\prod _{i,j}\left (-\frac {\partial }{\partial t_{ij}}\right )^{\delta _{ij}}\exp(-\sum _j\tilde H(t_{ij}|\mathbf u _{ij},Z_i)),\label{Lc}
\end{aligned}
\end{eqnarray}
where $\zeta=(a,b,\beta _{scale},\beta _{shape})$ is the  vector of parameters, $\delta _{ij}$ stands for censoring (1 if censored and 0, otherwise) and indices  $i$ and $j$ correspond to a cluster and a subject in that cluster, respectively.
We consider parameter estimation in the model~(\ref{Lc}) in the next section.
\subsection{Point  estimation of the parameters}
Marginal likelihood is calculated by marginalizing out the frailty $Z$ in the  conditional likelihood~(\ref{Lc}).  From a random effect, the marginal likelihood inherits only parameter $\sigma ^2$.
The $k$-dimensional vector of the unknown parameters $\zeta_{\sigma} =(\zeta,\sigma ^2)$  is  estimated by maximizing the marginal likelihood function (or its logarithm)
\begin{eqnarray}
\begin{aligned}
\mathcal L_m(\data|\zeta_{\sigma})=\mathbb E\mathcal L_c(\data|\zeta,Z_1,...,Z_{N_{cl}})=\prod _{i,j}\left (-\frac {\partial }{\partial t_{ij}}\right )^{\delta _{ij}}\left (1+\sigma ^2\sum _jH(t_{ij}|\mathbf u _{ij})\right )^{-1/\sigma ^2}.\label{Lm}
\end{aligned}
\end{eqnarray}

It is easy to see that the marginal likelihood function (\ref{Lm}) is a Laplace transform of the frailty distribution  calculated at the point $H(t|\mathbf u)$. The Gamma frailty results in a closed-form marginal likelihood.

\subsection{Confidence intervals for the parameters}
\subsubsection{Standard-error based confidence intervals}
Information about the standard errors of the parameter estimates can be obtained from the inverse of the Hessian matrix. Let $\mu $ be the $k$-vector parameter and $\hat \mu $ its maximum likelihood estimate (MLE). Assume that $[\mu _i^l,\mu _i^u]$ is the $1-\alpha $ confidence interval for the $j^{th}$ component $\hat \mu _j$, $j=1,...,k$. The coverage probability for component $j$ is defined by
\begin{eqnarray}
\begin{aligned}
P_{cov}(\mu _j)=\mathbb P(\mu _j\in [\mu _j^l,\mu _j^u]).\nonumber
\end{aligned}
\end{eqnarray}
This probability can be estimated from simulations as a proportion of cases when the true value of $\mu _j$ lies in interval $[\mu _j^l,\mu _j^u]$.

 The scale and the shape parameters $a$ and $b$ of the baseline hazard distribution and the Cox-regression parameters $\beta _{scale}$ and $\beta _{shape}$ do not lie on the boundary. But the frailty variance  $\sigma ^2$ can be equal to zero  corresponding to the model without frailty. If $\sigma ^2>0$ and $\sigma ^2$ is large compared with the standard error for ${\hat \sigma} ^2$,  the confidence intervals can be specified  using the asymptotic normality of MLE:
\begin{eqnarray}
\begin{aligned}
\sqrt n(\hat \mu - \mu)\xrightarrow {d}\mathcal N (\mathbf 0,\Sigma ),\nonumber
\end{aligned}
\end{eqnarray}
where  $n$ is the number of observations, $\mathcal N$ is the $k$-variate normal distribution with zero mean and  the $k\times k$ asymptotical variance-covariance matrix $\Sigma $  with elements $\Sigma _{l,m}$, $l,m=1,...,k$. In general, for any  values of $\sigma^2$, the confidence intervals can be calculated using a mixture of the truncated normal distribution and a point mass at zero for $\hat\sigma^2$, as in \cite{Bohnstedt}:
\begin{eqnarray}\label{eq:SE}
\begin{aligned}
\sqrt n(\hat \mu _j- \mu _j)&\xrightarrow {d}\Phi \left (\frac {\nu}{\kappa}\right )\mathcal {TN} _1(\mathbf \mu_{\nu},\Sigma _j,s,t)+\Phi \left (-\frac {\nu}{\kappa}\right )\mathcal {N} \left (-\frac {\Sigma _{j,k}}{\Sigma _{k,k}}\nu ,\Sigma _{k,k}-\frac {\Sigma _{j,k}^2}{\Sigma _{k,k}}\right ), \\
\sqrt n\hat \sigma ^2&\xrightarrow {d}\Phi \left (\frac {\nu}{\kappa}\right )\mathcal {TN} (\mathbf \nu,\kappa ^2,0,\infty)+\Phi \left (-\frac {\nu}{\kappa}\right )\chi_0^2,
\end{aligned}
\end{eqnarray}
where $\chi _0$ is a point mass at zero, $\kappa =\sqrt {\Sigma _{k,k}}$, $\nu=\sqrt n\sigma ^2$, $\mu _\nu=(0,\nu)^T$, $\Sigma _j=\left( \begin{smallmatrix} \Sigma _{j,j}&\Sigma _{j,k}\\ \Sigma _{k,j}&\Sigma _{k,k} \end{smallmatrix} \right)$, $j=1,...,k-1$, $\mathcal {TN}_1(x,V,s,t)$ is the marginal distribution of the first component of a truncated bivariate normal distribution with mean vector $x$, covariance matrix $V$, and the lower and upper truncation limits $s=(-\infty ,0)^T$ and $t=(\infty, \infty)^T$, respectively. We call these intervals standard-error-based.

\subsubsection{Profile likelihood based confidence intervals} The  standard-error based confidence intervals may have low coverage. The profile likelihood (PL) based confidence intervals are an attractive alternative. The profile likelihood  method inverts the likelihood ratio test to obtain a confidence interval for the parameter under study. Let $(\xi ,\zeta )$ be a vector of the unknown parameters of interest $\xi $ and the nuisance parameters  $\zeta $ in a statistical model. Denote by $Lik(\xi,\zeta)$ the likelihood function, and by $(\xi ^*,\zeta ^*)$ the maximum likelihood estimates. Define the profile likelihood function $Lik_1(\xi)$ by maximizing the function $Lik(\xi,\zeta)$ over the parameter $\zeta $, i.e. $Lik_1(\xi)=\max _\zeta Lik(\xi,\zeta)$. The point $\xi _0$ lies in the 95\% confidence interval for $\xi$ if $2(\ln Lik_1(\xi^*)-\ln Lik_1(\xi_0))\leq3.84$ (3.84 is the 95\% percentile of the $\chi ^2(1)$ distribution). The coverage probability of the 95\%  confidence interval for parameter $\xi$ can be estimated from simulations as a proportion of cases when the true value $\xi_0$ lies in the 95\% interval for $\xi$.

\section{Design of the simulations}
In the simulation study,  we investigate the quality of the parameter estimation for a double-Cox model with two covariates: a binary  "Success" and a continuous "Score", for both the Gompertz and the Weibull baseline hazards. The scale and shape parameters of these distributions were fixed at $a=20$, $b=1.5$ for Weibull and at  $a=0.0001$, $b=0.1$ for Gompertz hazards. The   variance of the gamma-distributed frailty $\sigma^2$ varied from zero to 5. 
   The choice of the  shape and scale parameters guaranteed a realistic life expectancy. 

 The binary covariate "Success" was generated from  the Bernoulli($p_{Success}$) distribution for each individual in a study. The covariate "Score" was generated from  the normal distribution with mean 0 and  variance 0.2 independently of the covariate "Success". The corresponding  Cox-regression vector-parameters $\beta _{shape}$ 
  and   $\beta_{scale}$
   were chosen independently of each other.  To make effects of scale and shape parameters comparable, the  Cox-regression coefficients for shape parameters were chosen to be one order smaller than those for scale.

 Apart of the parameters of the model $a$, $b$, $\beta _{scale}$, $\beta _{shape}$ and $\sigma ^2$, we also studied the effects of the additional structure parameters of the data such as the  sample size $N_{sample}$, the number of clusters $N_{cl}$, the proportion $p_{Success}$ of the baseline values for the binary covariate  and the censoring rate $p_{cens}$.
5000 repetitions were used for each of 1440 configurations of the parameters.

Additionally, to test the behaviour of the parameter estimates for an over-parametrised double-Cox model with zero shape,  we generated the data using a subset of the parameters, resulting in 16 configurations.
The full list  of the parameters used in simulations is provided in Table~\ref{tab:design}.
R statistical software \cite{rrr} was used for simulations.

\begin{table}[htb]
\caption{ \label{tab:design} {Values of parameters in the simulations}}
\begin{center}
\begingroup\fontsize{8pt}{10pt}\selectfont
\begin{tabular}{ll}
  \hline
\multicolumn{1}{c}{Parameter}&\multicolumn{1}{c}{Values}\\
\hline
Number of repetitions&5000\\
Sample size $N$ & 300*, 1000*, 10000\\
Number of clusters $N_{cl}$&10*, 100*\\
Censoring rate $p_{cens}$&0*, 0.4*, 0.8\\
Proportion $p_{Success}$ of 1's in the variable "Success"&0.25, 0.5*\\
$(\beta_{scale-success},\beta_{scale-score})$&(0.5,\;1), (-0.5,\;-1)* \\
$(\beta_{shape-success},\beta_{shape-score})$&(0.05,\;0.1), (-0.05,\;-0.1) \\
True variance of frailty $\sigma^2$&0*, 0.5, 1, 2*, 5\\
($a_{Gompertz}$,$b_{Gompertz}$)&(0.0001*,\;0.1*)\\
($a_{Weibull}$,$b_{Weibull}$)&(20*,\;1.5*)\\
$p_{Success}$&0.25, 0.5\\
"Success"&\it{Bernoulli($p_{Success}$)}\\
"Score"&$\mathcal{N}$(0,0.2)\\
\hline
{*Parameters of the additional simulations for the over-parametrised model.}
\end{tabular}
\endgroup
\end{center}
\end{table}

\subsection{Designing a simulation with a predefined proportion of censoring}
The time-to-failure  $X=\min (T,C)$, where conditionally independent (given frailty and covariates) random variables $T$ and $C$ are survival and censoring times, respectively. We assume that the censoring time is uniformly distributed on an interval [0,$\theta$] ($C\sim Uni(0,\theta)$) with probability density function $g(c|\theta )=\theta ^{-1}$.
 For an individual $i$ ,  $\text {P}(T\geq C|Z_i,\mathbf u_i)$ is the probability of being censored, and $\delta =Ind(C\leq T) $ is the censoring index. Then, (see \cite{Wan}) 
\begin{eqnarray}
\begin{aligned}
\mathbb P(\delta =&1|\theta ,Z_i,\mathbf u_i)=\mathbb P(C\leq T<\infty,0\leq C<\infty|Z_i,\mathbf u_i)=\int _0^\theta g(c|\theta)\int_c^\infty f(t|a,b,Z_i,\mathbf u_i)dtdc\nonumber \\
=&\theta^{-1}\int _0^\infty \int_c^\infty f(t|a,b,Z_i,\mathbf u_i)dtdc,
\end{aligned}
\end{eqnarray}
where 
$f(.)$ is the probability density function of the survival time. Applying this formula to the Weibull and the Gompertz double-Cox models, we obtain
\begin{eqnarray}
\begin{aligned}
\mathbb P(\delta=1|\theta ,Z_i,\mathbf u_i)=&\frac {Z_iae^{\beta _{scale}\mathbf u_i}}{\theta be^{\beta _{shape}\mathbf u_i}}\int _0^{(\theta /Z_iae^{\beta _{scale}\mathbf u_i})^{be^{\beta _{shape}\mathbf u_i}}}x^{b^{-1}e^{-\beta _{shape}\mathbf u_i}-1}e^{-x}dx \nonumber \\
=&\frac {Z_iae^{\beta _{scale}\mathbf u_i}}{\theta be^{\beta _{shape}\mathbf u_i}}\Gamma \left (b^{-1}e^{-\beta _{shape}\mathbf u_i},(\theta /Z_iae^{\beta _{scale}\mathbf u_i})^{be^{\beta _{shape}\mathbf u_i}}\right ),\nonumber
\end{aligned}
\end{eqnarray}
in the case of the Weibull model and
\begin{eqnarray}
\begin{aligned}
\mathbb P(\delta=1|\theta ,Z_i,\mathbf u_i)=&\frac {e^{Z_iae^{\beta _{scale}\mathbf u_i}/be^{\beta _{shape}\mathbf u_i}}}{\theta be^{\beta _{shape}\mathbf u_i}}\nonumber \\
\times &\left (E_1(Z_iae^{\beta _{scale}\mathbf u_i}/be^{\beta _{shape}\mathbf u_i})-E_1(Z_iae^{\beta _{scale}\mathbf u_i}e^{be^{\beta _{shape}\mathbf u_i}}/be^{\beta _{shape}\mathbf u_i})\right ), \nonumber \\
\end{aligned}
\end{eqnarray}
where $E_1(x)$ is the  exponential integral $E_1(x)=\int _x^\infty e^{-s}s^{-1}ds$, in the case of the Gompertz model.  \\
The population censoring rate is obtained by averaging the individual censoring rates with respect to random variables $Z$ and $\mathbf u$,
\begin{eqnarray}
\begin{aligned}
\mathbb P(\delta =1|\theta )=\mathbb E_{Z,\mathbf u}\mathbb P(\delta =1|\theta ,Z,\mathbf u)=\int _{D_u\times {D_z}}P(\delta =1|\theta ,Z,\mathbf u)p_Z(z)p_U(u)dzdu ,\nonumber
\end{aligned}
\end{eqnarray}
where $p_U(u)$ and $p_Z(z)$ is the probability density functions for independent random variables $\mathbf U$ and $Z$ defined in the domains $D_u$ and $D_z$, respectively. There is no  simple closed form for last integral. We will use the Monte-Carlo integration
\begin{eqnarray}
\begin{aligned}
P(\delta =1|\theta)&\approx N^{-1} \sum _{i=1}^N P(\delta =1|\theta ,Z_i,\mathbf u_i)   \label{eq1} \\
\end{aligned}
\end{eqnarray}
for independent random pairs $(Z_i,\mathbf u_i)$. This approximation converges to $P(\delta =1|\theta)$ as $N\to \infty$  in accordance with the law of large numbers.\\
Given the population censoring rates $p_{cens}$,  the unknown parameter $\theta $  is a numerical solution to  the equation
\begin{eqnarray}
\begin{aligned}
\mathbb P(\delta =1|\theta)=p_{cens}.\label{eq2}
\end{aligned}
\end{eqnarray}

In simulations, the censoring times $c_i$ for individuals $i=1,\cdots, N$,  are generated  from $Uni(0,\theta )$ distribution. For the survival times $t_i$, define  $s_i\sim U(0,1)$. The survival time $t_i$ for an individual $i$ is generated as
\begin{eqnarray}
\begin{aligned}
t_i=a\left( -\frac {\log(s_{i})}{Z_ie^{\beta _{scale}\mathbf u_i}}\right )^{1/be^{\beta _{shape}\mathbf u_i}} \nonumber \\
\end{aligned}
\end{eqnarray}
 in the case of the Weibull model and as
\begin{eqnarray}
\begin{aligned}
t_i= \frac {1}{be^{\beta _{shape}\mathbf u_i}}\log\left (-\frac {be^{\beta _{shape}\mathbf u_i}\log(s_{i})}{ae^{\beta _{scale}\mathbf u_i}Z_i}+1\right ) \nonumber \\
\end{aligned}
\end{eqnarray}
in the case of the Gompertz model. 
\\

\section{Simulation results}
\subsection{Effects of over-parametrization}
To compare the quality of estimation in the  over-parametrised  double-Cox model (shape and scale parameters) with that of the true single-Cox model (scale only), we  calculated two sets of the scale estimates for the true model:  using our own  software for the ML estimation under the single Cox model,  and using the function \lq parfm' from the R-package of the same name. We  also used our double-Cox function to estimated parameters for the over-parametrised  model with additional Cox-regression term for shape $b$ (double Cox). The results are given in Tables \ref{tab:A1} and \ref{tab:A2} in the Appendix. The scale estimates from our single-Cox program  and the \lq parfm' \cite{parfm} were practically identical for all combinations of the parameters $(N_{sample},\;N_{cl},\;p_{cens}\;,p_{Success},\;\sigma ^2_{true})$. For the binary covariate, relative error of estimation was within 2\% for $N=300$ and within 1\% for $N=1000$. For the continuous covariate, the relative error of estimation was within 5\% for $N=300$ and within 1\% for $N=1000$. The double-Cox estimation of scale had a slightly higher relative error, within 3\% for $N=300$, for the binary covariate, and practically the same error as the single-Cox for $N=1000$ and for the  continuous covariate.
Therefore, the estimates are robust against over-parametrization in the shape term.\\
\subsection{General simulation results} The estimates of unknown parameters $(a,b, \beta_{scale},\beta _{shape},\sigma ^2)$ were calculated maximizing the marginal likelihood for 5000 simulations and then averaged. Parameter $\theta$ of the uniform distribution $Uni(0,\theta)$ required to achieve a specified censoring proportion $p_{cens}$ was estimated from equations (\ref{eq1})-(\ref{eq2}) with $N =10^6$ Monte-Carlo simulations. 
We studied the bias of estimation for all parameters, and also coverage of the standard-error-based and the profile-likelihood-based confidence intervals at nominal 95\% level. Results are provided in Tables A.3 to A.12 and in 145 A4 Figures for bias and coverage in Appendices A to F. 

\subsection{Biases in the parameter estimation}
The estimation of the parameters "a" and "b" of the baseline distribution is almost unbiased for small values of the frailty variance $\sigma^2$. However, the bias in estimation of $a$ grows linearly with $\sigma^2$ and may be considerable when $N_{cl}=10$. This bias is positive for the Weibull, and negative for the Gompertz distribution.  The bias in estimation of both $a$ and $b$ decreases in the number of clusters. 

The estimation of the Cox regression scale parameters is almost unbiased for censoring rates up to 40\%, but is noticeably biased when the high censoring rate of 80\% is combined with a modest sample size. For $n=300$, the estimates of the shape parameters are considerably more biased than their scale counterparts,  especially when  $\sigma^2\leq 2$ or for high censoring rate. The sign of the bias depends on the sign of the coefficients.  

Smaller censoring rate increases the number of informative cases. In a sense, this effect is equivalent to an increase in the sample size. It seems that 200 of informative cases (corresponding to the sample size of 1000 and censoring rate of 0.8) are sufficient for  a relatively good estimation of the Cox-regression parameters with bias not exceeding 10\% of their true values.

 Although the model is identifiable, the shape and the scale parameters relating to the same covariates can compete for the likelihood in searching for the maximum likelihood estimates. This  can slow down the convergence of the estimates to their true values.

The estimation bias of $\hat \sigma^2$ is negative and declines practically linearly with the true value of $\sigma^2$, thus the relative error appears to be constant for larger sample sizes.   When the number of clusters  $N_{cl}=100$,
the bias of the Cox-regression parameters is not affected, but that of $\sigma^2$ is considerably reduced. 
This effect may be due to an increase in the number of underlying frailty values equal to the number of different clusters.

Overall, these findings are in agreement with the consistency of the ML estimates and well known underestimation  of  $\sigma ^2$.
The results are similar for the Weibull and the Gompertz model.

\subsubsection{Coverage of the standard-error based (SE) confidence intervals for the Cox regression parameters}
SE confidence intervals provide an asymptotically nominal coverage for the scale Cox-regression parameters,  but not the shape parameters. 
The coverage of the scale parameters suffers  under high censoring combined with smaller sample sizes, and can be as low as 70-80\% for both the Weibull and the Gompertz models. Worryingly, for some parameter combinations, the  coverage of the  shape parameters deteriorates with larger sample sizes, suggesting the use of a wrong limit distribution. Coverage of all parameters improves with the number of clusters. 

When $N_{cl}=10$, the coverage of the frailty variance $\sigma^2$ can be as low as 80\% for some values of $\sigma^2$. 
Coverage improves but may still be erratic when $N_{cl}=100$. Overall, we do not recommend the use of the SE confidence intervals.

 \subsubsection{Coverage of the profile-likelihood-based (PL) confidence intervals}
 PL confidence intervals provide a reasonable amount of coverage to all scale and shape Cox-regression parameters in  both Weibull and Gompertz models. The coverage generally decreases with higher censoring, but it still remains well above 90\% for all studied parameter combinations at a 95\% nominal confidence level, when the sample size $n=300$ and the censoring is as high as 80\%.
 The coverage improves for lower censoring rates and converges to the nominal level for larger sample sizes.
 
 The coverage of the baseline distribution parameters $a$ and $b$ is also acceptable. For 10 clusters, the coverage of $a$ is somewhat below nominal, at about 92\%, but it reaches nominal level for 100 clusters. 

The coverage of $\sigma^2$ is  usually above nominal when $\sigma^2=0$, and at about 92\% when   $\sigma^2>0$ even for large sample sizes when the number of clusters is 10. This parallels the negative bias in estimation of $\sigma^2$. The coverage  improves for larger number of clusters, and is almost nominal for all parameters when $N_{cl}=100$.

 Overall, the PL  method appears to be more robust and reliable and is recommended for use in practice.

\section{Discussion}
The  semi-parametric and parametric methods of survival analysis are  routinely used  in numerous  medical,  biological and demographic studies. Parametric maximum likelihood estimation guarantees the most efficient inferential procedures  when the true form of the underlying hazard function is known. To enable additional flexibility of the  parametric approach,  we included an additional Cox regression term modelling the shape of the baseline hazard function. 

We investigated the marginal likelihood-based estimation of the model parameters. This method works well for the Cox regression parameters, but underestimates the frailty variance and also may result in a biased estimate of the scale $a$ of the baseline hazard function. 

 In general, the underestimation  of the variance of the frailty in the survival models with random effect is a serious drawback of the maximum likelihood estimator. Unfortunately, the REML estimate of the frailty variance $\sigma^2$ is not available in the case of the gamma distributed random effect \cite{Duchateau}. Further research is required to develop the methodology  for unbiased estimation of the frailty variance for the distributions beyond the  log-normal.

In real-life applications that we studied so far, \cite{Begun2019}, \cite{HRT2022},  \cite{Ncube2022},  the frailty variance is usually rather low, and the number of clusters, such as the GP practices, is sufficiently high to make the marginal likelihood estimation of all the model parameters practically unbiased.

 We also considered interval estimation and demonstrated that the profile likelihood-based confidence intervals provide good coverage for all model parameters.
 
 
Overall, the double-Cox regression is a useful addition to the toolkit of survival analysis, and can be recommended for use in practice. 
 
 \section*{Acknowledgements}
The work by A. Begun and E. Kulinskaya was supported by the Economic and Social Research Council [grant number ES/L011859/1].

\bibliographystyle{vancouver} 
\bibliography{SimulationPaper}      %

\clearpage

\section*{Appendix}
\renewcommand{\theequation}{A.\arabic{equation}}
\setcounter{equation}{0}

\renewcommand{\thesection}{A.\arabic{section}}
\setcounter{section}{0}

\clearpage

\renewcommand{\thetable}{A.\arabic{table}}
\setcounter{table}{0}
\section*{Tables}

Tables \ref{tab:A1} and \ref{tab:A2}  compare the quality of estimation in the  over-parametrised  double-Cox model (shape and scale parameters) with that of the true single-Cox model (scale only). These tables provide two sets of the scale estimates for the true model:  using our own  software for the ML estimation under the single Cox model,  and using the function \lq parfm' from the R-package of the same name. Additionally, our double-Cox function was used to estimated parameters for the over-parametrised  model with additional Cox-regression term for shape $b$ (double Cox). 

Tables A.3 to A.12 provide range of the estimation bias for the Cox regression parameters and the frailty variance $\sigma^2$ across all combinations of parameters, separately for the Weibull and the Gompertz baseline hazards.

\clearpage
\begin{sidewaystable}[ht]
\centering
\begingroup\scriptsize
\begin{tabular}{cccccccccccc}
  \hline
\multicolumn{1}{c}{$n$}&\multicolumn{1}{c}{$\sigma ^2$}&\multicolumn{1}{c}{$N_{cl}$}&\multicolumn{1}{c}{\%cens}&\multicolumn{2}{c}{$\beta _{success-scale}$}&\multicolumn{1}{c}{$\beta _{success-shape}$}&\multicolumn{2}{c}{$\beta _{score-scale}$}&\multicolumn{1}{c}{$\beta _{score-shape}$}&\multicolumn{2}{c}{$\sigma ^2$}\\
\cline{5-12}
&&&&single Cox (parfm) & double Cox & double Cox & single Cox (parfm) & double Cox & double Cox & single Cox (parfm) & double Cox \\
\hline
\multirow{8}{*}{300}&\multirow{4}{*}{0}&\multirow{2}{*}{10}&0&-5.59e-03(-5.58e-03) & -1.00e-02 & 1.47e-03 & -1.04e-02(-1.04e-02) & -1.54e-02 & 6.62e-03 & 6.96e-06(4.99e-07) & 7.79e-06 \\
\cline{4-12}
&&&40&  -7.27e-03(-7.26e-03) & -9.43e-03 & -1.22e-03 & -1.40e-02(-1.40e-02) & -1.55e-02 & -1.10e-02 & 1.02e-05(6.64e-07) & 1.15e-05 \\
\cline{3-12}
&&\multirow{2}{*}{100}&0&  -9.43e-03(-9.42e-03) & -1.21e-02 & -1.21e-03 & -1.88e-02(-1.88e-02) & -2.17e-02 & 5.60e-03 & 6.50e-05(6.75e-06) & 6.79e-05 \\
\cline{4-12}
&&&40&  -1.32e-02(-1.32e-02) & -1.53e-02 & -3.28e-03 & -3.74e-02(-3.74e-02) & -3.96e-02 & -9.76e-03 & 1.12e-04(1.13e-05) & 1.32e-04 \\
\cline{2-12}
&\multirow{4}{*}{2}&\multirow{2}{*}{10}&0&  -3.86e-03(-3.86e-03) & -8.53e-03 & 2.39e-04 & -4.93e-03(-4.93e-03) & -7.32e-03 & 9.92e-05 & -3.03e-01(-3.03e-01) & -2.96e-01 \\
\cline{4-12}
&&&40&  -4.29e-03(-4.29e-03) & -8.37e-03 & 1.94e-04 & -1.17e-02(-1.17e-02) & -1.74e-02 & -2.15e-03 & -2.70e-01(-2.70e-01) & -2.60e-01 \\
\cline{3-12}
&&\multirow{2}{*}{100}&0&  -3.78e-03(-3.78e-03) & -8.76e-03 & 4.31e-04 & -1.41e-02(-1.41e-02) & -2.21e-02 & 3.80e-03 & -1.81e-02(-1.81e-02) & -8.56e-03 \\
\cline{4-12}
&&&40&  -7.19e-03(-7.19e-03) & -1.02e-02 & -2.75e-03 & 5.04e-03(5.04e-03) & -1.73e-04 & 3.16e-03 & -4.05e-03(-4.06e-03) & 1.06e-02 \\
\hline
\multirow{8}{*}{1000}&\multirow{4}{*}{0}&\multirow{2}{*}{10}&0&  -1.91e-03(-1.90e-03) & -2.89e-03 & -2.63e-05 & -5.68e-03(-5.68e-03) & -6.86e-03 & 2.08e-03 & 3.17e-06(3.40e-07) & 3.68e-06 \\
\cline{4-12}
&&&40&  -1.98e-03(-1.97e-03) & -2.51e-03 & -1.14e-03 & -1.07e-02(-1.07e-02) & -1.12e-02 & -5.27e-03 & 4.39e-06(3.96e-07) & 5.02e-06 \\
\cline{3-12}
&&\multirow{2}{*}{100}&0&  -9.62e-04(-9.57e-04) & -2.16e-03 & 2.52e-04 & -5.91e-03(-5.90e-03) & -6.35e-03 & 9.61e-04 & 3.07e-05(4.74e-06) & 3.32e-05 \\
\cline{4-12}
&&&40&  -4.97e-03(-4.97e-03) & -5.47e-03 & -1.13e-03 & -1.17e-02(-1.17e-02) & -1.18e-02 & -7.37e-03 & 5.42e-05(7.86e-06) & 6.21e-05 \\
\cline{2-12}
&\multirow{4}{*}{2}&\multirow{2}{*}{10}&0&  -3.19e-03(-3.19e-03) & -3.29e-03 & -5.07e-04 & -4.49e-03(-4.49e-03) & -3.39e-03 & -7.66e-04 & -2.99e-01(-2.99e-01) & -2.97e-01 \\
\cline{4-12}
&&&40&  -1.82e-03(-1.82e-03) & -3.32e-03 & 1.17e-04 & -1.69e-03(-1.69e-03) & -1.38e-03 & -1.52e-03 & -2.92e-01(-2.92e-01) & -2.89e-01 \\
\cline{3-12}
&&\multirow{2}{*}{100}&0&  -2.47e-03(-2.47e-03) & -2.53e-03 & -5.65e-04 & -1.27e-03(-1.27e-03) & -2.38e-03 & 2.28e-04 & -3.16e-02(-3.16e-02) & -2.90e-02 \\
\cline{4-12}
&&&40&  5.17e-04(5.17e-04) & -5.39e-04 & -1.79e-04 & -6.00e-03(-6.00e-03) & -7.18e-03 & 7.18e-04 & -2.05e-02(-2.05e-02) & -1.69e-02 \\
\end{tabular}
\endgroup
\caption{\label{tab:A1}{Bias in the over-parametrised Weibull model by sample size $n$, $\sigma ^2$, the number of clusters $N_{cl}$, and the percent of censored subjects. 
True values of the parameters: $\beta _{scale-Success}=-0.5$, $\beta _{scale-score}=-1$, $\beta _{shape-Success}=0$, $\beta _{shape-score}=0$, $p _{Success}=0.5$.}}
\end{sidewaystable}

\clearpage

\begin{sidewaystable}[ht]
\centering
\begingroup\scriptsize
\begin{tabular}{cccccccccccc}
  \hline
\multicolumn{1}{c}{$n$}&\multicolumn{1}{c}{$\sigma ^2$}&\multicolumn{1}{c}{$N_{cl}$}&\multicolumn{1}{c}{\%cens}&\multicolumn{2}{c}{$\beta _{success-scale}$}&\multicolumn{1}{c}{$\beta _{success-shape}$}&\multicolumn{2}{c}{$\beta _{score-scale}$}&\multicolumn{1}{c}{$\beta _{score-shape}$}&\multicolumn{2}{c}{$\sigma ^2$}\\
\cline{5-12}
&&&&single Cox (parfm) & double Cox & double Cox & single Cox (parfm) & double Cox & double Cox & single Cox (parfm) & double Cox \\
\hline
\multirow{8}{*}{300}&\multirow{4}{*}{0}&\multirow{2}{*}{10}&0&-3.60e-03(-3.60e-03) & 2.72e-03 & -1.30e-03 & -9.49e-03(-9.49e-03) & 2.42e-02 & -5.48e-03 & 7.41e-06(4.45e-07) & 8.09e-06 \\
\cline{4-12}
&&&40&  -1.28e-02(-1.27e-02) & 2.29e-02 & -5.19e-03 & -1.94e-02(-1.93e-02) & 5.14e-03 & -4.69e-03 & 1.03e-05(7.32e-07) & 1.16e-05 \\
\cline{3-12}
&&\multirow{2}{*}{100}&0&  -7.03e-03(-7.01e-03) & -2.81e-04 & -9.85e-04 & -1.50e-02(-1.50e-02) & 4.47e-02 & -8.93e-03 & 7.35e-05(6.55e-06) & 7.47e-05 \\
 \cline{4-12}
&&&40&  -1.29e-02(-1.29e-02) & -4.12e-03 & -1.66e-03 & -2.04e-02(-2.04e-02) & 2.68e-02 & -7.55e-03 & 8.75e-05(8.37e-06) & 7.75e-05 \\
\cline{2-12}
&\multirow{4}{*}{2}&\multirow{2}{*}{10}&0&  -2.54e-03(-2.54e-03) & -8.80e-03 & 3.75e-04 & -7.52e-03(-7.52e-03) & -5.63e-02 & 5.43e-03 & -3.04e-01(-3.04e-01) & -2.96e-01 \\
\cline{4-12}
&&&40&  -6.16e-03(-6.16e-03) & -1.92e-02 & 1.36e-03 & -2.31e-02(-2.31e-02) & -2.75e-02 & -1.64e-04 & -2.98e-01(-2.98e-01) & -2.88e-01 \\
\cline{3-12}
&&\multirow{2}{*}{100}&0&  -7.74e-03(-7.74e-03) & -1.43e-02 & 3.60e-04 & -1.29e-02(-1.29e-02) & -4.20e-02 & 2.80e-03 & -1.77e-02(-1.77e-02) & -8.75e-03 \\
\cline{4-12}
&&&40&  -9.15e-03(-9.15e-03) & -1.33e-02 & 1.32e-04 & -9.14e-03(-9.14e-03) & -7.04e-03 & -1.61e-03 & -1.50e-02(-1.50e-02) & -2.37e-03 \\
\hline
\multirow{8}{*}{1000}&\multirow{4}{*}{0}&\multirow{2}{*}{10}&0&  2.31e-05(2.72e-05) & -2.11e-03 & 8.34e-05 & -4.38e-03(-4.38e-03) & -2.52e-03 & -4.57e-04 & 3.56e-06(1.98e-07) & 4.00e-06 \\
\cline{4-12}
&&&40&  -2.43e-03(-2.42e-03) & -4.62e-03 & 2.26e-04 & -4.92e-03(-4.92e-03) & 2.47e-03 & -1.20e-03 & 4.64e-06(2.58e-07) & 5.32e-06 \\
\cline{3-12}
&&\multirow{2}{*}{100}&0&  -2.49e-03(-2.48e-03) & -7.33e-03 & 6.56e-04 & -6.02e-03(-6.02e-03) & -9.13e-04 & -6.66e-04 & 3.64e-05(3.21e-06) & 3.73e-05 \\
\cline{4-12}
&&&40&  -4.67e-03(-4.65e-03) & -4.63e-04 & -7.43e-04 & -1.03e-02(-1.03e-02) & 2.67e-02 & -5.78e-03 & 4.38e-05(3.90e-06) & 4.63e-05 \\
\cline{2-12}
&\multirow{4}{*}{2}&\multirow{2}{*}{10}&0&  -2.22e-03(-2.22e-03) & -8.25e-04 & -2.26e-04 & -4.19e-03(-4.19e-03) & 1.67e-04 & -6.40e-04 & -2.98e-01(-2.98e-01) & -2.96e-01 \\
\cline{4-12}
&&&40&  -1.11e-03(-1.11e-03) & -2.90e-03 & 1.03e-04 & 2.21e-03(2.21e-03) & 5.32e-04 & -8.38e-05 & -3.11e-01(-3.11e-01) & -3.08e-01 \\
\cline{3-12}
&&\multirow{2}{*}{100}&0&  -1.73e-03(-1.73e-03) & -6.96e-03 & 5.97e-04 & 1.60e-03(1.60e-03) & -3.28e-04 & 8.33e-05 & -3.20e-02(-3.20e-02) & -2.94e-02 \\
\cline{4-12}
&&&40&  -2.17e-04(-2.17e-04) & -5.29e-03 & 5.64e-04 & -3.65e-03(-3.65e-03) & 9.01e-03 & -2.00e-03 & -2.33e-02(-2.33e-02) & -2.00e-02 \\    \hline
\end{tabular}
\endgroup
\caption {\label{tab:A2}{Bias in the over-parametrised Gompertz model by sample size $n$, $\sigma ^2$, the number of clusters $N_{cl}$, and the percent of censored subjects.  True values: $\beta _{scale-Success}=-0.5$, $\beta _{scale-score}=-1$, $\beta _{shape-Success}=0$, $\beta _{shape-score}=0$, $p _{Success}=0.5$.}}
\end{sidewaystable}

\clearpage
\newpage
\begin{sidewaystable}[ht]
\centering
\begingroup\tiny
\begin{tabular}{rrrlcccccccccc}
  \hline
Sample size & \#of clusters & \%Cens &   &    &     &       &         &          &        &         &          &           &            \\
  \hline
 &  &  & True $\sigma ^2$ & 0 & 0 & 0.5 & 0.5 & 1 & 1 & 2 & 2 & 5 & 5 \\
   &  &  & \%Success & 25 & 50 & 25 & 50 & 25 & 50 & 25 & 50 & 25 & 50 \\
   \hline
  &   &   &   &    &     &       &         & Min-Max bias &        &         &          &           &            \\
   \hline
 &  & 0 &  & -0.014 - 0.012 & -0.011 - 0.007 & -0.015 - 0.012 & -0.011 - 0.005 & -0.012 - 0.008 & -0.011 - 0.006 & -0.02 - 0.01 & -0.011 - 0.01 & -0.024 - 0.02 & -0.021 - 0.02 \\
   & 10 & 40 &  & -0.018 - 0.016 & -0.01 - 0.012 & -0.003 - 0.015 & -0.011 - 0.01 & -0.009 - 0.015 & -0.012 - 0.013 & -0.015 - 0.019 & -0.014 - 0.014 & -0.03 - 0.025 & -0.024 - 0.024 \\
  300 &  & 80 &  & 0.03 - 0.067 & 0.012 - 0.066 & 0.039 - 0.104 & 0.021 - 0.072 & 0.042 - 0.087 & 0.024 - 0.079 & 0.045 - 0.114 & 0.02 - 0.09 & 0.099 - 0.209 & 0.028 - 0.12 \\
   &  & 0 &  & -0.016 - 0.015 & -0.016 - 0.01 & -0.009 - 0.01 & -0.009 - 0.007 & -0.012 - 0.009 & -0.013 - 0.007 & -0.015 - 0.005 & -0.005 - 0.007 & -0.024 - 0.007 & -0.015 - 0.005 \\
   & 100 & 40 &  & -0.02 - 0.026 & -0.014 - 0.023 & -0.01 - 0.013 & -0.011 - 0.013 & -0.017 - 0.016 & -0.012 - 0.01 & -0.017 - 0.015 & -0.015 - 0.009 & -0.025 - 0.005 & -0.019 - 0.014 \\
   &  & 80 &  & 0.023 - 0.11 & 0.007 - 0.099 & 0.026 - 0.092 & 0.014 - 0.07 & 0.021 - 0.081 & 0.011 - 0.068 & 0.034 - 0.089 & 0.015 - 0.07 & 0.008 - 0.102 & -0.002 - 0.058 \\
   &  & 0 &  & -0.004 - 0.005 & -0.002 - 0.003 & -0.002 - 0.004 & -0.005 - 0.004 & -0.003 - 0.003 & -0.004 - 0.005 & -0.01 - 0.008 & -0.009 - 0.005 & -0.013 - 0.014 & -0.015 - 0.017 \\
   & 10 & 40 &  & -0.003 - 0.006 & -0.004 - 0.004 & -0.005 - 0.006 & -0.007 - 0.002 & -0.004 - 0.007 & -0.004 - 0.006 & -0.006 - 0.005 & -0.008 - 0.007 & -0.019 - 0.018 & -0.017 - 0.014 \\
  1000 &  & 80 &  & 0.007 - 0.017 & 0.004 - 0.019 & 0.005 - 0.023 & -0.004 - 0.016 & 0.015 - 0.025 & 0.003 - 0.013 & 0.006 - 0.032 & -0.002 - 0.021 & 0.007 - 0.049 & -0.01 - 0.034 \\
   &  & 0 &  & -0.006 - 0.004 & -0.004 - 0.004 & -0.003 - 0.003 & -0.003 - 0.001 & -0.005 - 0.004 & -0.003 - 0.003 & -0.004 - 0.003 & -0.005 - 0.002 & -0.003 - 0.005 & -0.005 - 0.002 \\
   & 100 & 40 &  & -0.008 - 0.008 & -0.006 - 0.006 & -0.003 - 0.003 & -0.006 - 0.003 & -0.004 - 0.004 & -0.001 - 0.004 & -0.006 - 0.005 & -0.004 - 0.003 & -0.003 - 0.005 & -0.005 - 0.006 \\
   &  & 80 &  & 0.003 - 0.023 & 0.001 - 0.018 & 0.001 - 0.019 & 0.003 - 0.017 & 0.005 - 0.017 & 0.004 - 0.017 & 0.001 - 0.018 & 0.004 - 0.013 & 0.006 - 0.021 & -0.002 - 0.019 \\
   &  & 0 &  & 0 - 0 & 0 - 0 & -0.001 - 0.002 & -0.001 - 0.001 & -0.003 - 0.003 & -0.003 - 0.003 & -0.005 - 0.006 & -0.005 - 0.005 & -0.013 - 0.014 & -0.012 - 0.013 \\
   & 10 & 40 &  & 0 - 0.001 & 0 - 0.001 & -0.001 - 0.002 & -0.002 - 0.001 & -0.003 - 0.003 & -0.002 - 0.003 & -0.005 - 0.006 & -0.006 - 0.006 & -0.013 - 0.014 & -0.013 - 0.014 \\
  10000 &  & 80 &  & 0 - 0.003 & 0.001 - 0.002 & -0.002 - 0.002 & -0.001 - 0.004 & -0.001 - 0.006 & -0.002 - 0.005 & -0.005 - 0.009 & -0.006 - 0.007 & -0.011 - 0.016 & -0.012 - 0.016 \\
   &  & 0 &  & 0 - 0.001 & -0.001 - 0 & -0.001 - 0.001 & -0.001 - 0.001 & -0.001 - 0.001 & 0 - 0 & -0.001 - 0 & 0 - 0.001 & -0.002 - 0.002 & -0.002 - 0.002 \\
   & 100 & 40 &  & -0.001 - 0.001 & 0 - 0.001 & -0.001 - 0 & 0 - 0.001 & -0.001 - 0.001 & -0.001 - 0.001 & 0 - 0.001 & -0.001 - 0.001 & -0.001 - 0.002 & -0.002 - 0.001 \\
   &  & 80 &  & -0.003 - 0.004 & -0.002 - 0.005 & -0.001 - 0.002 & 0 - 0.002 & 0 - 0.002 & 0 - 0.001 & 0 - 0.003 & 0.001 - 0.002 & 0 - 0.003 & -0.001 - 0.001 \\
   \hline
\end{tabular}
\endgroup
\caption{Bias over all possible combinations of the Cox-regression parameters.  Parameter  $\beta _{Success-scale}$ .  Weibull  hazard function.}
\end{sidewaystable}
\begin{sidewaystable}[ht]
\centering
\begingroup\tiny
\begin{tabular}{rrrlcccccccccc}
  \hline
Sample size & \#of clusters & \%Cens &   &    &     &       &         &          &        &         &          &           &            \\
  \hline
 &  &  & True $\sigma ^2$ & 0 & 0 & 0.5 & 0.5 & 1 & 1 & 2 & 2 & 5 & 5 \\
   &  &  & \%Success & 25 & 50 & 25 & 50 & 25 & 50 & 25 & 50 & 25 & 50 \\
   \hline
  &   &   &   &    &     &       &         & Min-Max bias &        &         &          &           &            \\
   \hline
 &  & 0 &  & -0.117 - -0.055 & -0.024 - 0.013 & -0.106 - -0.038 & -0.018 - 0.019 & -0.064 - -0.026 & -0.008 - 0.022 & -0.05 - -0.03 & -0.015 - 0.013 & -0.032 - -0.007 & -0.008 - 0.009 \\
   & 10 & 40 &  & -0.174 - -0.088 & -0.01 - 0.015 & -0.141 - -0.069 & -0.01 - 0.02 & -0.164 - -0.064 & -0.008 - 0.025 & -0.087 - -0.061 & -0.01 - 0 & -0.055 - -0.032 & -0.016 - 0.014 \\
  300 &  & 80 &  & -0.422 - -0.205 & -0.048 - 0 & -0.399 - -0.131 & -0.11 - 0.05 & -0.447 - -0.145 & -0.042 - 0.089 & -0.718 - -0.128 & -0.025 - 0.075 & -0.615 - -0.121 & -0.06 - 0.101 \\
   &  & 0 &  & -0.123 - -0.071 & -0.027 - 0.022 & -0.101 - -0.055 & -0.007 - 0.02 & -0.085 - -0.043 & -0.004 - 0.02 & -0.063 - -0.014 & -0.01 - 0.009 & -0.028 - 0.004 & -0.005 - 0.007 \\
   & 100 & 40 &  & -0.141 - -0.092 & -0.041 - 0.021 & -0.193 - -0.099 & -0.01 - 0.08 & -0.162 - -0.071 & -0.023 - 0.019 & -0.085 - -0.039 & -0.016 - 0.023 & -0.06 - -0.024 & -0.022 - 0.03 \\
   &  & 80 &  & -0.388 - -0.197 & -0.035 - 0.036 & -0.37 - -0.188 & -0.074 - 0.015 & -0.416 - -0.132 & -0.087 - 0.045 & -0.38 - -0.165 & -0.044 - 0.078 & -0.303 - -0.113 & -0.059 - 0.035 \\
   &  & 0 &  & -0.035 - -0.025 & -0.017 - 0.002 & -0.026 - -0.006 & -0.007 - 0.006 & -0.025 - -0.01 & -0.006 - 0.007 & -0.019 - -0.005 & -0.007 - 0.012 & -0.012 - 0.001 & -0.006 - 0.003 \\
   & 10 & 40 &  & -0.052 - -0.026 & -0.003 - 0.004 & -0.045 - -0.022 & -0.009 - 0.008 & -0.025 - -0.012 & -0.013 - 0.006 & -0.02 - -0.002 & -0.002 - 0.008 & -0.02 - -0.006 & -0.006 - 0.003 \\
  1000 &  & 80 &  & -0.119 - -0.037 & -0.004 - 0.009 & -0.096 - -0.021 & -0.015 - 0.004 & -0.101 - -0.047 & -0.026 - 0.032 & -0.088 - -0.027 & -0.017 - 0.011 & -0.071 - -0.045 & -0.016 - 0.01 \\
   &  & 0 &  & -0.033 - -0.014 & -0.003 - 0.008 & -0.03 - -0.014 & -0.014 - 0.006 & -0.022 - -0.006 & -0.008 - 0.004 & -0.013 - -0.006 & -0.007 - 0.005 & -0.004 - 0.002 & -0.003 - 0.001 \\
   & 100 & 40 &  & -0.043 - -0.019 & -0.008 - -0.002 & -0.029 - -0.019 & -0.009 - 0.001 & -0.042 - -0.006 & -0.007 - 0.007 & -0.022 - -0.011 & -0.014 - 0.005 & -0.01 - 0 & -0.002 - 0.002 \\
   &  & 80 &  & -0.075 - -0.038 & -0.007 - 0.011 & -0.105 - -0.047 & -0.012 - 0.013 & -0.077 - -0.021 & -0.01 - 0.02 & -0.084 - -0.042 & -0.013 - 0.003 & -0.06 - -0.03 & -0.012 - 0.017 \\
   &  & 0 &  & -0.003 - -0.001 & 0 - 0.003 & -0.002 - 0 & -0.001 - 0 & -0.001 - 0 & -0.002 - 0.002 & -0.003 - 0 & -0.001 - 0 & -0.001 - 0.002 & 0 - 0.001 \\
   & 10 & 40 &  & -0.005 - -0.001 & -0.003 - 0.004 & -0.006 - 0 & -0.001 - 0.003 & -0.005 - -0.001 & -0.002 - 0.001 & -0.002 - 0 & -0.002 - 0 & -0.003 - 0 & -0.001 - 0 \\
  10000 &  & 80 &  & -0.005 - 0.001 & -0.002 - 0.003 & -0.009 - -0.001 & -0.005 - 0.001 & -0.009 - -0.002 & 0 - 0.005 & -0.011 - -0.001 & -0.004 - 0.004 & -0.01 - 0.001 & -0.005 - 0.001 \\
   &  & 0 &  & -0.003 - 0 & -0.002 - 0.005 & -0.002 - -0.002 & -0.002 - 0.001 & -0.002 - 0.001 & 0 - 0.002 & -0.002 - 0 & 0 - 0 & -0.001 - 0.001 & -0.001 - 0.001 \\
   & 100 & 40 &  & -0.01 - 0 & -0.002 - 0.001 & -0.004 - 0.003 & -0.001 - 0.002 & -0.006 - -0.001 & 0 - 0.002 & -0.002 - -0.001 & 0 - 0.001 & -0.002 - 0 & 0 - 0.001 \\
   &  & 80 &  & -0.01 - -0.003 & -0.006 - 0.003 & -0.007 - -0.003 & -0.004 - 0.002 & -0.01 - -0.003 & -0.003 - -0.001 & -0.007 - 0 & -0.001 - 0.004 & -0.009 - 0 & -0.002 - 0.003 \\
   \hline
\end{tabular}
\endgroup
\caption{Bias over all possible combinations of the Cox-regression parameters.  Parameter  $\beta _{Success-scale}$ .  Gompertz  hazard function.}
\end{sidewaystable}
\begin{sidewaystable}[ht]
\centering
\begingroup\tiny
\begin{tabular}{rrrlcccccccccc}
  \hline
Sample size & \#of clusters & \%Cens &   &    &     &       &         &          &        &         &          &           &            \\
  \hline
 &  &  & True $\sigma ^2$ & 0 & 0 & 0.5 & 0.5 & 1 & 1 & 2 & 2 & 5 & 5 \\
   &  &  & \%Success & 25 & 50 & 25 & 50 & 25 & 50 & 25 & 50 & 25 & 50 \\
   \hline
  &   &   &   &    &     &       &         & Min-Max bias &        &         &          &           &            \\
   \hline
 &  & 0 &  & -0.011 - 0.014 & -0.014 - 0.016 & -0.014 - 0.017 & -0.021 - 0.012 & -0.017 - 0.02 & -0.019 - 0.017 & -0.023 - 0.016 & -0.024 - 0.024 & -0.046 - 0.038 & -0.032 - 0.046 \\
   & 10 & 40 &  & -0.014 - 0.03 & -0.024 - 0.025 & -0.032 - 0.018 & -0.014 - 0.019 & -0.031 - 0.014 & -0.015 - 0.019 & -0.035 - 0.028 & -0.02 - 0.029 & -0.035 - 0.042 & -0.031 - 0.045 \\
  300 &  & 80 &  & -0.095 - 0.098 & -0.109 - 0.118 & -0.057 - 0.094 & -0.093 - 0.109 & -0.062 - 0.081 & -0.085 - 0.132 & -0.073 - 0.118 & -0.079 - 0.116 & -0.126 - 0.251 & -0.181 - 0.096 \\
   &  & 0 &  & -0.025 - 0.016 & -0.026 - 0.022 & -0.006 - 0.011 & -0.013 - 0.005 & -0.016 - 0.013 & -0.004 - 0.016 & -0.019 - 0.009 & -0.01 - 0.019 & -0.018 - 0.015 & -0.023 - 0.025 \\
   & 100 & 40 &  & -0.04 - 0.032 & -0.036 - 0.039 & -0.022 - 0.011 & -0.009 - 0.024 & -0.026 - 0.017 & -0.021 - 0.018 & -0.014 - 0.023 & -0.029 - 0.027 & -0.024 - 0.016 & -0.028 - 0.027 \\
   &  & 80 &  & -0.108 - 0.095 & -0.092 - 0.164 & -0.095 - 0.077 & -0.064 - 0.074 & -0.076 - 0.072 & -0.06 - 0.085 & -0.085 - 0.089 & -0.057 - 0.111 & -0.085 - 0.111 & -0.078 - 0.101 \\
   &  & 0 &  & -0.005 - 0.003 & -0.004 - 0.006 & -0.006 - 0.003 & -0.01 - 0.004 & -0.006 - 0.009 & -0.012 - 0.002 & -0.018 - 0.014 & -0.012 - 0.014 & -0.025 - 0.03 & -0.029 - 0.034 \\
   & 10 & 40 &  & -0.007 - 0.01 & -0.009 - 0.011 & -0.004 - 0.011 & -0.009 - 0.01 & -0.01 - 0.015 & -0.011 - 0.01 & -0.012 - 0.011 & -0.011 - 0.007 & -0.034 - 0.036 & -0.028 - 0.034 \\
  1000 &  & 80 &  & -0.023 - 0.021 & -0.012 - 0.021 & -0.012 - 0.009 & -0.021 - 0.029 & -0.01 - 0.025 & -0.013 - 0.004 & -0.034 - 0.039 & -0.02 - 0.026 & -0.06 - 0.068 & -0.049 - 0.042 \\
   &  & 0 &  & -0.006 - 0.009 & -0.008 - 0.006 & -0.004 - 0.006 & -0.004 - 0.009 & -0.01 - 0.003 & -0.009 - 0.004 & 0 - 0.007 & -0.004 - 0.007 & -0.014 - 0.005 & -0.003 - 0.007 \\
   & 100 & 40 &  & -0.009 - 0.015 & -0.014 - 0.016 & -0.003 - 0.01 & -0.002 - 0.006 & -0.009 - 0.003 & -0.009 - 0.011 & -0.006 - 0.008 & -0.005 - 0.003 & -0.005 - 0.009 & -0.008 - 0.012 \\
   &  & 80 &  & -0.019 - 0.026 & -0.021 - 0.028 & -0.022 - 0.029 & -0.023 - 0.03 & -0.025 - 0.02 & -0.028 - 0.03 & -0.018 - 0.023 & -0.015 - 0.026 & -0.023 - 0.011 & -0.02 - 0.018 \\
   &  & 0 &  & 0 - 0.001 & -0.001 - 0.001 & -0.002 - 0.004 & -0.002 - 0.002 & -0.006 - 0.008 & -0.007 - 0.008 & -0.011 - 0.012 & -0.011 - 0.012 & -0.026 - 0.026 & -0.027 - 0.026 \\
   & 10 & 40 &  & -0.002 - 0.001 & -0.001 - 0 & -0.003 - 0.003 & -0.003 - 0.004 & -0.007 - 0.006 & -0.005 - 0.008 & -0.011 - 0.01 & -0.012 - 0.011 & -0.027 - 0.026 & -0.025 - 0.028 \\
  10000 &  & 80 &  & -0.009 - 0.003 & -0.003 - 0.003 & -0.002 - 0.005 & -0.003 - 0.008 & -0.006 - 0.006 & -0.01 - 0.008 & -0.011 - 0.012 & -0.013 - 0.012 & -0.027 - 0.029 & -0.029 - 0.024 \\
   &  & 0 &  & -0.001 - 0.002 & -0.001 - 0.001 & -0.003 - 0 & -0.002 - 0.001 & -0.002 - 0 & 0 - 0.002 & -0.001 - 0.002 & -0.001 - 0.002 & -0.005 - 0.005 & -0.003 - 0.003 \\
   & 100 & 40 &  & -0.002 - 0.001 & -0.001 - 0.002 & 0.001 - 0.002 & -0.001 - 0.001 & -0.001 - 0.002 & -0.001 - 0.001 & -0.002 - 0 & 0 - 0.001 & -0.005 - 0.003 & -0.001 - 0.003 \\
   &  & 80 &  & -0.007 - 0.005 & -0.002 - 0.009 & -0.003 - 0.003 & -0.001 - 0.004 & -0.001 - 0.001 & -0.004 - 0.003 & -0.005 - 0.003 & -0.005 - -0.002 & -0.005 - 0.003 & -0.002 - 0.002 \\
   \hline
\end{tabular}
\endgroup
\caption{Bias over all possible combinations of the Cox-regression parameters.  Parameter  $\beta _{score-scale}$ .  Weibull  hazard function.}
\end{sidewaystable}
\begin{sidewaystable}[ht]
\centering
\begingroup\tiny
\begin{tabular}{rrrlcccccccccc}
  \hline
Sample size & \#of clusters & \%Cens &   &    &     &       &         &          &        &         &          &           &            \\
  \hline
 &  &  & True $\sigma ^2$ & 0 & 0 & 0.5 & 0.5 & 1 & 1 & 2 & 2 & 5 & 5 \\
   &  &  & \%Success & 25 & 50 & 25 & 50 & 25 & 50 & 25 & 50 & 25 & 50 \\
   \hline
  &   &   &   &    &     &       &         & Min-Max bias &        &         &          &           &            \\
   \hline
 &  & 0 &  & -0.062 - 0.032 & -0.02 - 0.04 & -0.014 - 0.014 & -0.013 - 0.037 & -0.029 - 0.037 & -0.012 - 0.051 & -0.04 - 0.016 & -0.049 - 0.012 & -0.033 - 0.016 & -0.026 - 0.034 \\
   & 10 & 40 &  & -0.027 - 0.05 & -0.07 - 0.123 & -0.022 - 0.054 & -0.009 - 0.067 & -0.033 - 0.064 & -0.016 - -0.003 & -0.04 - 0.045 & -0.021 - 0.025 & -0.025 - 0.028 & -0.025 - 0.021 \\
  300 &  & 80 &  & -0.163 - 0.185 & -0.194 - 0.256 & -0.027 - 0.158 & -0.181 - 0.136 & -0.13 - 0.117 & -0.189 - 0.068 & -0.038 - 0.096 & -0.068 - 0.15 & -0.019 - 0.106 & -0.065 - 0.055 \\
   &  & 0 &  & -0.058 - 0.036 & -0.03 - 0.057 & 0.005 - 0.056 & -0.056 - 0.021 & -0.029 - 0.038 & -0.035 - 0.037 & -0.02 - 0 & 0.002 - 0.013 & -0.02 - 0.03 & -0.027 - 0.015 \\
   & 100 & 40 &  & -0.02 - 0.063 & -0.088 - 0.055 & -0.024 - 0.023 & -0.071 - 0.116 & -0.039 - 0.031 & -0.036 - 0.071 & -0.048 - 0.016 & -0.027 - 0.061 & -0.002 - 0.038 & -0.022 - 0.044 \\
   &  & 80 &  & -0.254 - 0.279 & -0.191 - 0.086 & -0.205 - 0.262 & -0.119 - 0.12 & -0.123 - 0.127 & -0.218 - 0.137 & -0.076 - 0.155 & -0.082 - 0.08 & -0.097 - 0.023 & -0.125 - 0.081 \\
   &  & 0 &  & 0.002 - 0.026 & -0.013 - 0.012 & -0.005 - 0.006 & -0.013 - 0.014 & -0.003 - 0.016 & -0.004 - 0.025 & -0.005 - 0.005 & -0.032 - 0.004 & -0.01 - 0.01 & -0.003 - 0.011 \\
   & 10 & 40 &  & -0.013 - 0.01 & -0.003 - 0.007 & -0.015 - 0.017 & -0.003 - 0.013 & -0.016 - 0.011 & -0.003 - 0.007 & -0.008 - 0.012 & -0.015 - 0.008 & -0.002 - 0.02 & -0.003 - 0.006 \\
  1000 &  & 80 &  & -0.103 - 0.067 & -0.012 - 0.071 & -0.059 - 0.076 & -0.05 - 0.03 & -0.045 - 0.055 & -0.033 - 0.049 & -0.026 - 0.051 & -0.016 - 0.031 & -0.017 - 0.039 & -0.014 - 0.027 \\
   &  & 0 &  & -0.007 - 0.01 & -0.013 - 0.008 & -0.002 - 0.013 & -0.003 - 0.025 & -0.02 - 0.007 & -0.004 - 0.006 & -0.015 - 0.008 & -0.005 - 0.005 & -0.007 - 0.013 & -0.008 - -0.003 \\
   & 100 & 40 &  & -0.016 - 0.014 & -0.015 - 0.023 & -0.015 - 0.003 & -0.006 - 0.013 & -0.026 - 0.007 & -0.004 - 0.016 & -0.01 - -0.001 & -0.019 - 0.012 & -0.011 - 0.009 & -0.009 - 0.004 \\
   &  & 80 &  & -0.014 - 0.055 & -0.08 - 0.076 & -0.04 - 0.071 & -0.057 - 0.05 & -0.038 - 0.034 & -0.041 - 0.063 & -0.052 - 0.058 & -0.022 - 0.052 & -0.039 - 0.022 & -0.052 - 0.003 \\
   &  & 0 &  & -0.005 - 0.008 & -0.008 - 0.011 & -0.001 - 0.002 & -0.005 - 0.003 & -0.002 - 0.001 & -0.009 - 0.003 & -0.002 - 0.004 & -0.001 - 0.004 & 0 - 0.003 & -0.003 - 0 \\
   & 10 & 40 &  & -0.006 - 0.002 & -0.003 - 0.007 & 0 - 0.001 & -0.006 - 0.007 & -0.005 - 0.006 & -0.004 - 0.002 & -0.005 - 0.01 & -0.002 - 0.003 & -0.004 - 0.002 & -0.002 - 0.002 \\
  10000 &  & 80 &  & -0.001 - 0.017 & -0.005 - 0 & -0.008 - 0.013 & -0.005 - 0.003 & -0.004 - 0.013 & -0.006 - 0.003 & -0.008 - 0.004 & -0.009 - 0.006 & -0.004 - 0.013 & -0.008 - 0.006 \\
   &  & 0 &  & -0.004 - 0.007 & 0.001 - 0.005 & -0.003 - 0.002 & -0.002 - 0.007 & -0.002 - 0.002 & -0.004 - 0.006 & 0.001 - 0.001 & -0.002 - 0 & 0 - 0.002 & -0.001 - 0.001 \\
   & 100 & 40 &  & -0.004 - 0.003 & 0 - 0.002 & -0.003 - 0.012 & -0.003 - 0.005 & -0.004 - 0.003 & -0.001 - 0.008 & -0.001 - 0.005 & -0.008 - 0.002 & -0.001 - 0.004 & -0.003 - 0.004 \\
   &  & 80 &  & -0.014 - 0.009 & -0.002 - 0.009 & -0.003 - 0.006 & -0.004 - 0.004 & -0.009 - -0.006 & -0.012 - 0.019 & -0.007 - 0.016 & 0.002 - 0.007 & -0.002 - 0.006 & -0.003 - 0.007 \\
   \hline
\end{tabular}
\endgroup
\caption{Bias over all possible combinations of the Cox-regression parameters.  Parameter  $\beta _{score-scale}$ .  Gompertz  hazard function.}
\end{sidewaystable}
\begin{sidewaystable}[ht]
\centering
\begingroup\tiny
\begin{tabular}{rrrlcccccccccc}
  \hline
Sample size & \#of clusters & \%Cens &   &    &     &       &         &          &        &         &          &           &            \\
  \hline
 &  &  & True $\sigma ^2$ & 0 & 0 & 0.5 & 0.5 & 1 & 1 & 2 & 2 & 5 & 5 \\
   &  &  & \%Success & 25 & 50 & 25 & 50 & 25 & 50 & 25 & 50 & 25 & 50 \\
   \hline
  &   &   &   &    &     &       &         & Min-Max bias &        &         &          &           &            \\
   \hline
 &  & 0 &  & 0.008 - 0.01 & -0.001 - 0.002 & 0.005 - 0.011 & -0.001 - 0.001 & 0.005 - 0.01 & 0 - 0.002 & 0.003 - 0.006 & 0 - 0.001 & 0.001 - 0.003 & -0.001 - 0 \\
   & 10 & 40 &  & 0.007 - 0.011 & -0.004 - 0 & 0.007 - 0.01 & 0 - 0.001 & 0.006 - 0.01 & -0.001 - 0.001 & 0.004 - 0.009 & -0.002 - 0.001 & 0.003 - 0.007 & 0 - 0.003 \\
  300 &  & 80 &  & 0.014 - 0.037 & -0.002 - 0.003 & 0.022 - 0.026 & -0.001 - 0.011 & 0.009 - 0.029 & 0.001 - 0.009 & 0.016 - 0.036 & 0.001 - 0.004 & 0.02 - 0.041 & -0.006 - 0.032 \\
   &  & 0 &  & 0.005 - 0.011 & -0.002 - 0.004 & 0.007 - 0.011 & -0.001 - 0.002 & 0.003 - 0.01 & -0.005 - -0.001 & 0.003 - 0.006 & -0.001 - 0 & 0.001 - 0.003 & 0 - 0.001 \\
   & 100 & 40 &  & 0.008 - 0.014 & -0.006 - 0.005 & 0.002 - 0.012 & -0.002 - 0.004 & 0.006 - 0.011 & -0.002 - 0 & 0.005 - 0.009 & -0.002 - 0 & 0.004 - 0.006 & -0.003 - 0.001 \\
   &  & 80 &  & 0.019 - 0.033 & -0.004 - 0.013 & 0.01 - 0.027 & -0.002 - 0.008 & 0.015 - 0.024 & -0.001 - 0.004 & 0.015 - 0.028 & -0.005 - 0.004 & 0.019 - 0.033 & -0.005 - 0.003 \\
   &  & 0 &  & 0.002 - 0.004 & 0 - 0 & 0.001 - 0.003 & 0 - 0.001 & 0.001 - 0.003 & -0.001 - 0 & 0.001 - 0.003 & 0 - 0 & 0 - 0.001 & 0 - 0 \\
   & 10 & 40 &  & 0.002 - 0.005 & -0.002 - 0.001 & 0.001 - 0.003 & -0.001 - 0 & 0.002 - 0.004 & 0 - 0.002 & 0.002 - 0.003 & -0.001 - 0.001 & 0.001 - 0.003 & -0.001 - 0 \\
  1000 &  & 80 &  & 0.002 - 0.007 & 0 - 0.002 & 0.002 - 0.008 & -0.002 - 0.005 & 0.005 - 0.01 & -0.003 - 0.001 & 0.004 - 0.007 & -0.003 - 0 & 0 - 0.011 & -0.003 - 0.003 \\
   &  & 0 &  & 0.002 - 0.005 & -0.001 - 0.001 & 0.002 - 0.003 & -0.001 - 0.001 & 0.001 - 0.003 & -0.001 - 0 & 0 - 0.001 & 0 - 0 & 0 - 0.001 & 0 - 0.001 \\
   & 100 & 40 &  & 0.002 - 0.006 & -0.002 - 0.002 & 0.001 - 0.004 & -0.001 - 0 & 0.001 - 0.003 & -0.001 - 0.002 & 0.001 - 0.003 & -0.001 - 0.001 & 0.001 - 0.002 & -0.001 - 0.001 \\
   &  & 80 &  & 0.003 - 0.008 & -0.002 - 0.001 & 0.002 - 0.006 & -0.002 - 0.002 & 0.001 - 0.006 & -0.001 - 0.003 & 0.003 - 0.009 & -0.002 - 0.002 & 0.004 - 0.006 & -0.003 - 0.003 \\
   &  & 0 &  & 0 - 0.001 & -0.001 - 0 & 0 - 0 & 0 - 0 & 0 - 0 & -0.001 - 0 & 0 - 0 & 0 - 0 & 0 - 0 & 0 - 0 \\
   & 10 & 40 &  & 0 - 0.001 & 0 - 0 & 0 - 0 & 0 - 0 & 0 - 0.001 & 0 - 0.001 & 0 - 0.001 & 0 - 0 & 0 - 0 & 0 - 0 \\
  10000 &  & 80 &  & 0 - 0.001 & 0 - 0 & 0 - 0.001 & -0.002 - 0.001 & 0 - 0.001 & 0 - 0.001 & 0 - 0.001 & -0.001 - 0 & 0 - 0.003 & -0.001 - 0.001 \\
   &  & 0 &  & 0 - 0.001 & 0 - 0 & 0 - 0 & 0 - 0 & 0 - 0 & 0 - 0 & 0 - 0 & 0 - 0 & 0 - 0 & 0 - 0 \\
   & 100 & 40 &  & 0 - 0.001 & 0 - 0 & 0 - 0.001 & 0 - 0 & 0 - 0.001 & 0 - 0 & 0 - 0.001 & 0 - 0 & 0 - 0 & 0 - 0 \\
   &  & 80 &  & -0.001 - 0.001 & -0.001 - 0.001 & -0.001 - 0 & -0.001 - 0.001 & -0.001 - 0.001 & -0.001 - 0.001 & 0 - 0.002 & -0.001 - 0 & 0 - 0.002 & -0.001 - 0 \\
   \hline
\end{tabular}
\endgroup
\caption{Bias over all possible combinations of the Cox-regression parameters.  Parameter  $\beta _{Success-shape}$ .  Weibull  hazard function.}
\end{sidewaystable}
\begin{sidewaystable}[ht]
\centering
\begingroup\tiny
\begin{tabular}{rrrlcccccccccc}
  \hline
Sample size & \#of clusters & \%Cens &   &    &     &       &         &          &        &         &          &           &            \\
  \hline
 &  &  & True $\sigma ^2$ & 0 & 0 & 0.5 & 0.5 & 1 & 1 & 2 & 2 & 5 & 5 \\
   &  &  & \%Success & 25 & 50 & 25 & 50 & 25 & 50 & 25 & 50 & 25 & 50 \\
   \hline
  &   &   &   &    &     &       &         & Min-Max bias &        &         &          &           &            \\
   \hline
 &  & 0 &  & 0.007 - 0.014 & -0.001 - 0.002 & 0.004 - 0.011 & -0.002 - 0.001 & 0.004 - 0.007 & -0.003 - 0.001 & 0.004 - 0.006 & -0.001 - 0.001 & 0.002 - 0.003 & 0 - 0.001 \\
   & 10 & 40 &  & 0.014 - 0.019 & -0.004 - 0.002 & 0.01 - 0.015 & -0.002 - 0.002 & 0.01 - 0.019 & -0.003 - 0.003 & 0.009 - 0.011 & -0.001 - 0.001 & 0.005 - 0.006 & 0 - 0.001 \\
  300 &  & 80 &  & 0.024 - 0.034 & -0.007 - 0.01 & 0.017 - 0.032 & -0.002 - 0.008 & 0.018 - 0.036 & -0.007 - 0 & 0.016 - 0.036 & -0.003 - -0.001 & 0.016 - 0.027 & -0.004 - 0.002 \\
   &  & 0 &  & 0.007 - 0.013 & -0.002 - 0.002 & 0.005 - 0.012 & -0.002 - 0.002 & 0.006 - 0.009 & -0.004 - 0.001 & 0.002 - 0.006 & -0.001 - 0.001 & 0.001 - 0.002 & 0 - 0.001 \\
   & 100 & 40 &  & 0.012 - 0.015 & 0 - 0.003 & 0.014 - 0.022 & -0.01 - -0.001 & 0.009 - 0.018 & -0.001 - 0.002 & 0.004 - 0.01 & -0.002 - 0.002 & 0 - 0.007 & -0.002 - 0.002 \\
   &  & 80 &  & 0.02 - 0.033 & -0.014 - 0.012 & 0.024 - 0.029 & -0.001 - 0.005 & 0.016 - 0.03 & -0.003 - 0.005 & 0.019 - 0.028 & -0.005 - 0.003 & 0.015 - 0.024 & 0 - 0.011 \\
   &  & 0 &  & 0.003 - 0.005 & 0 - 0.002 & 0.001 - 0.003 & -0.001 - 0.001 & 0.001 - 0.004 & -0.001 - 0.001 & 0.001 - 0.002 & -0.001 - 0.001 & 0 - 0.001 & 0 - 0 \\
   & 10 & 40 &  & 0.004 - 0.006 & -0.001 - 0.001 & 0.002 - 0.005 & -0.001 - 0.001 & 0.002 - 0.004 & -0.001 - 0.001 & 0.001 - 0.003 & -0.001 - 0 & 0.001 - 0.002 & 0 - 0.001 \\
  1000 &  & 80 &  & 0.005 - 0.01 & -0.004 - 0.002 & 0.003 - 0.009 & 0 - 0.001 & 0.005 - 0.009 & -0.004 - 0.003 & 0.003 - 0.009 & 0 - 0 & 0.004 - 0.008 & 0 - 0.001 \\
   &  & 0 &  & 0.003 - 0.004 & -0.002 - 0.001 & 0.002 - 0.004 & -0.001 - 0.002 & 0.001 - 0.003 & 0 - 0.001 & 0.001 - 0.002 & 0 - 0.001 & 0 - 0.001 & 0 - 0 \\
   & 100 & 40 &  & 0.003 - 0.004 & 0 - 0.001 & 0.002 - 0.004 & 0 - 0.001 & 0.001 - 0.005 & -0.001 - 0.001 & 0.001 - 0.003 & 0 - 0.002 & 0 - 0.002 & -0.001 - 0 \\
   &  & 80 &  & 0.003 - 0.01 & -0.006 - 0.005 & 0.006 - 0.01 & -0.001 - 0.001 & 0.002 - 0.008 & -0.002 - 0.002 & 0.006 - 0.008 & -0.001 - 0.001 & 0.003 - 0.006 & -0.002 - 0 \\
   &  & 0 &  & 0 - 0.001 & 0 - 0 & 0 - 0 & 0 - 0 & 0 - 0 & 0 - 0 & 0 - 0 & 0 - 0 & 0 - 0 & 0 - 0 \\
   & 10 & 40 &  & 0 - 0.001 & -0.001 - 0 & 0 - 0.001 & 0 - 0 & 0 - 0.001 & 0 - 0 & 0 - 0 & 0 - 0 & 0 - 0 & 0 - 0 \\
  10000 &  & 80 &  & 0 - 0 & -0.001 - 0.001 & 0 - 0.001 & 0 - 0.001 & 0 - 0.001 & -0.001 - 0 & 0 - 0.001 & 0 - 0 & 0 - 0.001 & 0 - 0.001 \\
   &  & 0 &  & 0 - 0 & -0.001 - 0 & 0 - 0 & 0 - 0 & 0 - 0 & 0 - 0 & 0 - 0 & 0 - 0 & 0 - 0 & 0 - 0 \\
   & 100 & 40 &  & 0 - 0.001 & 0 - 0 & 0 - 0.001 & 0 - 0 & 0 - 0.001 & 0 - 0 & 0 - 0 & 0 - 0 & 0 - 0 & 0 - 0 \\
   &  & 80 &  & 0 - 0.001 & -0.001 - 0.001 & 0 - 0.001 & 0 - 0.001 & 0 - 0.001 & 0 - 0.001 & 0 - 0.001 & -0.001 - 0 & 0 - 0.001 & -0.001 - 0 \\
   \hline
\end{tabular}
\endgroup
\caption{Bias over all possible combinations of the Cox-regression parameters.  Parameter  $\beta _{Success-shape}$ .  Gompertz  hazard function.}
\end{sidewaystable}
\begin{sidewaystable}[ht]
\centering
\begingroup\tiny
\begin{tabular}{rrrlcccccccccc}
  \hline
Sample size & \#of clusters & \%Cens &   &    &     &       &         &          &        &         &          &           &            \\
  \hline
 &  &  & True $\sigma ^2$ & 0 & 0 & 0.5 & 0.5 & 1 & 1 & 2 & 2 & 5 & 5 \\
   &  &  & \%Success & 25 & 50 & 25 & 50 & 25 & 50 & 25 & 50 & 25 & 50 \\
   \hline
  &   &   &   &    &     &       &         & Min-Max bias &        &         &          &           &            \\
   \hline
 &  & 0 &  & -0.002 - 0.002 & 0.001 - 0.007 & -0.006 - 0.005 & -0.001 - 0.007 & -0.005 - 0.006 & -0.001 - 0.003 & -0.005 - 0.004 & -0.003 - 0.002 & -0.001 - 0 & -0.003 - 0.002 \\
   & 10 & 40 &  & -0.004 - 0.025 & -0.01 - 0.011 & -0.009 - 0.004 & -0.016 - 0.012 & -0.011 - 0.004 & -0.008 - 0.003 & -0.004 - 0.003 & -0.005 - 0.003 & 0 - 0.007 & -0.002 - 0.002 \\
  300 &  & 80 &  & -0.029 - 0.021 & -0.024 - 0.02 & -0.022 - 0.014 & -0.033 - 0.031 & -0.019 - 0.024 & -0.028 - 0.033 & -0.022 - 0.027 & -0.02 - 0.02 & -0.057 - 0.025 & -0.031 - -0.001 \\
   &  & 0 &  & -0.002 - 0 & 0 - 0.008 & -0.006 - 0.004 & -0.003 - 0.003 & -0.006 - 0.003 & -0.005 - 0 & -0.002 - 0.005 & -0.001 - 0.003 & -0.002 - 0.001 & -0.002 - 0 \\
   & 100 & 40 &  & -0.016 - 0.015 & -0.013 - 0.018 & -0.011 - 0.007 & -0.007 - 0.004 & -0.005 - 0.001 & -0.005 - 0.002 & 0 - 0.002 & -0.008 - 0.002 & -0.008 - 0.006 & -0.003 - 0.004 \\
   &  & 80 &  & -0.032 - 0.021 & -0.035 - 0.028 & -0.03 - 0.022 & -0.018 - 0.014 & -0.03 - 0.015 & -0.029 - 0.022 & -0.018 - 0.028 & -0.006 - 0.024 & -0.032 - 0.037 & -0.028 - 0.027 \\
   &  & 0 &  & -0.001 - 0.002 & -0.003 - 0.002 & 0 - 0.002 & -0.001 - 0.005 & -0.002 - 0 & -0.001 - 0.003 & -0.001 - 0.001 & -0.001 - 0.002 & -0.001 - 0.001 & -0.001 - 0 \\
   & 10 & 40 &  & -0.003 - 0.007 & -0.005 - 0.001 & -0.002 - 0.002 & 0 - 0.004 & -0.003 - 0.002 & -0.003 - 0.003 & -0.004 - 0.003 & -0.001 - 0.001 & -0.002 - 0.002 & -0.002 - 0.002 \\
  1000 &  & 80 &  & -0.011 - 0.008 & -0.004 - 0.007 & -0.005 - -0.002 & -0.004 - 0.008 & 0 - 0.008 & -0.007 - -0.001 & -0.007 - 0.01 & -0.004 - 0.006 & -0.006 - 0.006 & -0.006 - 0.006 \\
   &  & 0 &  & -0.004 - 0.001 & -0.001 - 0.003 & -0.002 - 0.004 & -0.003 - 0 & -0.003 - 0.003 & 0 - 0.002 & -0.002 - 0 & -0.002 - 0.001 & -0.001 - 0.001 & -0.001 - 0.001 \\
   & 100 & 40 &  & -0.004 - 0.007 & -0.002 - 0.006 & -0.01 - 0.003 & -0.006 - 0.002 & -0.004 - 0.003 & -0.005 - 0.004 & -0.004 - 0.002 & 0 - 0.004 & -0.001 - 0.003 & -0.001 - 0.001 \\
   &  & 80 &  & -0.007 - 0.008 & -0.007 - 0.006 & -0.007 - 0.008 & -0.006 - 0.01 & -0.004 - 0.004 & -0.005 - 0.009 & -0.003 - 0.005 & -0.005 - 0.009 & -0.004 - 0.004 & -0.002 - 0.001 \\
   &  & 0 &  & 0 - 0 & -0.001 - 0.001 & -0.001 - 0.001 & 0 - 0.001 & 0 - 0 & 0 - 0 & -0.001 - 0 & 0 - 0 & 0 - 0 & 0 - 0 \\
   & 10 & 40 &  & -0.001 - 0.002 & 0 - 0.002 & -0.001 - 0 & -0.001 - 0.001 & -0.001 - 0.001 & -0.001 - 0.001 & 0 - 0 & 0 - 0.001 & -0.001 - 0 & -0.001 - 0 \\
  10000 &  & 80 &  & -0.004 - 0.001 & -0.002 - 0.001 & 0 - 0.002 & -0.001 - 0.003 & -0.002 - 0.001 & 0 - 0.001 & 0.001 - 0.002 & -0.002 - 0.002 & 0.001 - 0.001 & -0.002 - 0.001 \\
   &  & 0 &  & -0.001 - 0 & 0 - 0 & -0.001 - 0.001 & 0 - 0.001 & 0 - 0.001 & -0.001 - 0 & 0 - 0 & 0 - 0 & 0 - 0 & 0 - 0 \\
   & 100 & 40 &  & 0 - 0.001 & -0.001 - 0 & -0.001 - 0 & -0.001 - 0.001 & 0 - 0.001 & -0.002 - 0.001 & 0 - 0 & -0.001 - 0 & 0 - 0.001 & -0.001 - 0 \\
   &  & 80 &  & -0.003 - 0.002 & -0.002 - 0.004 & -0.002 - 0.001 & -0.001 - 0.002 & -0.001 - 0.002 & -0.001 - 0.001 & 0 - 0.001 & -0.003 - 0 & -0.002 - 0.001 & -0.002 - 0 \\
   \hline
\end{tabular}
\endgroup
\caption{Bias over all possible combinations of the Cox-regression parameters.  Parameter  $\beta _{score-shape}$ .  Weibull  hazard function.}
\end{sidewaystable}
\begin{sidewaystable}[ht]
\centering
\begingroup\tiny
\begin{tabular}{rrrlcccccccccc}
  \hline
Sample size & \#of clusters & \%Cens &   &    &     &       &         &          &        &         &          &           &            \\
  \hline
 &  &  & True $\sigma ^2$ & 0 & 0 & 0.5 & 0.5 & 1 & 1 & 2 & 2 & 5 & 5 \\
   &  &  & \%Success & 25 & 50 & 25 & 50 & 25 & 50 & 25 & 50 & 25 & 50 \\
   \hline
  &   &   &   &    &     &       &         & Min-Max bias &        &         &          &           &            \\
   \hline
 &  & 0 &  & -0.006 - 0.012 & -0.008 - 0.005 & -0.004 - 0.004 & -0.007 - 0.003 & -0.006 - 0.005 & -0.006 - 0 & 0 - 0.004 & 0 - 0.005 & -0.001 - 0.002 & -0.002 - 0.002 \\
   & 10 & 40 &  & -0.011 - 0.007 & -0.021 - 0.012 & -0.006 - 0.004 & -0.011 - 0.003 & -0.01 - 0.007 & -0.002 - 0.004 & -0.005 - 0.008 & -0.005 - 0.002 & -0.001 - 0.002 & -0.002 - 0 \\
  300 &  & 80 &  & -0.038 - 0.035 & -0.054 - 0.045 & -0.028 - 0.015 & -0.028 - 0.042 & -0.026 - 0.026 & -0.019 - 0.042 & -0.024 - 0.014 & -0.025 - 0.021 & -0.025 - 0.007 & -0.015 - 0.017 \\
   &  & 0 &  & -0.008 - 0.012 & -0.012 - 0.007 & -0.007 - -0.002 & -0.005 - 0.01 & -0.006 - 0.007 & -0.004 - 0.006 & -0.002 - 0.002 & -0.002 - 0 & -0.001 - 0 & 0 - 0.001 \\
   & 100 & 40 &  & -0.012 - 0.008 & -0.012 - 0.017 & -0.004 - 0.002 & -0.018 - 0.012 & -0.007 - 0.01 & -0.009 - 0.004 & -0.003 - 0.01 & -0.013 - 0.003 & -0.006 - 0.002 & -0.002 - 0.001 \\
   &  & 80 &  & -0.058 - 0.053 & -0.029 - 0.052 & -0.049 - 0.041 & -0.031 - 0.027 & -0.026 - 0.027 & -0.026 - 0.039 & -0.036 - 0.026 & -0.025 - 0.024 & -0.012 - 0.026 & -0.023 - 0.027 \\
   &  & 0 &  & -0.003 - -0.001 & -0.002 - 0.003 & -0.001 - 0.001 & -0.001 - 0.001 & -0.002 - 0 & -0.003 - 0 & 0 - 0 & -0.001 - 0.003 & -0.001 - 0.001 & -0.001 - 0 \\
   & 10 & 40 &  & -0.002 - 0.002 & 0 - 0.002 & -0.002 - 0.003 & -0.003 - 0.001 & -0.001 - 0.002 & -0.002 - 0.001 & -0.002 - 0.002 & -0.002 - 0.003 & -0.002 - -0.001 & -0.001 - 0.001 \\
  1000 &  & 80 &  & -0.017 - 0.023 & -0.016 - 0.006 & -0.015 - 0.014 & -0.007 - 0.009 & -0.011 - 0.01 & -0.01 - 0.008 & -0.012 - 0.001 & -0.007 - 0.004 & -0.006 - 0.004 & -0.007 - 0.002 \\
   &  & 0 &  & -0.003 - 0.002 & -0.002 - 0.001 & -0.002 - 0 & -0.003 - 0.001 & -0.001 - 0.002 & -0.001 - 0 & -0.001 - 0.002 & -0.001 - 0.001 & -0.001 - 0 & 0 - 0.001 \\
   & 100 & 40 &  & -0.003 - 0.004 & -0.005 - 0.004 & -0.001 - 0.003 & -0.002 - 0.002 & -0.002 - 0.003 & -0.003 - 0.001 & 0 - 0.001 & -0.001 - 0.002 & 0 - 0.001 & -0.001 - 0.001 \\
   &  & 80 &  & -0.014 - 0.008 & -0.016 - 0.02 & -0.013 - 0.01 & -0.011 - 0.014 & -0.008 - 0.009 & -0.014 - 0.011 & -0.014 - 0.012 & -0.01 - 0.007 & -0.004 - 0.008 & -0.003 - 0.008 \\
   &  & 0 &  & -0.001 - 0.001 & -0.002 - 0.001 & 0 - 0 & -0.001 - 0.001 & 0 - 0 & 0 - 0.001 & 0 - 0 & -0.001 - 0 & 0 - 0 & 0 - 0 \\
   & 10 & 40 &  & 0 - 0.001 & -0.001 - 0.001 & 0 - 0 & -0.001 - 0.001 & -0.001 - 0.001 & 0 - 0.001 & -0.001 - 0.001 & 0 - 0 & 0 - 0 & 0 - 0 \\
  10000 &  & 80 &  & -0.004 - 0 & 0 - 0.001 & -0.003 - 0.002 & -0.001 - 0.002 & -0.002 - 0.001 & -0.001 - 0.002 & 0 - 0.002 & -0.001 - 0.002 & -0.002 - 0.001 & -0.001 - 0.001 \\
   &  & 0 &  & -0.001 - 0.001 & 0 - 0 & 0 - 0 & -0.001 - 0 & 0 - 0 & -0.001 - 0.001 & 0 - 0 & 0 - 0 & 0 - 0 & 0 - 0 \\
   & 100 & 40 &  & 0 - 0.001 & 0 - 0 & -0.002 - 0 & -0.001 - 0.001 & -0.001 - 0.001 & -0.001 - 0 & -0.001 - 0 & 0 - 0.001 & 0 - 0 & 0 - 0 \\
   &  & 80 &  & -0.002 - 0.003 & -0.002 - 0.001 & -0.001 - 0.001 & -0.001 - 0.001 & 0 - 0.002 & -0.004 - 0.003 & -0.003 - 0.001 & -0.001 - 0 & -0.001 - 0.001 & -0.001 - 0.001 \\
   \hline
\end{tabular}
\endgroup
\caption{Bias over all possible combinations of the Cox-regression parameters.  Parameter  $\beta _{score-shape}$ .  Gompertz  hazard function.}
\end{sidewaystable}
\begin{sidewaystable}[ht]
\centering
\begingroup\tiny
\begin{tabular}{rrrlcccccccccc}
  \hline
Sample size & \#of clusters & \%Cens &   &    &     &       &         &          &        &         &          &           &            \\
  \hline
 &  &  & True $\sigma ^2$ & 0 & 0 & 0.5 & 0.5 & 1 & 1 & 2 & 2 & 5 & 5 \\
   &  &  & \%Success & 25 & 50 & 25 & 50 & 25 & 50 & 25 & 50 & 25 & 50 \\
   \hline
  &   &   &   &    &     &       &         & Min-Max bias &        &         &          &           &            \\
   \hline
 &  & 0 &  & 0 - 0 & 0 - 0 & -0.094 - -0.093 & -0.094 - -0.093 & -0.168 - -0.167 & -0.168 - -0.168 & -0.295 - -0.294 & -0.295 - -0.294 & -0.587 - -0.586 & -0.589 - -0.584 \\
   & 10 & 40 &  & 0 - 0 & 0 - 0 & -0.094 - -0.093 & -0.094 - -0.093 & -0.163 - -0.161 & -0.164 - -0.162 & -0.26 - -0.258 & -0.261 - -0.258 & -0.453 - -0.448 & -0.452 - -0.449 \\
  300 &  & 80 &  & 0 - 0 & 0 - 0 & -0.209 - -0.202 & -0.212 - -0.205 & -0.227 - -0.219 & -0.225 - -0.211 & -0.306 - -0.293 & -0.306 - -0.295 & -2.414 - -0.621 & -0.971 - -0.609 \\
   &  & 0 &  & 0 - 0 & 0 - 0 & -0.018 - -0.017 & -0.018 - -0.017 & -0.013 - -0.013 & -0.013 - -0.012 & -0.009 - -0.008 & -0.009 - -0.008 & 0.023 - 0.024 & 0.021 - 0.024 \\
   & 100 & 40 &  & 0 - 0 & 0 - 0 & -0.027 - -0.024 & -0.028 - -0.024 & -0.013 - -0.011 & -0.013 - -0.011 & 0.01 - 0.012 & 0.009 - 0.012 & 0.051 - 0.059 & 0.048 - 0.056 \\
   &  & 80 &  & 0 - 0 & 0 - 0 & -0.363 - -0.334 & -0.356 - -0.347 & -0.216 - -0.177 & -0.211 - -0.198 & -0.033 - -0.014 & -0.037 - -0.016 & 0.119 - 0.135 & 0.126 - 0.139 \\
   &  & 0 &  & 0 - 0 & 0 - 0 & -0.091 - -0.091 & -0.091 - -0.091 & -0.169 - -0.169 & -0.169 - -0.169 & -0.297 - -0.297 & -0.297 - -0.297 & -0.652 - -0.65 & -0.652 - -0.649 \\
   & 10 & 40 &  & 0 - 0 & 0 - 0 & -0.092 - -0.092 & -0.093 - -0.092 & -0.171 - -0.171 & -0.172 - -0.171 & -0.291 - -0.29 & -0.291 - -0.29 & -0.556 - -0.551 & -0.553 - -0.551 \\
  1000 &  & 80 &  & 0 - 0 & 0 - 0 & -0.104 - -0.103 & -0.103 - -0.102 & -0.175 - -0.173 & -0.175 - -0.172 & -0.289 - -0.283 & -0.288 - -0.285 & -0.592 - -0.585 & -0.597 - -0.575 \\
   &  & 0 &  & 0 - 0 & 0 - 0 & -0.009 - -0.009 & -0.009 - -0.009 & -0.014 - -0.014 & -0.014 - -0.014 & -0.029 - -0.029 & -0.029 - -0.029 & -0.046 - -0.046 & -0.047 - -0.046 \\
   & 100 & 40 &  & 0 - 0 & 0 - 0 & -0.011 - -0.011 & -0.011 - -0.011 & -0.01 - -0.009 & -0.01 - -0.009 & -0.018 - -0.015 & -0.017 - -0.016 & -0.017 - -0.014 & -0.016 - -0.013 \\
   &  & 80 &  & 0 - 0 & 0 - 0 & -0.028 - -0.025 & -0.028 - -0.026 & -0.024 - -0.023 & -0.025 - -0.023 & -0.029 - -0.025 & -0.029 - -0.025 & -0.012 - -0.001 & -0.013 - -0.009 \\
   &  & 0 &  & 0 - 0 & 0 - 0 & -0.095 - -0.095 & -0.095 - -0.095 & -0.171 - -0.171 & -0.171 - -0.171 & -0.31 - -0.31 & -0.31 - -0.31 & -0.677 - -0.676 & -0.677 - -0.676 \\
   & 10 & 40 &  & 0 - 0 & 0 - 0 & -0.09 - -0.09 & -0.09 - -0.09 & -0.175 - -0.174 & -0.175 - -0.174 & -0.297 - -0.296 & -0.298 - -0.296 & -0.585 - -0.583 & -0.586 - -0.583 \\
  10000 &  & 80 &  & 0 - 0 & 0 - 0 & -0.091 - -0.09 & -0.091 - -0.09 & -0.173 - -0.172 & -0.173 - -0.173 & -0.286 - -0.285 & -0.287 - -0.285 & -0.548 - -0.544 & -0.55 - -0.545 \\
   &  & 0 &  & 0 - 0 & 0 - 0 & -0.009 - -0.009 & -0.009 - -0.009 & -0.009 - -0.009 & -0.009 - -0.009 & -0.027 - -0.027 & -0.027 - -0.027 & -0.053 - -0.052 & -0.053 - -0.052 \\
   & 100 & 40 &  & 0 - 0 & 0 - 0 & -0.011 - -0.011 & -0.011 - -0.011 & -0.015 - -0.014 & -0.015 - -0.014 & -0.025 - -0.025 & -0.025 - -0.025 & -0.039 - -0.037 & -0.039 - -0.038 \\
   &  & 80 &  & 0 - 0 & 0 - 0 & -0.013 - -0.012 & -0.012 - -0.012 & -0.015 - -0.014 & -0.014 - -0.014 & -0.025 - -0.024 & -0.025 - -0.024 & -0.036 - -0.032 & -0.035 - -0.032 \\
   \hline
\end{tabular}
\endgroup
\caption{Bias over all possible combinations of the Cox-regression parameters.  Parameter  $\sigma ^2$ .  Weibull  hazard function.}
\end{sidewaystable}
\begin{sidewaystable}[ht]
\centering
\begingroup\tiny
\begin{tabular}{rrrlcccccccccc}
  \hline
Sample size & \#of clusters & \%Cens &   &    &     &       &         &          &        &         &          &           &            \\
  \hline
 &  &  & True $\sigma ^2$ & 0 & 0 & 0.5 & 0.5 & 1 & 1 & 2 & 2 & 5 & 5 \\
   &  &  & \%Success & 25 & 50 & 25 & 50 & 25 & 50 & 25 & 50 & 25 & 50 \\
   \hline
  &   &   &   &    &     &       &         & Min-Max bias &        &         &          &           &            \\
   \hline
 &  & 0 &  & 0 - 0 & 0 - 0 & -0.094 - -0.093 & -0.094 - -0.093 & -0.168 - -0.167 & -0.168 - -0.167 & -0.296 - -0.294 & -0.295 - -0.294 & -0.6 - -0.599 & -0.6 - -0.599 \\
   & 10 & 40 &  & 0 - 0 & 0 - 0 & -0.098 - -0.096 & -0.099 - -0.097 & -0.173 - -0.171 & -0.172 - -0.171 & -0.286 - -0.284 & -0.285 - -0.284 & -0.497 - -0.491 & -0.497 - -0.491 \\
  300 &  & 80 &  & 0 - 0 & 0 - 0 & -0.212 - -0.206 & -0.211 - -0.197 & -0.191 - -0.181 & -0.189 - -0.185 & -0.237 - -0.229 & -0.236 - -0.23 & -0.327 - -0.306 & -0.332 - -0.313 \\
   &  & 0 &  & 0 - 0 & 0 - 0 & -0.018 - -0.018 & -0.018 - -0.017 & -0.014 - -0.013 & -0.014 - -0.013 & -0.01 - -0.008 & -0.009 - -0.009 & 0.021 - 0.022 & 0.019 - 0.02 \\
   & 100 & 40 &  & 0 - 0 & 0 - 0 & -0.04 - -0.038 & -0.038 - -0.036 & -0.023 - -0.021 & -0.024 - -0.022 & -0.002 - 0 & -0.003 - -0.001 & 0.05 - 0.052 & 0.05 - 0.055 \\
   &  & 80 &  & 0 - 0 & 0 - 0 & -0.362 - -0.353 & -0.366 - -0.336 & -0.226 - -0.173 & -0.211 - -0.185 & -0.006 - 0.027 & -0.001 - 0.019 & 0.247 - 0.267 & 0.253 - 0.291 \\
   &  & 0 &  & 0 - 0 & 0 - 0 & -0.091 - -0.091 & -0.091 - -0.091 & -0.169 - -0.169 & -0.169 - -0.169 & -0.297 - -0.297 & -0.297 - -0.297 & -0.66 - -0.66 & -0.66 - -0.66 \\
   & 10 & 40 &  & 0 - 0 & 0 - 0 & -0.093 - -0.093 & -0.094 - -0.093 & -0.175 - -0.175 & -0.175 - -0.175 & -0.309 - -0.308 & -0.309 - -0.308 & -0.629 - -0.627 & -0.629 - -0.625 \\
  1000 &  & 80 &  & 0 - 0 & 0 - 0 & -0.1 - -0.099 & -0.1 - -0.098 & -0.168 - -0.165 & -0.167 - -0.165 & -0.273 - -0.27 & -0.275 - -0.271 & -0.521 - -0.516 & -0.52 - -0.518 \\
   &  & 0 &  & 0 - 0 & 0 - 0 & -0.009 - -0.009 & -0.009 - -0.009 & -0.014 - -0.014 & -0.014 - -0.014 & -0.029 - -0.029 & -0.029 - -0.029 & -0.051 - -0.05 & -0.051 - -0.05 \\
   & 100 & 40 &  & 0 - 0 & 0 - 0 & -0.013 - -0.013 & -0.013 - -0.013 & -0.012 - -0.011 & -0.012 - -0.011 & -0.02 - -0.019 & -0.02 - -0.019 & -0.027 - -0.024 & -0.026 - -0.025 \\
   &  & 80 &  & 0 - 0 & 0 - 0 & -0.024 - -0.021 & -0.023 - -0.019 & -0.015 - -0.014 & -0.016 - -0.014 & -0.009 - -0.005 & -0.01 - -0.005 & 0.026 - 0.031 & 0.024 - 0.029 \\
   &  & 0 &  & 0 - 0 & 0 - 0 & -0.095 - -0.095 & -0.095 - -0.095 & -0.171 - -0.171 & -0.171 - -0.171 & -0.311 - -0.31 & -0.311 - -0.31 & -0.682 - -0.682 & -0.682 - -0.682 \\
   & 10 & 40 &  & 0 - 0 & 0 - 0 & -0.09 - -0.09 & -0.09 - -0.09 & -0.175 - -0.175 & -0.175 - -0.175 & -0.306 - -0.306 & -0.306 - -0.306 & -0.649 - -0.646 & -0.649 - -0.646 \\
  10000 &  & 80 &  & 0 - 0 & 0 - 0 & -0.09 - -0.09 & -0.09 - -0.09 & -0.173 - -0.173 & -0.173 - -0.173 & -0.289 - -0.287 & -0.289 - -0.287 & -0.551 - -0.546 & -0.549 - -0.546 \\
   &  & 0 &  & 0 - 0 & 0 - 0 & -0.009 - -0.009 & -0.009 - -0.009 & -0.009 - -0.009 & -0.009 - -0.009 & -0.027 - -0.027 & -0.027 - -0.027 & -0.055 - -0.055 & -0.055 - -0.055 \\
   & 100 & 40 &  & 0 - 0 & 0 - 0 & -0.011 - -0.011 & -0.011 - -0.011 & -0.014 - -0.014 & -0.014 - -0.014 & -0.027 - -0.027 & -0.027 - -0.027 & -0.058 - -0.058 & -0.058 - -0.057 \\
   &  & 80 &  & 0 - 0 & 0 - 0 & -0.013 - -0.012 & -0.013 - -0.012 & -0.013 - -0.012 & -0.013 - -0.012 & -0.019 - -0.018 & -0.019 - -0.018 & -0.032 - -0.031 & -0.034 - -0.029 \\
   \hline
\end{tabular}
\endgroup
\caption{Bias over all possible combinations of the Cox-regression parameters.  Parameter  $\sigma ^2$ .  Gompertz  hazard function.}
\end{sidewaystable}

\clearpage
\section*{Figures}

\begin{itemize}
\item A: Biases in estimation of the scale   and shape  parameters of the  baseline distribution
\item B: Biases in estimation of the Cox regression parameters and of the frailty variance
\item C: Coverage of the profile likelihood confidence intervals for  the scale   and shape  parameters of the  baseline distribution
\item D: Coverage of the profile likelihood confidence intervals for the Cox regression parameters and  the frailty variance
\item E: Coverage of the standard error based confidence intervals for  the scale   and shape  parameters of the  baseline distribution
\item  F: Coverage of the standard-error-based confidence intervals for the Cox regression parameters and  the frailty variance.
\end{itemize}
\clearpage

\setcounter{figure}{0}
\setcounter{section}{0}
\renewcommand{\thefigure}{A.\arabic{figure}}

\section*{A: Biases in estimation of the scale   and shape  parameters of the  baseline distribution}

Each figure corresponds to a particular  baseline distribution  (Weibull or Gompertz), a value of the probability of success $p_{success}$ (= 0.25 or 0.5), a value for the number of clusters $N_{cl}$ (=10, 100) and a particular choice of the signs of the Cox regression parameters (+ + + +, + - + -, - + - + and - - - -).\\

The absolute values of the Cox regression parameters are held constant at $\beta_{success-scale}$ = 0.5 , $\beta_{success-shape}$= 0.05 , $\beta_{score-scale}$= 1 , $\beta_{score-shape}$= 0.1.

For each combination of a censoring proportion  (= 0, 40\%, 80\%), a panel plots, versus the frailty variance $\sigma^2$ (= 0, 0.5, 1, 2, 3, 4, 5), the difference between the estimated and the true value of a scale parameter $a$  and between the estimated and true value of a shape parameter $b$ of the baseline Weibull or Gompertz distribution for three sample sizes (300, 1000 and 10000) . \\

\clearpage

 \begin{figure}[ht]
	\centering
	\includegraphics[scale=1]{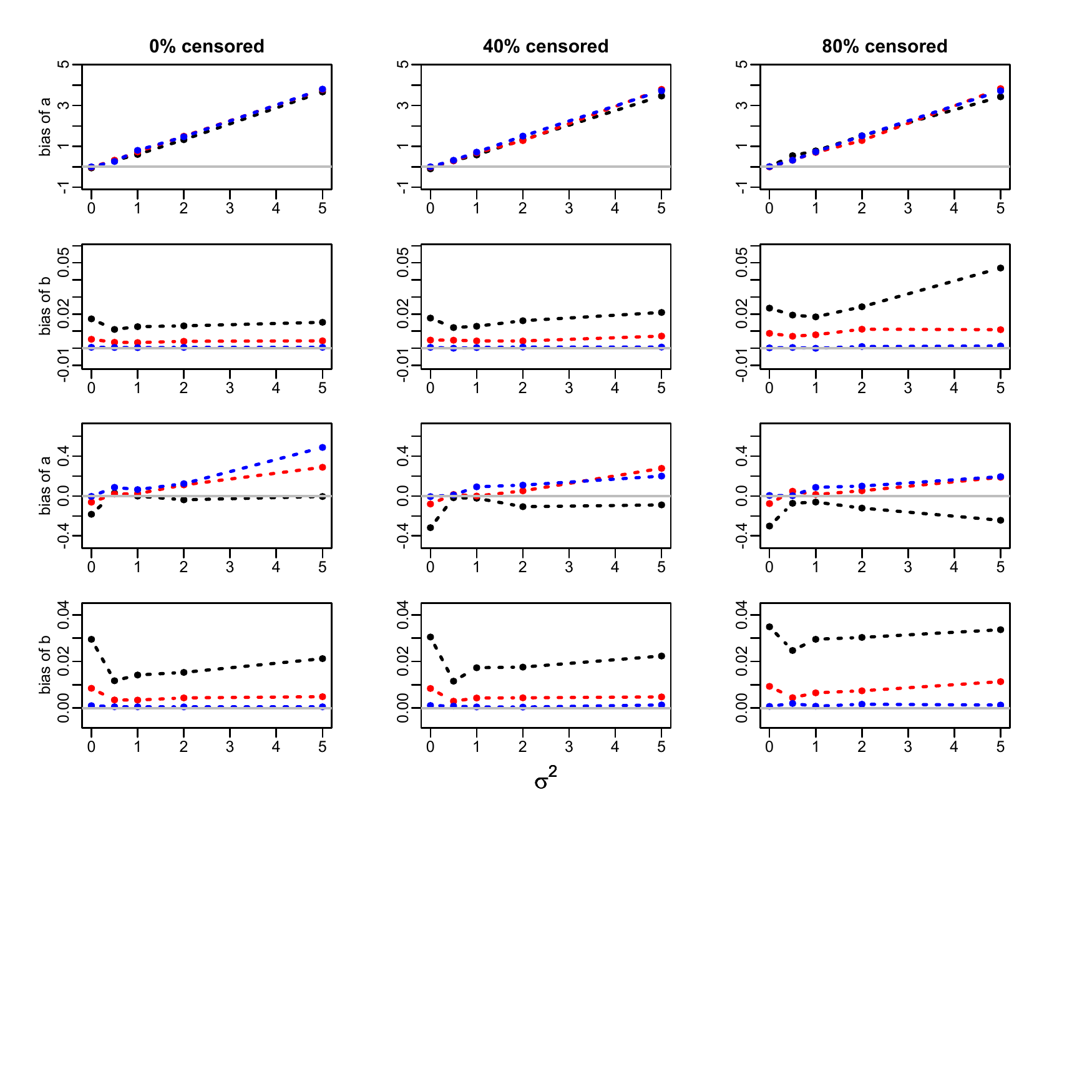}
	\caption{Bias of the estimation of the $a$ and $b$ parameters in the Weibull model.
Success proportion= 0.25.  Sample sizes: 300 (black), 1000 (red) and 10000 (blue).
True values: $\beta_{success-scale}$ = 0.5 , $\beta_{success-shape}$= 0.05 , $\beta_{score-scale}$= 1 , $\beta_{score-shape}$= 0.1.
Top two rows: 10 clusters; bottom two rows: 100 clusters.}
	\label{Bias10_100clustersWeibull1}
\end{figure}

\begin{figure}[ht]
	\centering
	\includegraphics[scale=1]{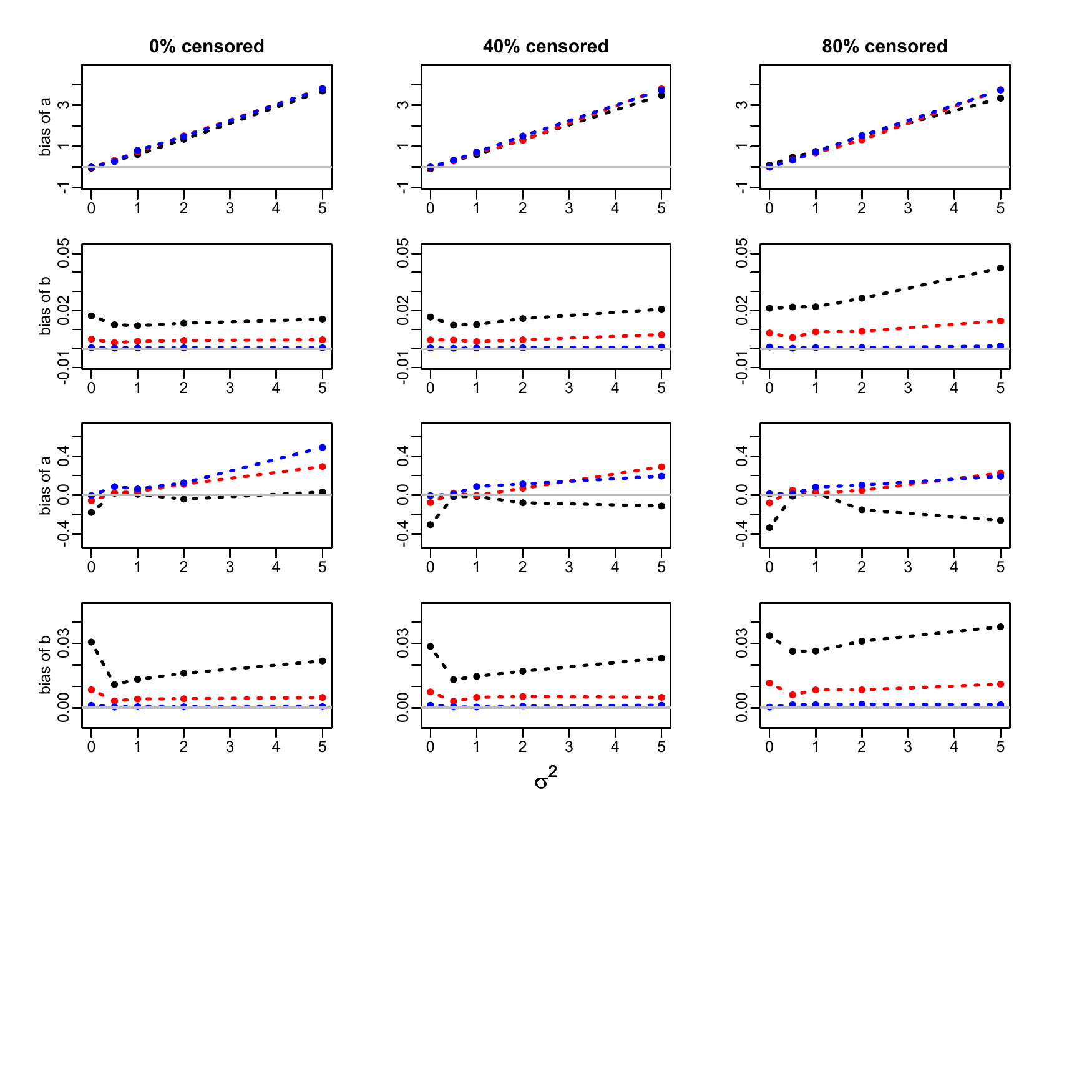}
	\caption{Bias of the estimation of the $a$ and $b$ parameters  in the Weibull model.
Success proportion= 0.25.  Sample sizes: 300 (black), 1000 (red) and 10000 (blue).
True values: $\beta_{success-scale}$ = 0.5 , $\beta_{success-shape}$= -0.05 , $\beta_{score-scale}$= 1 , $\beta_{score-shape}$= - 0.1.
Top two rows: 10 clusters; bottom two rows: 100 clusters.}
	\label{Bias10_100clustersWeibull2}
\end{figure}

\begin{figure}[ht]
	\centering
	\includegraphics[scale=1]{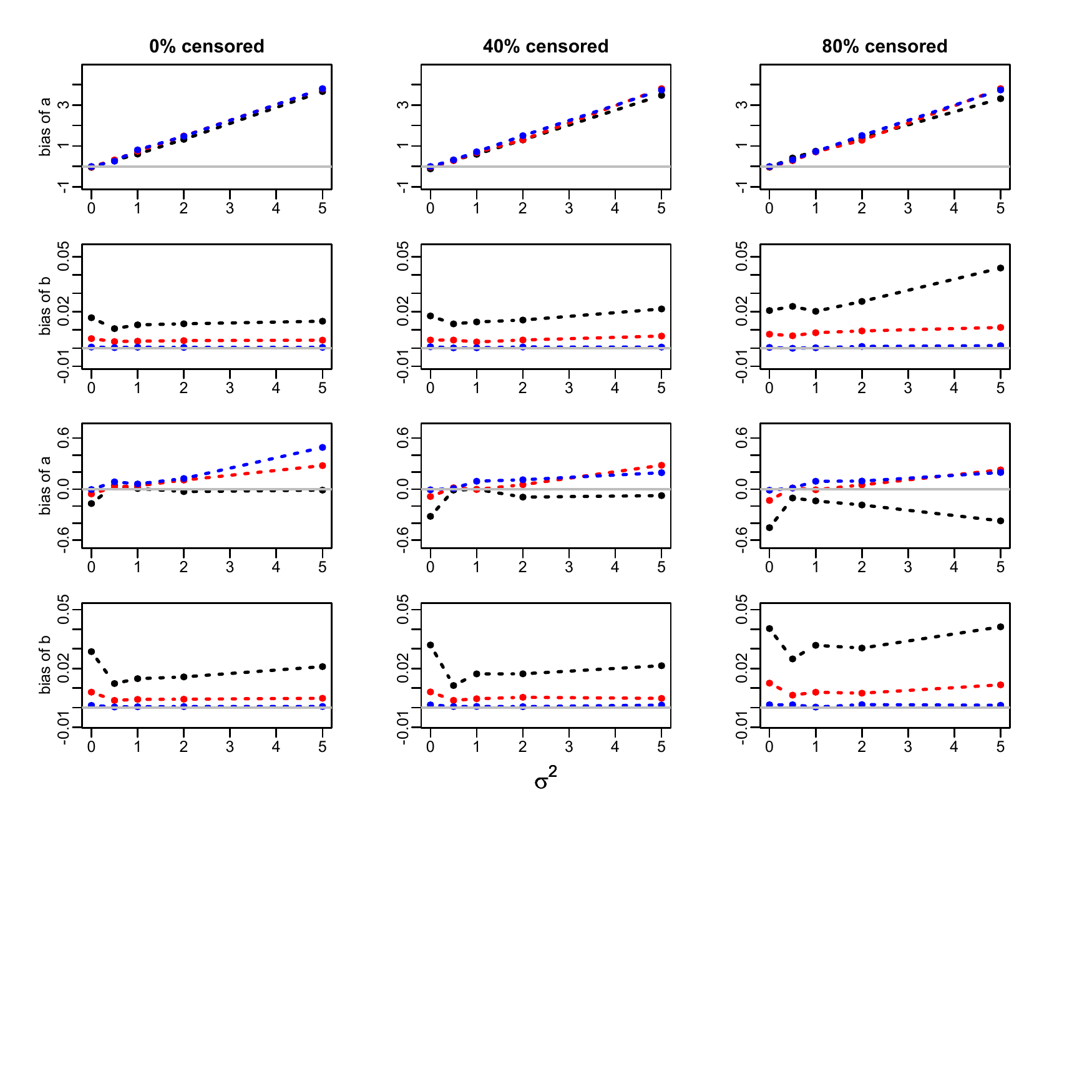}
	\caption{Bias of the estimation of the $a$ and $b$ parameters  in the Weibull model.
Success proportion= 0.25.  Sample sizes: 300 (black), 1000 (red) and 10000 (blue).
True values: $\beta_{success-scale}$ = -0.5 , $\beta_{success-shape}$= 0.05 , $\beta_{score-scale}$= -1 , $\beta_{score-shape}$= 0.1.
Top two rows: 10 clusters; bottom two rows: 100 clusters.}
	\label{Bias10_100clustersWeibull3}
\end{figure}

\begin{figure}[ht]
	\centering
	\includegraphics[scale=1]{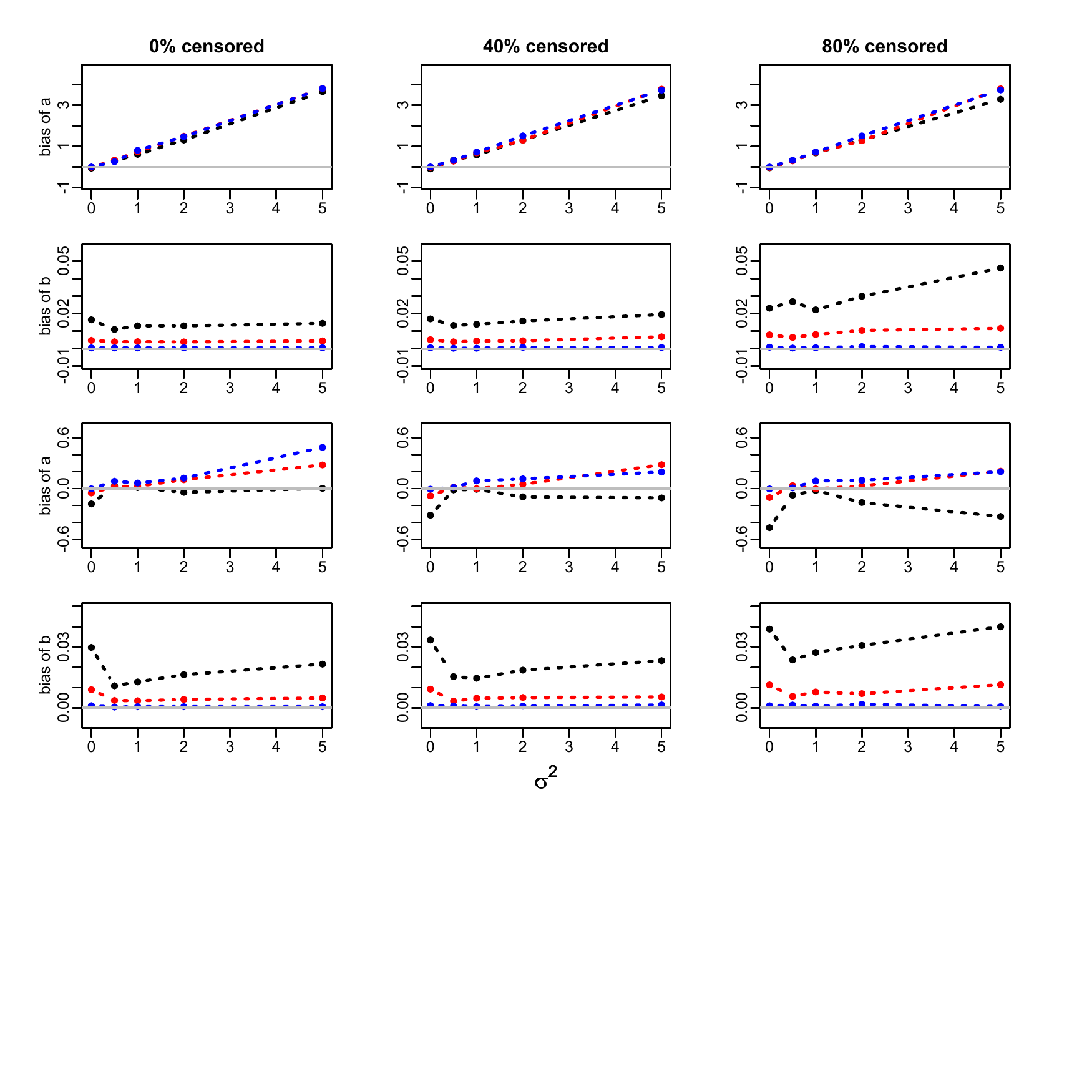}
	\caption{Bias of the estimation of the $a$ and $b$ parameters  in the Weibull model.
Success proportion= 0.25.  Sample sizes: 300 (black), 1000 (red) and 10000 (blue).
True values: $\beta_{success-scale}$ = -0.5 , $\beta_{success-shape}$= -0.05 , $\beta_{score-scale}$= -1 , $\beta_{score-shape}$= -0.1.
Top two rows: 10 clusters; bottom two rows: 100 clusters.}
	\label{Bias10_100clustersWeibull4}
\end{figure}

\begin{figure}[ht]
	\centering
	\includegraphics[scale=1]{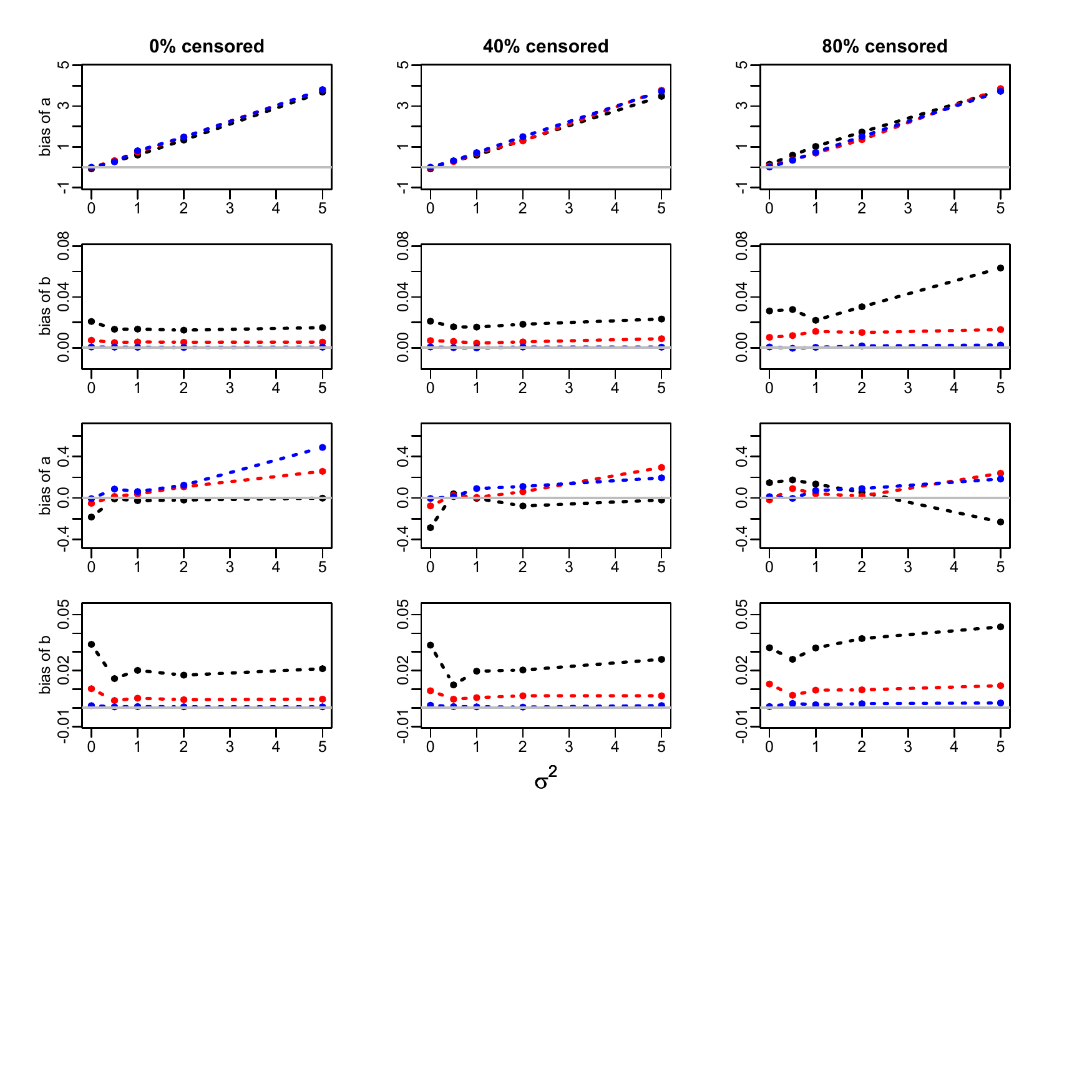}
	\caption{Bias of the estimation of the $a$ and $b$ parameters  in the Weibull model.
Success proportion= 0.5.  Sample sizes: 300 (black), 1000 (red) and 10000 (blue).
True values: $\beta_{success-scale}$ = 0.5 , $\beta_{success-shape}$= 0.05 , $\beta_{score-scale}$= 1 , $\beta_{score-shape}$= 0.1.
Top two rows: 10 clusters; bottom two rows: 100 clusters.}
	\label{Bias10_100clustersWeibull5}
\end{figure}

\begin{figure}[ht]
	\centering
	\includegraphics[scale=1]{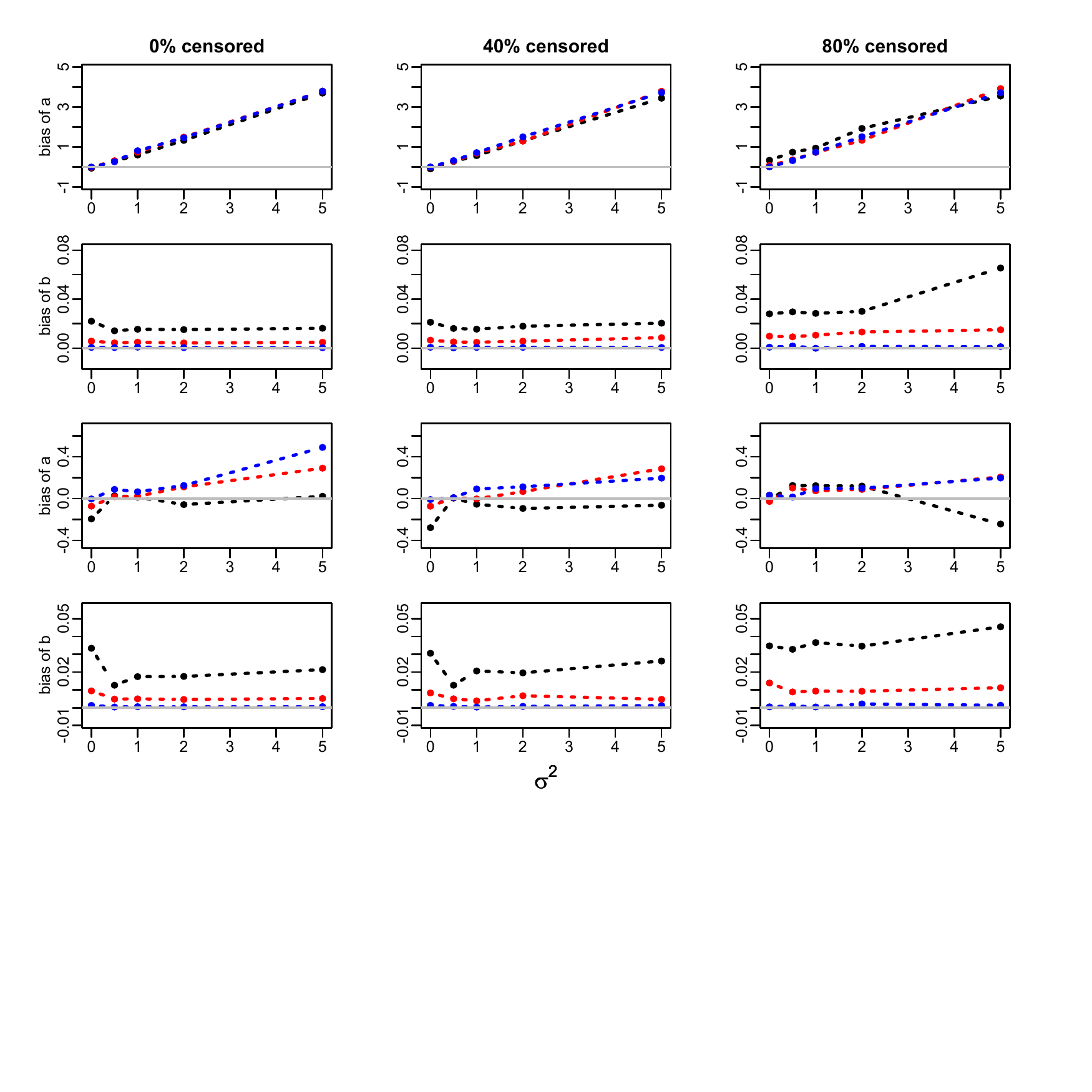}
	\caption{Bias of the estimation of the $a$ and $b$ parameters  in the Weibull model.
Success proportion= 0.5.  Sample sizes: 300 (black), 1000 (red) and 10000 (blue).
True values: $\beta_{success-scale}$ = 0.5 , $\beta_{success-shape}$= -0.05 , $\beta_{score-scale}$= 1 , $\beta_{score-shape}$= - 0.1.
Top two rows: 10 clusters; bottom two rows: 100 clusters.}
	\label{Bias10_100clustersWeibull6}
\end{figure}

\begin{figure}[ht]
	\centering
	\includegraphics[scale=1]{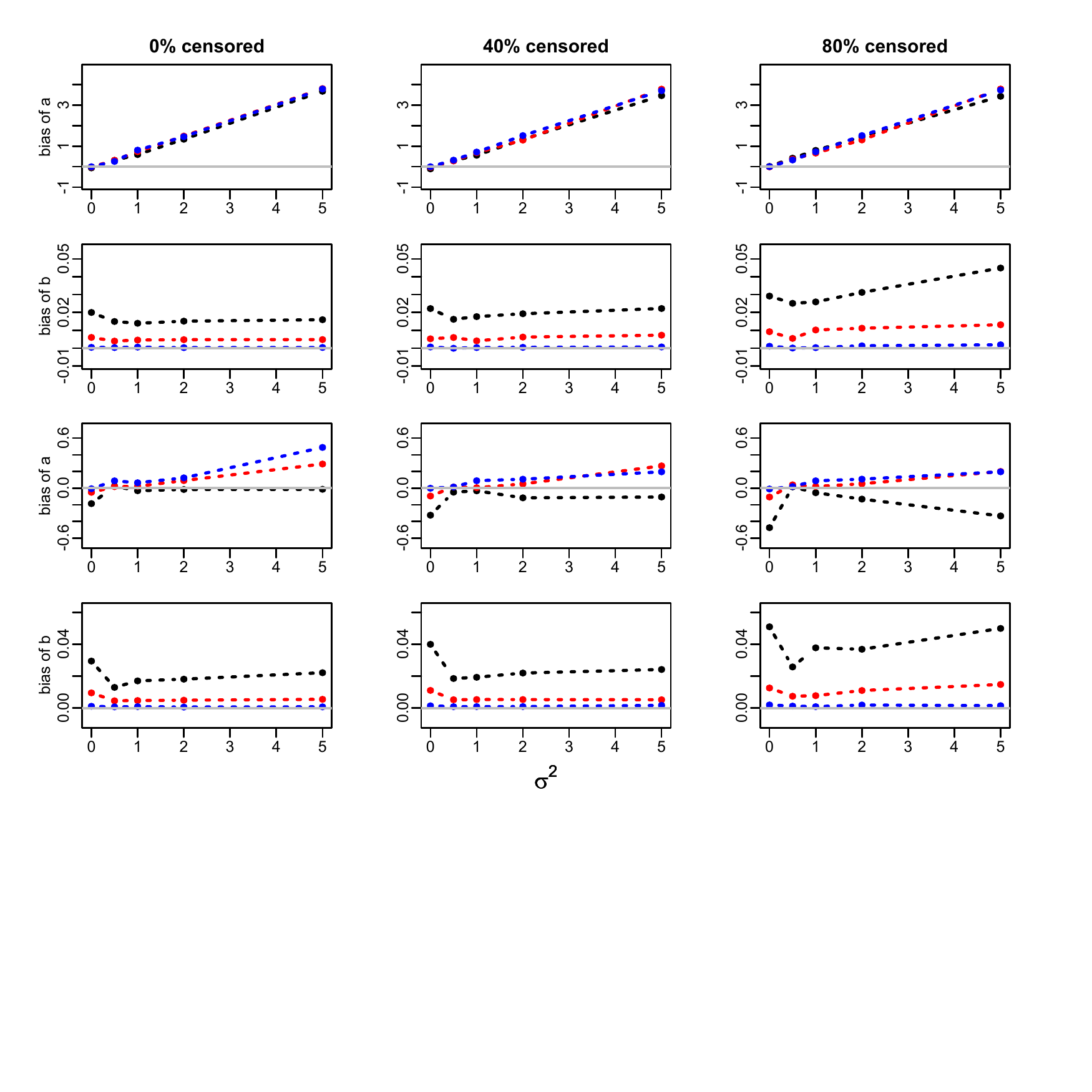}
	\caption{Bias of the estimation of the $a$ and $b$ parameters in the Weibull model.
Success proportion= 0.5.  Sample sizes: 300 (black), 1000 (red) and 10000 (blue).
True values: $\beta_{success-scale}$ = -0.5 , $\beta_{success-shape}$= 0.05 , $\beta_{score-scale}$= -1 , $\beta_{score-shape}$= 0.1.
Top two rows: 10 clusters; bottom two rows: 100 clusters.}
	\label{Bias10_100clustersWeibull7}
\end{figure}

\begin{figure}[ht]
	\centering
	\includegraphics[scale=1]{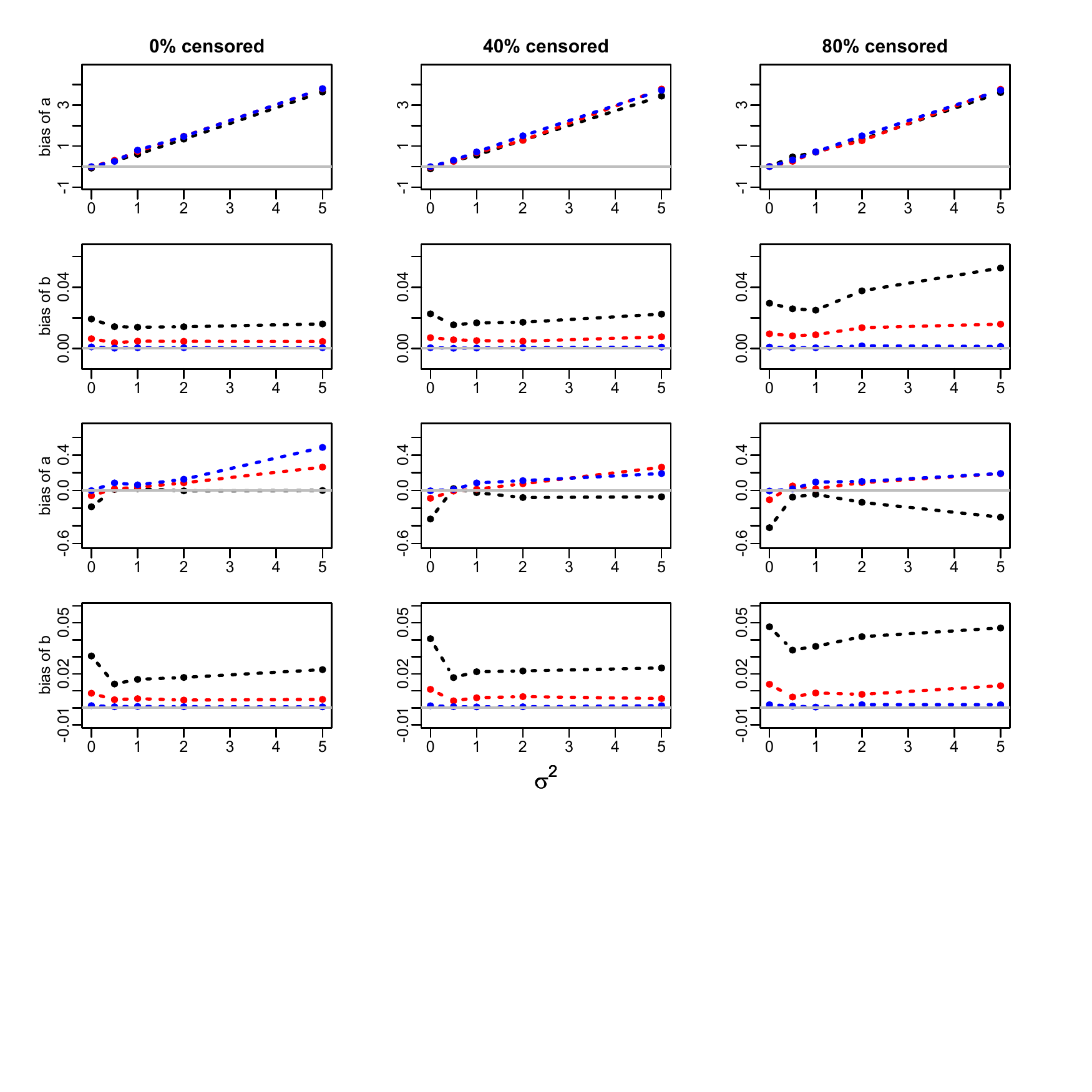}
	\caption{Bias of the estimation of the $a$ and $b$ parameters  in the Weibull model.
Success proportion= 0.5.  Sample sizes: 300 (black), 1000 (red) and 10000 (blue).
True values: $\beta_{success-scale}$ = -0.5 , $\beta_{success-shape}$= -0.05 , $\beta_{score-scale}$= -1 , $\beta_{score-shape}$= -0.1.
Top two rows: 10 clusters; bottom two rows: 100 clusters.}
	\label{Bias10_100clustersWeibull8}
\end{figure}


\begin{figure}[ht]
	\centering
	\includegraphics[scale=1]{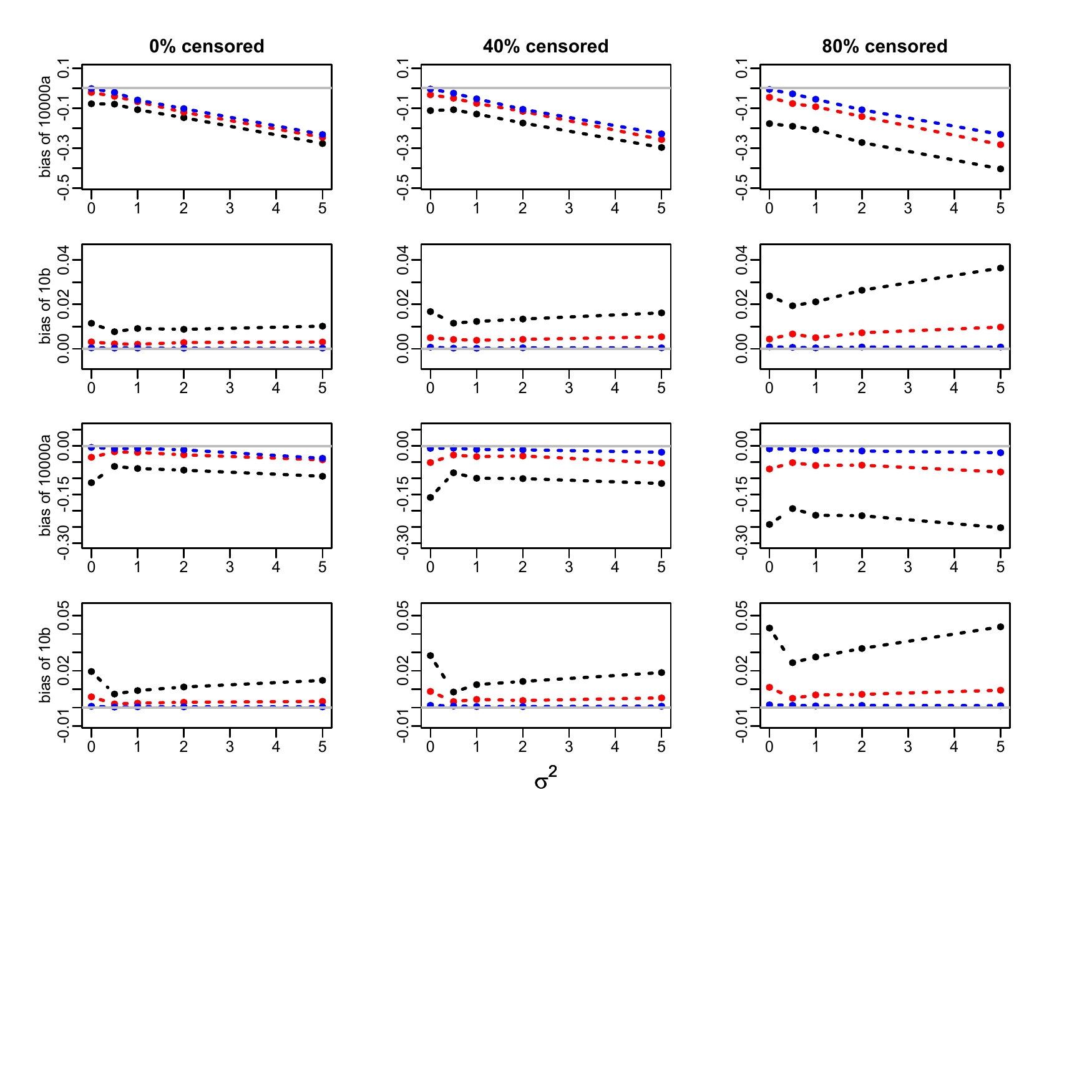}
	\caption{Bias of the estimation of the $a$ and $b$ parameters in the Gompertz model.
Success proportion= 0.25.  Sample sizes: 300 (black), 1000 (red) and 10000 (blue).
True values: $\beta_{success-scale}$ = 0.5 , $\beta_{success-shape}$= 0.05 , $\beta_{score-scale}$= 1 , $\beta_{score-shape}$= 0.1.
Top two rows: 10 clusters; bottom two rows: 100 clusters.}
	\label{Bias10_100clustersGompertz1}
\end{figure}

\begin{figure}[ht]
	\centering
	\includegraphics[scale=1]{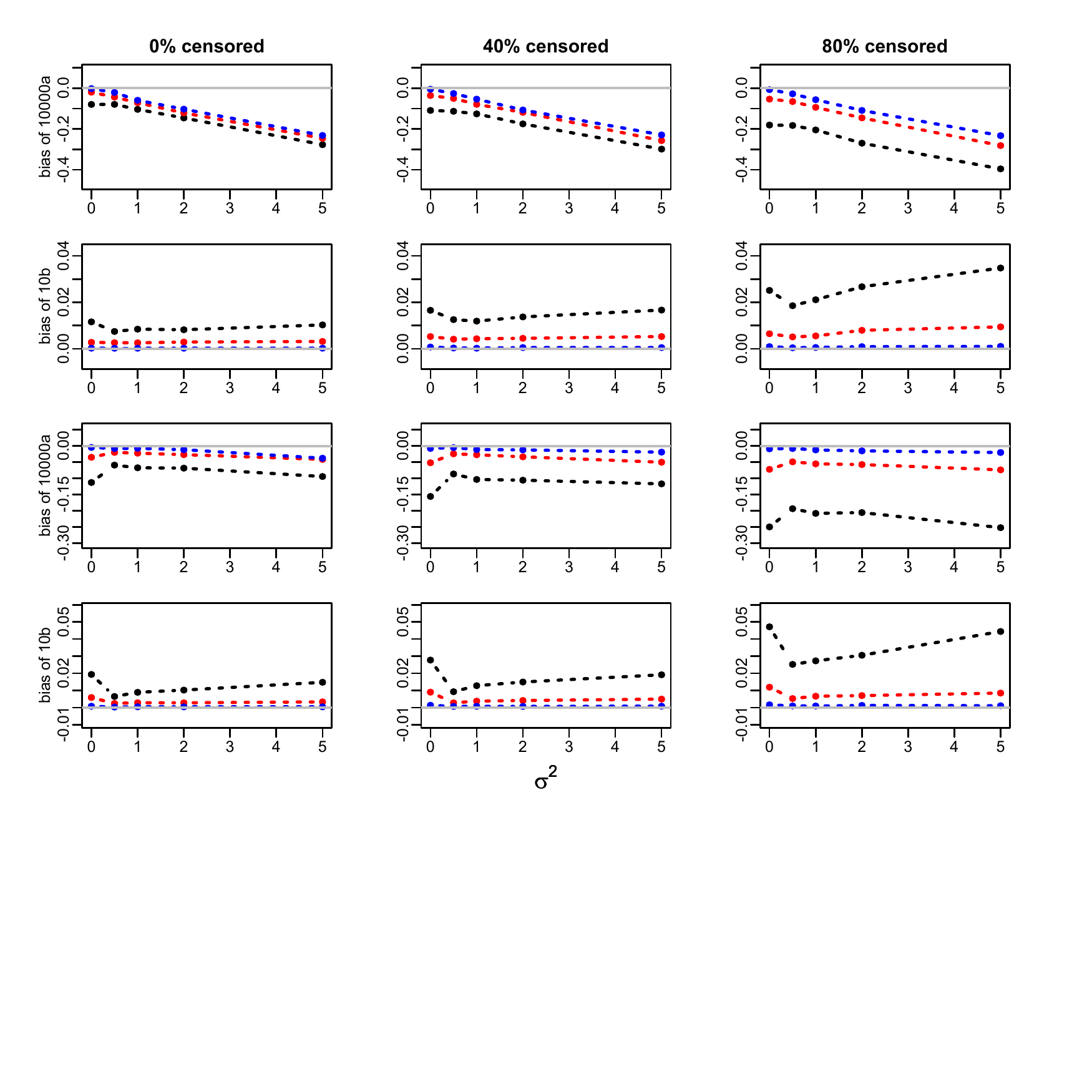}
	\caption{Bias of the estimation of the $a$ and $b$ parameters  in the Gompertz model.
Success proportion= 0.25.  Sample sizes: 300 (black), 1000 (red) and 10000 (blue).
True values: $\beta_{success-scale}$ = 0.5 , $\beta_{success-shape}$= -0.05 , $\beta_{score-scale}$= 1 , $\beta_{score-shape}$= - 0.1.
Top two rows: 10 clusters; bottom two rows: 100 clusters.}
	\label{Bias10_100clustersGompertz2}
\end{figure}

\begin{figure}[ht]
	\centering
	\includegraphics[scale=1]{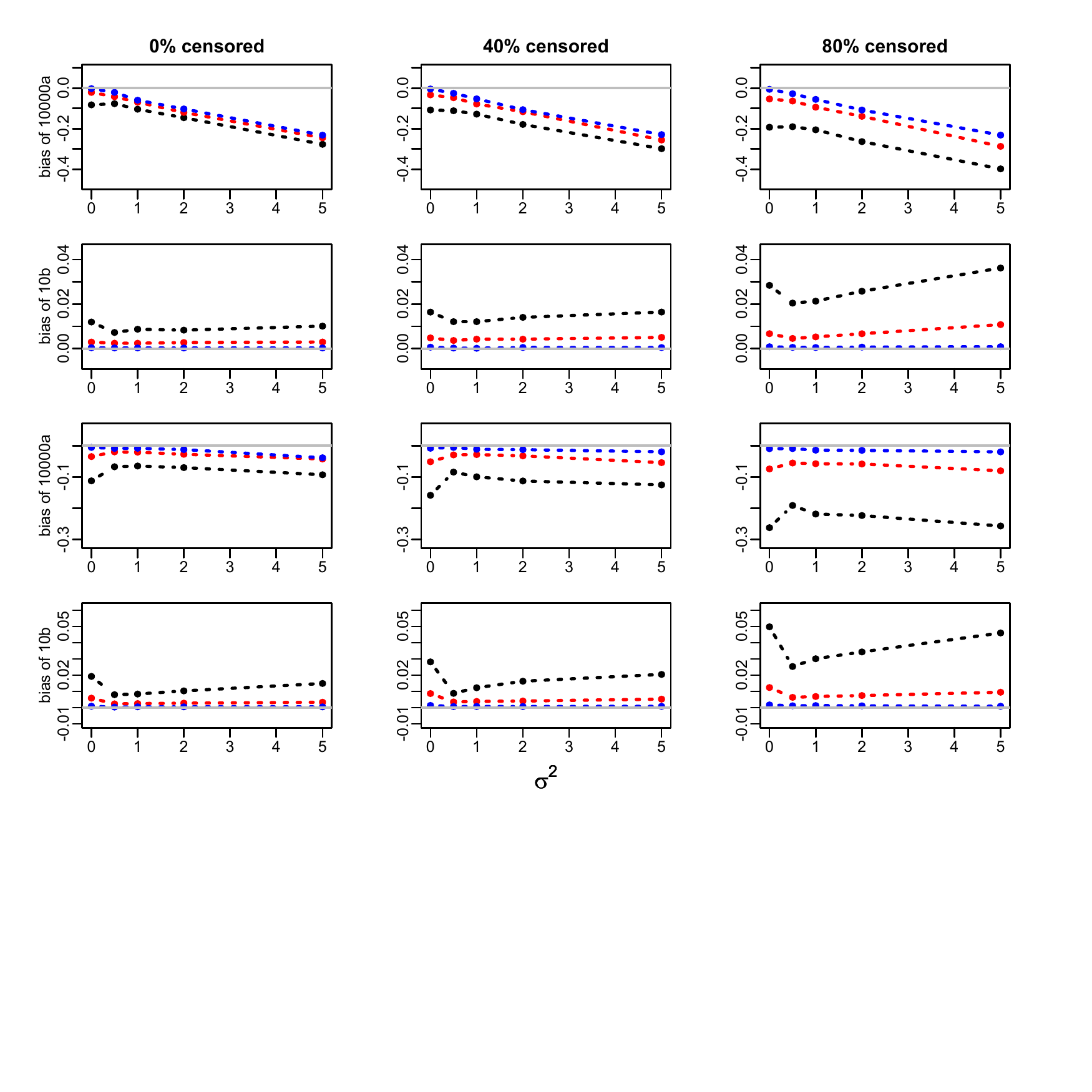}
	\caption{Bias of the estimation of the $a$ and $b$ parameters  in the Gompertz model.
Success proportion= 0.25.  Sample sizes: 300 (black), 1000 (red) and 10000 (blue).
True values: $\beta_{success-scale}$ = -0.5 , $\beta_{success-shape}$= 0.05 , $\beta_{score-scale}$= -1 , $\beta_{score-shape}$= 0.1.
Top two rows: 10 clusters; bottom two rows: 100 clusters.}
	\label{Bias10_100clustersGompertz3}
\end{figure}

\begin{figure}[ht]
	\centering
	\includegraphics[scale=1]{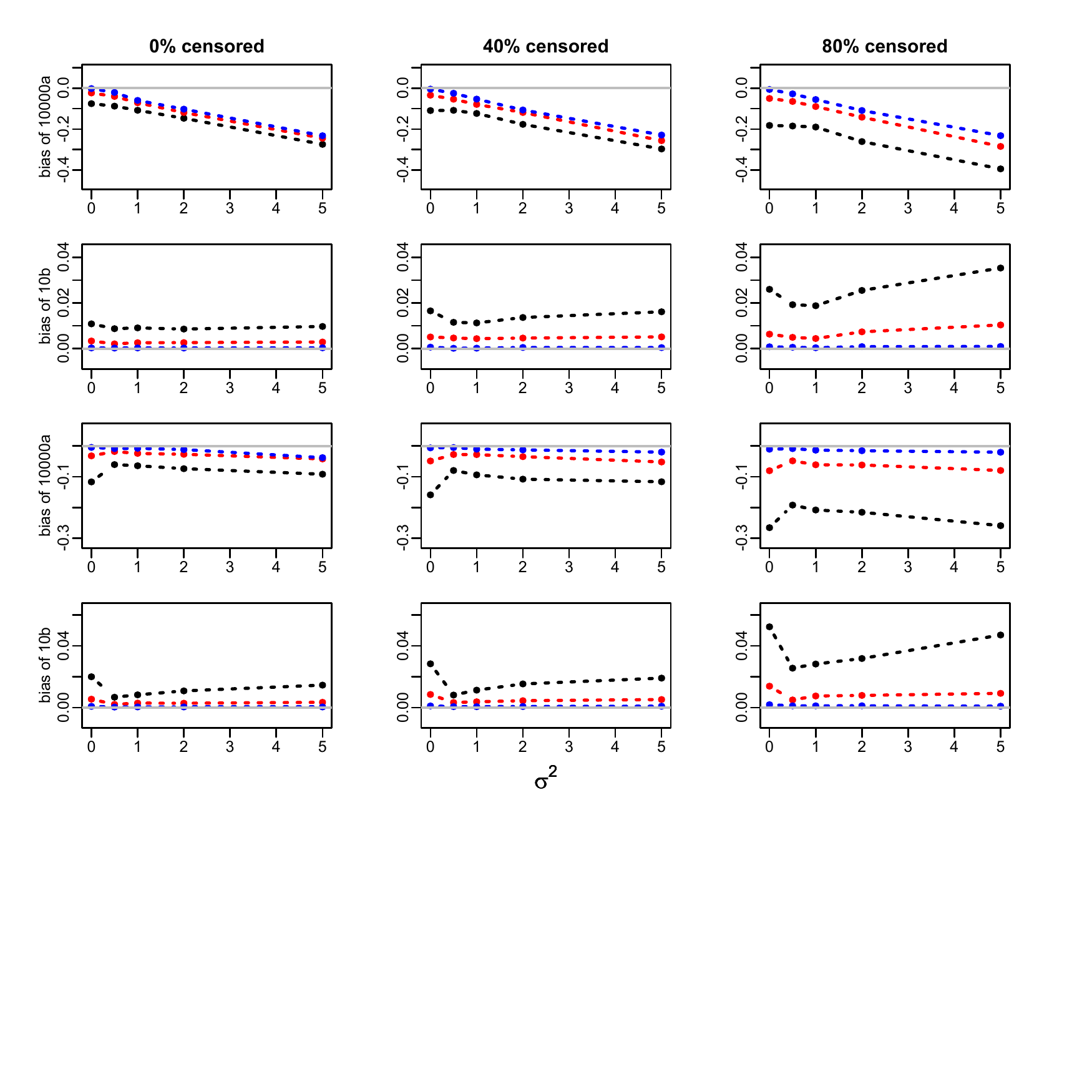}
	\caption{Bias of the estimation of the $a$ and $b$ parameters  in the Gompertz model.
Success proportion= 0.25.  Sample sizes: 300 (black), 1000 (red) and 10000 (blue).
True values: $\beta_{success-scale}$ = -0.5 , $\beta_{success-shape}$= -0.05 , $\beta_{score-scale}$= -1 , $\beta_{score-shape}$= -0.1.
Top two rows: 10 clusters; bottom two rows: 100 clusters.}
	\label{Bias10_100clustersGompertz4}
\end{figure}

\begin{figure}[ht]
	\centering
	\includegraphics[scale=1]{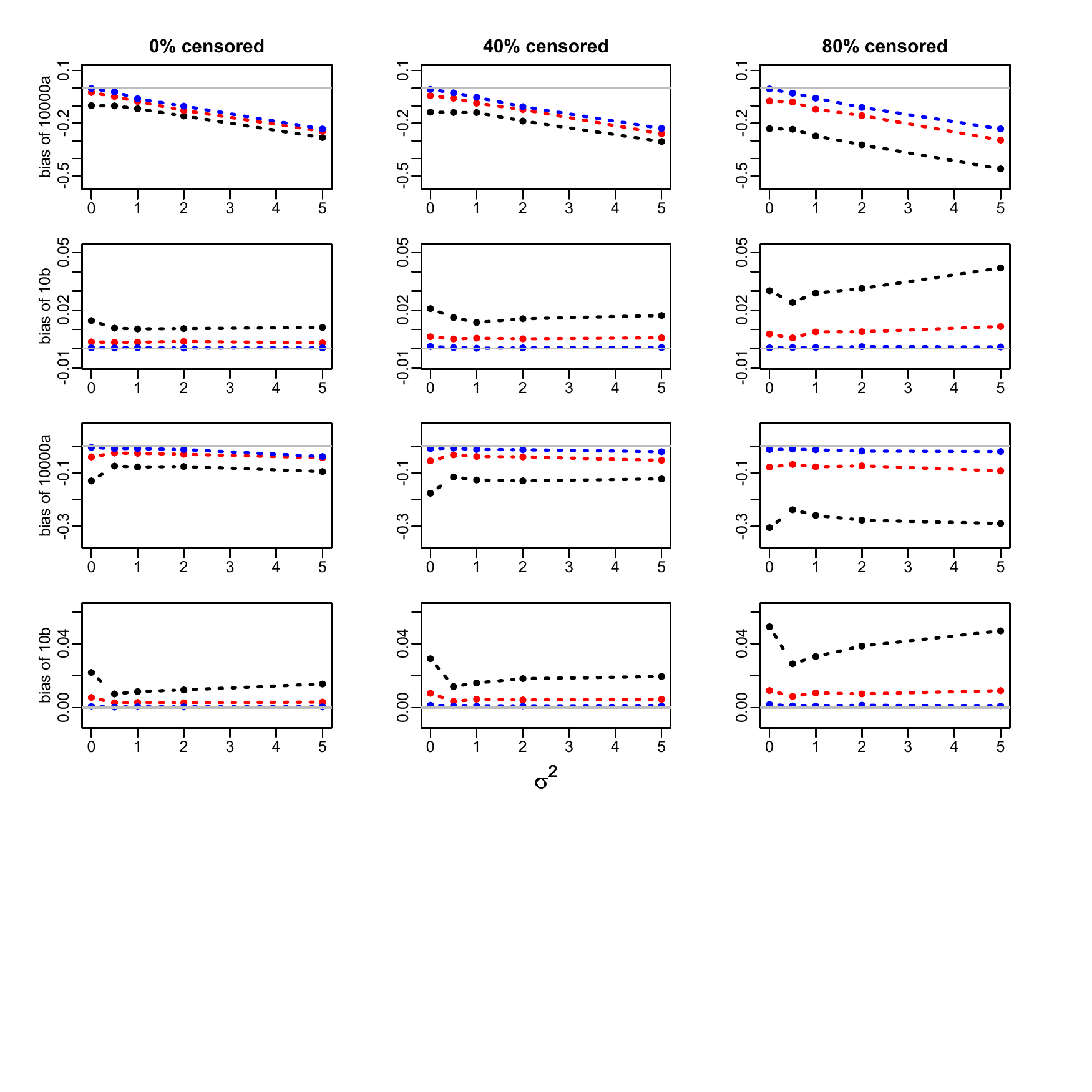}
	\caption{Bias of the estimation of the $a$ and $b$ parameters  in the Gompertz model.
Success proportion= 0.5.  Sample sizes: 300 (black), 1000 (red) and 10000 (blue).
True values: $\beta_{success-scale}$ = 0.5 , $\beta_{success-shape}$= 0.05 , $\beta_{score-scale}$= 1 , $\beta_{score-shape}$= 0.1.
Top two rows: 10 clusters; bottom two rows: 100 clusters.}
	\label{Bias10_100clustersGompertz5}
\end{figure}

\begin{figure}[ht]
	\centering
	\includegraphics[scale=1]{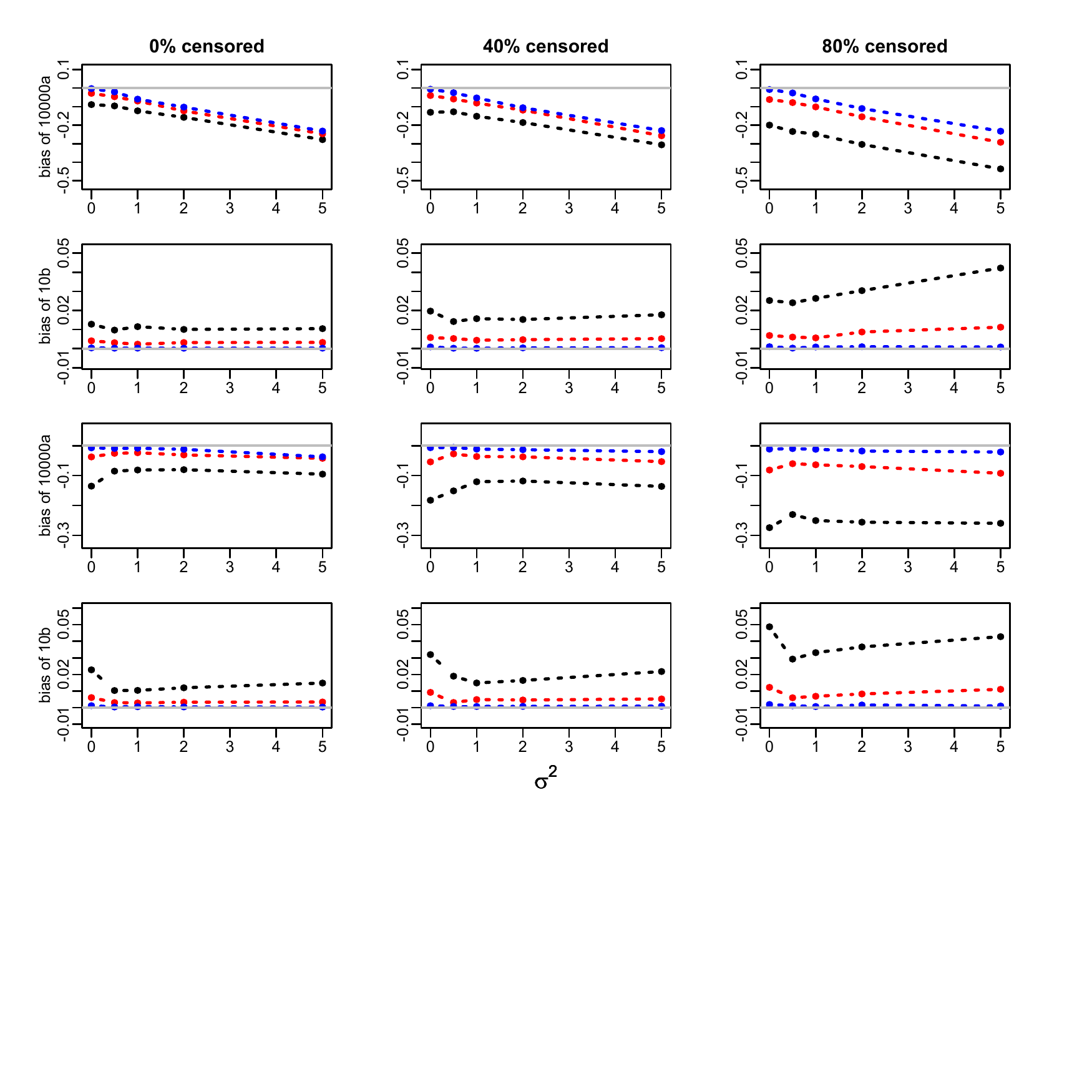}
	\caption{Bias of the estimation of the $a$ and $b$ parameters  in the Gompertz model.
Success proportion= 0.5.  Sample sizes: 300 (black), 1000 (red) and 10000 (blue).
True values: $\beta_{success-scale}$ = 0.5 , $\beta_{success-shape}$= -0.05 , $\beta_{score-scale}$= 1 , $\beta_{score-shape}$= - 0.1.
Top two rows: 10 clusters; bottom two rows: 100 clusters.}
	\label{Bias10_100clustersGompertz6}
\end{figure}

\begin{figure}[ht]
	\centering
	\includegraphics[scale=1]{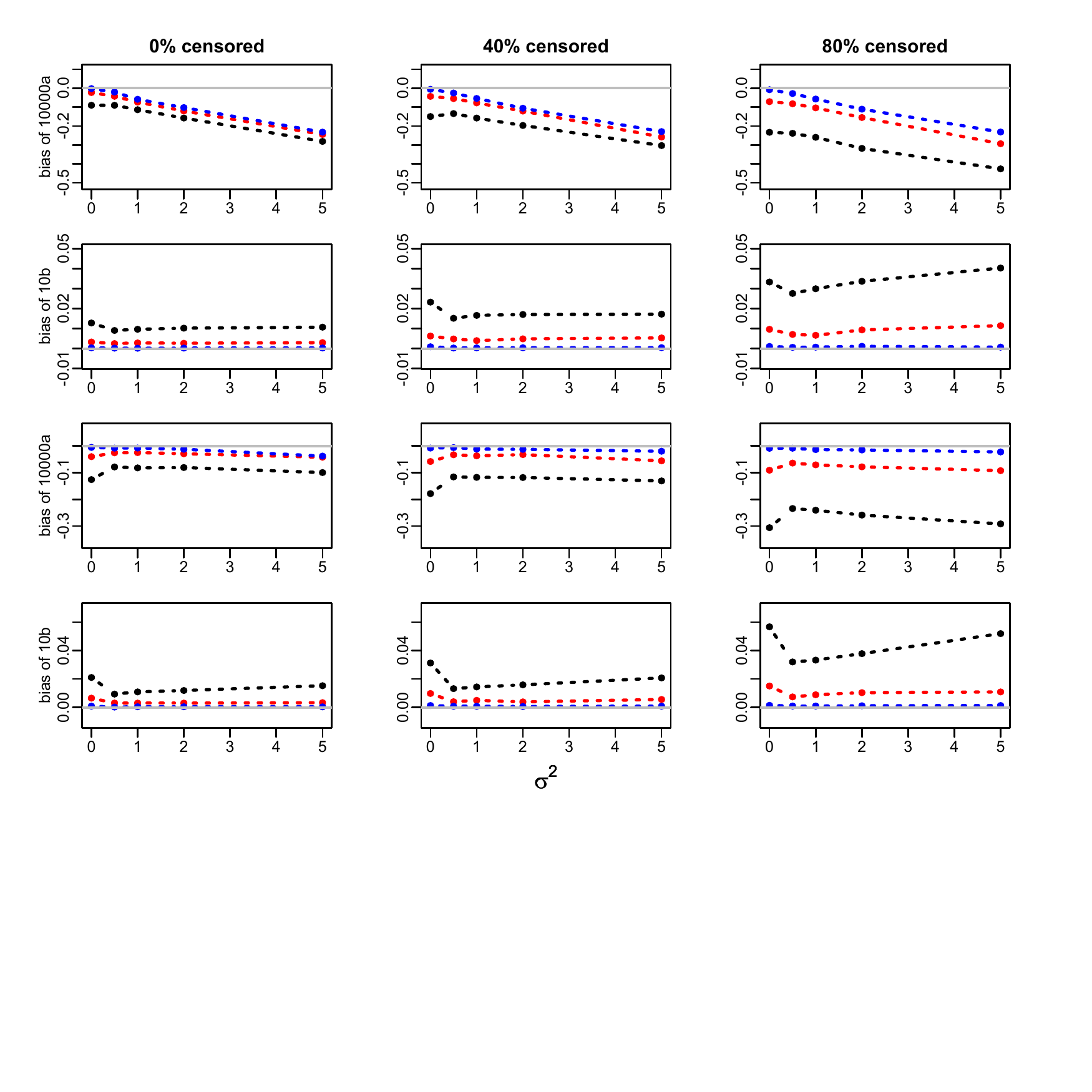}
	\caption{Bias of the estimation of the $a$ and $b$ parameters in the Gompertz model.
Success proportion= 0.5.  Sample sizes: 300 (black), 1000 (red) and 10000 (blue).
True values: $\beta_{success-scale}$ = -0.5 , $\beta_{success-shape}$= 0.05 , $\beta_{score-scale}$= -1 , $\beta_{score-shape}$= 0.1.
Top two rows: 10 clusters; bottom two rows: 100 clusters.}
	\label{Bias10_100clustersGompertz7}
\end{figure}

\begin{figure}[ht]
	\centering
	\includegraphics[scale=1]{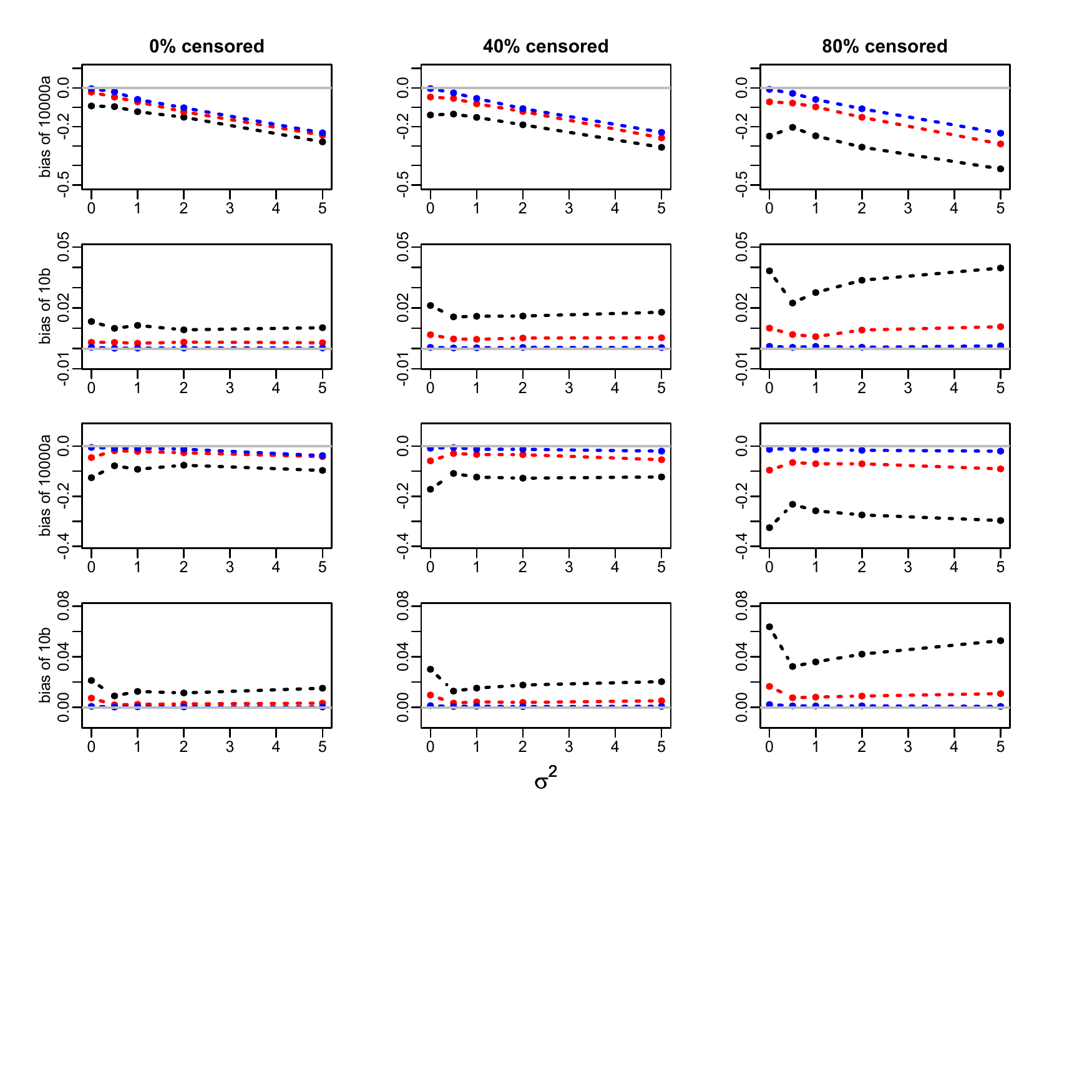}
	\caption{Bias of the estimation of the $a$ and $b$ parameters  in the Gompertz model.
Success proportion= 0.5.  Sample sizes: 300 (black), 1000 (red) and 10000 (blue).
True values: $\beta_{success-scale}$ = -0.5 , $\beta_{success-shape}$= -0.05 , $\beta_{score-scale}$= -1 , $\beta_{score-shape}$= -0.1.
Top two rows: 10 clusters; bottom two rows: 100 clusters.}
	\label{Bias10_100clustersGompertz8}
\end{figure}

\clearpage

\setcounter{figure}{0}
\setcounter{section}{0}
\renewcommand{\thefigure}{B.\arabic{figure}}

\section*{B: Biases in estimation of the Cox regression parameters and of the frailty variance}

Each figure corresponds to a particular  baseline distribution  (Weibull or Gompertz), a value of the probability of success $p_{success}$ (= 0.25 or 0.5), a value for the number of clusters $N_{cl}$ (=10, 100) and a particular choice of the signs of the Cox regression parameters (+ + + +, + - + -, - + - + and - - - -).\\

The absolute values of the Cox regression parameters are held constant at $\beta_{success-scale}$ = 0.5 , $\beta_{success-shape}$= 0.05 , $\beta_{score-scale}$= 1 , $\beta_{score-shape}$= 0.1. 

For each combination of a censoring proportion  (= 0, 40\%, 80\%), a panel plots, versus the frailty variance $\sigma^2$ (= 0, 0.5, 1, 2, 3, 4, 5), the difference between the estimated and the true value of a Cox regression parameter and between the estimated and true value of $\sigma^2$ for three sample sizes (300, 1000 and 10000) . \\

\clearpage

 \begin{figure}[ht]
	\centering
	\includegraphics[scale=1]{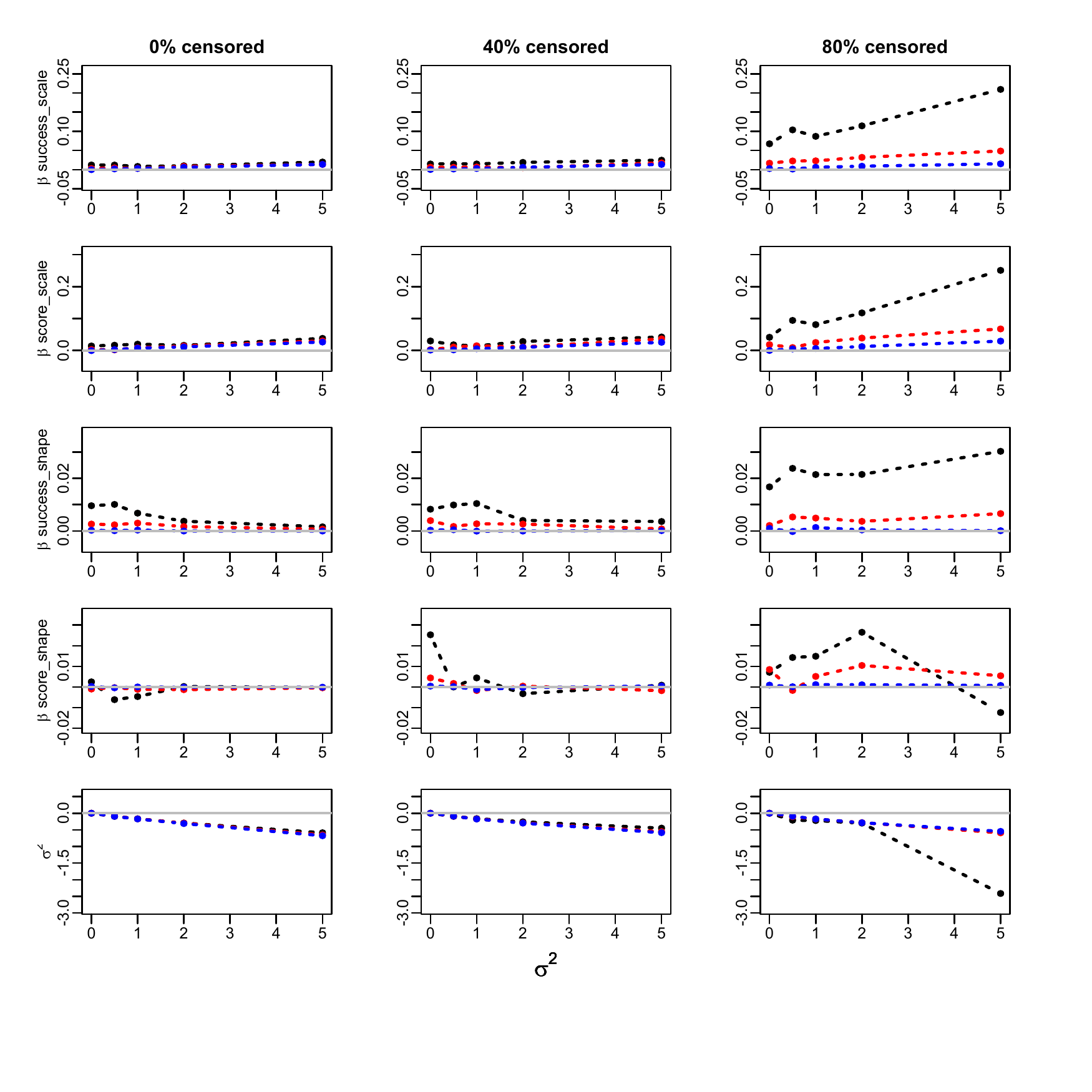}
	\caption{Bias of the estimation of the Cox regression parameters and $\sigma^2$ in the Weibull model.
Success proportion= 0.25. 10 clusters. Sample sizes: 300 (black), 1000 (red) and 10000 (blue).
True values: $\beta_{success-scale}$ = 0.5 , $\beta_{success-shape}$= 0.05 , $\beta_{score-scale}$= 1 , $\beta_{score-shape}$= 0.1}
	\label{Bias10clustersWeibull1}
\end{figure}

\begin{figure}[ht]
	\centering
	\includegraphics[scale=1]{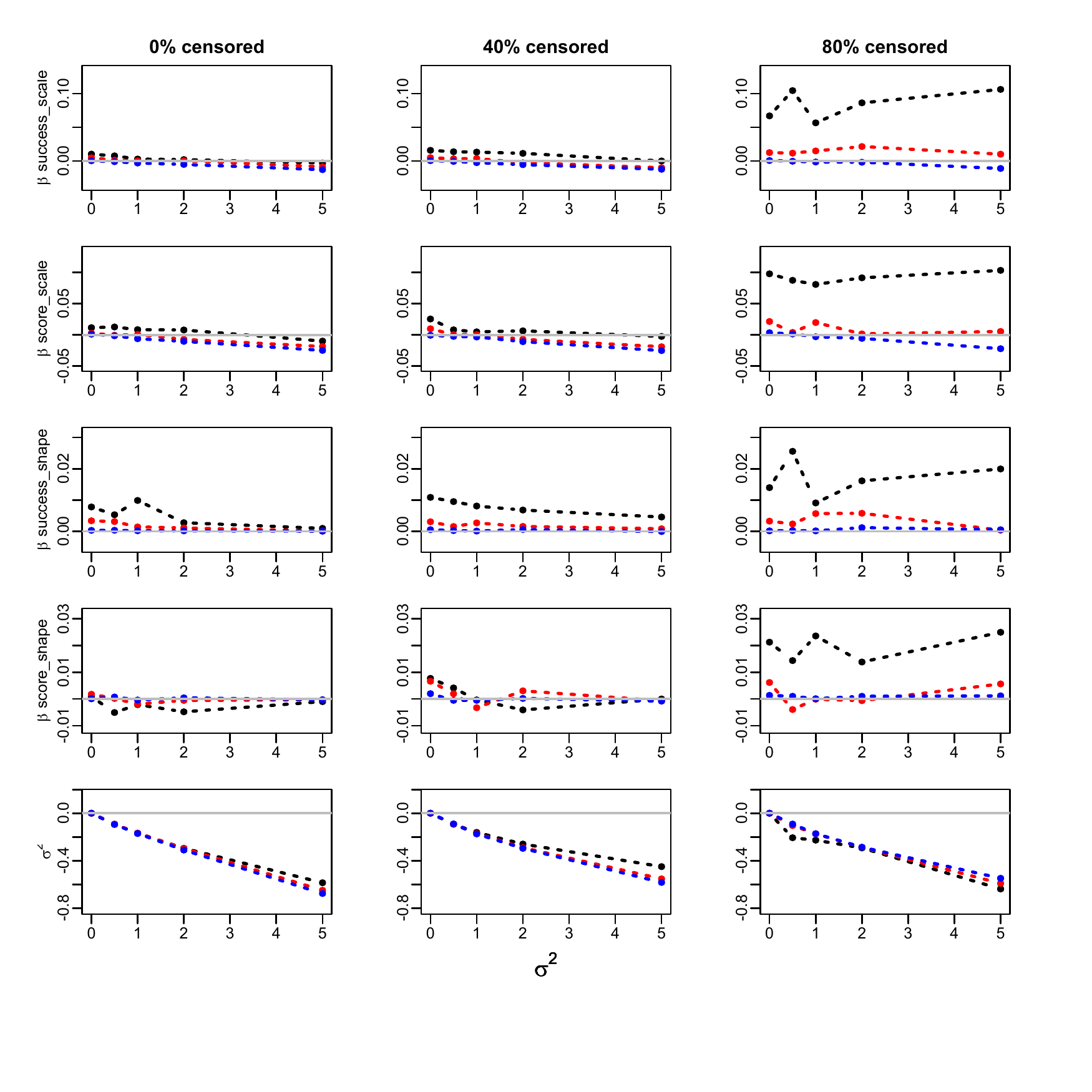}
	\caption{Bias of the estimation of the Cox regression parameters and $\sigma^2$ in the Weibull model.
Success proportion= 0.25. 10 clusters. Sample sizes: 300 (black), 1000 (red) and 10000 (blue).
True values: $\beta_{success-scale}$ = 0.5 , $\beta_{success-shape}$= -0.05 , $\beta_{score-scale}$= 1 , $\beta_{score-shape}$= - 0.1}
	\label{Bias10clustersWeibull2}
\end{figure}

\begin{figure}[ht]
	\centering
	\includegraphics[scale=1]{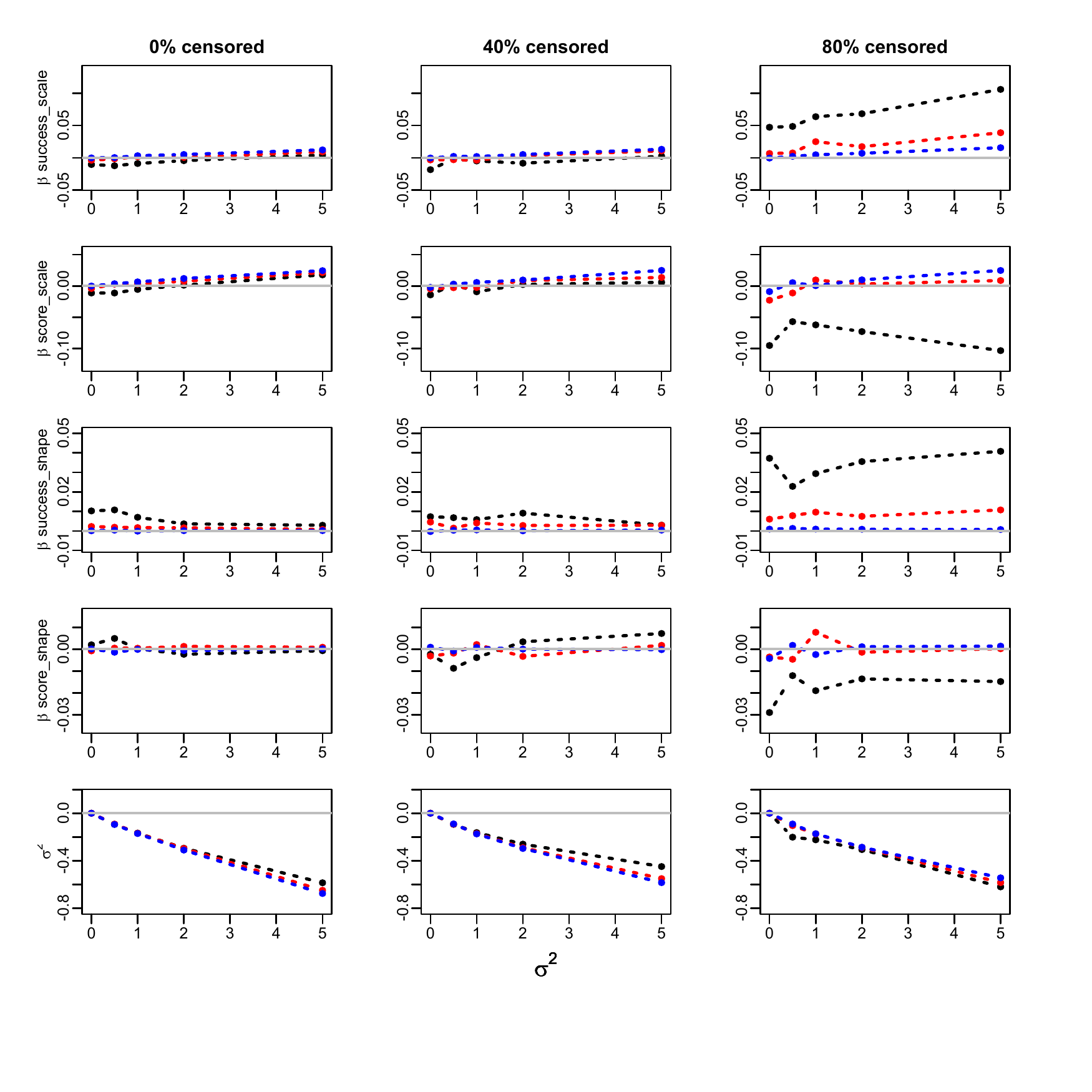}
	\caption{Bias of the estimation of the Cox regression parameters and $\sigma^2$ in the Weibull model.
Success proportion= 0.25. 10 clusters. Sample sizes: 300 (black), 1000 (red) and 10000 (blue).
True values: $\beta_{success-scale}$ = -0.5 , $\beta_{success-shape}$= 0.05 , $\beta_{score-scale}$= -1 , $\beta_{score-shape}$= 0.1}
	\label{Bias10clustersWeibull3}
\end{figure}

\begin{figure}[ht]
	\centering
	\includegraphics[scale=1]{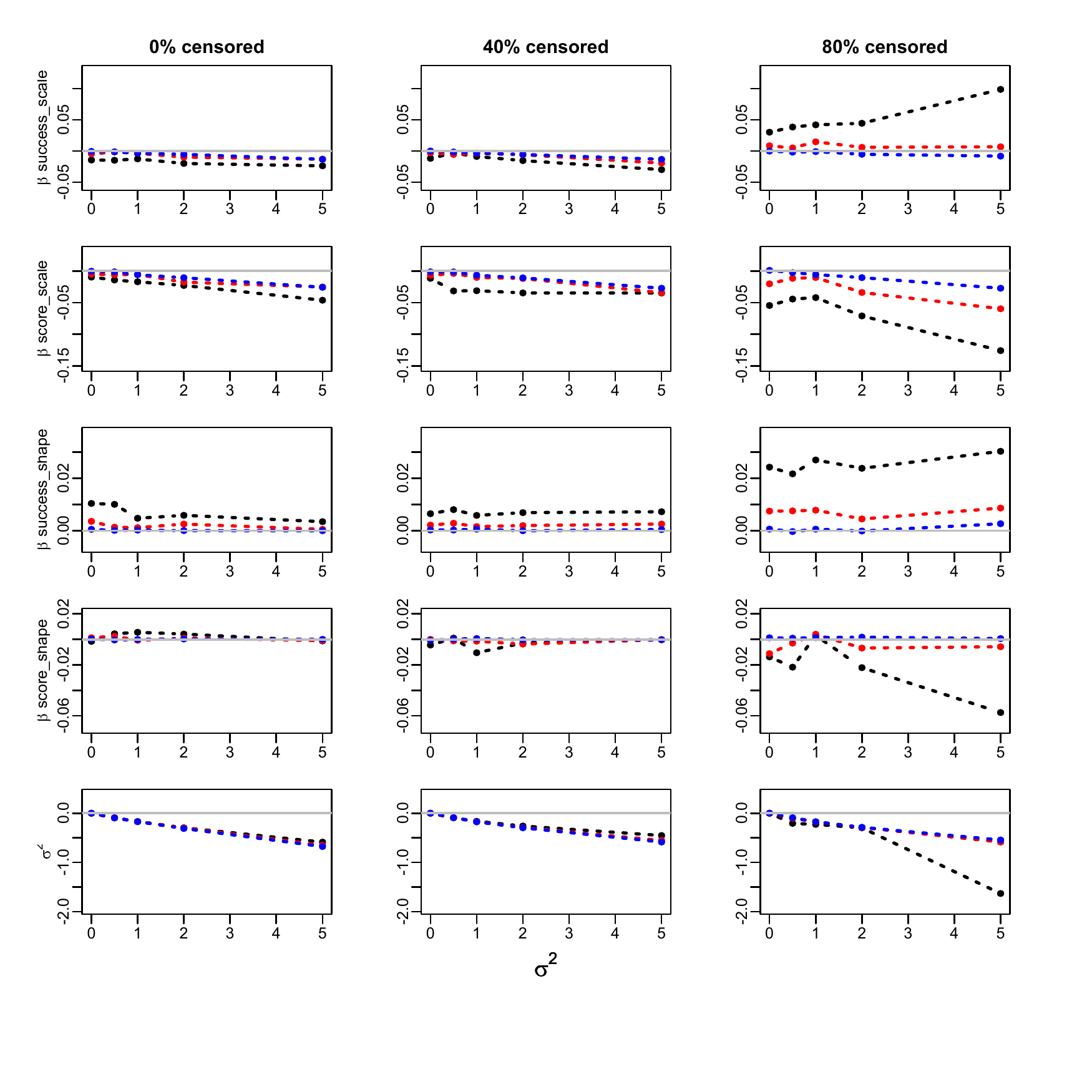}
	\caption{Bias of the estimation of the Cox regression parameters and $\sigma^2$ in the Weibull model.
Success proportion= 0.25. 10 clusters. Sample sizes: 300 (black), 1000 (red) and 10000 (blue).
True values: $\beta_{success-scale}$ = -0.5 , $\beta_{success-shape}$= -0.05 , $\beta_{score-scale}$= -1 , $\beta_{score-shape}$= -0.1}
	\label{Bias10clustersWeibull4}
\end{figure}

\begin{figure}[ht]
	\centering
	\includegraphics[scale=1]{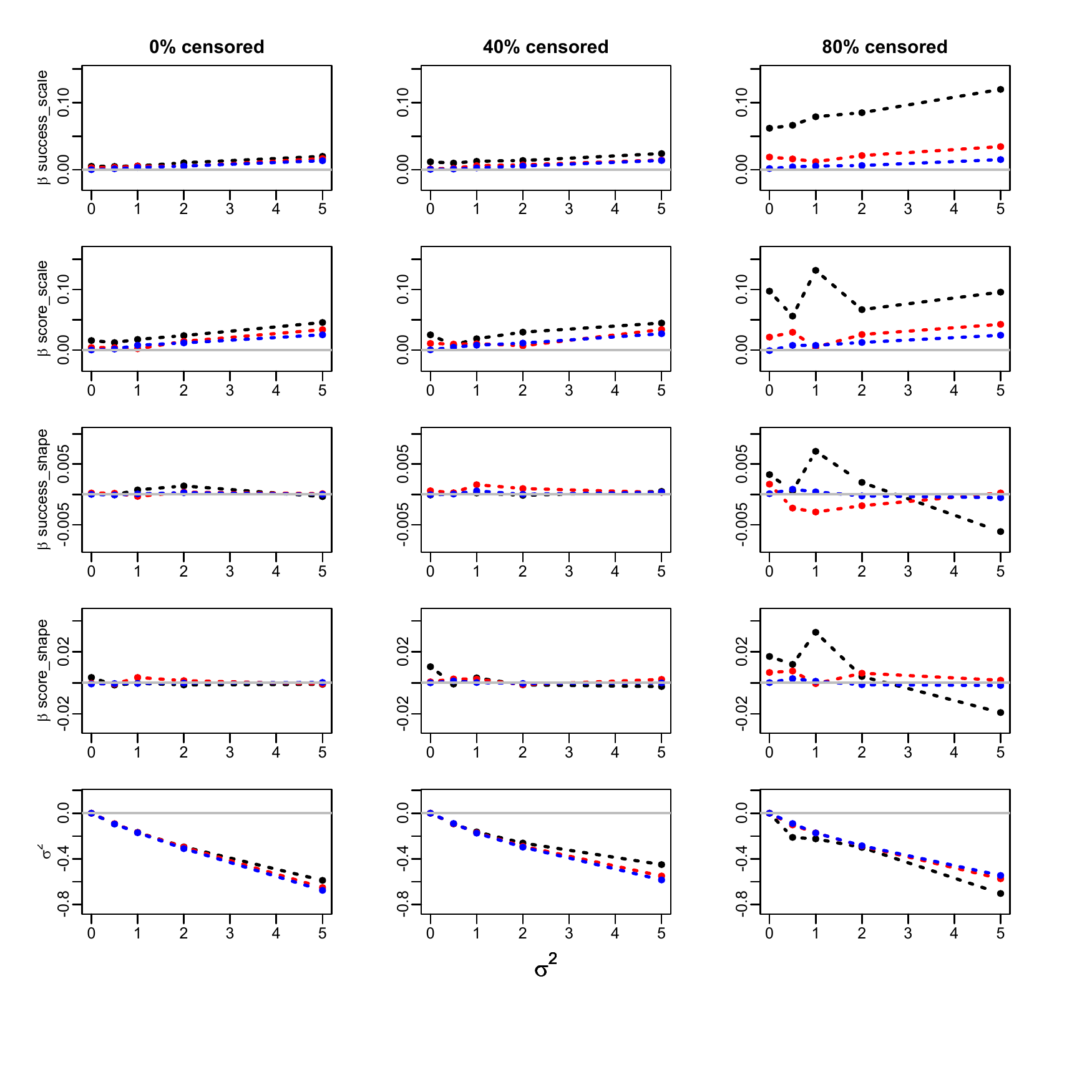}
	\caption{Bias of the estimation of the Cox regression parameters and $\sigma^2$ in the Weibull model.
Success proportion= 0.5. 10 clusters. Sample sizes: 300 (black), 1000 (red) and 10000 (blue).
True values: $\beta_{success-scale}$ = 0.5 , $\beta_{success-shape}$= 0.05 , $\beta_{score-scale}$= 1 , $\beta_{score-shape}$= 0.1}
	\label{Bias10clustersWeibull5}
\end{figure}

\begin{figure}[ht]
	\centering
	\includegraphics[scale=1]{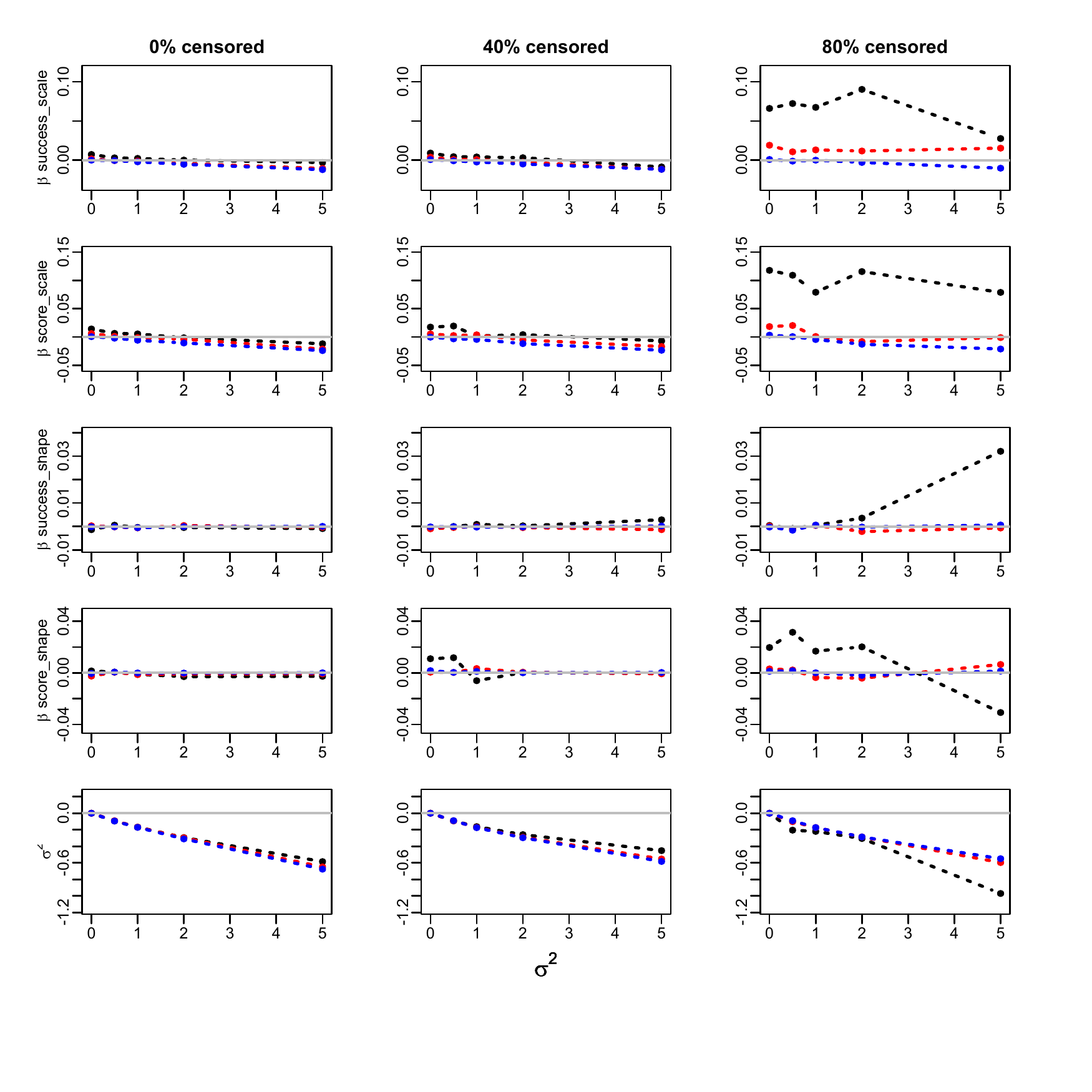}
	\caption{Bias of the estimation of the Cox regression parameters and $\sigma^2$ in the Weibull model.
Success proportion= 0.5. 10 clusters. Sample sizes: 300 (black), 1000 (red) and 10000 (blue).
True values: $\beta_{success-scale}$ = 0.5 , $\beta_{success-shape}$= -0.05 , $\beta_{score-scale}$= 1 , $\beta_{score-shape}$= - 0.1}
	\label{Bias10clustersWeibull6}
\end{figure}

\begin{figure}[ht]
	\centering
	\includegraphics[scale=1]{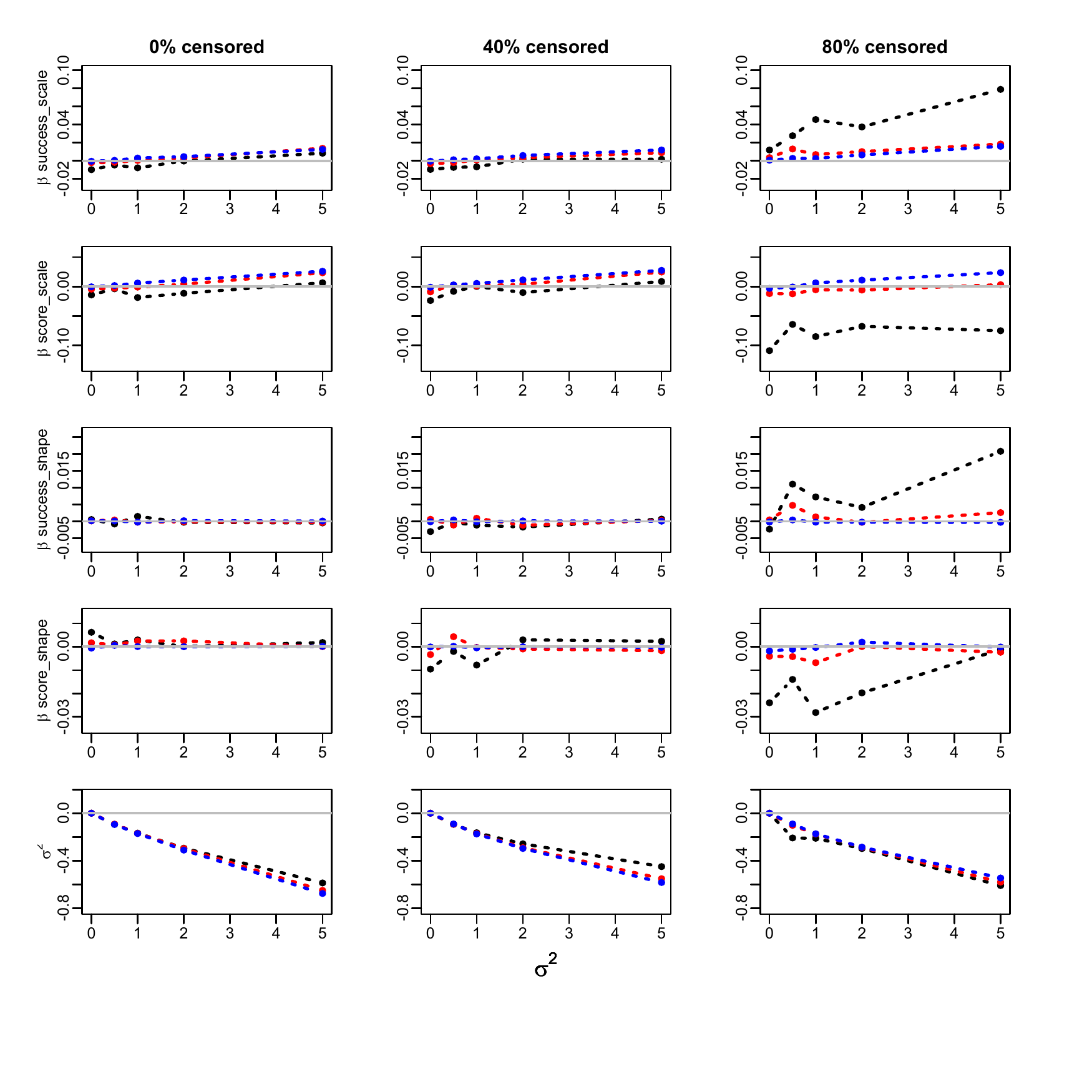}
	\caption{Bias of the estimation of the Cox regression parameters and $\sigma^2$ in the Weibull model.
Success proportion= 0.5. 10 clusters. Sample sizes: 300 (black), 1000 (red) and 10000 (blue).
True values: $\beta_{success-scale}$ = -0.5 , $\beta_{success-shape}$= 0.05 , $\beta_{score-scale}$= -1 , $\beta_{score-shape}$= 0.1}
	\label{Bias10clustersWeibull7}
\end{figure}

\begin{figure}[ht]
	\centering
	\includegraphics[scale=1]{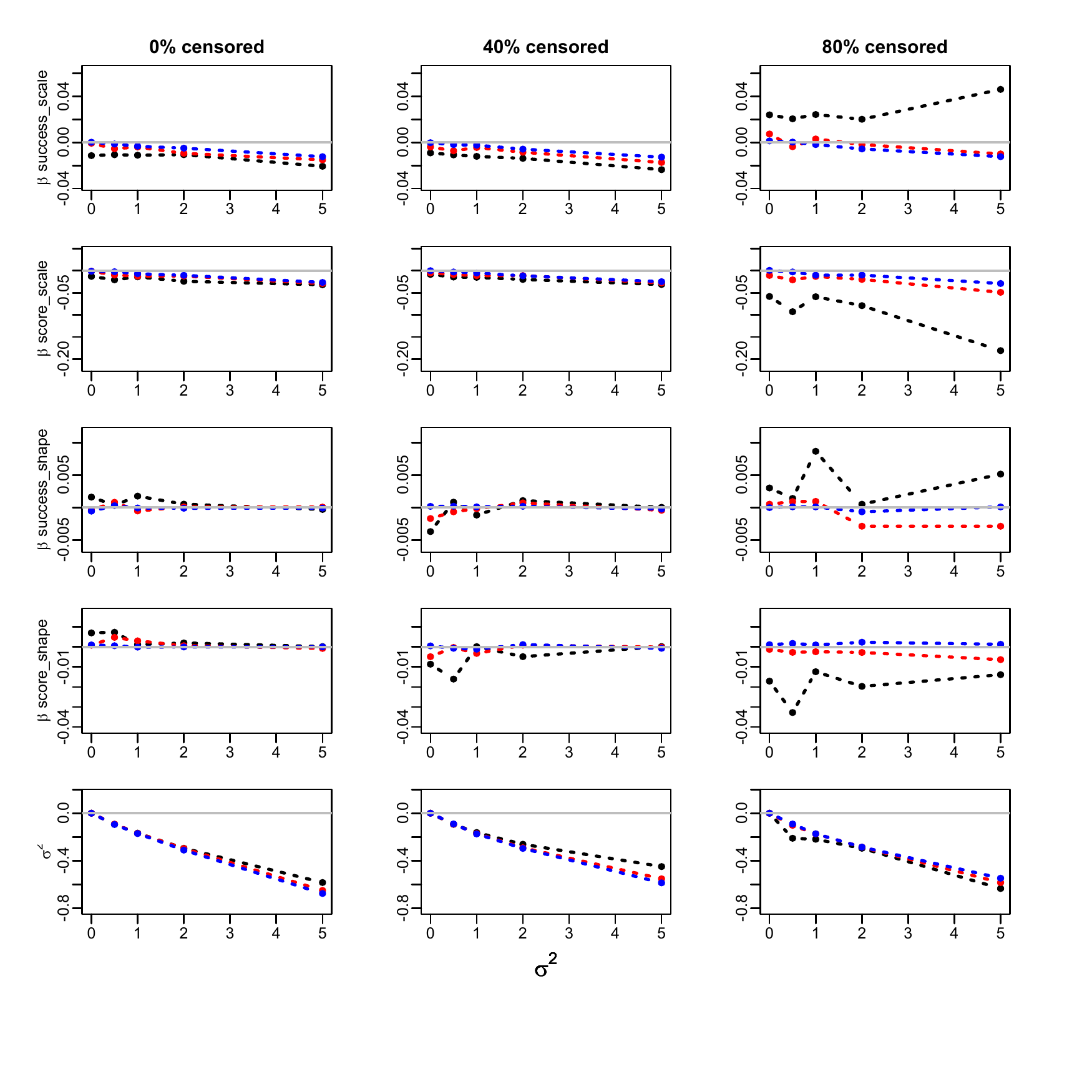}
	\caption{Bias of the estimation of the Cox regression parameters and $\sigma^2$ in the Weibull model.
Success proportion= 0.5. 10 clusters. Sample sizes: 300 (black), 1000 (red) and 10000 (blue).
True values: $\beta_{success-scale}$ = -0.5 , $\beta_{success-shape}$= -0.05 , $\beta_{score-scale}$= -1 , $\beta_{score-shape}$= -0.1}
	\label{Bias10clustersWeibull8}
\end{figure}

 \begin{figure}[ht]
	\centering
	\includegraphics[scale=1]{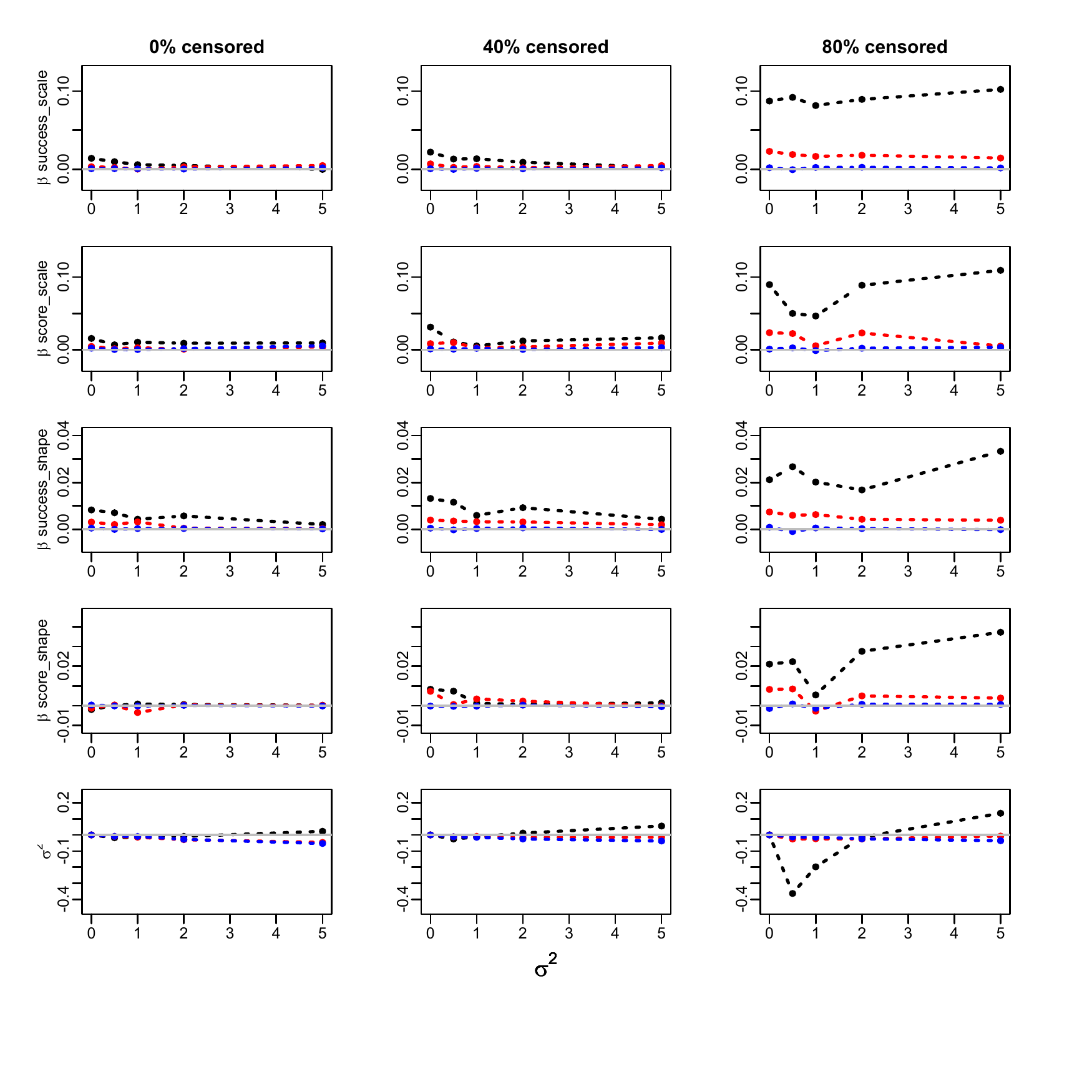}
	\caption{Bias of the estimation of the Cox regression parameters and $\sigma^2$ in the Weibull model.
Success proportion= 0.25. 100 clusters. Sample sizes: 300 (black), 1000 (red) and 10000 (blue).
True values: $\beta_{success-scale}$ = 0.5 , $\beta_{success-shape}$= 0.05 , $\beta_{score-scale}$= 1 , $\beta_{score-shape}$= 0.1}
	\label{Bias100clustersWeibull1}
\end{figure}

\begin{figure}[ht]
	\centering
	\includegraphics[scale=1]{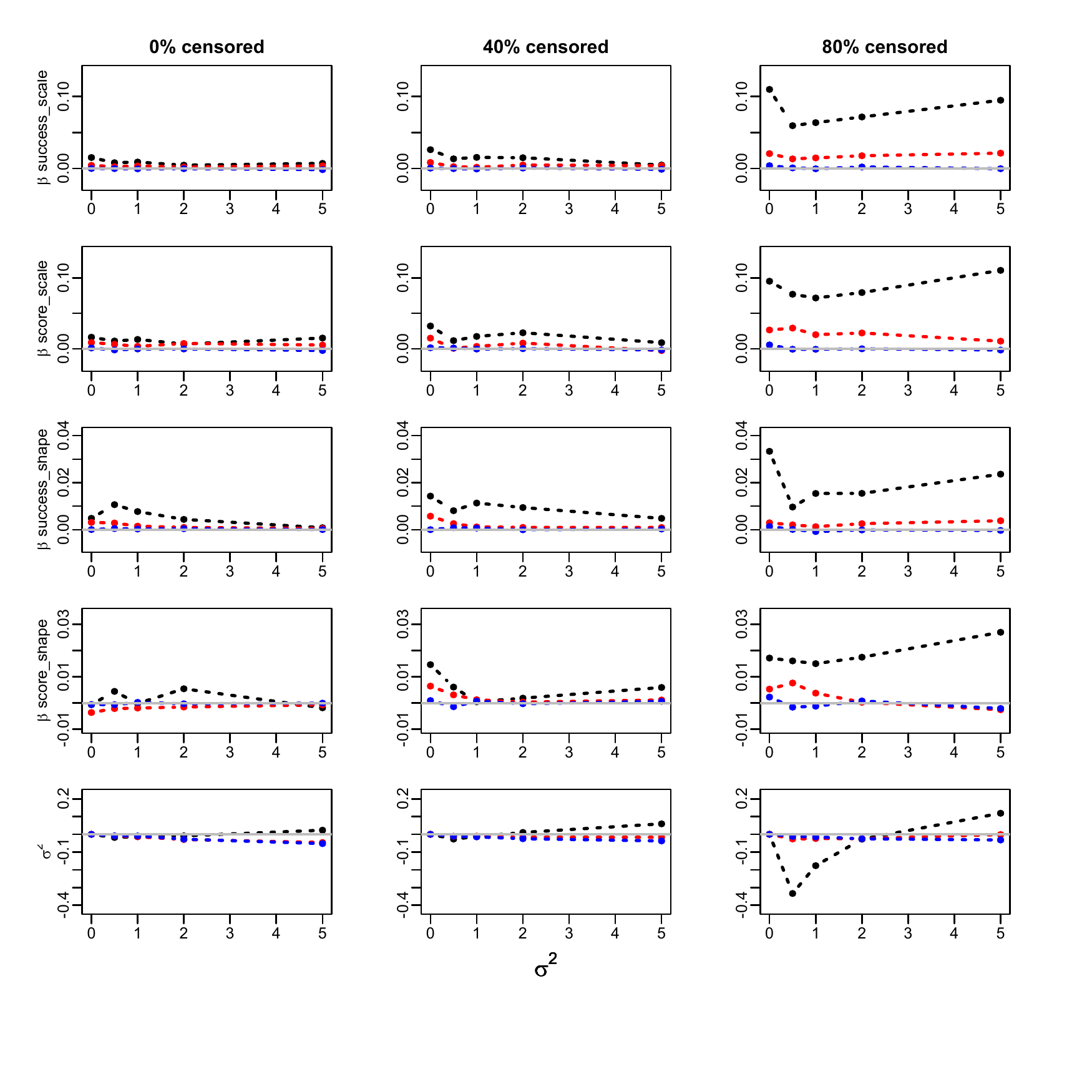}
	\caption{Bias of the estimation of the Cox regression parameters and $\sigma^2$ in the Weibull model.
Success proportion= 0.25. 100 clusters. Sample sizes: 300 (black), 1000 (red) and 10000 (blue).
True values: $\beta_{success-scale}$ = 0.5 , $\beta_{success-shape}$= -0.05 , $\beta_{score-scale}$= 1 , $\beta_{score-shape}$= - 0.1}
	\label{Bias100clustersWeibull2}
\end{figure}

\begin{figure}[ht]
	\centering
	\includegraphics[scale=1]{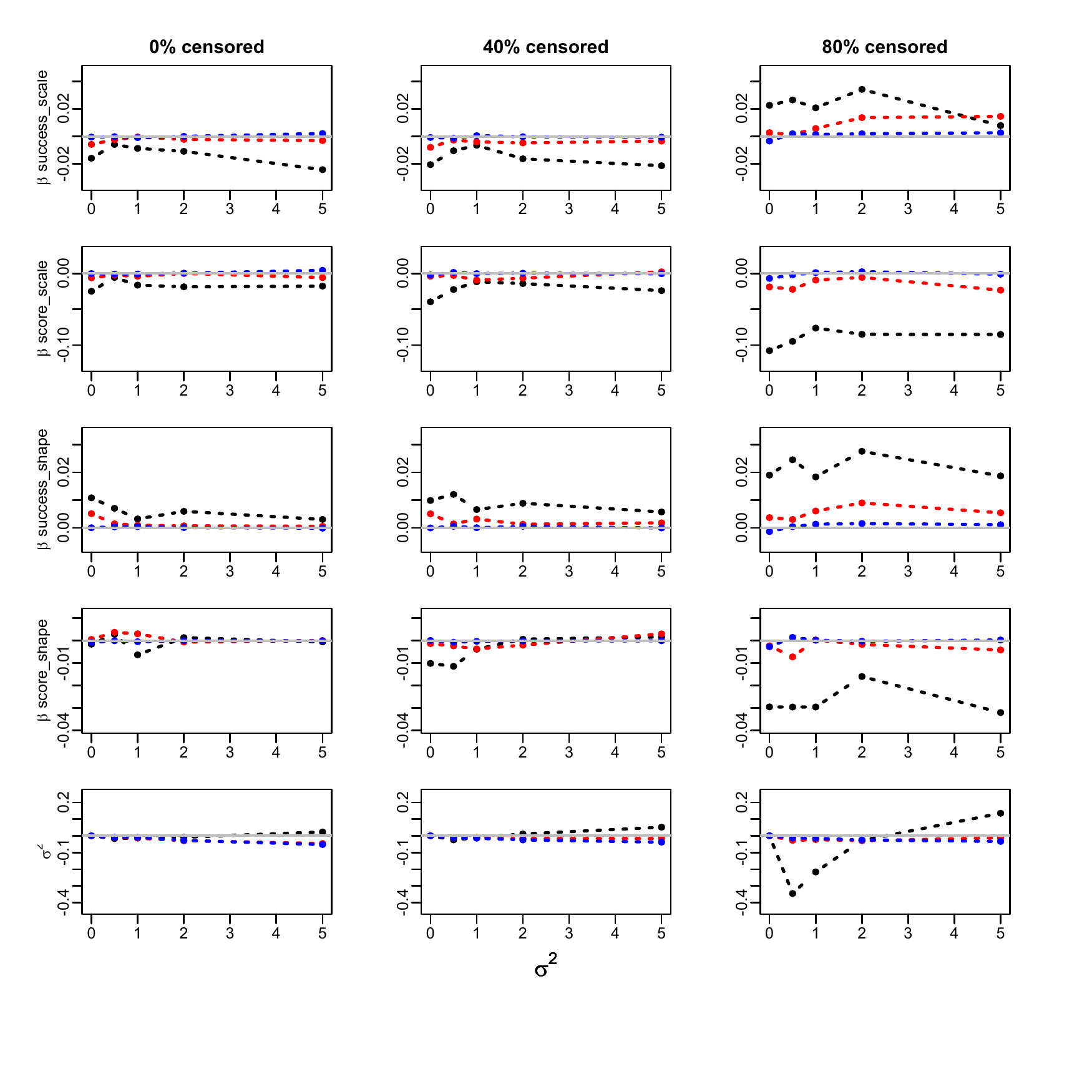}
	\caption{Bias of the estimation of the Cox regression parameters and $\sigma^2$ in the Weibull model.
Success proportion= 0.25. 100 clusters. Sample sizes: 300 (black), 1000 (red) and 10000 (blue).
True values: $\beta_{success-scale}$ = -0.5 , $\beta_{success-shape}$= 0.05 , $\beta_{score-scale}$= -1 , $\beta_{score-shape}$= 0.1}
	\label{Bias100clustersWeibull3}
\end{figure}

\begin{figure}[ht]
	\centering
	\includegraphics[scale=1]{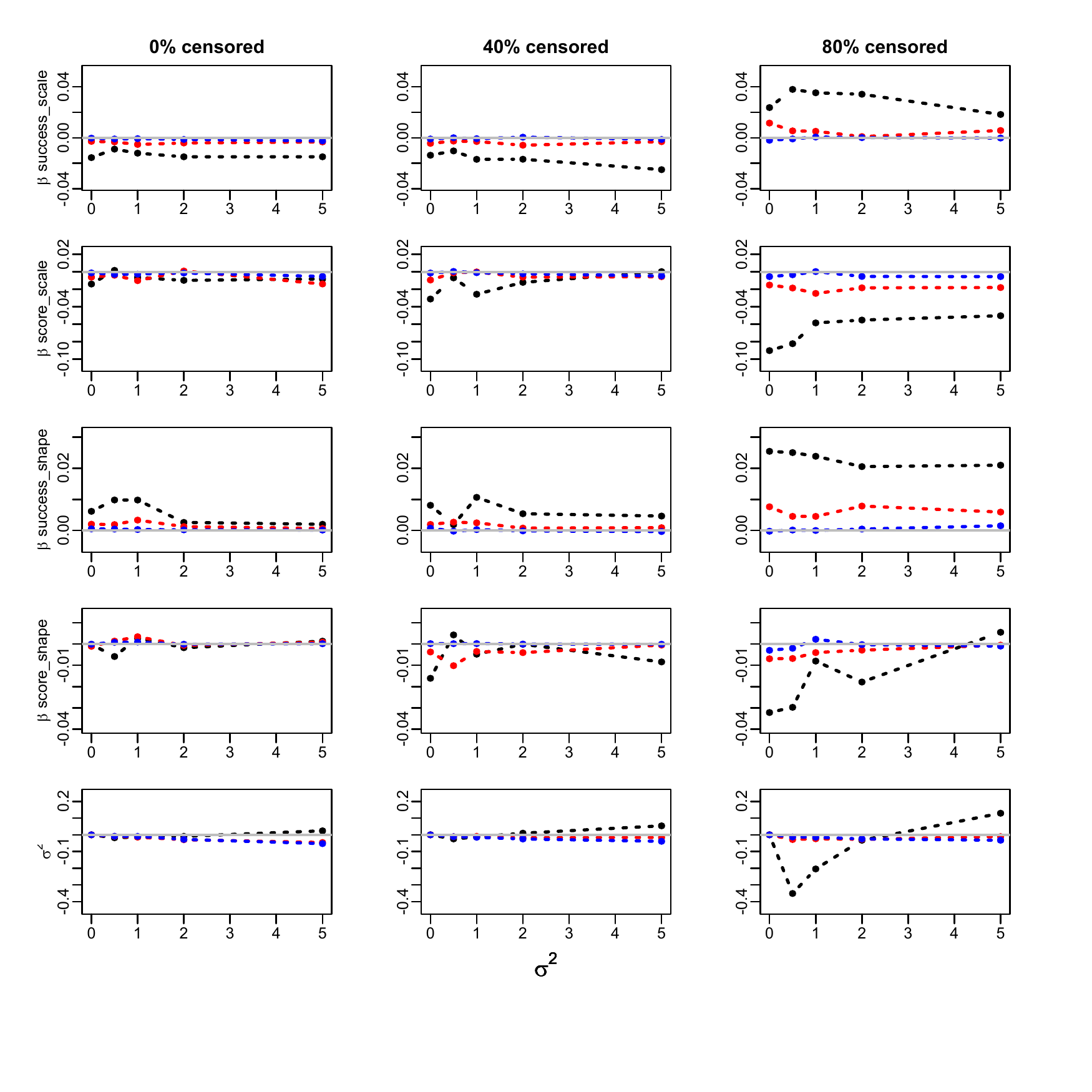}
	\caption{Bias of the estimation of the Cox regression parameters and $\sigma^2$ in the Weibull model.
Success proportion= 0.25. 100 clusters. Sample sizes: 300 (black), 1000 (red) and 10000 (blue).
True values: $\beta_{success-scale}$ = -0.5 , $\beta_{success-shape}$= -0.05 , $\beta_{score-scale}$= -1 , $\beta_{score-shape}$= -0.1}
	\label{Bias100clustersWeibull4}
\end{figure}

\begin{figure}[ht]
	\centering
	\includegraphics[scale=1]{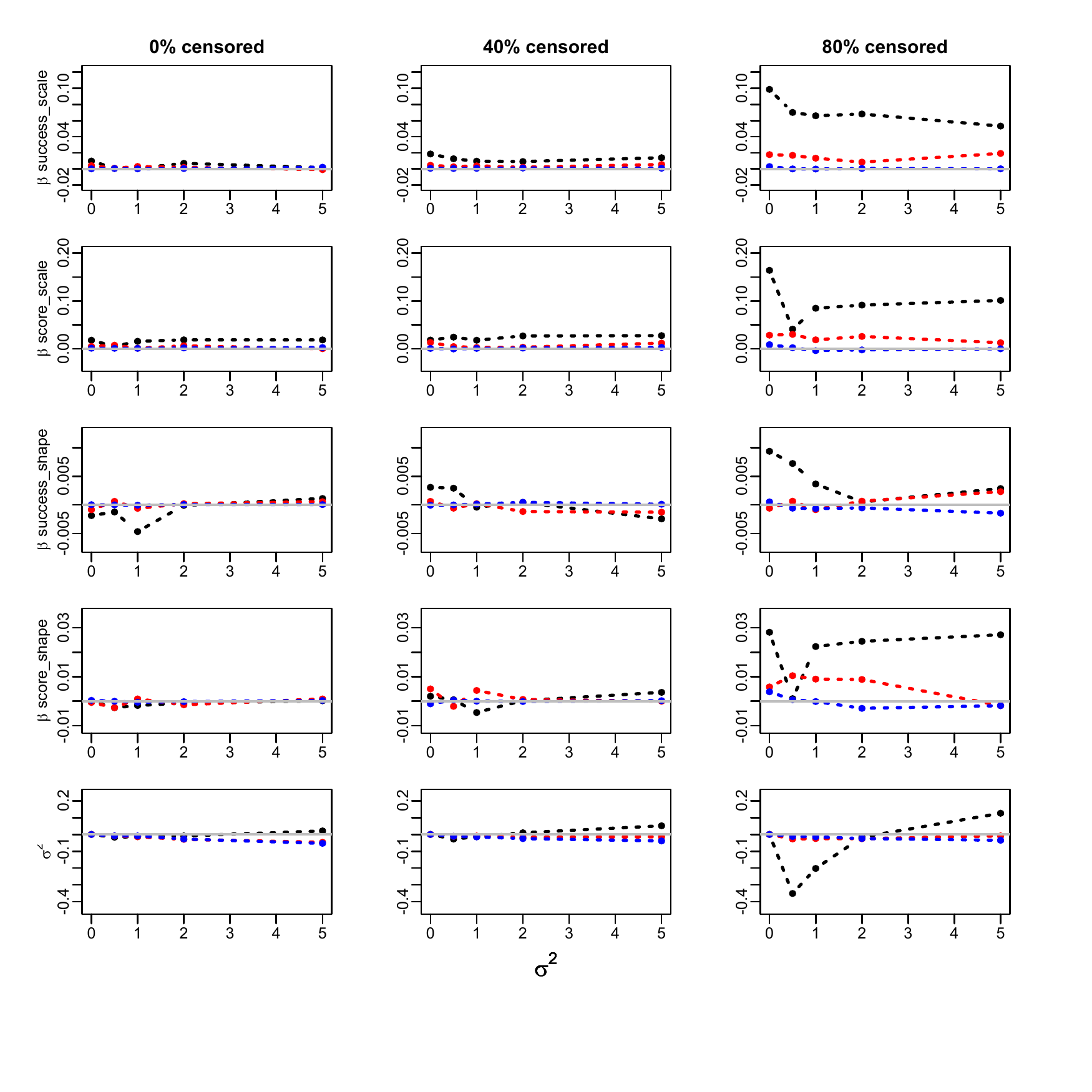}
	\caption{Bias of the estimation of the Cox regression parameters and $\sigma^2$ in the Weibull model.
Success proportion= 0.5. 100 clusters. Sample sizes: 300 (black), 1000 (red) and 10000 (blue).
True values: $\beta_{success-scale}$ = 0.5 , $\beta_{success-shape}$= 0.05 , $\beta_{score-scale}$= 1 , $\beta_{score-shape}$= 0.1}
	\label{Bias100clustersWeibull5}
\end{figure}

\begin{figure}[ht]
	\centering
	\includegraphics[scale=1]{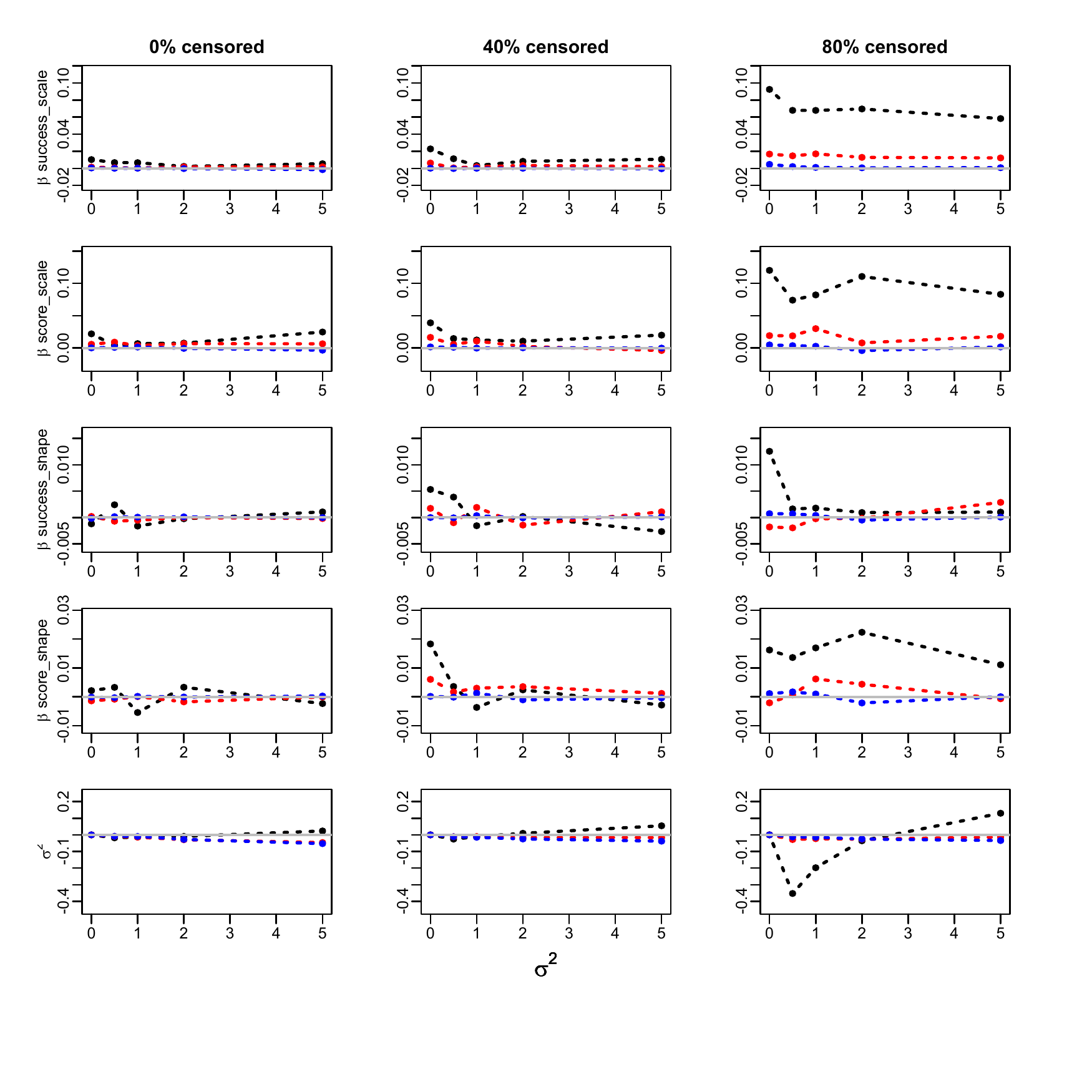}
	\caption{Bias of the estimation of the Cox regression parameters and $\sigma^2$ in the Weibull model.
Success proportion= 0.5. 100 clusters. Sample sizes: 300 (black), 1000 (red) and 10000 (blue).
True values: $\beta_{success-scale}$ = 0.5 , $\beta_{success-shape}$= -0.05 , $\beta_{score-scale}$= 1 , $\beta_{score-shape}$= - 0.1}
	\label{Bias100clustersWeibull6}
\end{figure}

\begin{figure}[ht]
	\centering
	\includegraphics[scale=1]{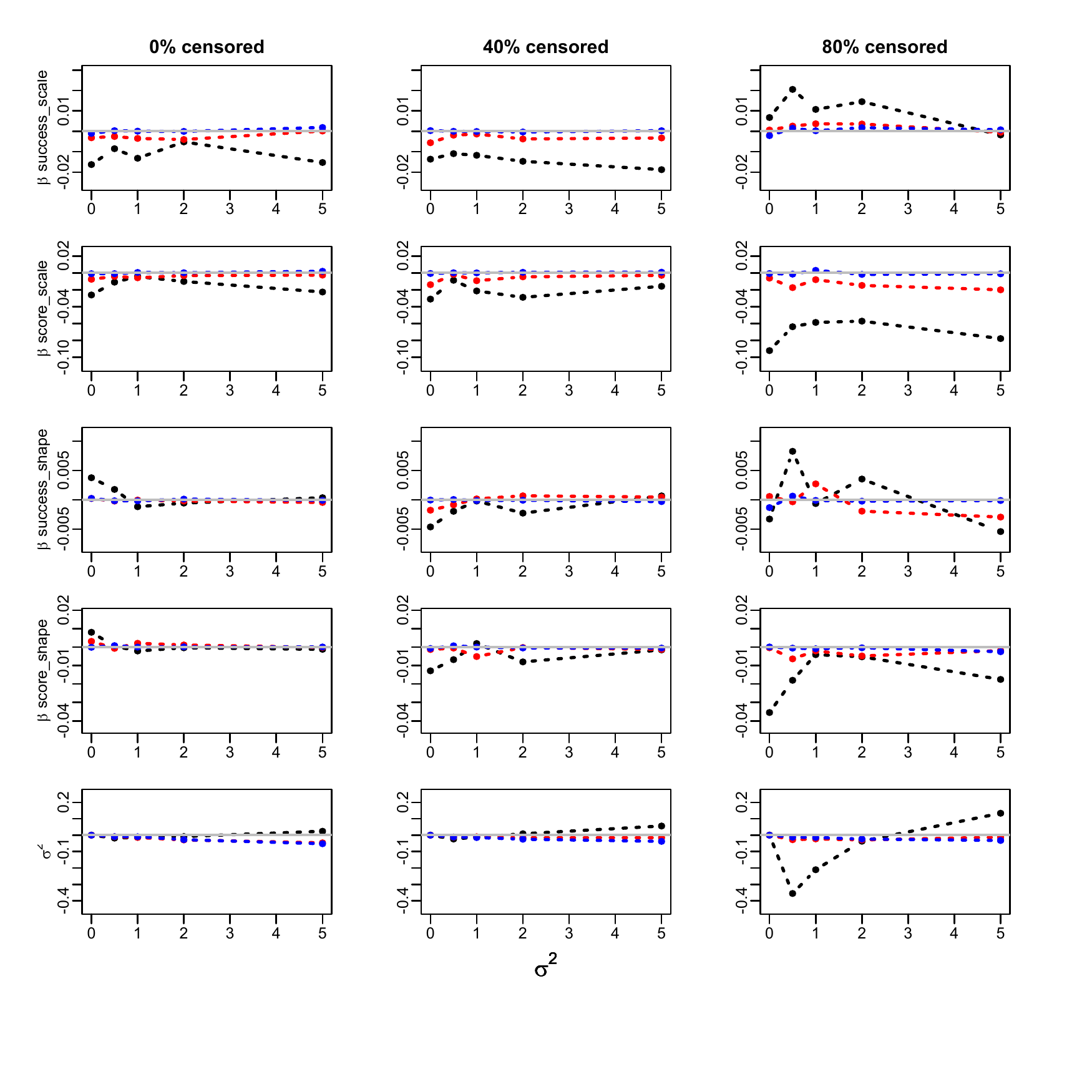}
	\caption{Bias of the estimation of the Cox regression parameters and $\sigma^2$ in the Weibull model.
Success proportion= 0.5. 100 clusters. Sample sizes: 300 (black), 1000 (red) and 10000 (blue).
True values: $\beta_{success-scale}$ = -0.5 , $\beta_{success-shape}$= 0.05 , $\beta_{score-scale}$= -1 , $\beta_{score-shape}$= 0.1}
	\label{Bias100clustersWeibull7}
\end{figure}

\begin{figure}[ht]
	\centering
	\includegraphics[scale=1]{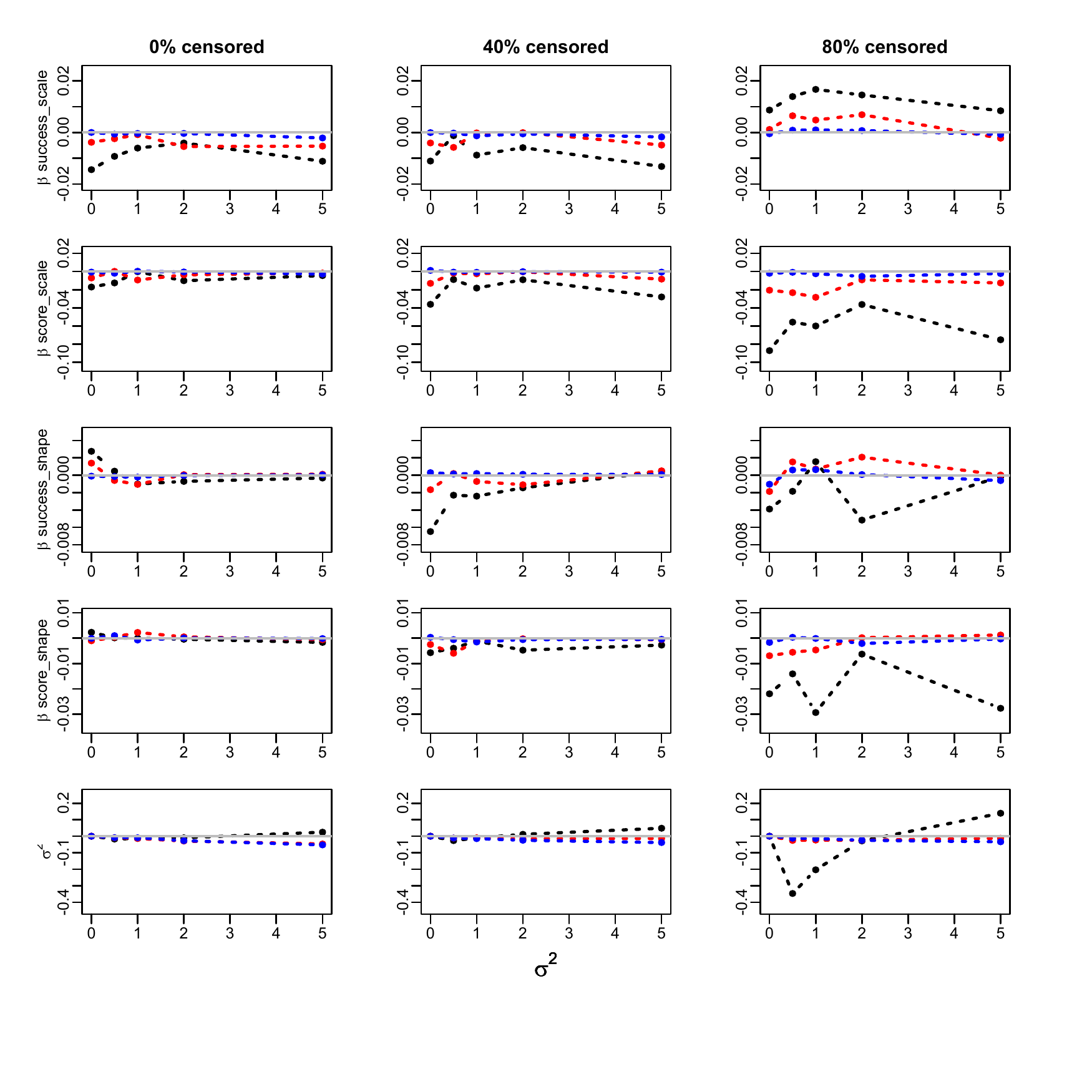}
	\caption{Bias of the estimation of the Cox regression parameters and $\sigma^2$ in the Weibull model.
Success proportion= 0.5. 100 clusters. Sample sizes: 300 (black), 1000 (red) and 10000 (blue).
True values: $\beta_{success-scale}$ = -0.5 , $\beta_{success-shape}$= -0.05 , $\beta_{score-scale}$= -1 , $\beta_{score-shape}$= -0.1}
	\label{Bias100clustersWeibull8}
\end{figure}


 \begin{figure}[ht]
	\centering
	\includegraphics[scale=1]{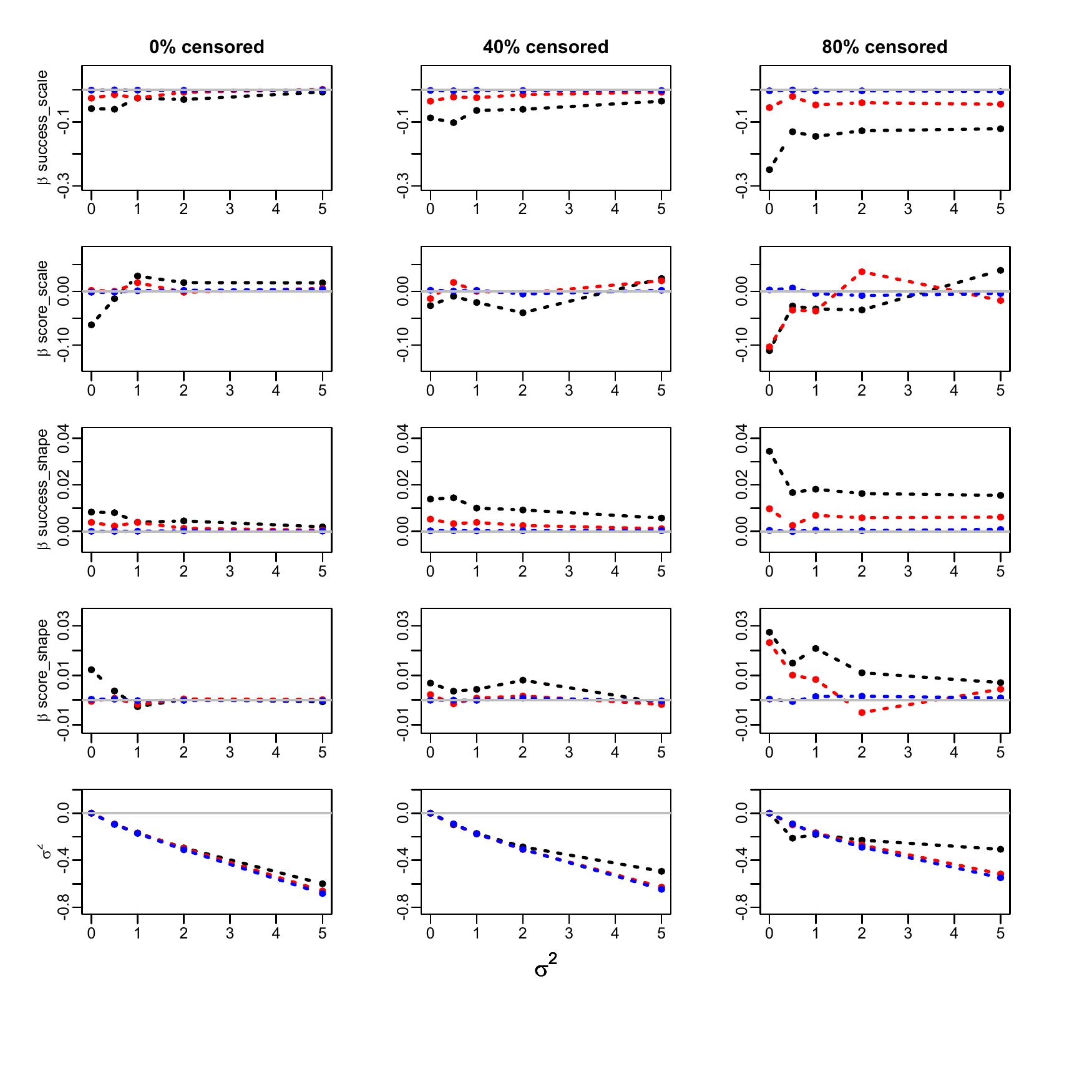}
	\caption{Bias of the estimation of the Cox regression parameters and $\sigma^2$ in the Gompertz model.
Success proportion= 0.25. 10 clusters. Sample sizes: 300 (black), 1000 (red) and 10000 (blue).
True values: $\beta_{success-scale}$ = 0.5 , $\beta_{success-shape}$= 0.05 , $\beta_{score-scale}$= 1 , $\beta_{score-shape}$= 0.1}
	\label{Bias10clustersGompertz1}
\end{figure}

\begin{figure}[ht]
	\centering
	\includegraphics[scale=1]{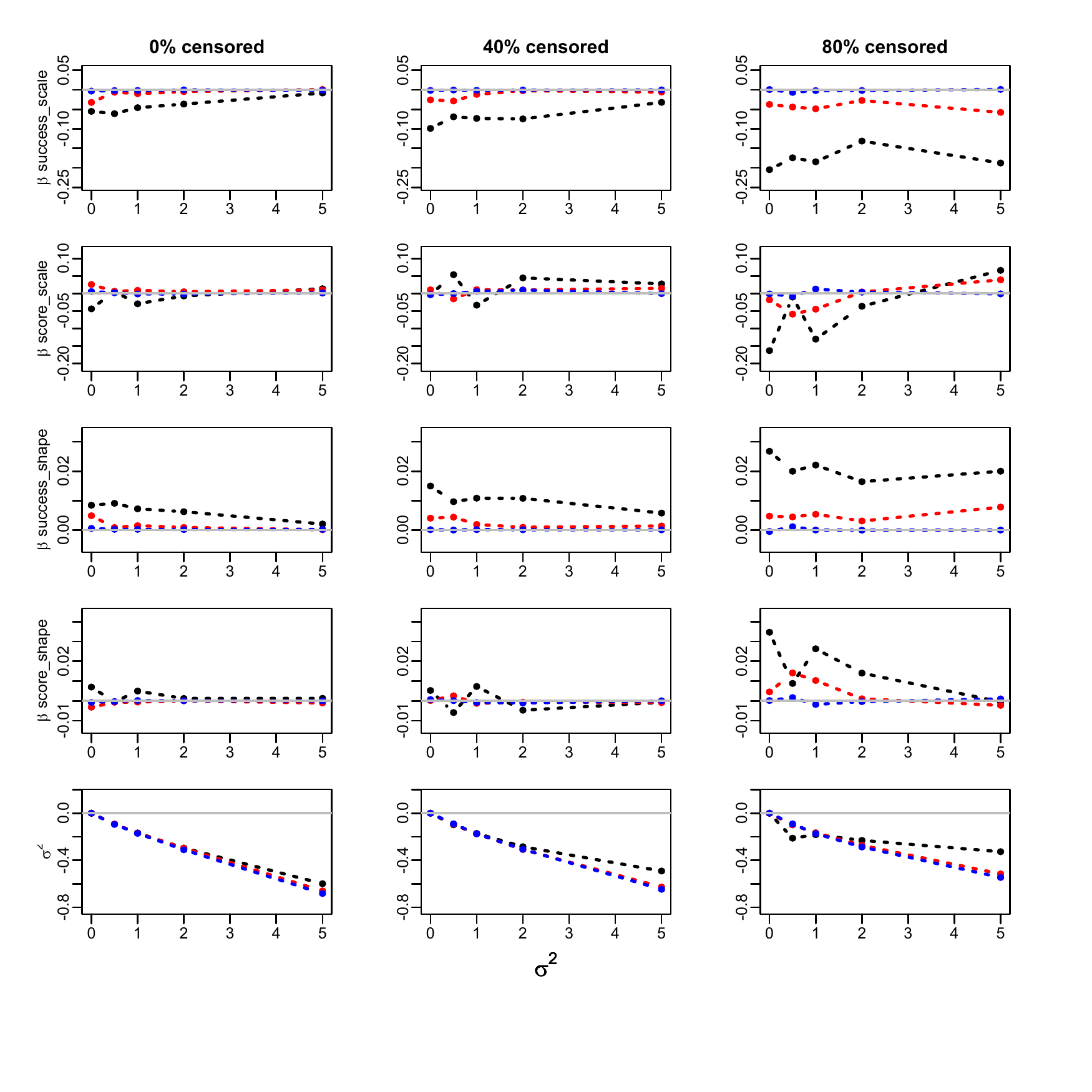}
	\caption{Bias of the estimation of the Cox regression parameters and $\sigma^2$ in the Gompertz model.
Success proportion= 0.25. 10 clusters. Sample sizes: 300 (black), 1000 (red) and 10000 (blue).
True values: $\beta_{success-scale}$ = 0.5 , $\beta_{success-shape}$= -0.05 , $\beta_{score-scale}$= 1 , $\beta_{score-shape}$= - 0.1}
	\label{Bias10clustersGompertz2}
\end{figure}

\begin{figure}[ht]
	\centering
	\includegraphics[scale=1]{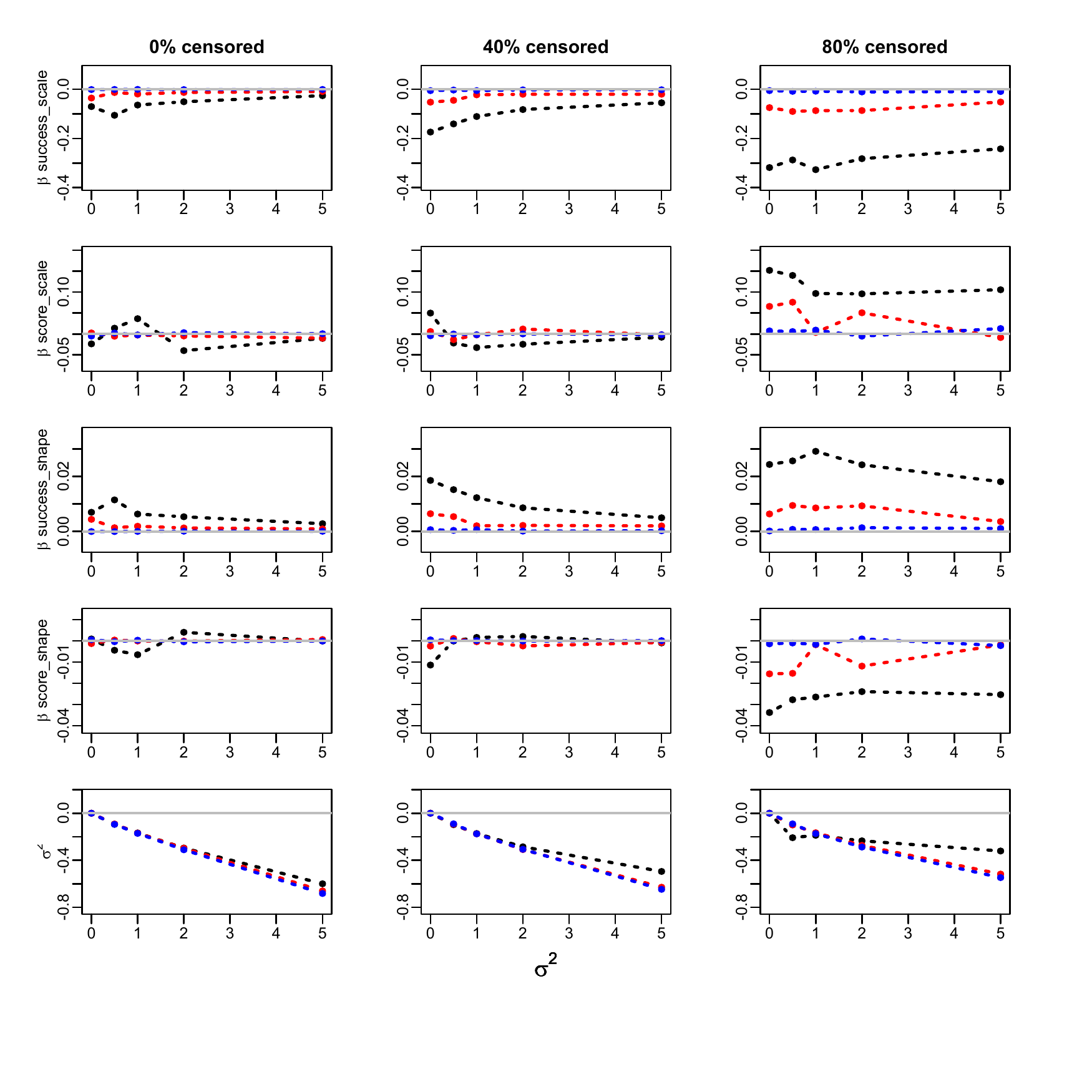}
	\caption{Bias of the estimation of the Cox regression parameters and $\sigma^2$ in the Gompertz model.
Success proportion= 0.25. 10 clusters. Sample sizes: 300 (black), 1000 (red) and 10000 (blue).
True values: $\beta_{success-scale}$ = -0.5 , $\beta_{success-shape}$= 0.05 , $\beta_{score-scale}$= -1 , $\beta_{score-shape}$= 0.1}
	\label{Bias10clustersGompertz3}
\end{figure}

\begin{figure}[ht]
	\centering
	\includegraphics[scale=1]{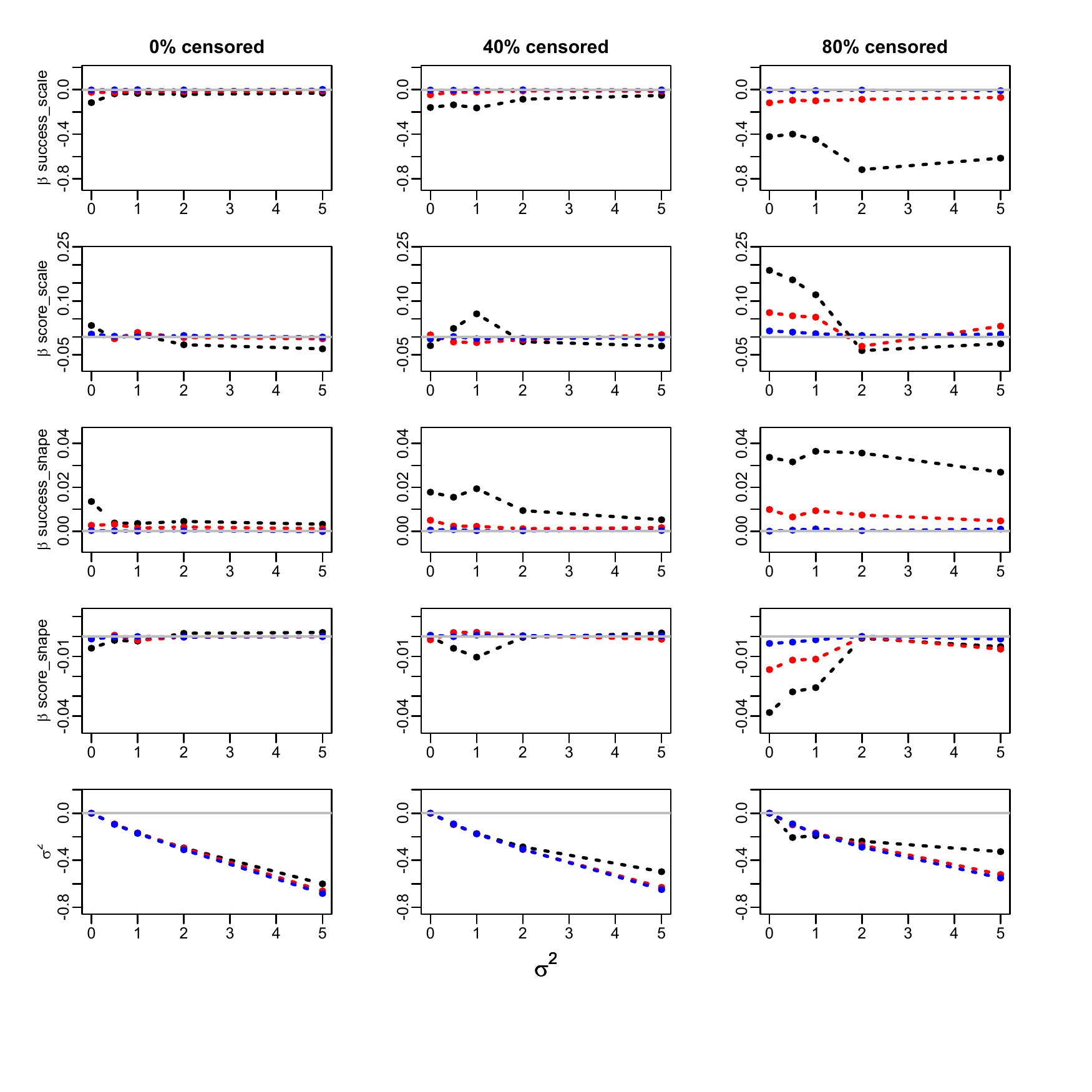}
	\caption{Bias of the estimation of the Cox regression parameters and $\sigma^2$ in the Gompertz model.
Success proportion= 0.25. 10 clusters. Sample sizes: 300 (black), 1000 (red) and 10000 (blue).
True values: $\beta_{success-scale}$ = -0.5 , $\beta_{success-shape}$= -0.05 , $\beta_{score-scale}$= -1 , $\beta_{score-shape}$= -0.1}
	\label{Bias10clustersGompertz4}
\end{figure}

\begin{figure}[ht]
	\centering
	\includegraphics[scale=1]{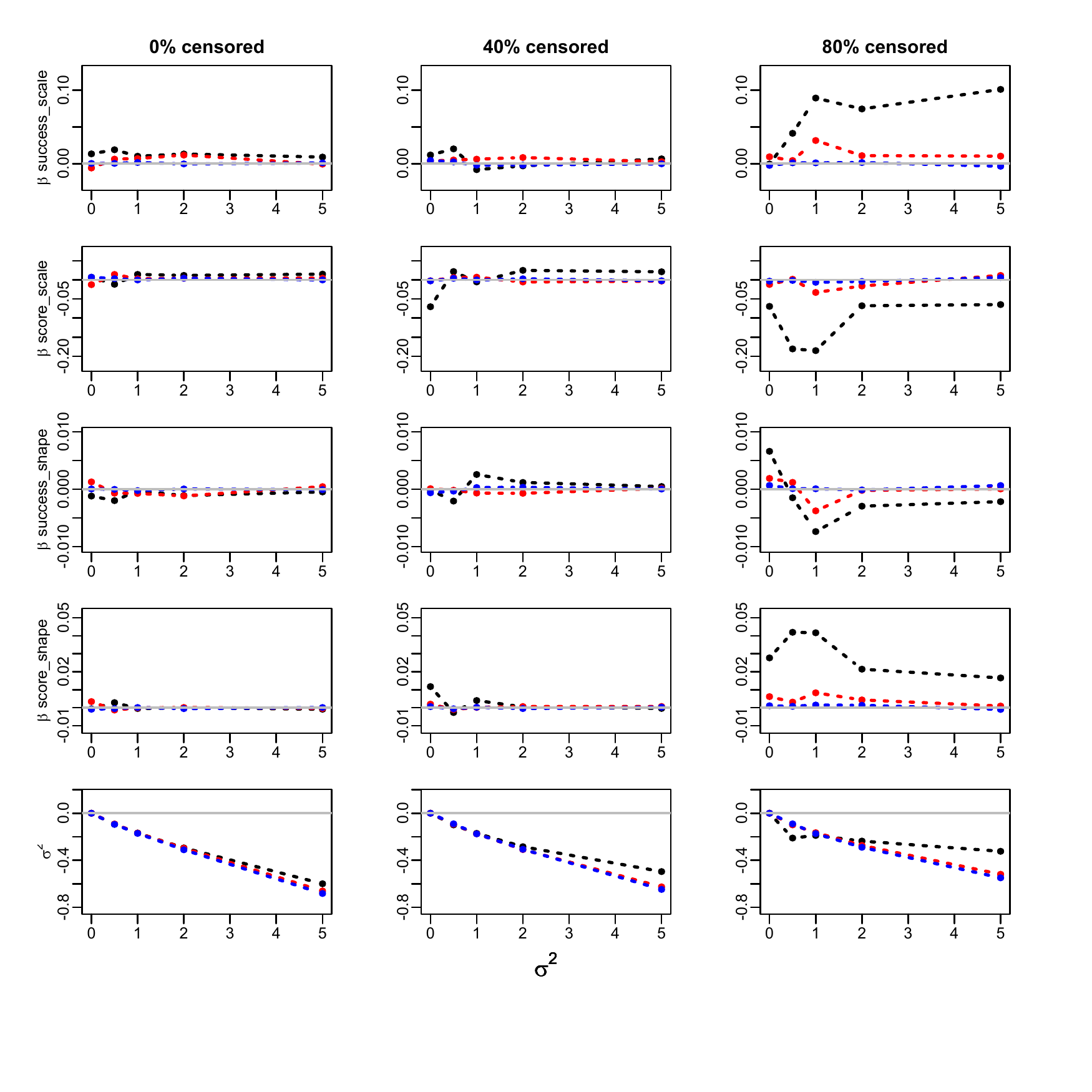}
	\caption{Bias of the estimation of the Cox regression parameters and $\sigma^2$ in the Gompertz model.
Success proportion= 0.5. 10 clusters. Sample sizes: 300 (black), 1000 (red) and 10000 (blue).
True values: $\beta_{success-scale}$ = 0.5 , $\beta_{success-shape}$= 0.05 , $\beta_{score-scale}$= 1 , $\beta_{score-shape}$= 0.1}
	\label{Bias10clustersGompertz5}
\end{figure}

\begin{figure}[ht]
	\centering
	\includegraphics[scale=1]{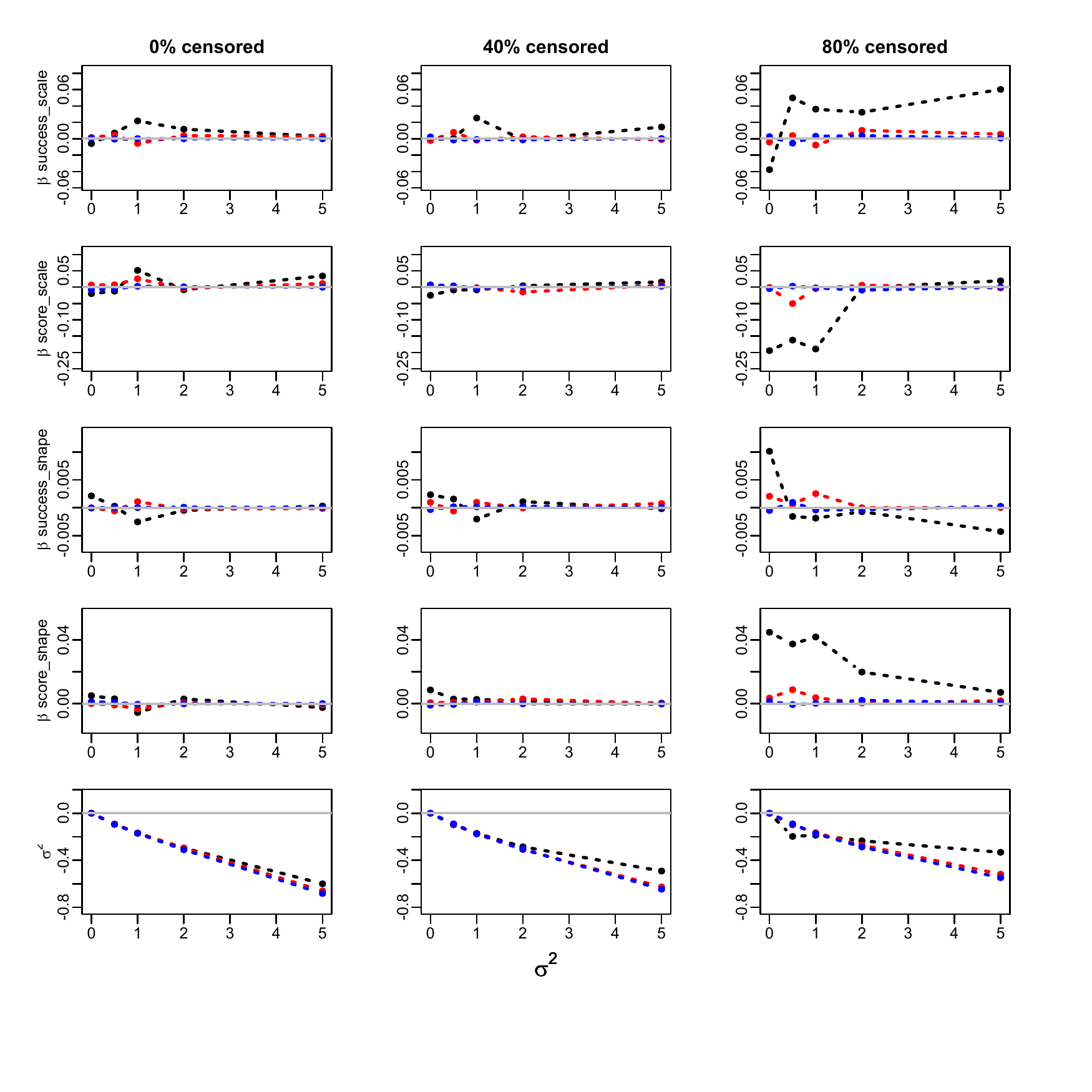}
	\caption{Bias of the estimation of the Cox regression parameters and $\sigma^2$ in the Weibull model.
Success proportion= 0.5. 10 clusters. Sample sizes: 300 (black), 1000 (red) and 10000 (blue).
True values: $\beta_{success-scale}$ = 0.5 , $\beta_{success-shape}$= -0.05 , $\beta_{score-scale}$= 1 , $\beta_{score-shape}$= - 0.1}
	\label{Bias10clustersGompertz6}
\end{figure}

\begin{figure}[ht]
	\centering
	\includegraphics[scale=1]{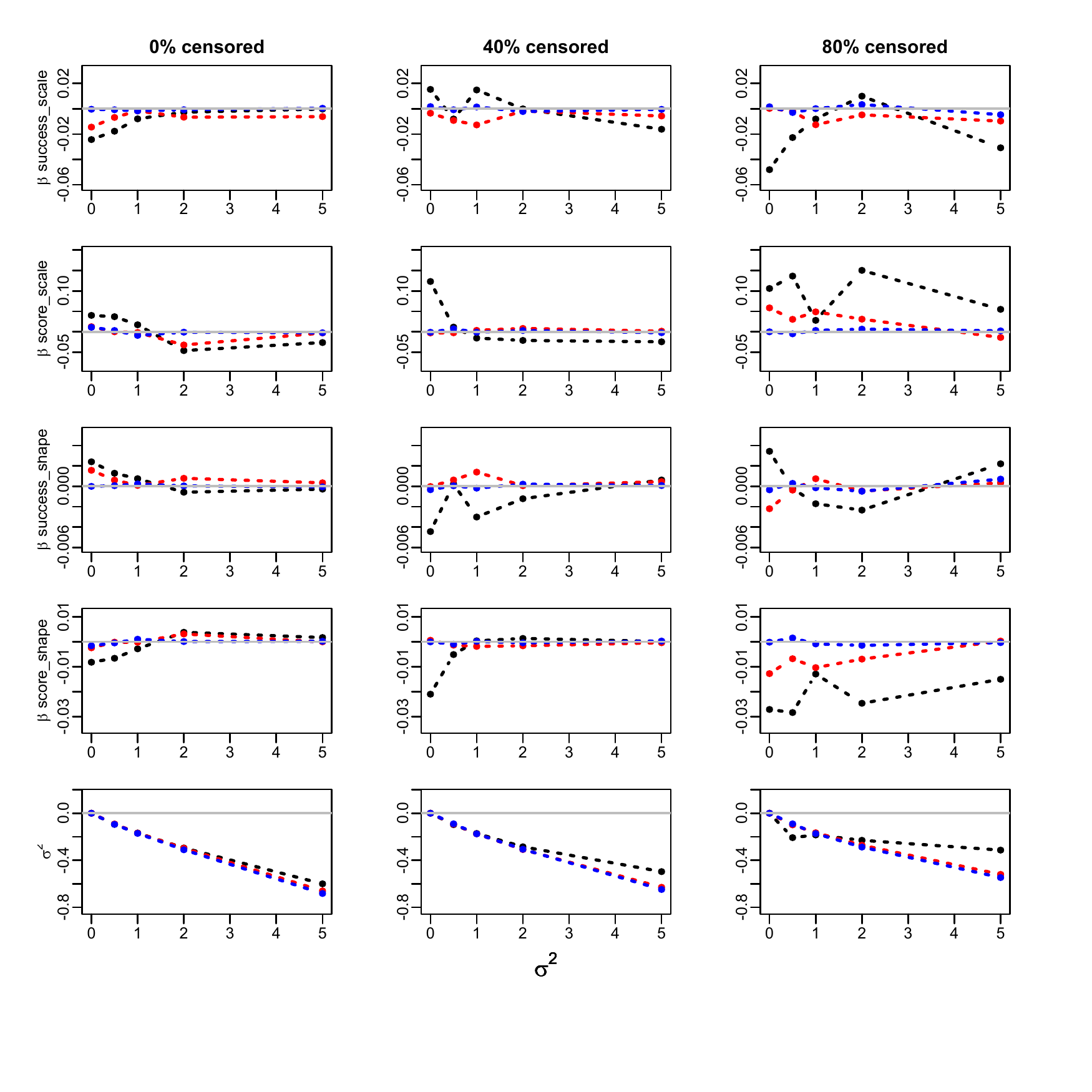}
	\caption{Bias of the estimation of the Cox regression parameters and $\sigma^2$ in the Gompertz model.
Success proportion= 0.5. 10 clusters. Sample sizes: 300 (black), 1000 (red) and 10000 (blue).
True values: $\beta_{success-scale}$ = -0.5 , $\beta_{success-shape}$= 0.05 , $\beta_{score-scale}$= -1 , $\beta_{score-shape}$= 0.1}
	\label{Bias10clustersGompertz7}
\end{figure}

\begin{figure}[ht]
	\centering
	\includegraphics[scale=1]{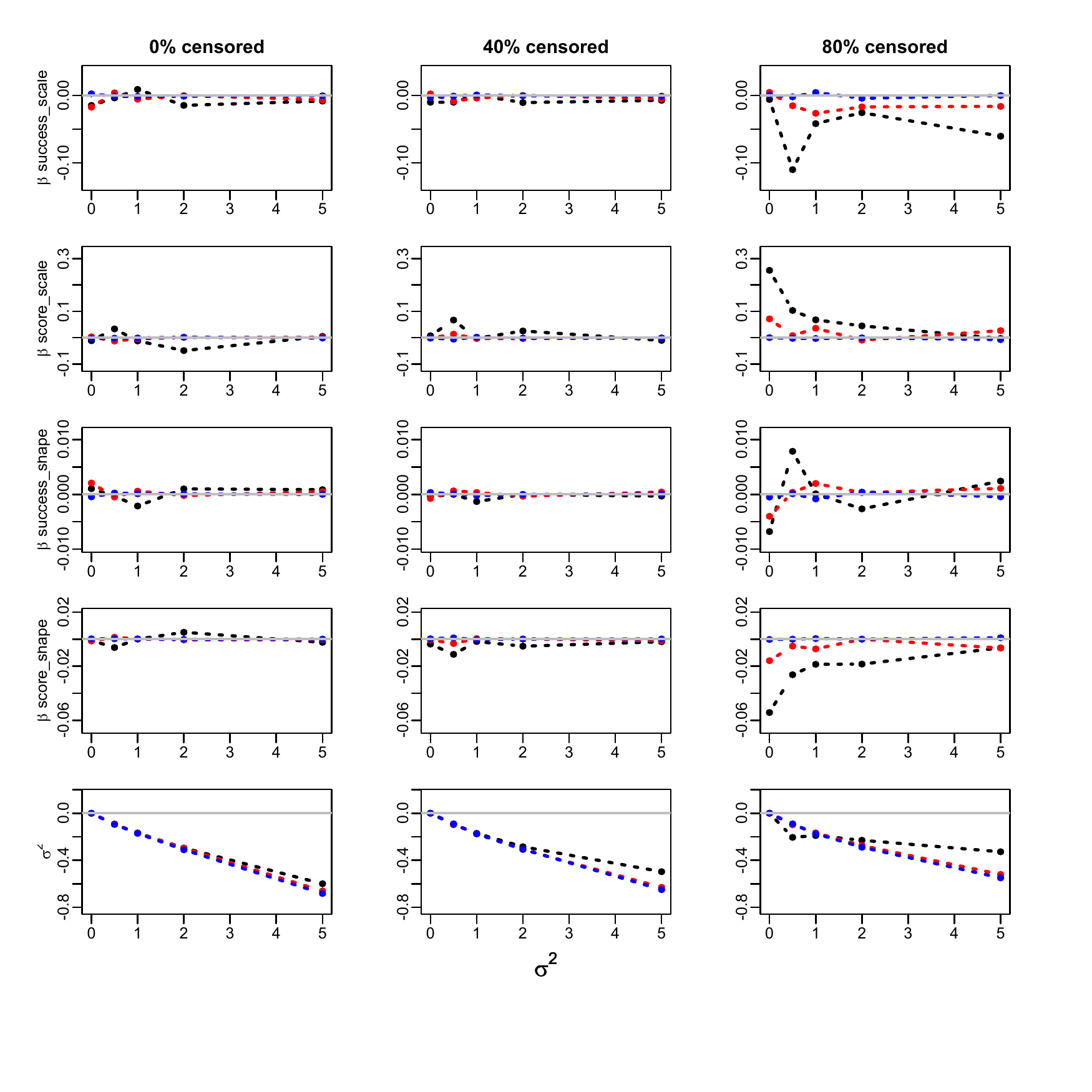}
	\caption{Bias of the estimation of the Cox regression parameters and $\sigma^2$ in the Gompertz model.
Success proportion= 0.5. 10 clusters. Sample sizes: 300 (black), 1000 (red) and 10000 (blue).
True values: $\beta_{success-scale}$ = -0.5 , $\beta_{success-shape}$= -0.05 , $\beta_{score-scale}$= -1 , $\beta_{score-shape}$= -0.1}
	\label{Bias10clustersGompertz8}
\end{figure}

 \begin{figure}[ht]
	\centering
	\includegraphics[scale=1]{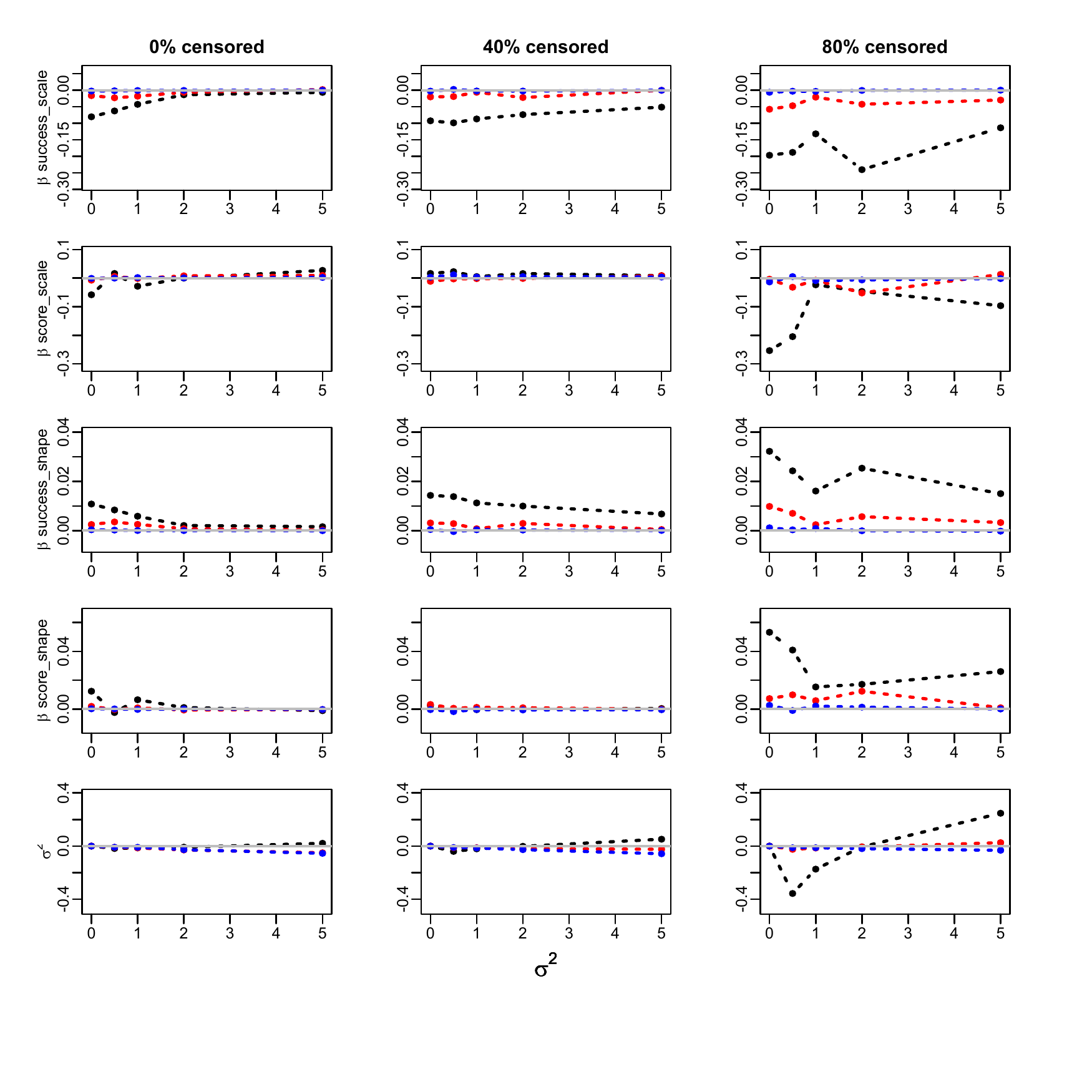}
	\caption{Bias of the estimation of the Cox regression parameters and $\sigma^2$ in the Gompertz model.
Success proportion= 0.25. 100 clusters. Sample sizes: 300 (black), 1000 (red) and 10000 (blue).
True values: $\beta_{success-scale}$ = 0.5 , $\beta_{success-shape}$= 0.05 , $\beta_{score-scale}$= 1 , $\beta_{score-shape}$= 0.1}
	\label{Bias100clustersGompertz1}
\end{figure}

\begin{figure}[ht]
	\centering
	\includegraphics[scale=1]{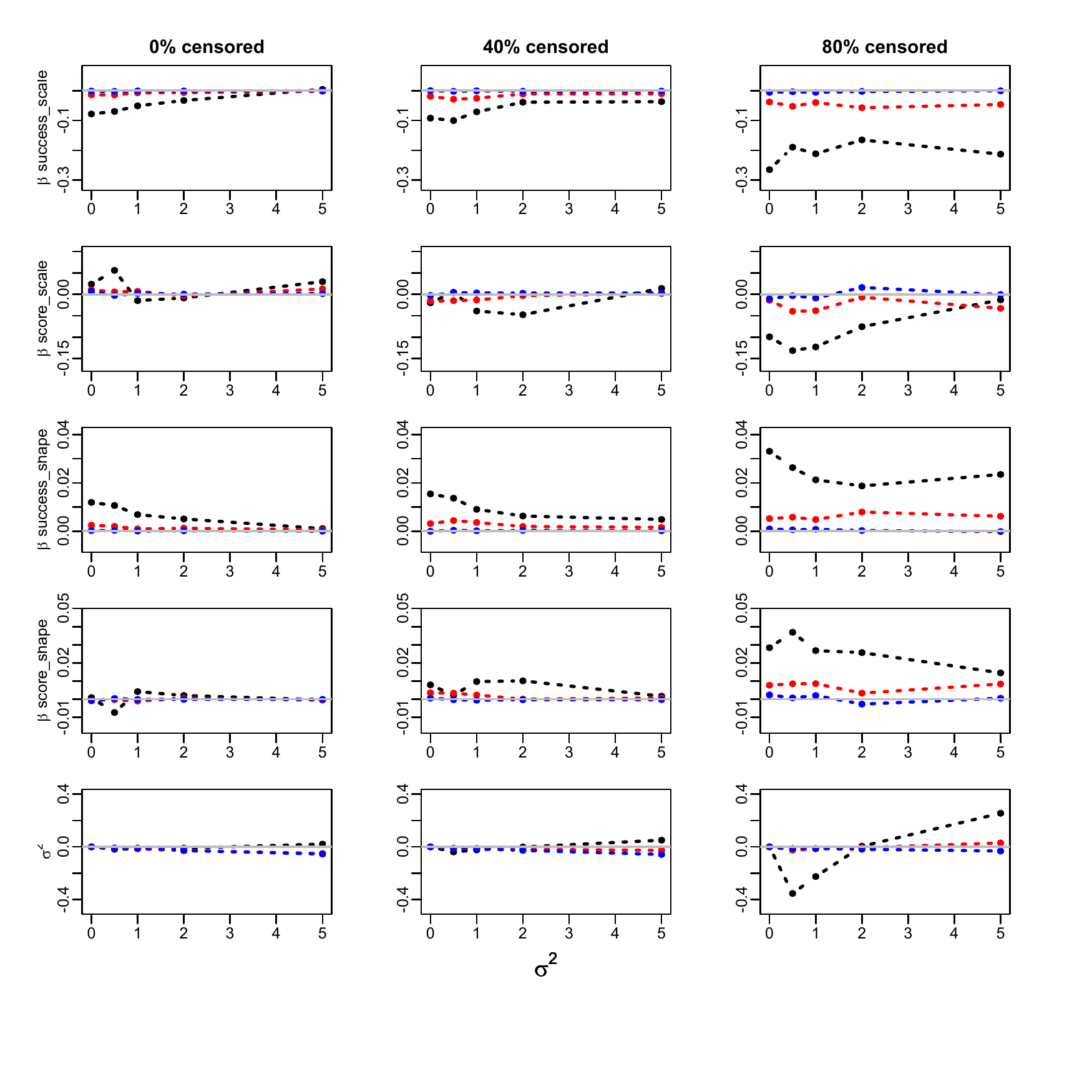}
	\caption{Bias of the estimation of the Cox regression parameters and $\sigma^2$ in the Gompertz model.
Success proportion= 0.25. 100 clusters. Sample sizes: 300 (black), 1000 (red) and 10000 (blue).
True values: $\beta_{success-scale}$ = 0.5 , $\beta_{success-shape}$= -0.05 , $\beta_{score-scale}$= 1 , $\beta_{score-shape}$= - 0.1}
	\label{Bias100clustersGompertz2}
\end{figure}

\begin{figure}[ht]
	\centering
	\includegraphics[scale=1]{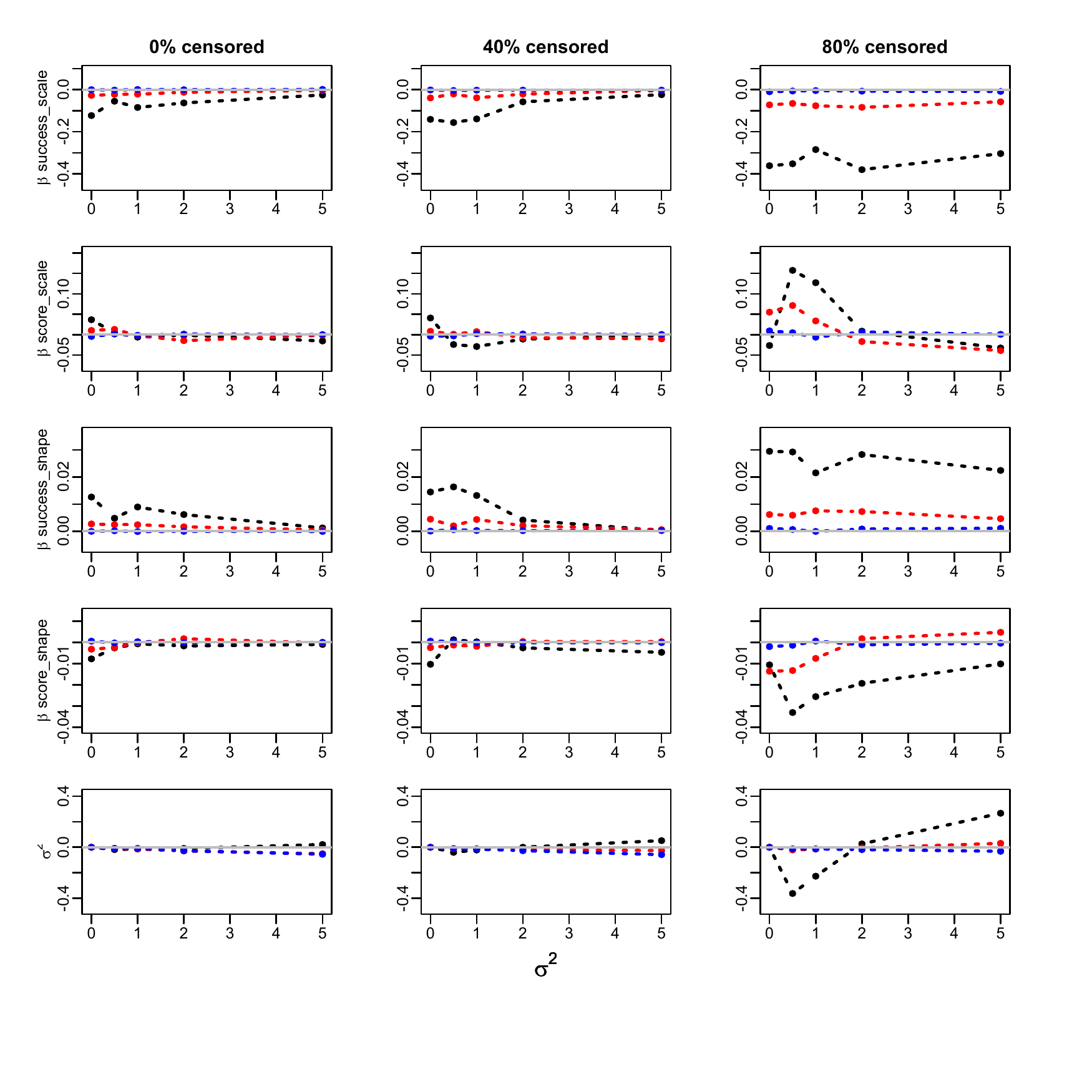}
	\caption{Bias of the estimation of the Cox regression parameters and $\sigma^2$ in the Gompertz model.
Success proportion= 0.25. 100 clusters. Sample sizes: 300 (black), 1000 (red) and 10000 (blue).
True values: $\beta_{success-scale}$ = -0.5 , $\beta_{success-shape}$= 0.05 , $\beta_{score-scale}$= -1 , $\beta_{score-shape}$= 0.1}
	\label{Bias100clustersGompertz3}
\end{figure}

\begin{figure}[ht]
	\centering
	\includegraphics[scale=1]{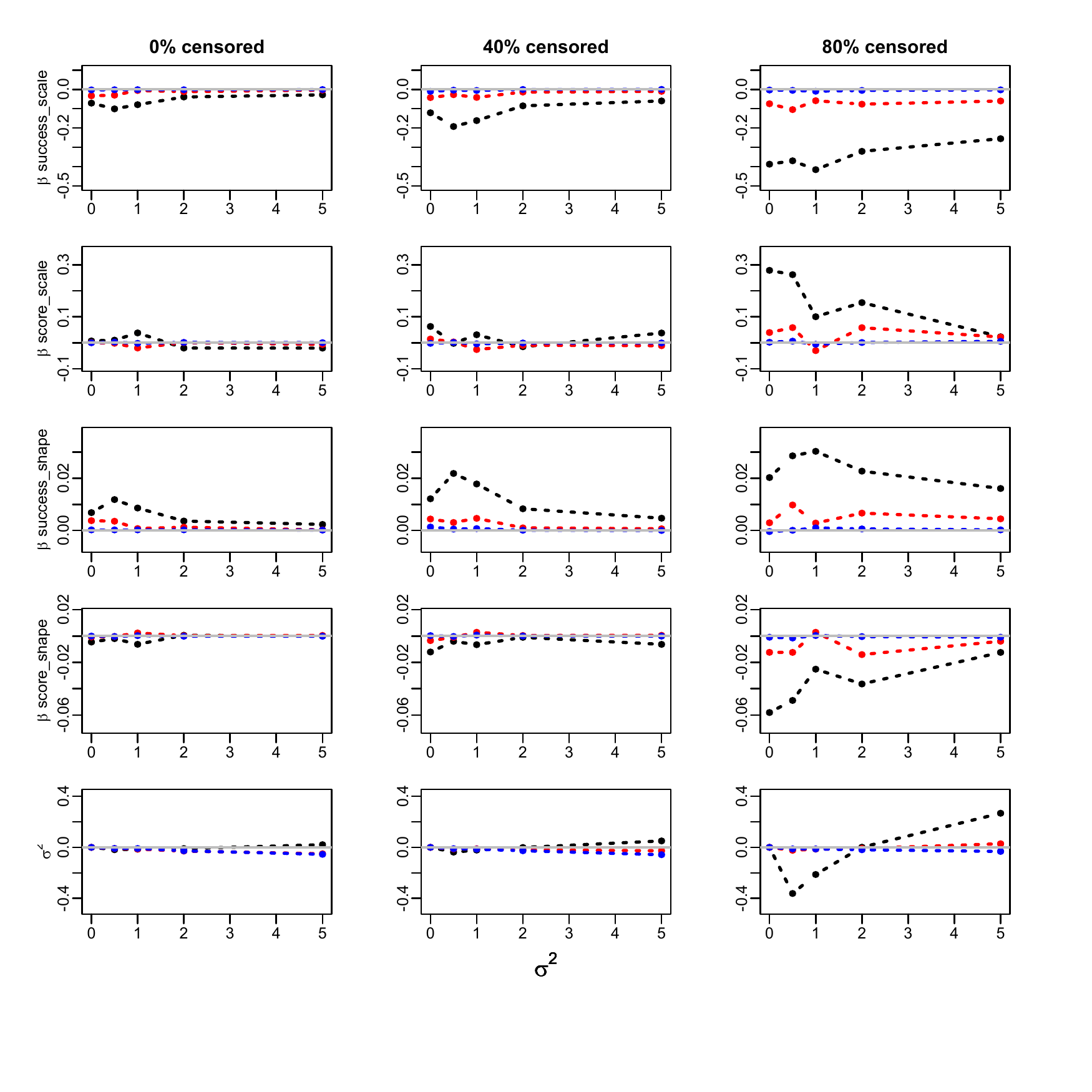}
	\caption{Bias of the estimation of the Cox regression parameters and $\sigma^2$ in the Gompertz model.
Success proportion= 0.25. 100 clusters. Sample sizes: 300 (black), 1000 (red) and 10000 (blue).
True values: $\beta_{success-scale}$ = -0.5 , $\beta_{success-shape}$= -0.05 , $\beta_{score-scale}$= -1 , $\beta_{score-shape}$= -0.1}
	\label{Bias100clustersGompertz4}
\end{figure}

\begin{figure}[ht]
	\centering
	\includegraphics[scale=1]{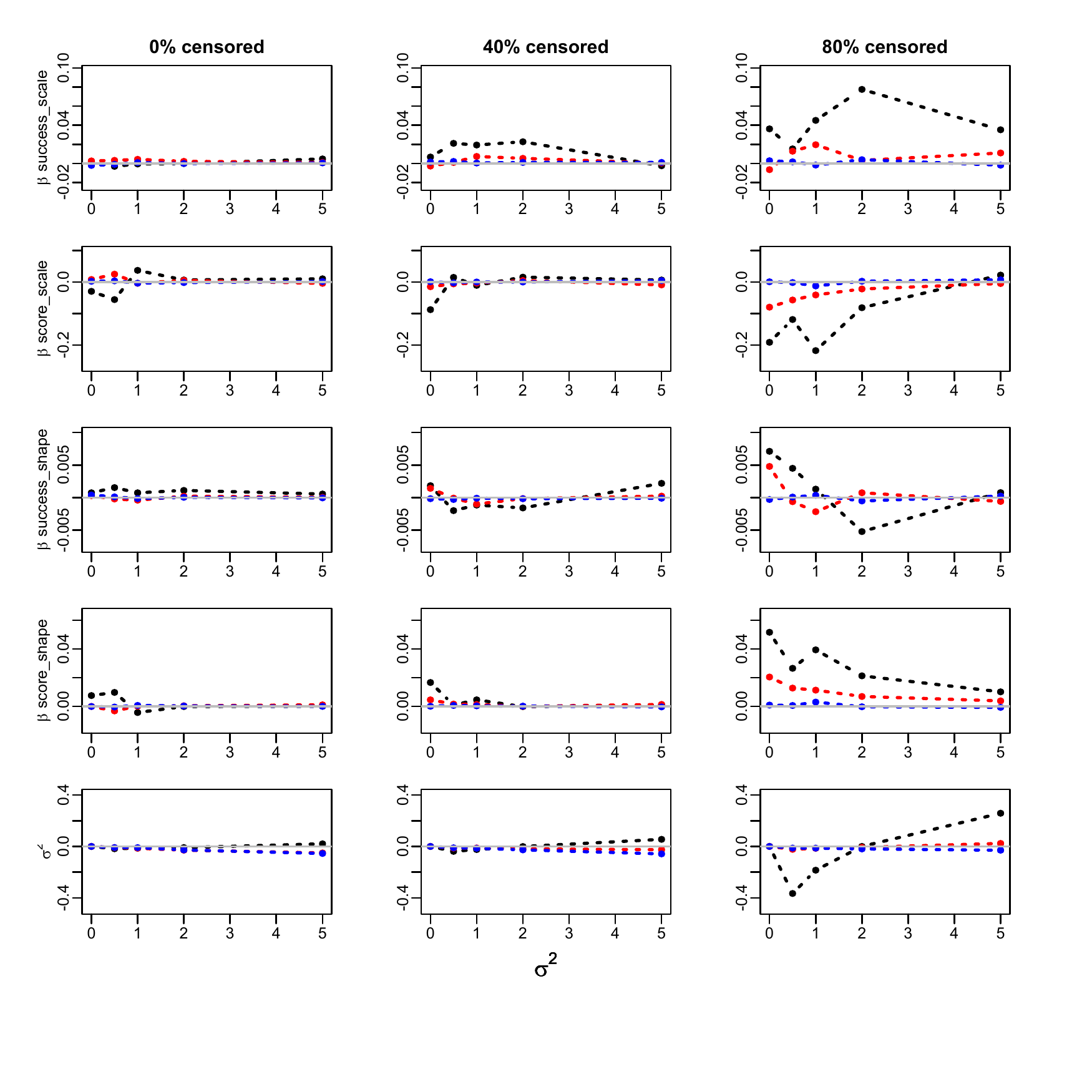}
	\caption{Bias of the estimation of the Cox regression parameters and $\sigma^2$ in the Gompertz model.
Success proportion= 0.5. 100 clusters. Sample sizes: 300 (black), 1000 (red) and 10000 (blue).
True values: $\beta_{success-scale}$ = 0.5 , $\beta_{success-shape}$= 0.05 , $\beta_{score-scale}$= 1 , $\beta_{score-shape}$= 0.1}
	\label{Bias100clustersGompertz5}
\end{figure}

\begin{figure}[ht]
	\centering
	\includegraphics[scale=1]{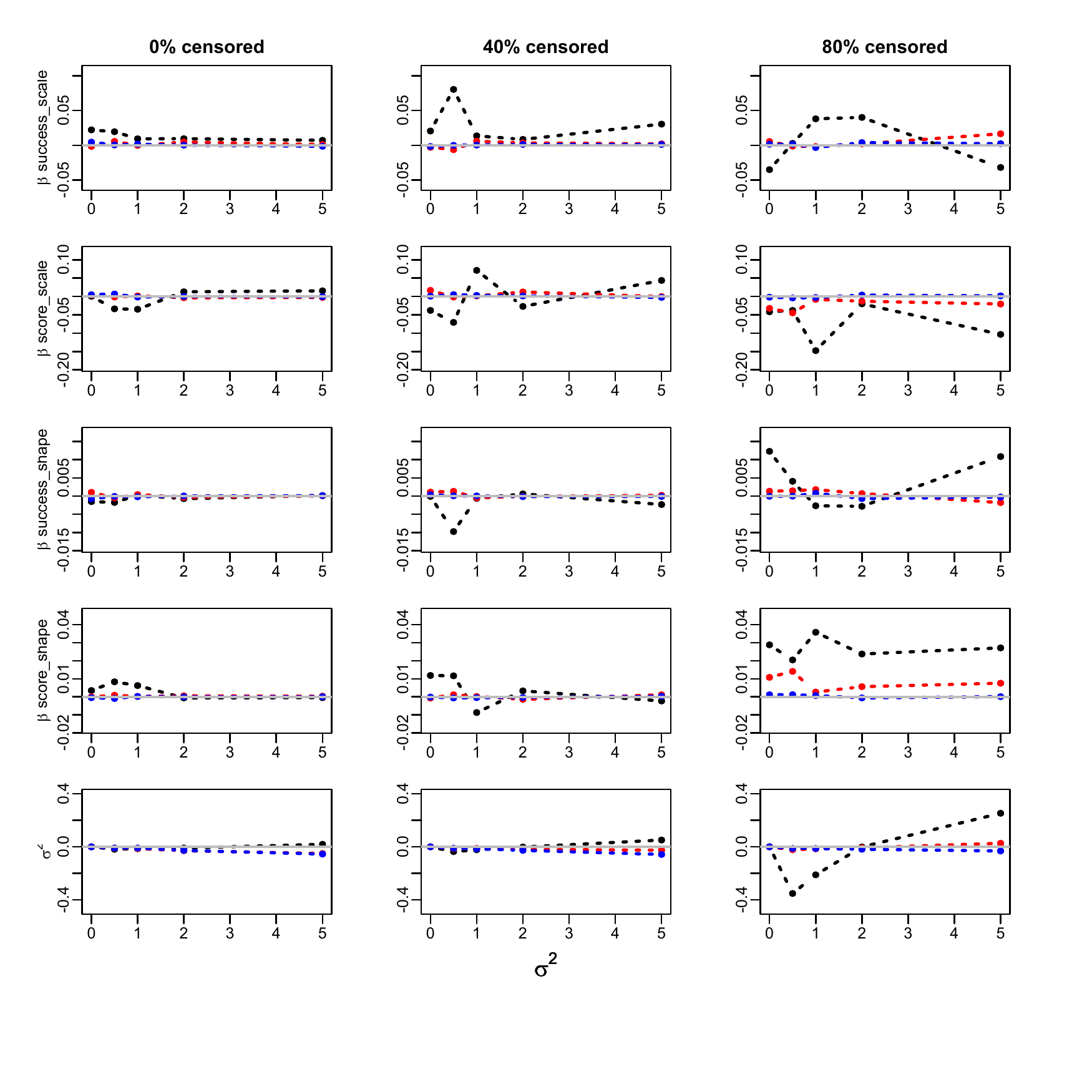}
	\caption{Bias of the estimation of the Cox regression parameters and $\sigma^2$ in the Gompertz model.
Success proportion= 0.5. 100 clusters. Sample sizes: 300 (black), 1000 (red) and 10000 (blue).
True values: $\beta_{success-scale}$ = 0.5 , $\beta_{success-shape}$= -0.05 , $\beta_{score-scale}$= 1 , $\beta_{score-shape}$= - 0.1}
	\label{Bias100clustersGompertz6}
\end{figure}

\begin{figure}[ht]
	\centering
	\includegraphics[scale=1]{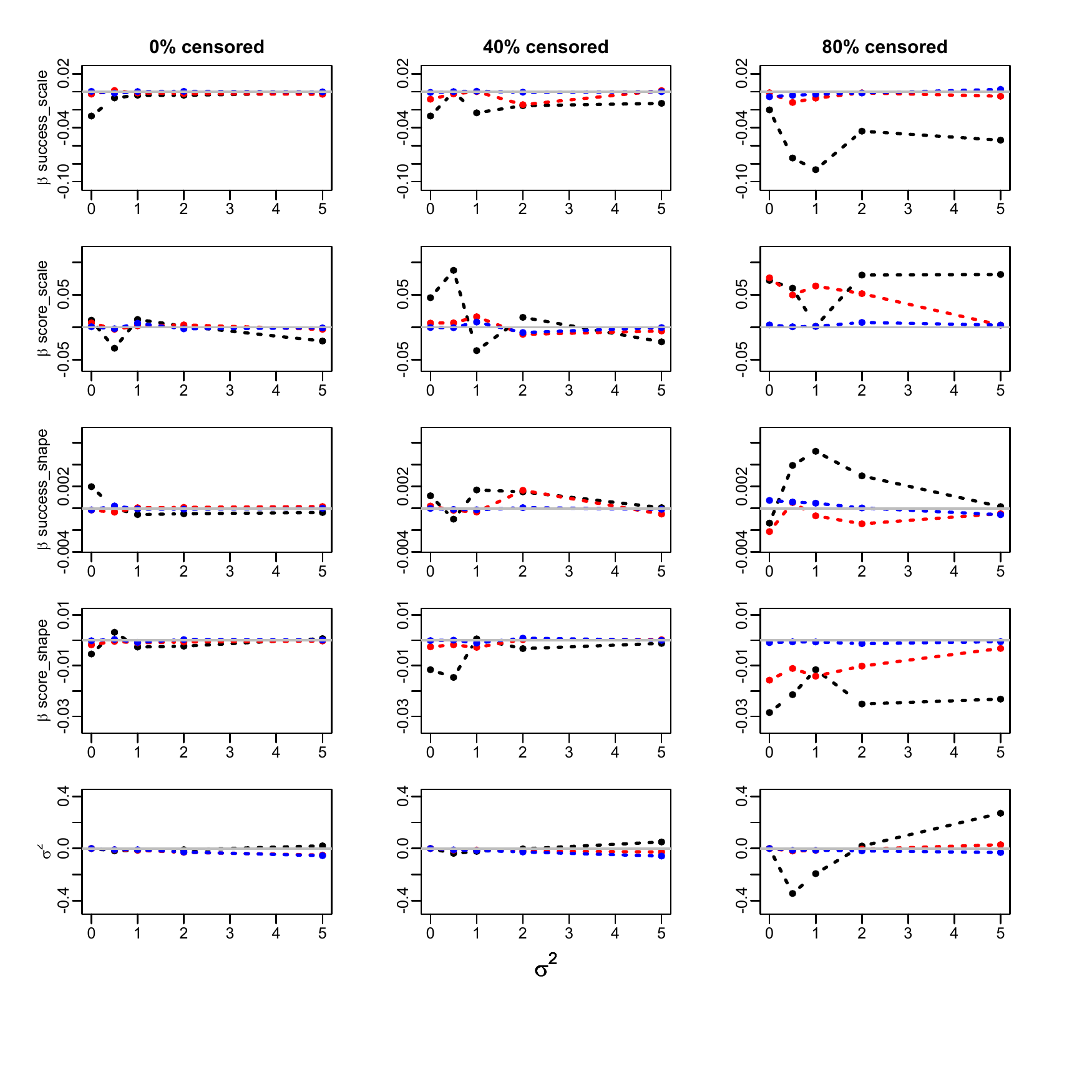}
	\caption{Bias of the estimation of the Cox regression parameters and $\sigma^2$ in the Gompertz model.
Success proportion= 0.5. 100 clusters. Sample sizes: 300 (black), 1000 (red) and 10000 (blue).
True values: $\beta_{success-scale}$ = -0.5 , $\beta_{success-shape}$= 0.05 , $\beta_{score-scale}$= -1 , $\beta_{score-shape}$= 0.1}
	\label{Bias100clustersGompertz7}
\end{figure}

\begin{figure}[ht]
	\centering
	\includegraphics[scale=1]{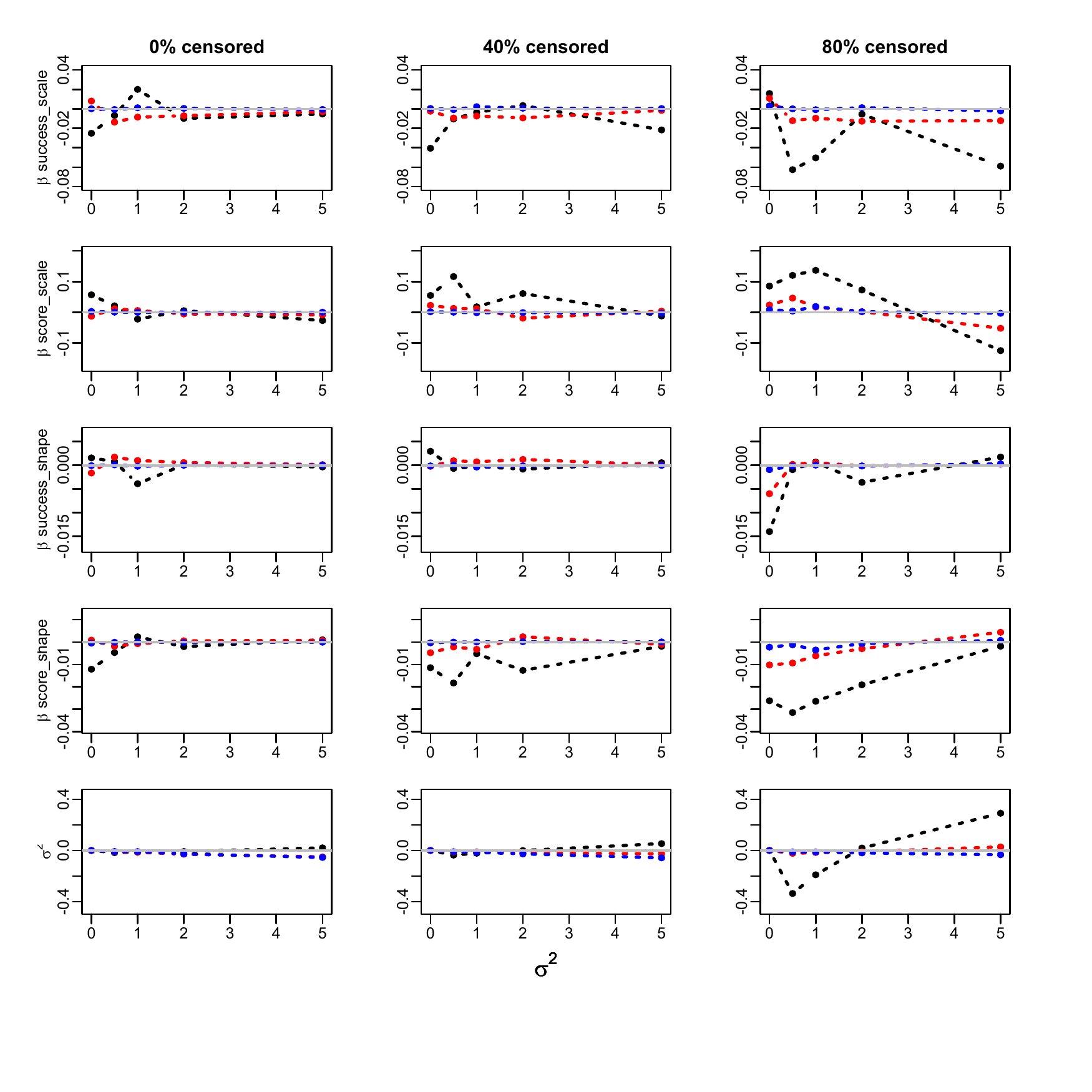}
	\caption{Bias of the estimation of the Cox regression parameters and $\sigma^2$ in the Gompertz model.
Success proportion= 0.5. 100 clusters. Sample sizes: 300 (black), 1000 (red) and 10000 (blue).
True values: $\beta_{success-scale}$ = -0.5 , $\beta_{success-shape}$= -0.05 , $\beta_{score-scale}$= -1 , $\beta_{score-shape}$= -0.1}
	\label{Bias100clustersGompertz8}
\end{figure}

\clearpage

\setcounter{figure}{0}
\setcounter{section}{0}
\renewcommand{\thefigure}{C.\arabic{figure}}

\section*{C: Coverage of the profile likelihood confidence intervals for  the scale   and shape  parameters of the  baseline distribution}

Each figure corresponds to a particular  baseline distribution  (Weibull or Gompertz), a value of the probability of success $p_{success}$ (= 0.25 or 0.5), a value for the number of clusters $N_{cl}$ (=10, 100) and a particular choice of the signs of the Cox regression parameters (+ + + +, + - + -, - + - + and - - - -).\\

The absolute values of the Cox regression parameters are held constant at $\beta_{success-scale}$ = 0.5 , $\beta_{success-shape}$= 0.05 , $\beta_{score-scale}$= 1 , $\beta_{score-shape}$= 0.1.

For each combination of a censoring proportion  (= 0, 40\%, 80\%), a panel plots, versus the frailty variance $\sigma^2$ (= 0, 0.5, 1, 2, 3, 4, 5), the empirical coverage of the true value of the  scale parameter $a$  or the true value of  the shape parameter $b$  by a PL confidence interval at 95\% nominal level, for three sample sizes (300, 1000 and 10000) . \\

\clearpage

 \begin{figure}[ht]
	\centering
	\includegraphics[scale=1]{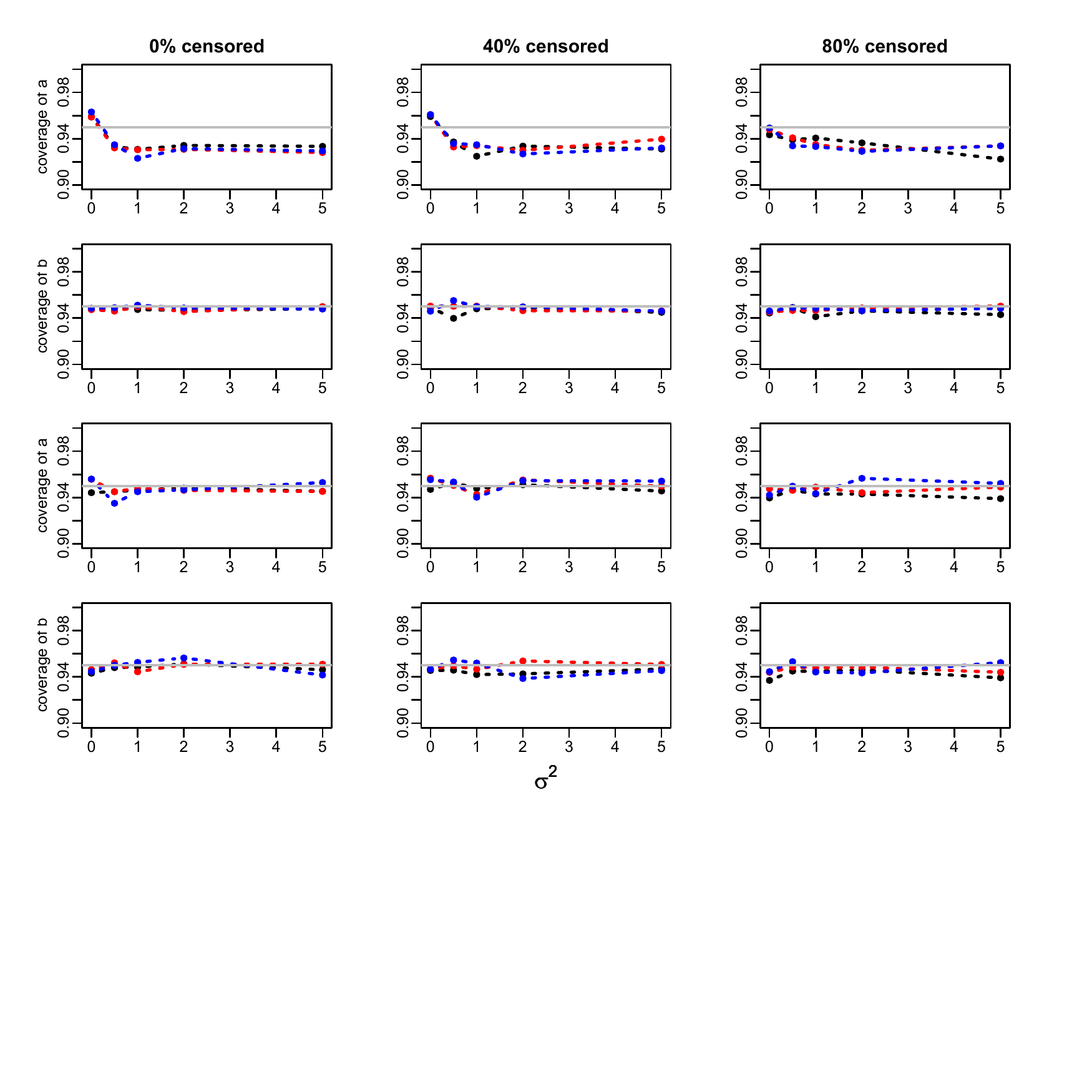}
	\caption{Coverage of the profile likelihood based confidence intervals for the $a$ and $b$ parameters in the Weibull model.
Success proportion= 0.25.  Sample sizes: 300 (black), 1000 (red) and 10000 (blue).
True values: $\beta_{success-scale}$ = 0.5 , $\beta_{success-shape}$= 0.05 , $\beta_{score-scale}$= 1 , $\beta_{score-shape}$= 0.1. 
Top two rows: 10 clusters; bottom two rows: 100 clusters.}
	\label{CoveragePL10_100clustersWeibull1}
\end{figure}

\begin{figure}[ht]
	\centering
	\includegraphics[scale=1]{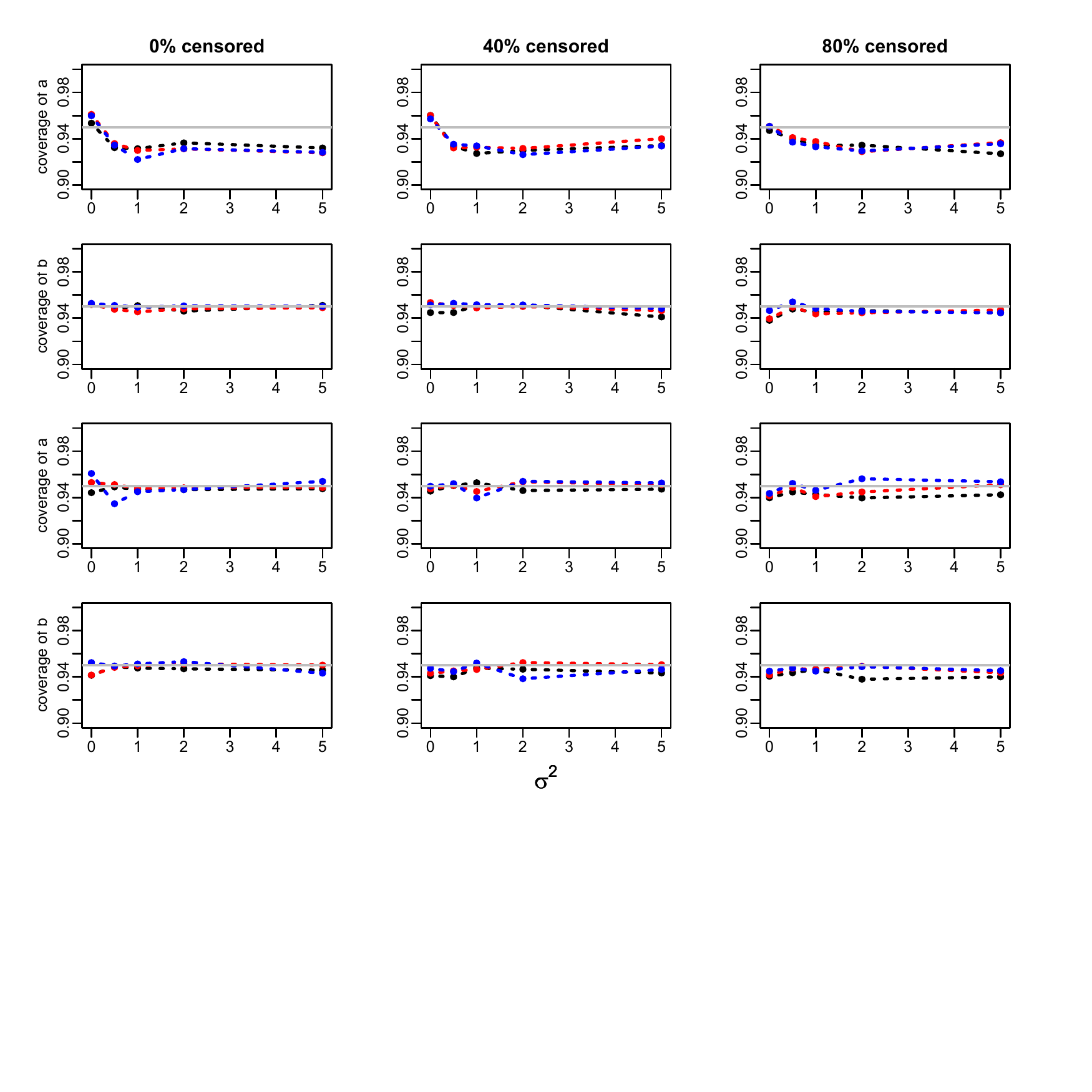}
	\caption{Coverage of the profile likelihood based confidence intervals for the $a$ and $b$ parameters  in the Weibull model.
Success proportion= 0.25.  Sample sizes: 300 (black), 1000 (red) and 10000 (blue).
True values: $\beta_{success-scale}$ = 0.5 , $\beta_{success-shape}$= -0.05 , $\beta_{score-scale}$= 1 , $\beta_{score-shape}$= - 0.1.
Top two rows: 10 clusters; bottom two rows: 100 clusters.}
	\label{CoveragePL10_100clustersWeibull2}
\end{figure}

\begin{figure}[ht]
	\centering
	\includegraphics[scale=1]{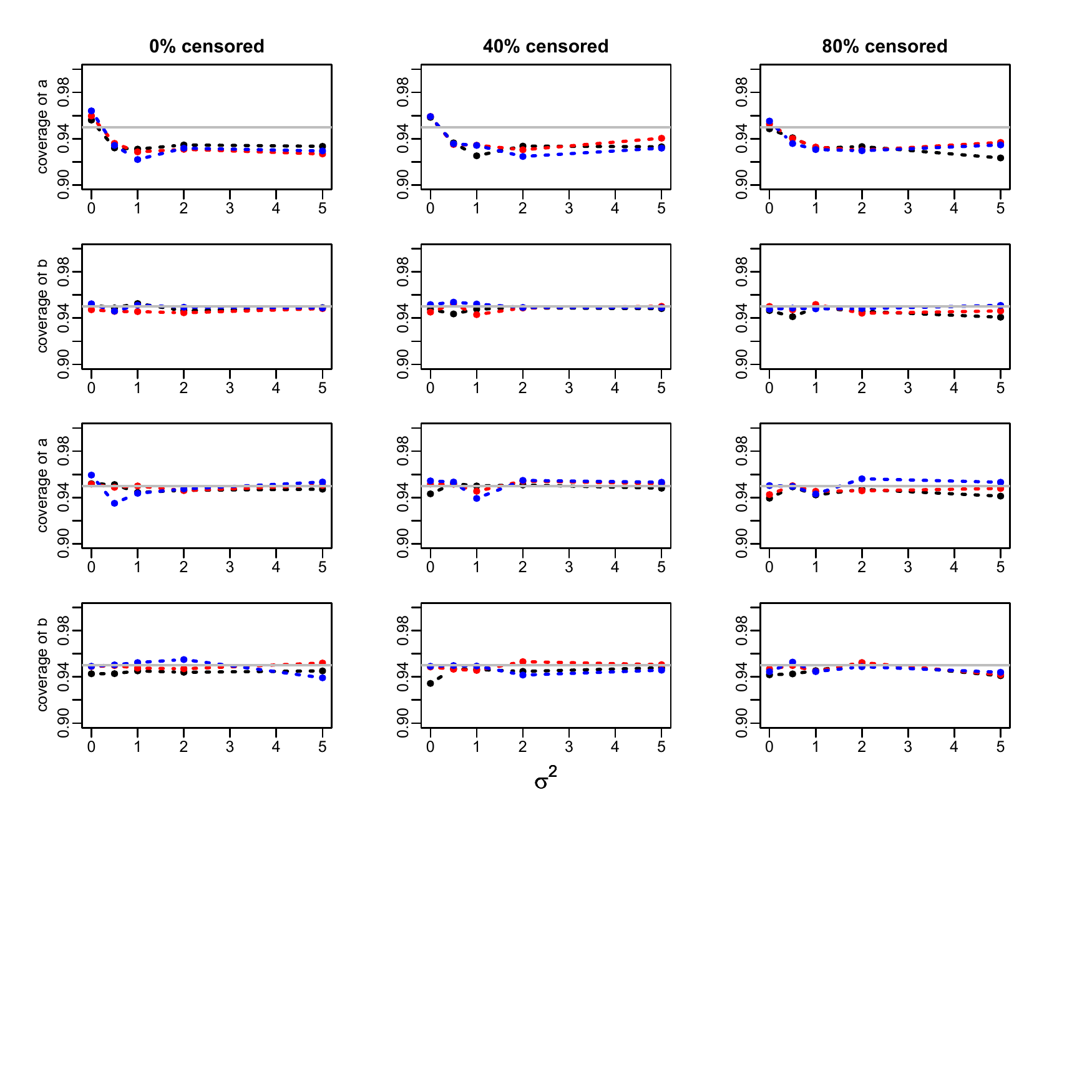}
	\caption{Coverage of the profile likelihood based confidence intervals for the $a$ and $b$ parameters  in the Weibull model.
Success proportion= 0.25.  Sample sizes: 300 (black), 1000 (red) and 10000 (blue).
True values: $\beta_{success-scale}$ = -0.5 , $\beta_{success-shape}$= 0.05 , $\beta_{score-scale}$= -1 , $\beta_{score-shape}$= 0.1.
Top two rows: 10 clusters; bottom two rows: 100 clusters.}
	\label{CoveragePL10_100clustersWeibull3}
\end{figure}

\begin{figure}[ht]
	\centering
	\includegraphics[scale=1]{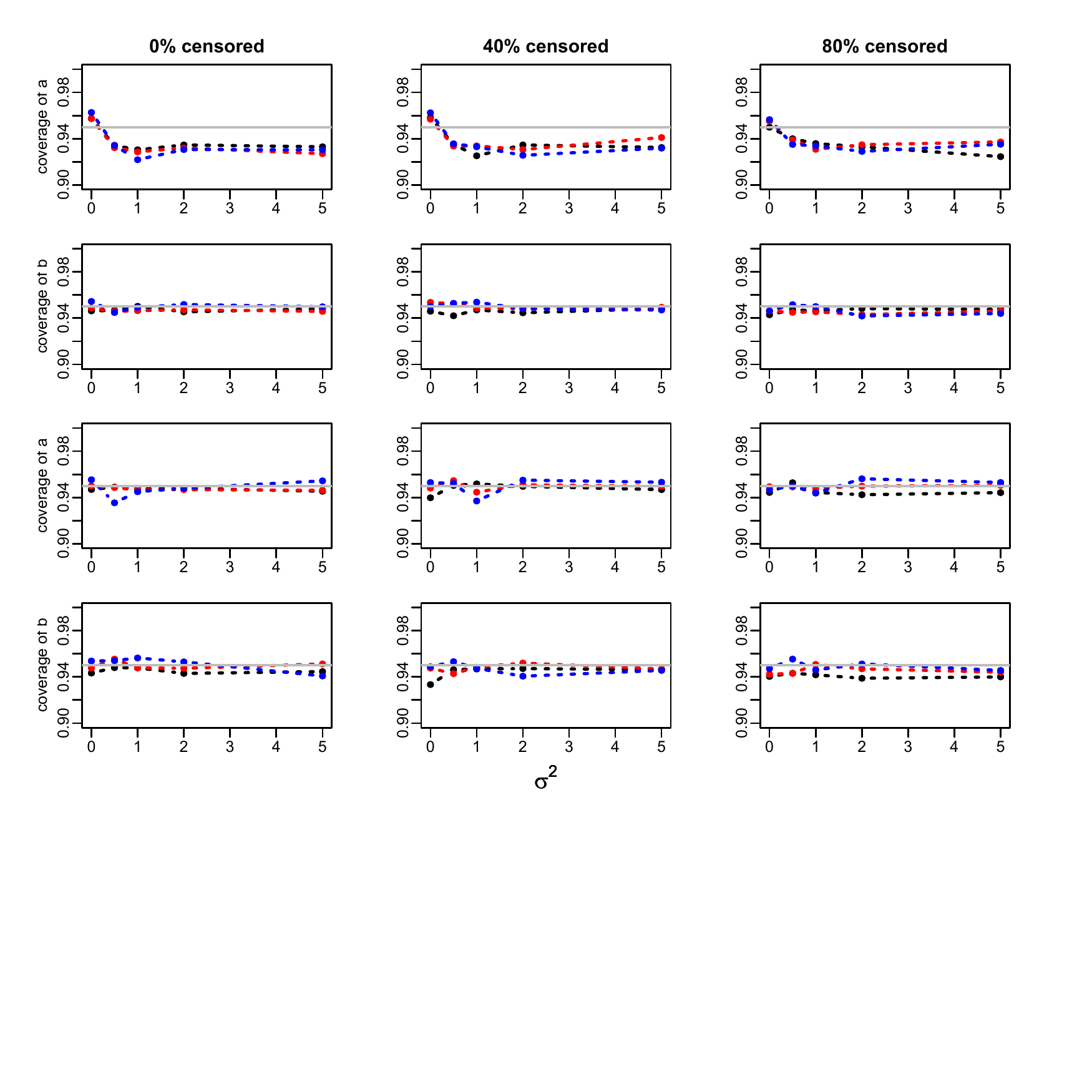}
	\caption{Coverage of the profile likelihood based confidence intervals for the $a$ and $b$ parameters  in the Weibull model.
Success proportion= 0.25. Sample sizes: 300 (black), 1000 (red) and 10000 (blue).
True values: $\beta_{success-scale}$ = -0.5 , $\beta_{success-shape}$= -0.05 , $\beta_{score-scale}$= -1 , $\beta_{score-shape}$= -0.1.
Top two rows: 10 clusters; bottom two rows: 100 clusters.}
	\label{CoveragePL10_100clustersWeibull4}
\end{figure}

\begin{figure}[ht]
	\centering
	\includegraphics[scale=1]{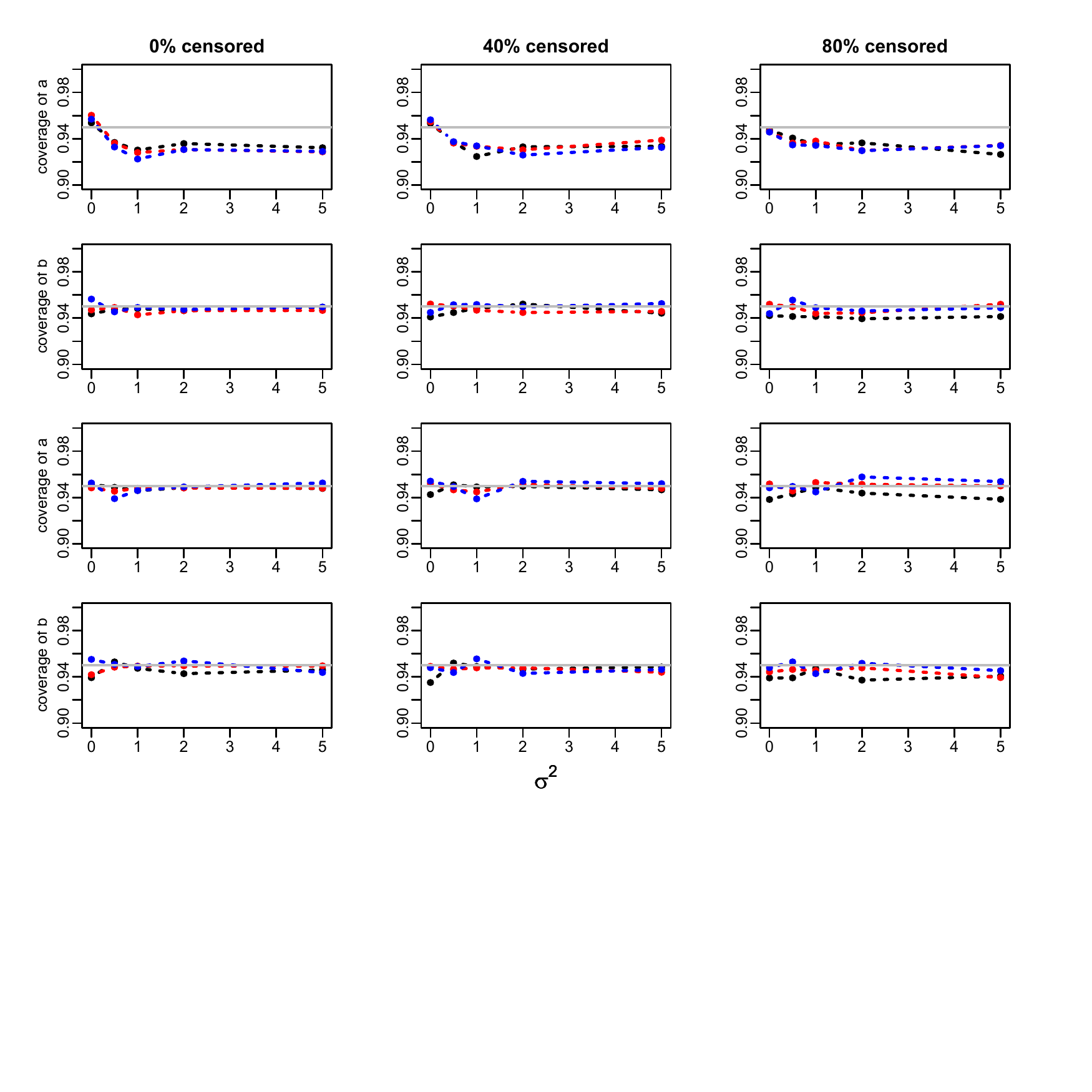}
	\caption{Coverage of the profile likelihood based confidence intervals for the $a$ and $b$ parameters  in the Weibull model.
Success proportion= 0.5.  Sample sizes: 300 (black), 1000 (red) and 10000 (blue).
True values: $\beta_{success-scale}$ = 0.5 , $\beta_{success-shape}$= 0.05 , $\beta_{score-scale}$= 1 , $\beta_{score-shape}$= 0.1.
Top two rows: 10 clusters; bottom two rows: 100 clusters.}
	\label{CoveragePL10_100clustersWeibull5}
\end{figure}

\begin{figure}[ht]
	\centering
	\includegraphics[scale=1]{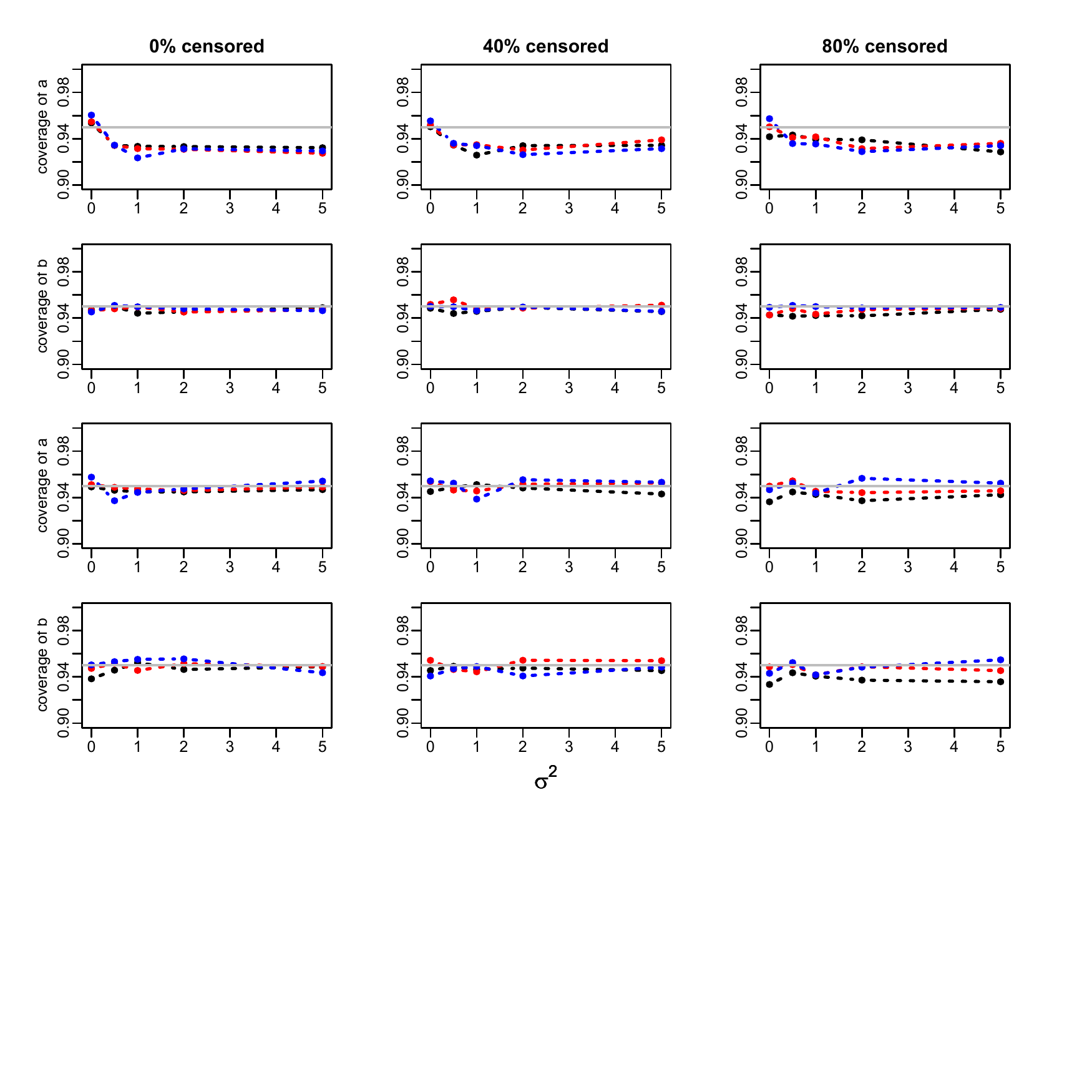}
	\caption{Coverage of the profile likelihood based confidence intervals for the $a$ and $b$ parameters  in the Weibull model.
Success proportion= 0.5.  Sample sizes: 300 (black), 1000 (red) and 10000 (blue).
True values: $\beta_{success-scale}$ = 0.5 , $\beta_{success-shape}$= -0.05 , $\beta_{score-scale}$= 1 , $\beta_{score-shape}$= - 0.1.
Top two rows: 10 clusters; bottom two rows: 100 clusters.}
	\label{CoveragePL10_100clustersWeibull6}
\end{figure}

\begin{figure}[ht]
	\centering
	\includegraphics[scale=1]{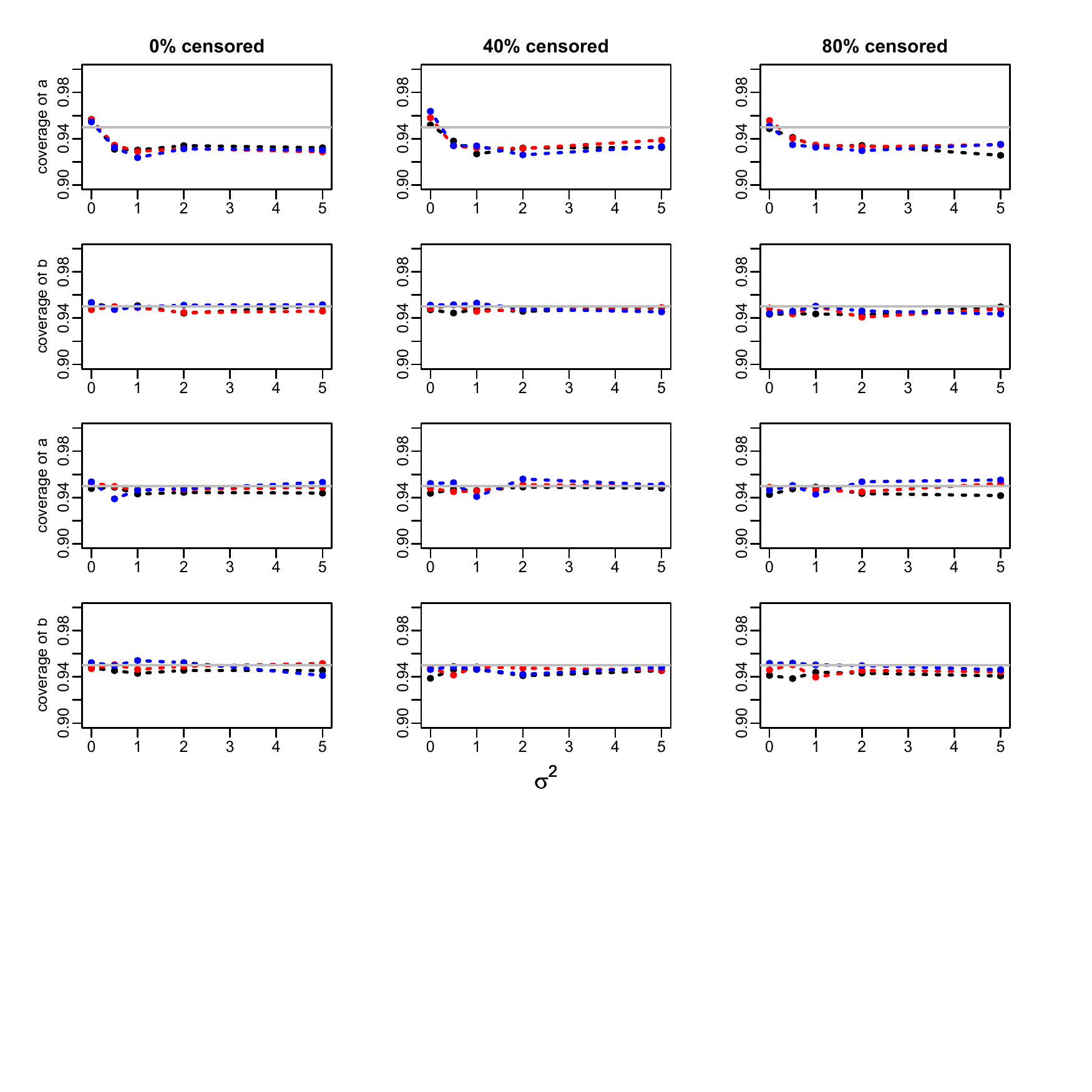}
	\caption{Coverage of the profile likelihood based confidence intervals for the $a$ and $b$ parameters in the Weibull model.
Success proportion= 0.5.  Sample sizes: 300 (black), 1000 (red) and 10000 (blue).
True values: $\beta_{success-scale}$ = -0.5 , $\beta_{success-shape}$= 0.05 , $\beta_{score-scale}$= -1 , $\beta_{score-shape}$= 0.1.
Top two rows: 10 clusters; bottom two rows: 100 clusters.}
	\label{CoveragePL10_100clustersWeibull7}
\end{figure}

\begin{figure}[ht]
	\centering
	\includegraphics[scale=1]{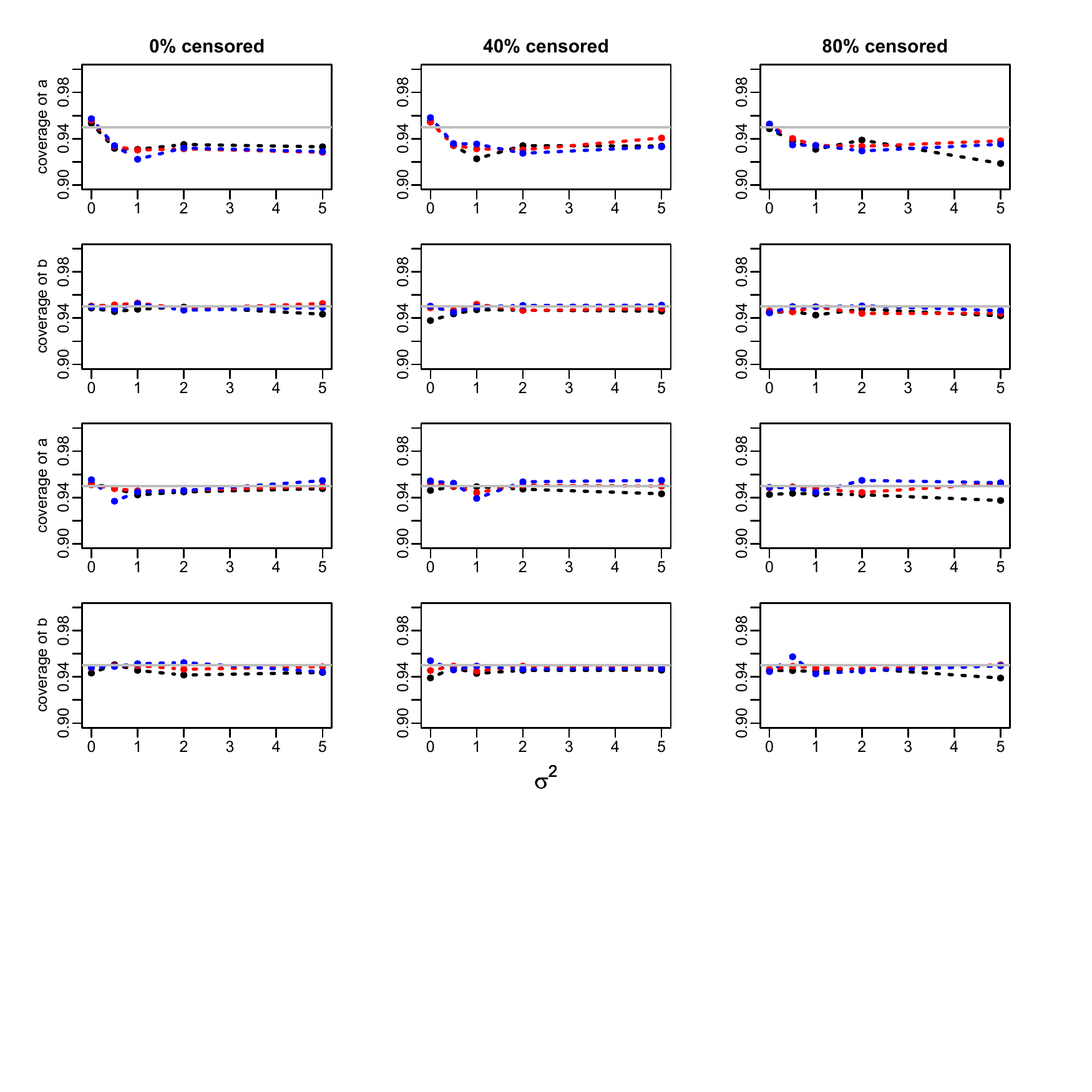}
	\caption{Coverage of the profile likelihood based confidence intervals for the $a$ and $b$ parameters  in the Weibull model.
Success proportion= 0.5.  Sample sizes: 300 (black), 1000 (red) and 10000 (blue).
True values: $\beta_{success-scale}$ = -0.5 , $\beta_{success-shape}$= -0.05 , $\beta_{score-scale}$= -1 , $\beta_{score-shape}$= -0.1.
Top two rows: 10 clusters; bottom two rows: 100 clusters.}
\label{CoveragePL10_100clustersWeibull8}
\end{figure}


\begin{figure}[ht]
	\centering
	\includegraphics[scale=1]{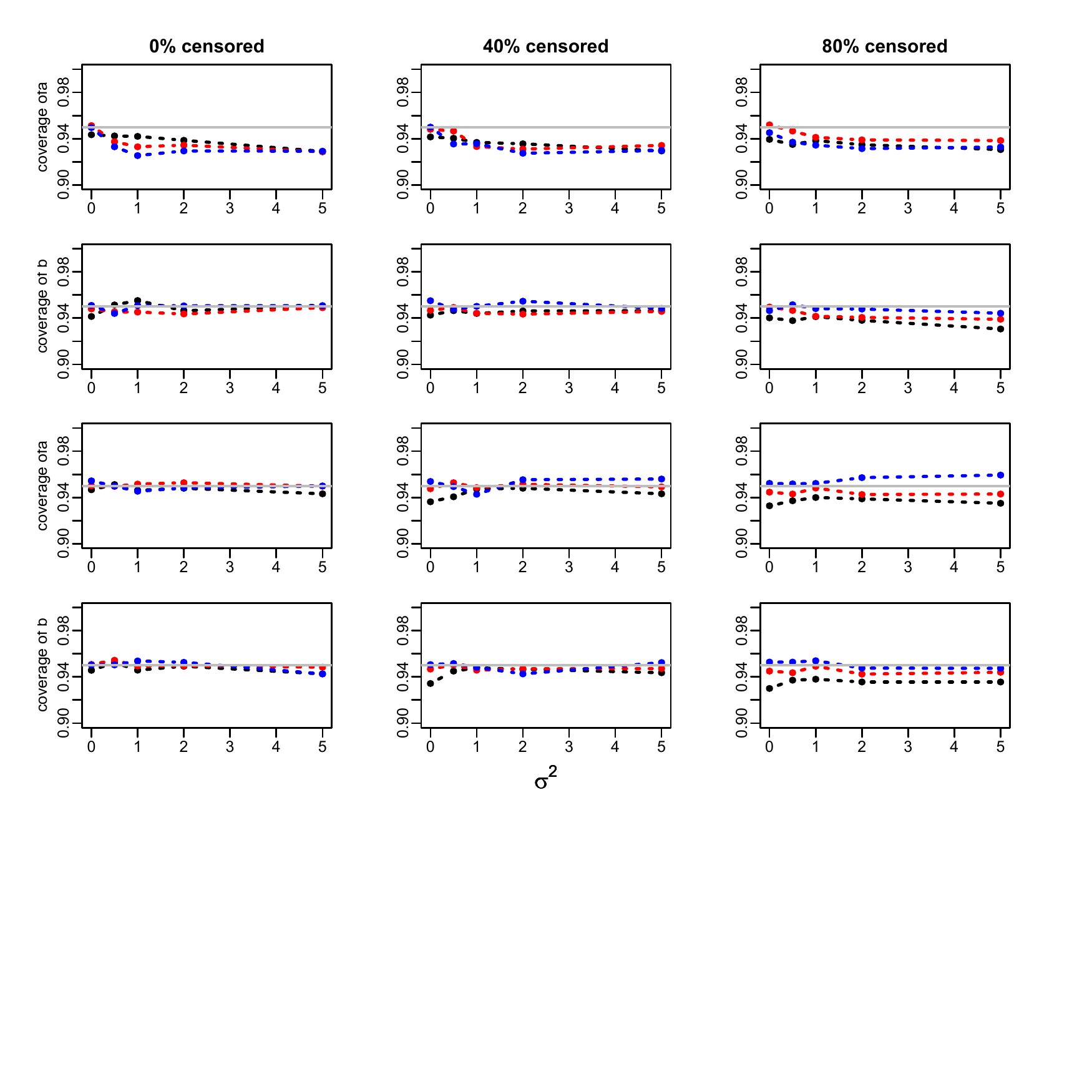}
	\caption{Coverage of the profile likelihood based confidence intervals for the $a$ and $b$ parameters in the Gompertz model.
Success proportion= 0.25.  Sample sizes: 300 (black), 1000 (red) and 10000 (blue).
True values: $\beta_{success-scale}$ = 0.5 , $\beta_{success-shape}$= 0.05 , $\beta_{score-scale}$= 1 , $\beta_{score-shape}$= 0.1.
Top two rows: 10 clusters; bottom two rows: 100 clusters.}
	\label{CoveragePL10_100clustersGompertz1}
\end{figure}

\begin{figure}[ht]
	\centering
	\includegraphics[scale=1]{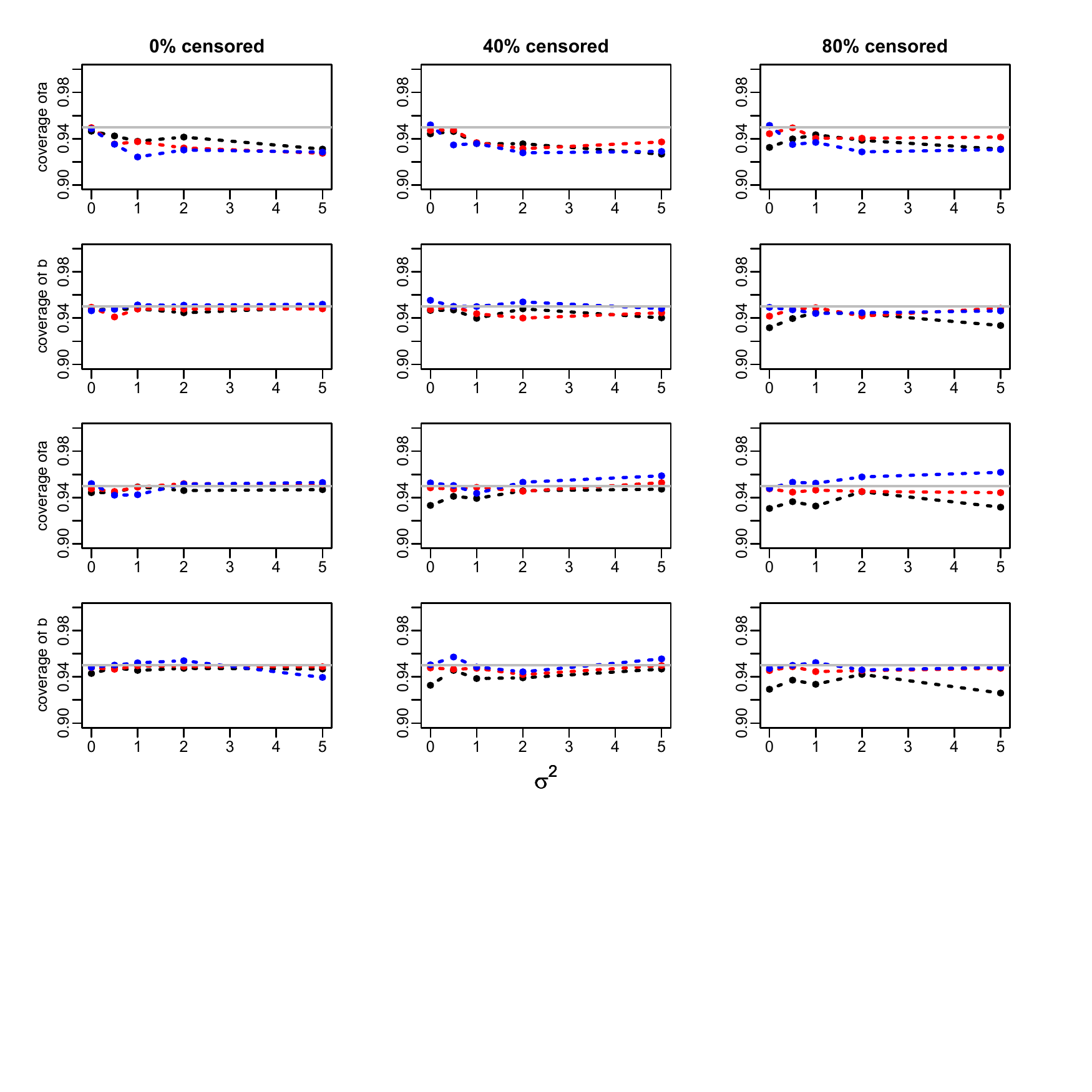}
	\caption{Coverage of the profile likelihood based confidence intervals for the $a$ and $b$ parameters  in the Gompertz model.
Success proportion= 0.25.  Sample sizes: 300 (black), 1000 (red) and 10000 (blue).
True values: $\beta_{success-scale}$ = 0.5 , $\beta_{success-shape}$= -0.05 , $\beta_{score-scale}$= 1 , $\beta_{score-shape}$= - 0.1.
Top two rows: 10 clusters; bottom two rows: 100 clusters.}
	\label{CoveragePL10_100clustersGompertz2}
\end{figure}

\begin{figure}[ht]
	\centering
	\includegraphics[scale=1]{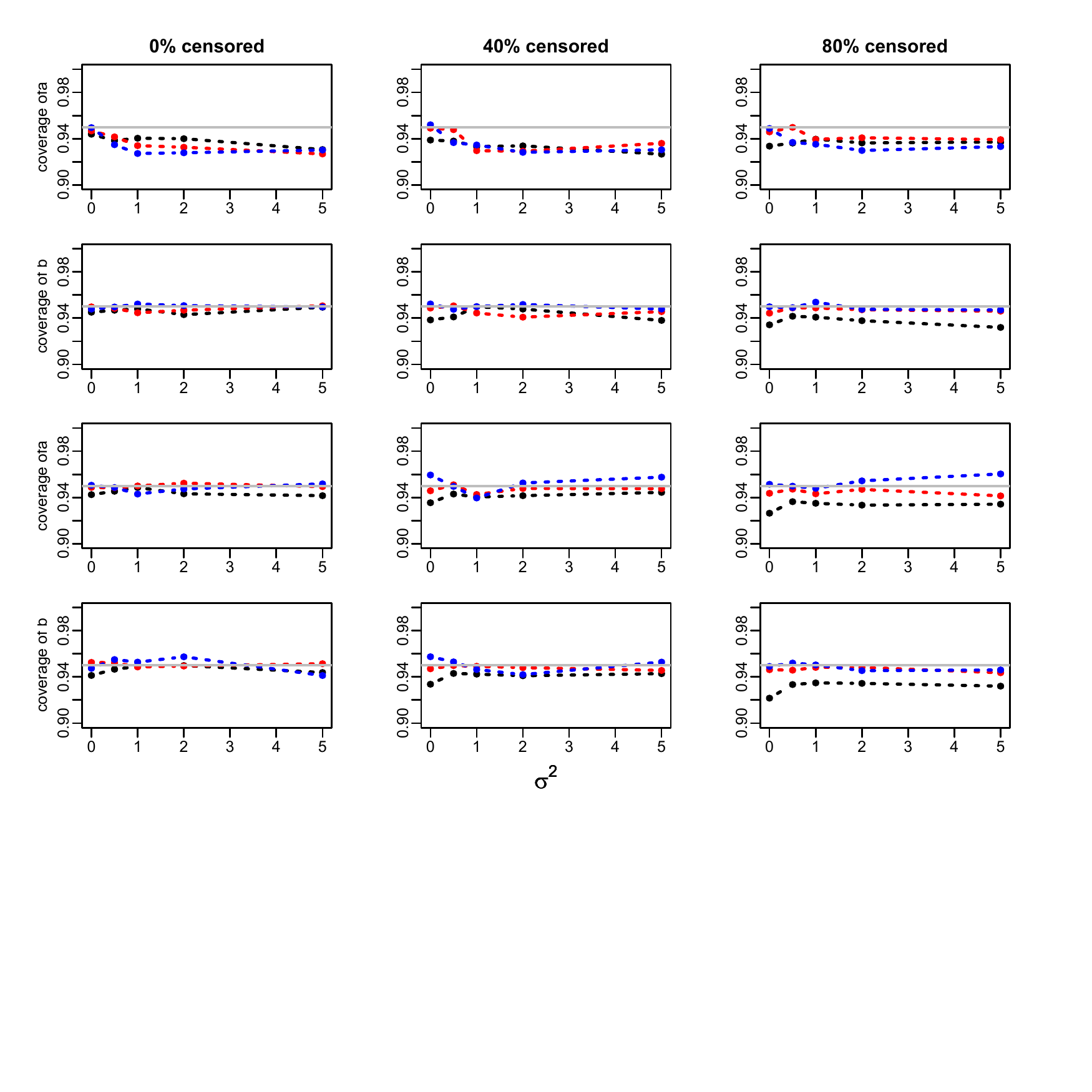}
	\caption{Coverage of the profile likelihood based confidence intervals for the $a$ and $b$ parameters  in the Gompertz model.
Success proportion= 0.25.  Sample sizes: 300 (black), 1000 (red) and 10000 (blue).
True values: $\beta_{success-scale}$ = -0.5 , $\beta_{success-shape}$= 0.05 , $\beta_{score-scale}$= -1 , $\beta_{score-shape}$= 0.1.
Top two rows: 10 clusters; bottom two rows: 100 clusters.}
	\label{CoveragePL10_100clustersGompertz3}
\end{figure}

\begin{figure}[ht]
	\centering
	\includegraphics[scale=1]{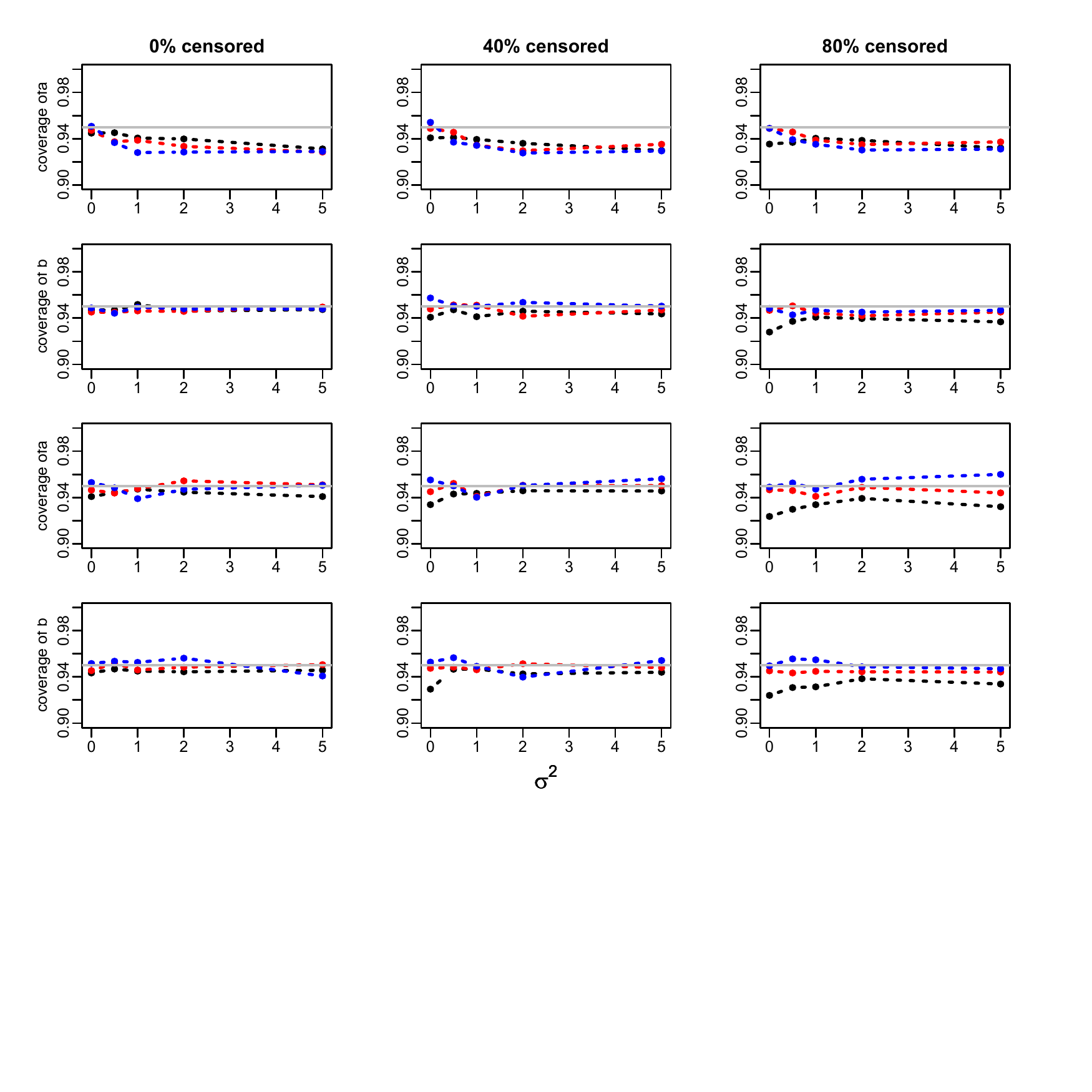}
	\caption{Coverage of the profile likelihood based confidence intervals for the $a$ and $b$ parameters  in the Gompertz model.
Success proportion= 0.25.  Sample sizes: 300 (black), 1000 (red) and 10000 (blue).
True values: $\beta_{success-scale}$ = -0.5 , $\beta_{success-shape}$= -0.05 , $\beta_{score-scale}$= -1 , $\beta_{score-shape}$= -0.1.
Top two rows: 10 clusters; bottom two rows: 100 clusters.}
	\label{CoveragePL10_100clustersGompertz4}
\end{figure}

\begin{figure}[ht]
	\centering
	\includegraphics[scale=1]{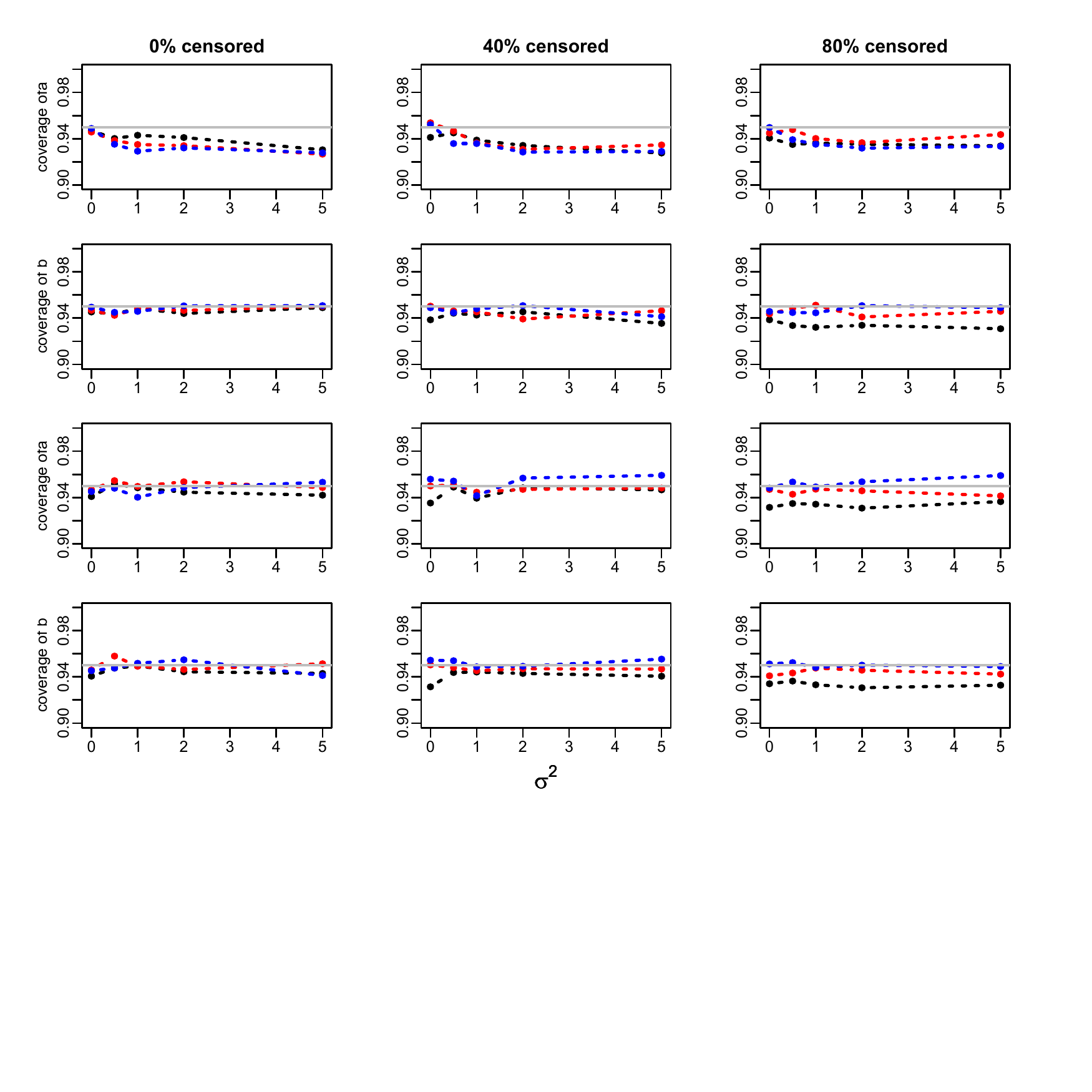}
	\caption{Coverage of the profile likelihood based confidence intervals for the $a$ and $b$ parameters  in the Gompertz model.
Success proportion= 0.5.  Sample sizes: 300 (black), 1000 (red) and 10000 (blue).
True values: $\beta_{success-scale}$ = 0.5 , $\beta_{success-shape}$= 0.05 , $\beta_{score-scale}$= 1 , $\beta_{score-shape}$= 0.1.
Top two rows: 10 clusters; bottom two rows: 100 clusters.}
	\label{CoveragePL10_100clustersGompertz5}
\end{figure}

\begin{figure}[ht]
	\centering
	\includegraphics[scale=1]{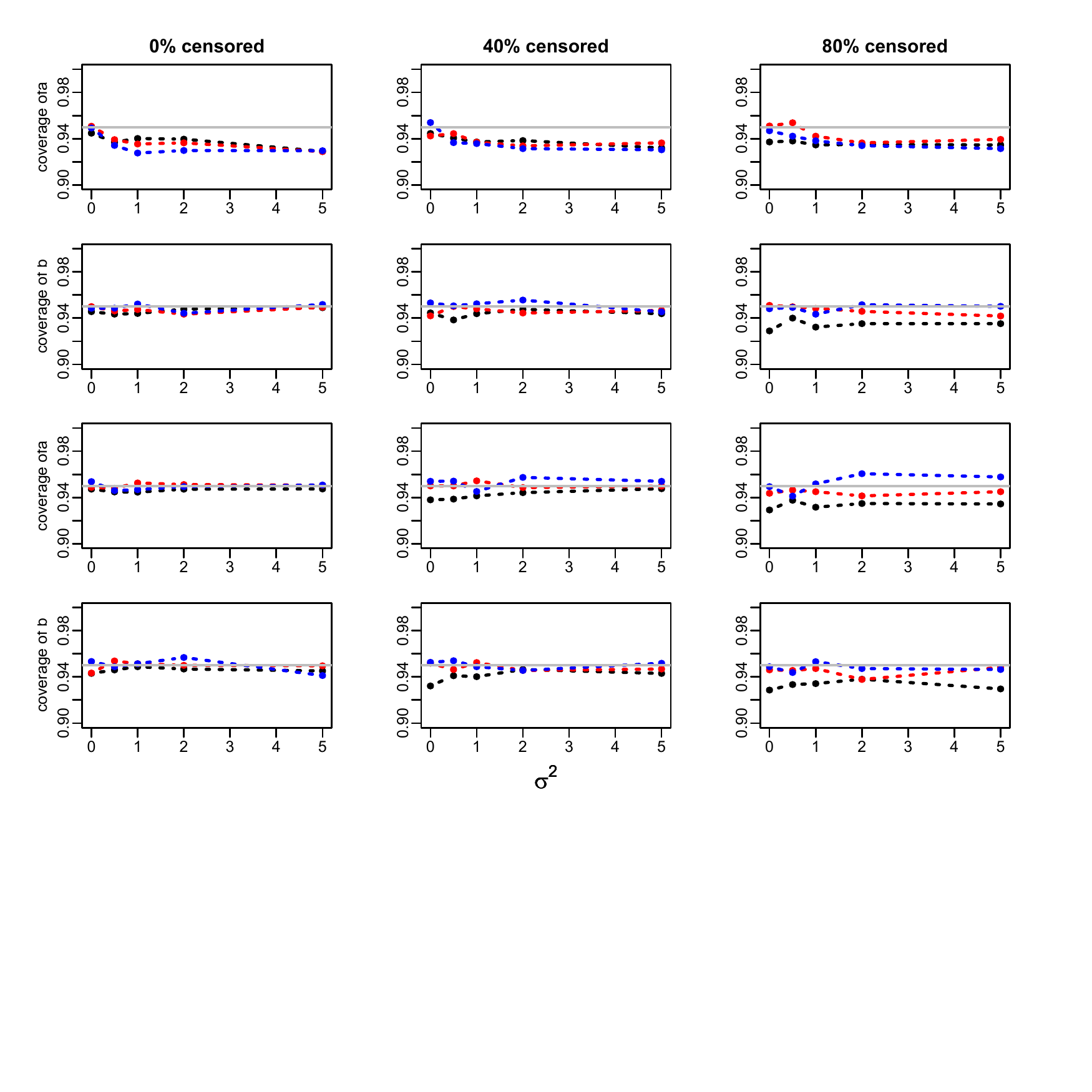}
	\caption{Coverage of the profile likelihood based confidence intervals for the $a$ and $b$ parameters  in the Gompertz model.
Success proportion= 0.5. Sample sizes: 300 (black), 1000 (red) and 10000 (blue).
True values: $\beta_{success-scale}$ = 0.5 , $\beta_{success-shape}$= -0.05 , $\beta_{score-scale}$= 1 , $\beta_{score-shape}$= - 0.1.
Top two rows: 10 clusters; bottom two rows: 100 clusters.}
	\label{CoveragePL10_100clustersGompertz6}
\end{figure}

\begin{figure}[ht]
	\centering
	\includegraphics[scale=1]{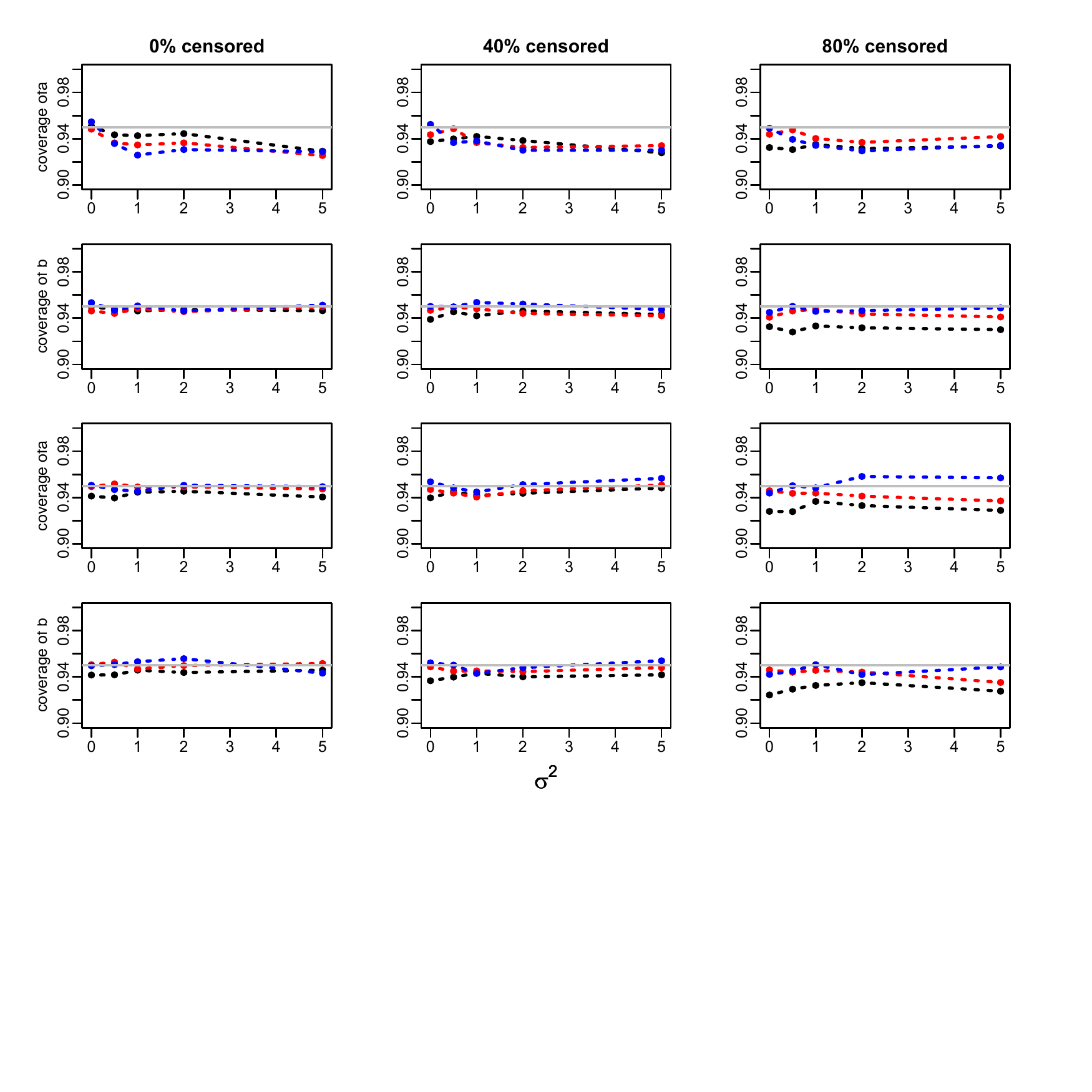}
	\caption{Coverage of the profile likelihood based confidence intervals for the $a$ and $b$ parameters in the Gompertz model.
Success proportion= 0.5.  Sample sizes: 300 (black), 1000 (red) and 10000 (blue).
True values: $\beta_{success-scale}$ = -0.5 , $\beta_{success-shape}$= 0.05 , $\beta_{score-scale}$= -1 , $\beta_{score-shape}$= 0.1.
Top two rows: 10 clusters; bottom two rows: 100 clusters.}
	\label{CoveragePL10_100clustersGompertz7}
\end{figure}

\begin{figure}[ht]
	\centering
	\includegraphics[scale=1]{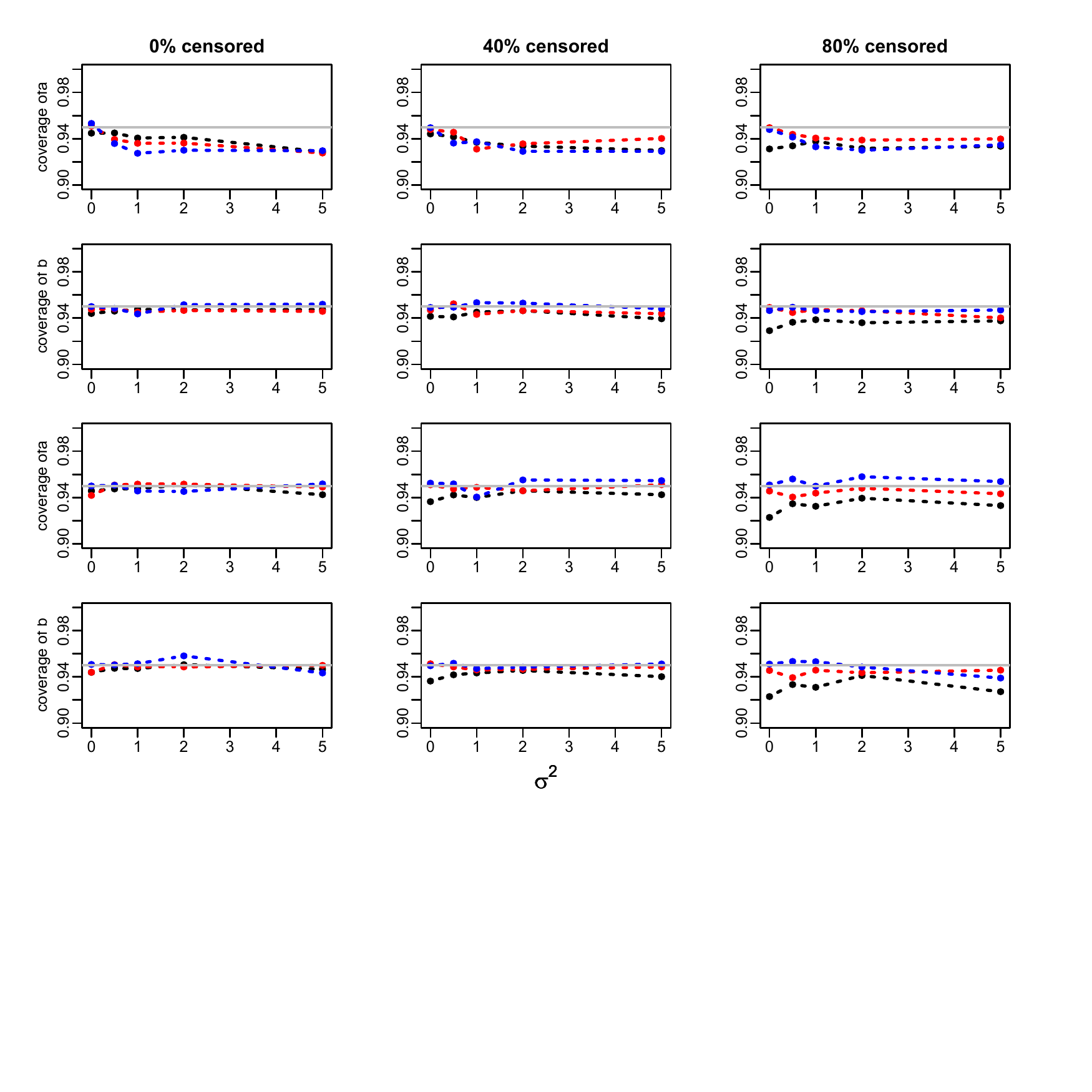}
	\caption{Coverage of the profile likelihood based confidence intervals for the $a$ and $b$ parameters  in the Gompertz model.
Success proportion= 0.5.  Sample sizes: 300 (black), 1000 (red) and 10000 (blue).
True values: $\beta_{success-scale}$ = -0.5 , $\beta_{success-shape}$= -0.05 , $\beta_{score-scale}$= -1 , $\beta_{score-shape}$= -0.1.
Top two rows: 10 clusters; bottom two rows: 100 clusters.}
	\label{CoveragePL10_100clustersGompertz8}
\end{figure}

\clearpage

\setcounter{figure}{0}
\setcounter{section}{0}
\renewcommand{\thefigure}{D.\arabic{figure}}

\section*{D: Coverage of the profile likelihood confidence intervals for the Cox regression parameters and  the frailty variance}

Each figure corresponds to a particular  baseline distribution  (Weibull or Gompertz), a value of the probability of success $p_{success}$ (= 0.25 or 0.5), a value for the number of clusters $N_{cl}$ (=10, 100) and a particular choice of the signs of the Cox regression parameters (+ + + +, + - + -, - + - + and - - - -).\\

The absolute values of the Cox regression parameters are held constant at $\beta_{success-scale}$ = 0.5 , $\beta_{success-shape}$= 0.05 , $\beta_{score-scale}$= 1 , $\beta_{score-shape}$= 0.1. 

For each combination of a censoring proportion  (= 0, 40\%, 80\%), a panel plots, versus the frailty variance $\sigma^2$ (= 0, 0.5, 1, 2, 3, 4, 5), the empirical coverage of the true value of a Cox regression parameter or the true value of $\sigma^2$ by a PL confidence interval at 95\% nominal level, for three sample sizes (300, 1000 and 10000) . \\

\clearpage


\begin{figure}[ht]
	\centering
	\includegraphics[scale=1]{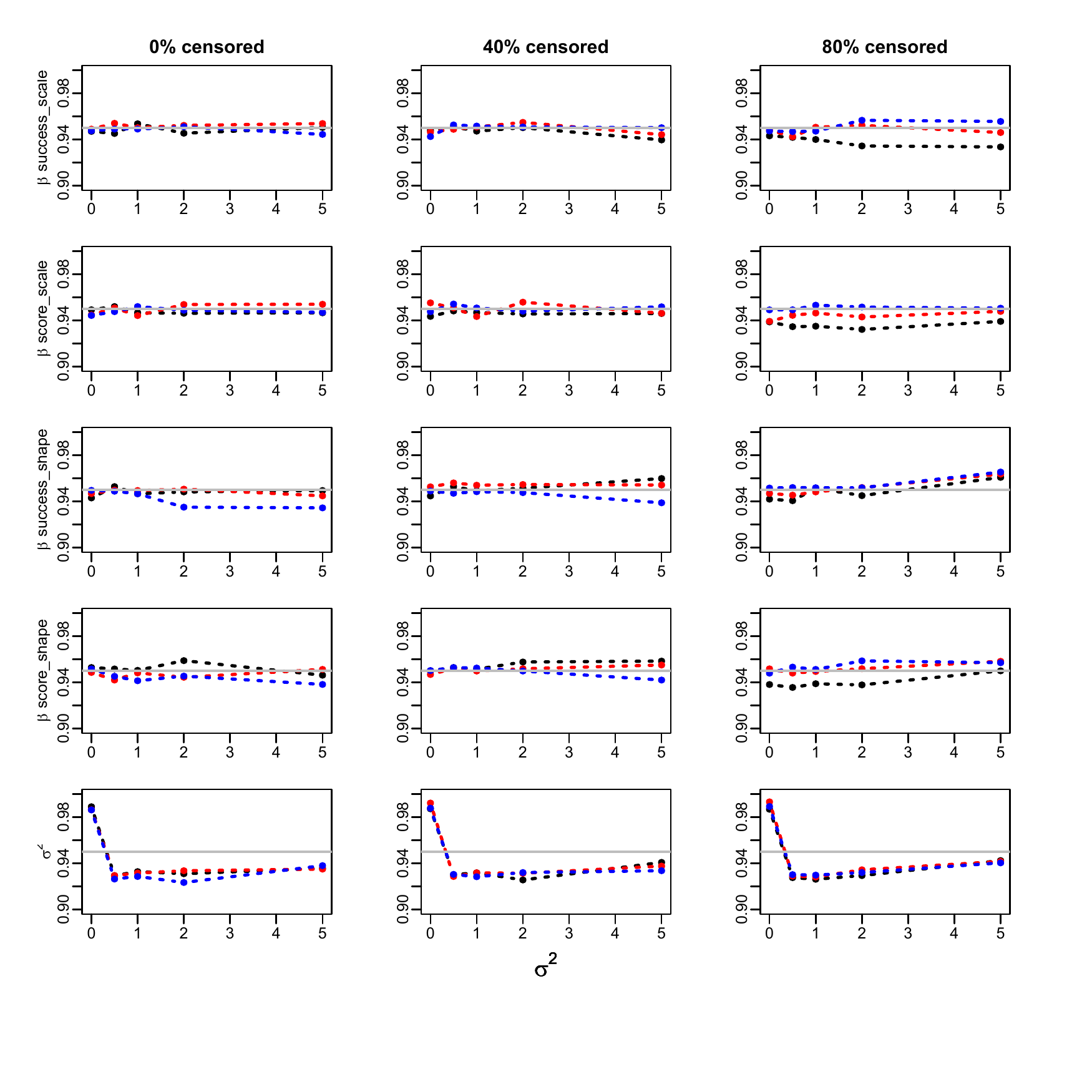}
	\caption{Coverage of the profile likelihood based confidence intervals for the Cox regression parameters and $\sigma^2$ at nominal 95\% level. Weibull model.
Success proportion= 0.25. 10 clusters. Sample sizes: 300 (black), 1000 (red) and 10000 (blue).
True values: $\beta_{success-scale}$ = 0.5 , $\beta_{success-shape}$= 0.05 , $\beta_{score-scale}$= 1 , $\beta_{score-shape}$= 0.1}
	\label{CoveragePL10clustersWeibull1}
\end{figure}

\begin{figure}[ht]
	\centering
	\includegraphics[scale=1]{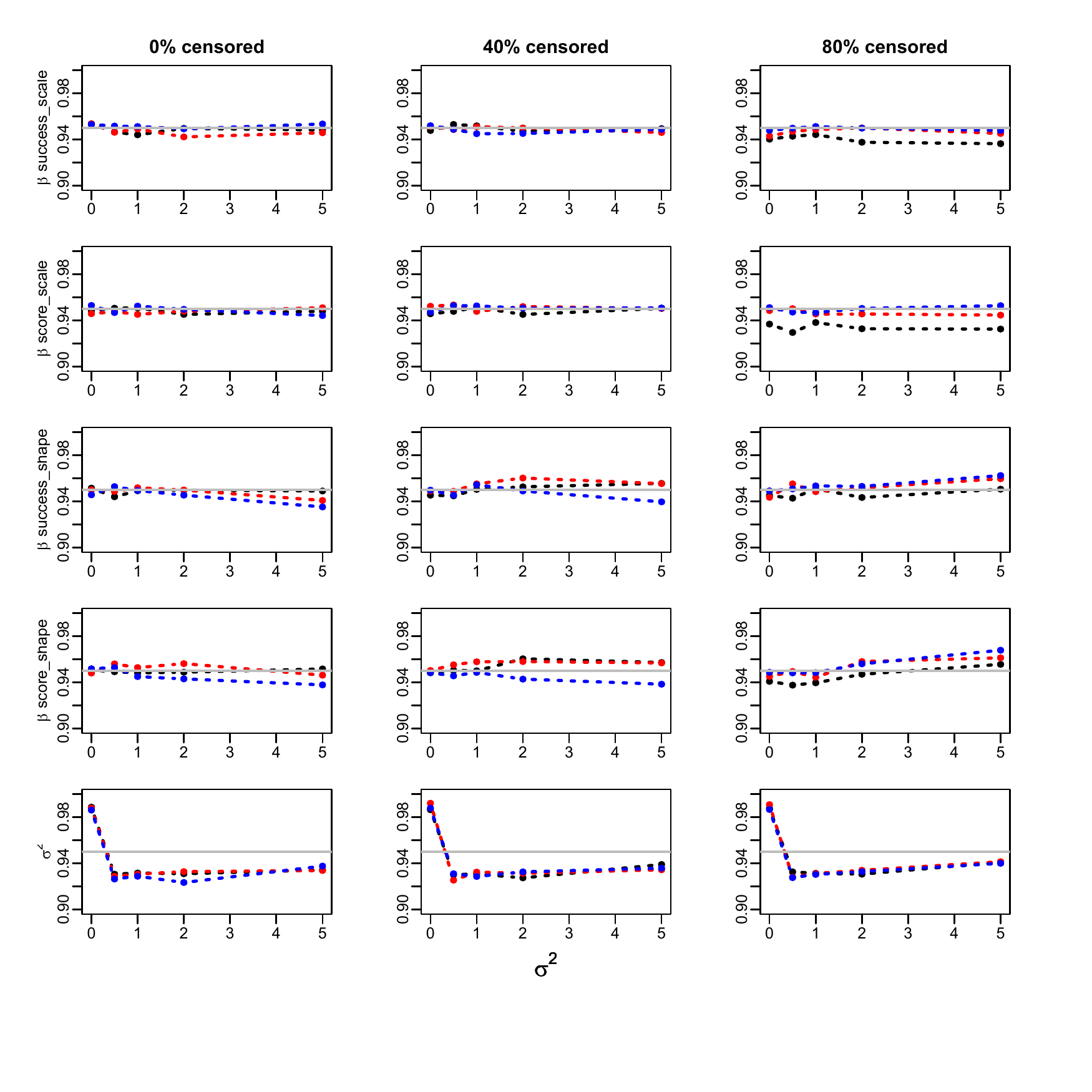}
	\caption{Coverage of the profile likelihood based confidence intervals for the Cox regression parameters and $\sigma^2$ at nominal 95\% level.  Weibull model.
Success proportion= 0.25. 10 clusters. Sample sizes: 300 (black), 1000 (red) and 10000 (blue).
True values: $\beta_{success-scale}$ = 0.5 , $\beta_{success-shape}$= -0.05 , $\beta_{score-scale}$= 1 , $\beta_{score-shape}$= - 0.1}
	\label{CoveragePL10clustersWeibull2}
\end{figure}

\begin{figure}[ht]
	\centering
	\includegraphics[scale=1]{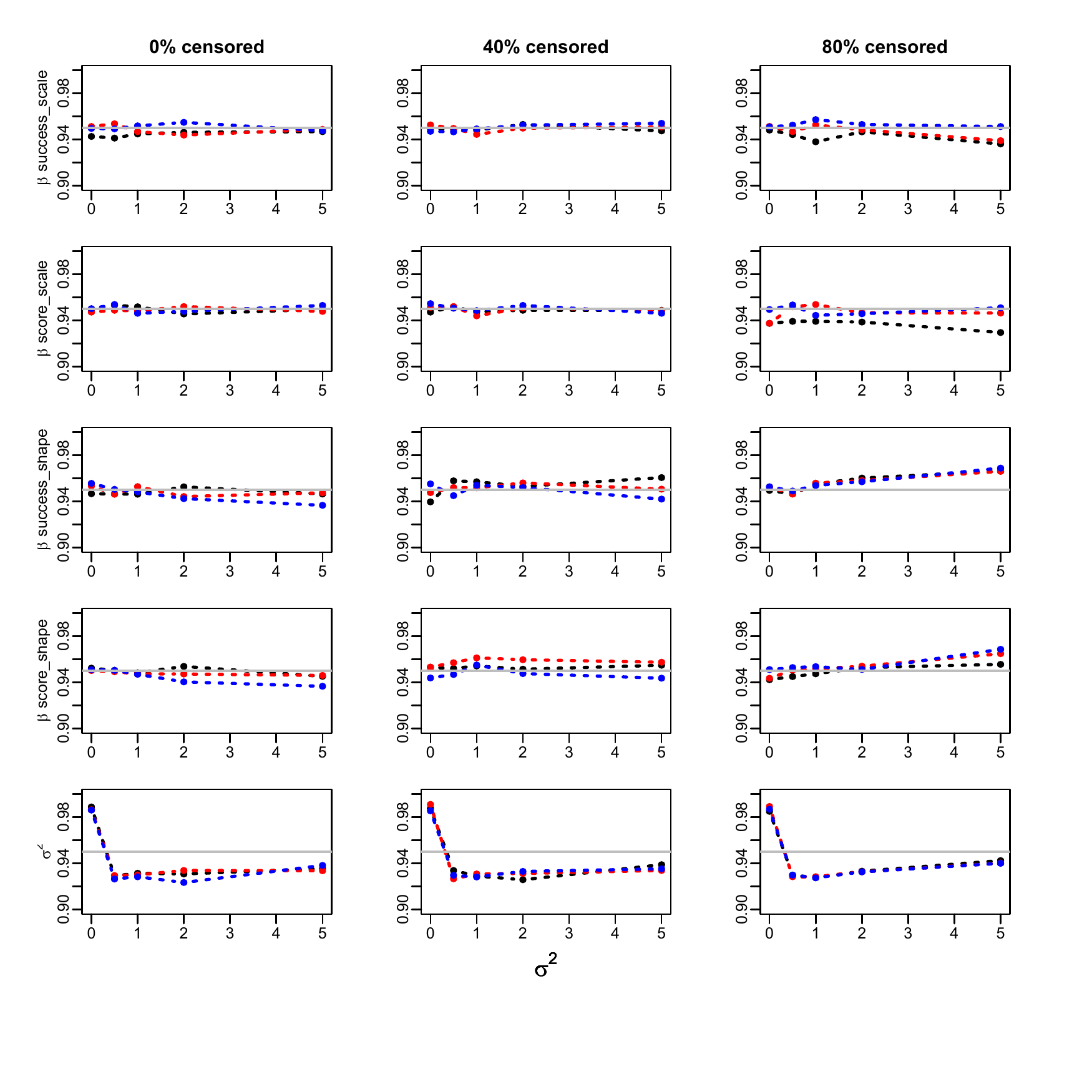}
	\caption{Coverage of the profile likelihood based confidence intervals for the Cox regression parameters and $\sigma^2$ at nominal 95\% level. Weibull model.
Success proportion= 0.25. 10 clusters. Sample sizes: 300 (black), 1000 (red) and 10000 (blue).
True values: $\beta_{success-scale}$ = -0.5 , $\beta_{success-shape}$= 0.05 , $\beta_{score-scale}$= -1 , $\beta_{score-shape}$= 0.1}
	\label{CoveragePL10clustersWeibull3}
\end{figure}

\begin{figure}[ht]
	\centering
	\includegraphics[scale=1]{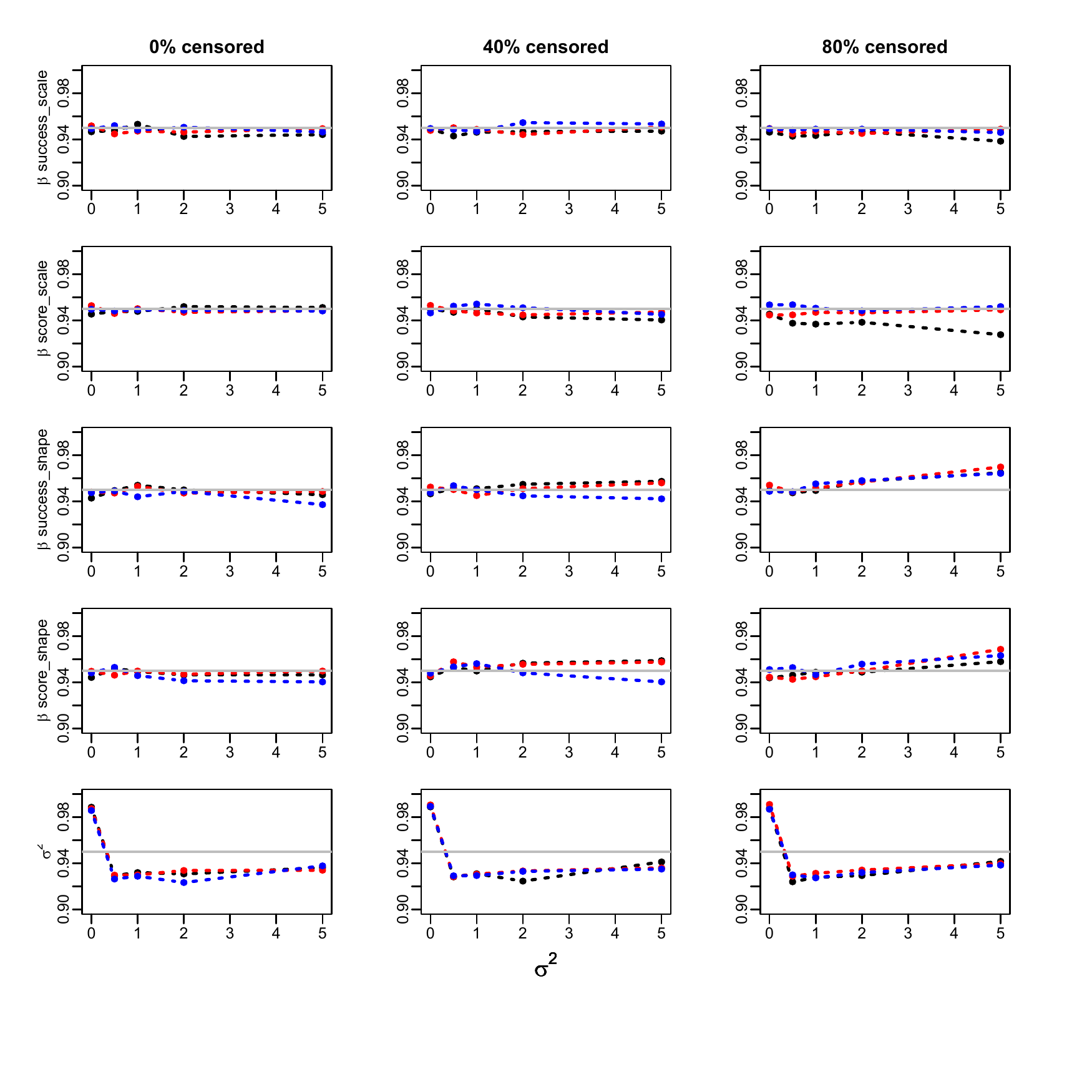}
	\caption{Coverage of the profile likelihood based confidence intervals for the Cox regression parameters and $\sigma^2$ at nominal 95\% level. Weibull model.
Success proportion= 0.25. 10 clusters. Sample sizes: 300 (black), 1000 (red) and 10000 (blue).
True values: $\beta_{success-scale}$ = -0.5 , $\beta_{success-shape}$= -0.05 , $\beta_{score-scale}$= -1 , $\beta_{score-shape}$= -0.1}
	\label{CoveragePL10clustersWeibull4}
\end{figure}

\begin{figure}[ht]
	\centering
	\includegraphics[scale=1]{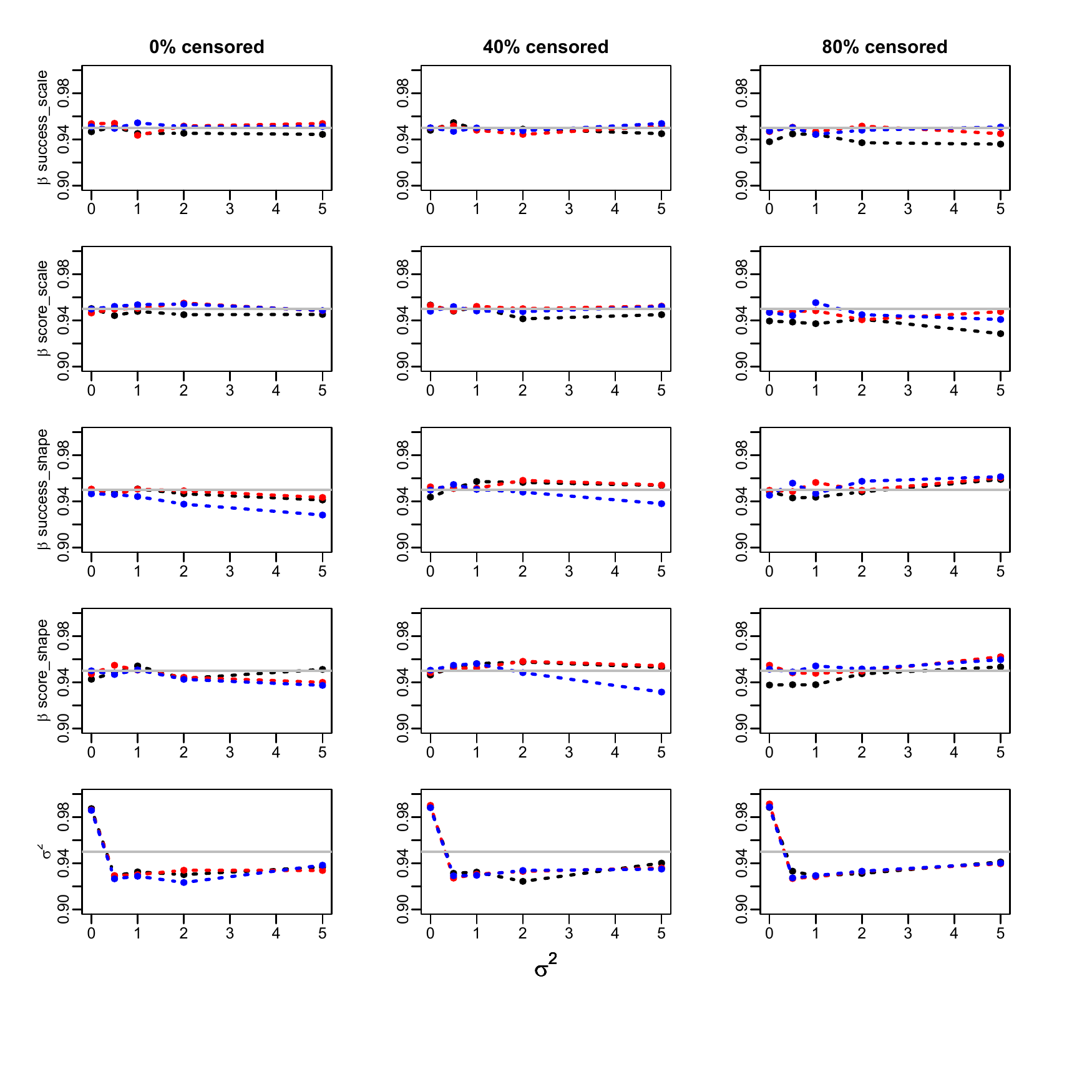}
	\caption{Coverage of the profile likelihood based confidence intervals for the Cox regression parameters and $\sigma^2$ at nominal 95\% level. Weibull model.
Success proportion= 0.5. 10 clusters. Sample sizes: 300 (black), 1000 (red) and 10000 (blue).
True values: $\beta_{success-scale}$ = 0.5 , $\beta_{success-shape}$= 0.05 , $\beta_{score-scale}$= 1 , $\beta_{score-shape}$= 0.1}
	\label{CoveragePL10clustersWeibull5}
\end{figure}

\begin{figure}[ht]
	\centering
	\includegraphics[scale=1]{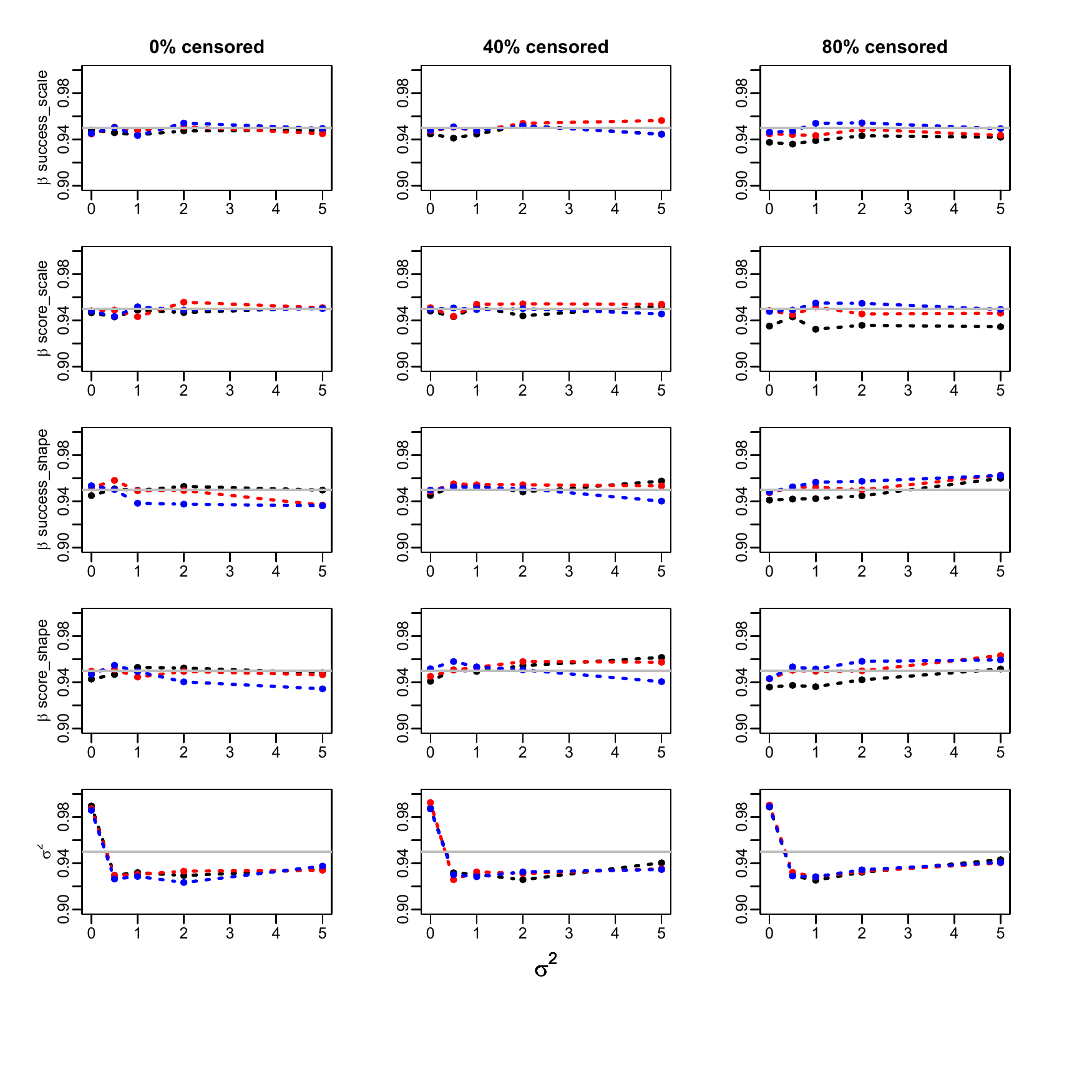}
	\caption{Coverage of the profile likelihood based confidence intervals for the Cox regression parameters and $\sigma^2$ at nominal 95\% level. Weibull model.
Success proportion= 0.5. 10 clusters. Sample sizes: 300 (black), 1000 (red) and 10000 (blue).
True values: $\beta_{success-scale}$ = 0.5 , $\beta_{success-shape}$= -0.05 , $\beta_{score-scale}$= 1 , $\beta_{score-shape}$= - 0.1}
	\label{CoveragePL10clustersWeibull6}
\end{figure}

\begin{figure}[ht]
	\centering
	\includegraphics[scale=1]{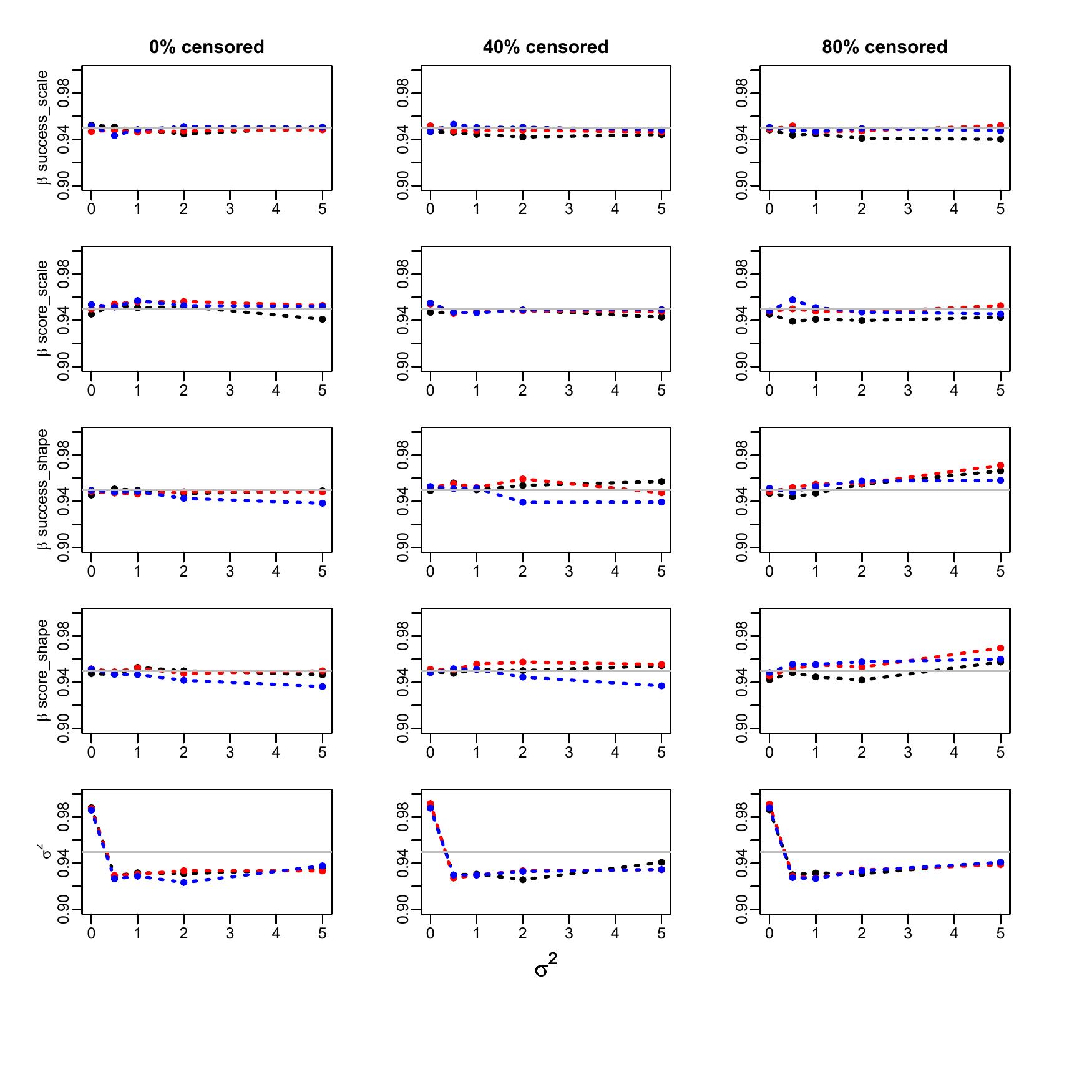}
	\caption{Coverage of the profile likelihood based confidence intervals for the Cox regression parameters and $\sigma^2$ at nominal 95\% level. Weibull model.
Success proportion= 0.5. 10 clusters. Sample sizes: 300 (black), 1000 (red) and 10000 (blue).
True values: $\beta_{success-scale}$ = -0.5 , $\beta_{success-shape}$= 0.05 , $\beta_{score-scale}$= -1 , $\beta_{score-shape}$= 0.1}
	\label{CoveragePL10clustersWeibull7}
\end{figure}

\begin{figure}[ht]
	\centering
	\includegraphics[scale=1]{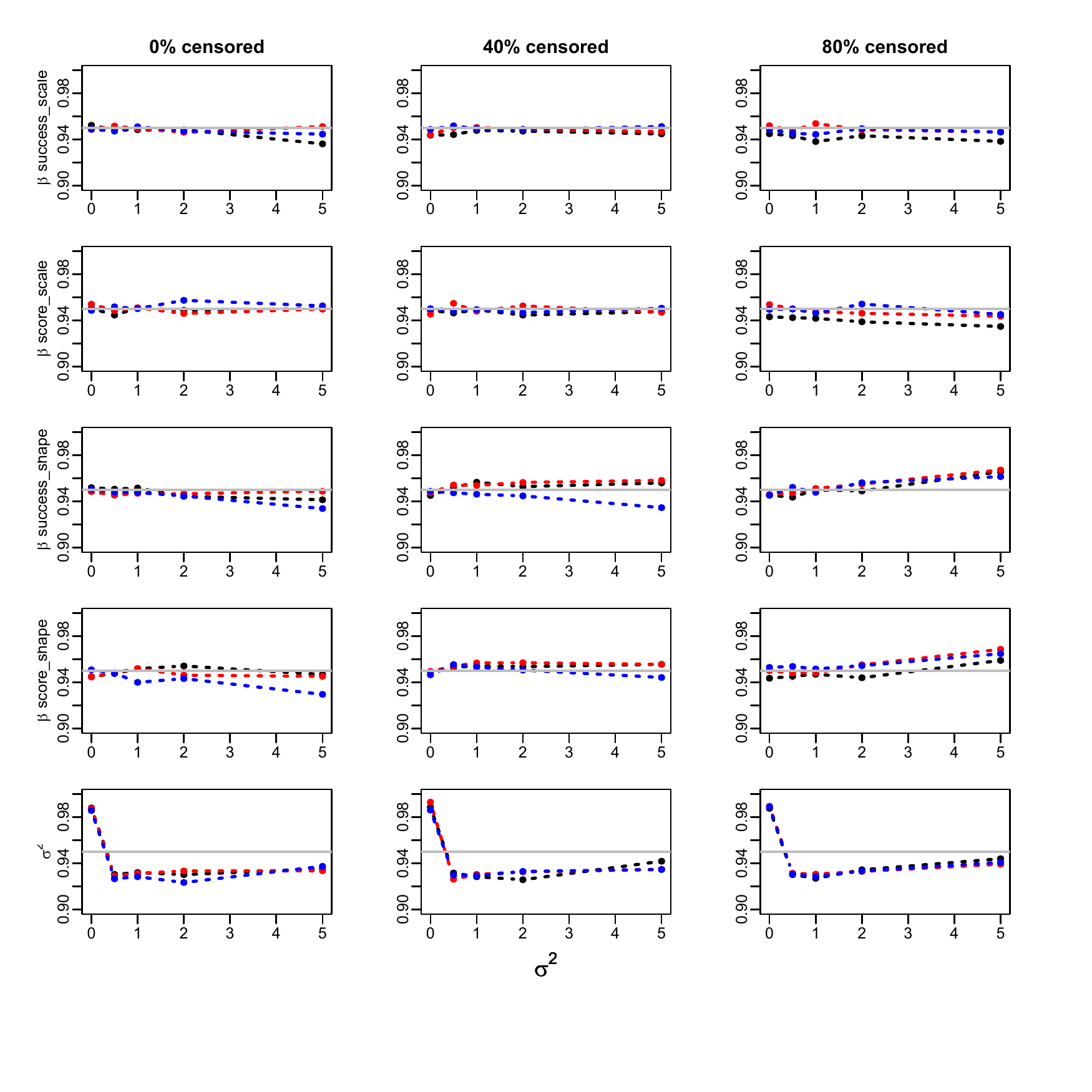}
	\caption{Coverage of the profile likelihood based confidence intervals for the Cox regression parameters and $\sigma^2$ at nominal 95\% level. Weibull model.
Success proportion= 0.5. 10 clusters. Sample sizes: 300 (black), 1000 (red) and 10000 (blue).
True values: $\beta_{success-scale}$ = -0.5 , $\beta_{success-shape}$= -0.05 , $\beta_{score-scale}$= -1 , $\beta_{score-shape}$= -0.1}
	\label{CoveragePL10clustersWeibull8}
\end{figure}

 \begin{figure}[ht]
	\centering
	\includegraphics[scale=1]{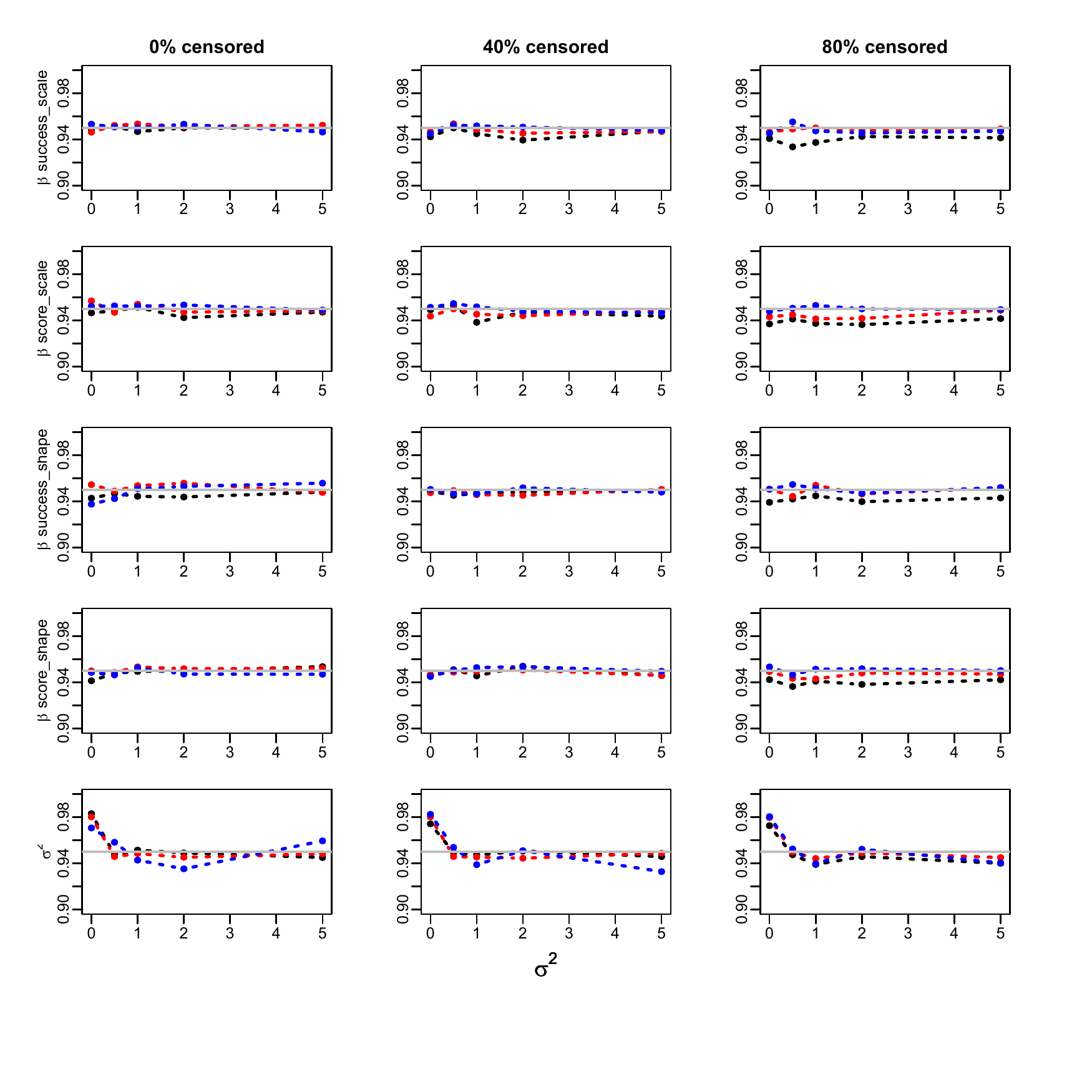}
	\caption{Coverage of the profile likelihood based confidence intervals for the Cox regression parameters and $\sigma^2$ at nominal 95\% level. Weibull model.
Success proportion= 0.25. 100 clusters. Sample sizes: 300 (black), 1000 (red) and 10000 (blue).
True values: $\beta_{success-scale}$ = 0.5 , $\beta_{success-shape}$= 0.05 , $\beta_{score-scale}$= 1 , $\beta_{score-shape}$= 0.1}
	\label{CoveragePL100clustersWeibull1}
\end{figure}

\begin{figure}[ht]
	\centering
	\includegraphics[scale=1]{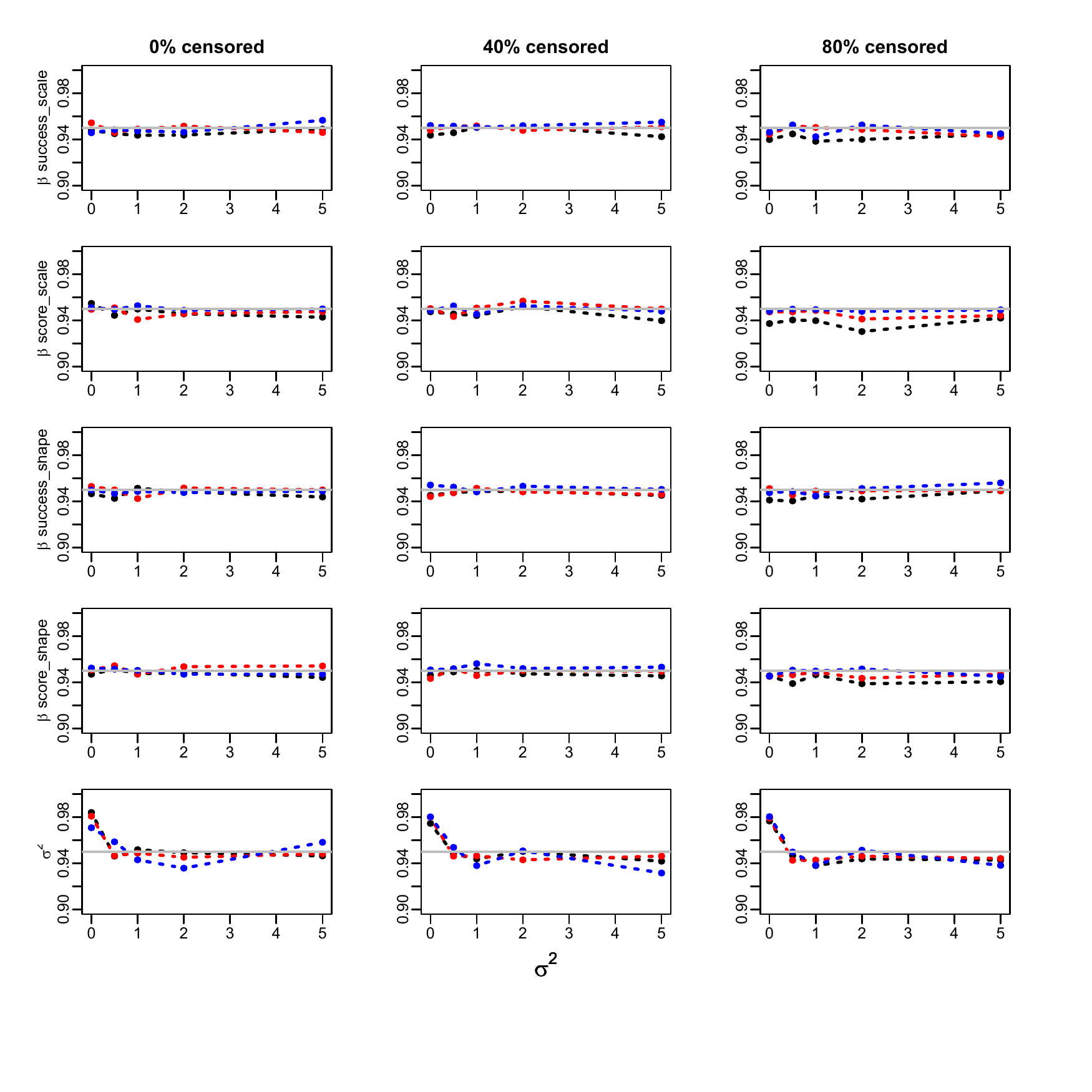}
	\caption{Coverage of the profile likelihood based confidence intervals for the Cox regression parameters and $\sigma^2$ at nominal 95\% level. Weibull model.
Success proportion= 0.25. 100 clusters. Sample sizes: 300 (black), 1000 (red) and 10000 (blue).
True values: $\beta_{success-scale}$ = 0.5 , $\beta_{success-shape}$= -0.05 , $\beta_{score-scale}$= 1 , $\beta_{score-shape}$= - 0.1}
	\label{CoveragePL100clustersWeibull2}
\end{figure}

\begin{figure}[ht]
	\centering
	\includegraphics[scale=1]{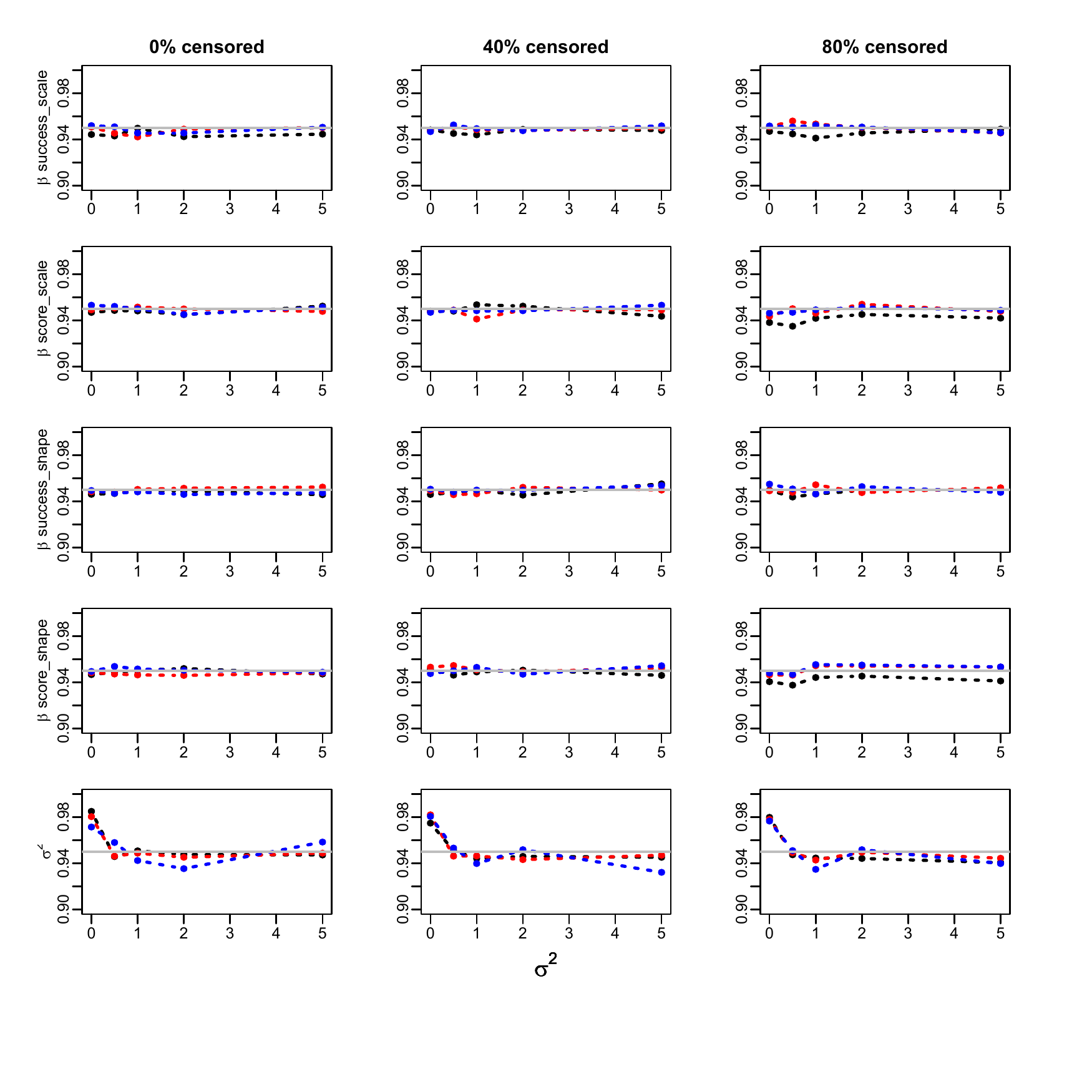}
	\caption{Coverage of the profile likelihood based confidence intervals for the Cox regression parameters and $\sigma^2$ at nominal 95\% level. Weibull model.
Success proportion= 0.25. 100 clusters. Sample sizes: 300 (black), 1000 (red) and 10000 (blue).
True values: $\beta_{success-scale}$ = -0.5 , $\beta_{success-shape}$= 0.05 , $\beta_{score-scale}$= -1 , $\beta_{score-shape}$= 0.1}
	\label{CoveragePL100clustersWeibull3}
\end{figure}

\begin{figure}[ht]
	\centering
	\includegraphics[scale=1]{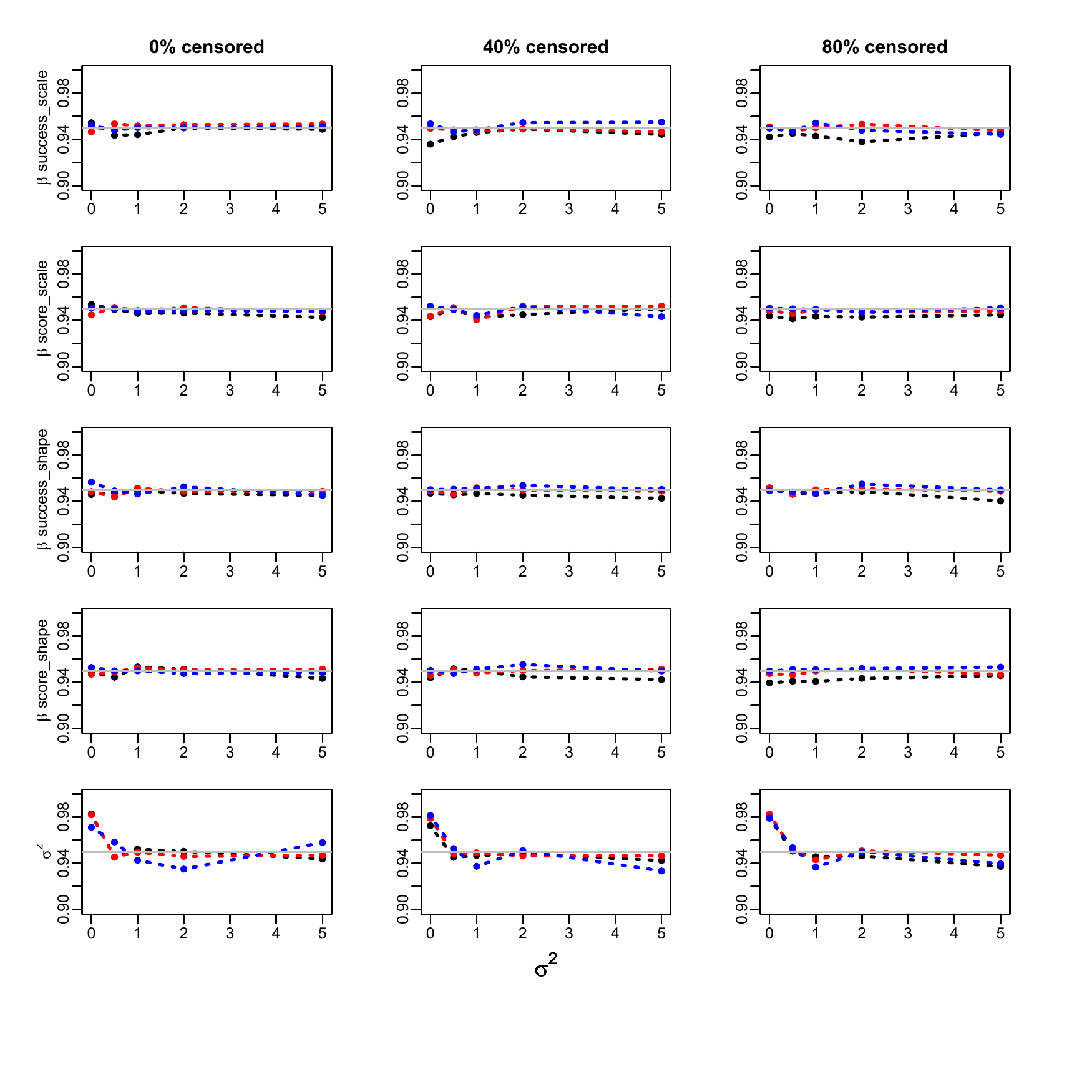}
	\caption{Coverage of the profile likelihood based confidence intervals for the Cox regression parameters and $\sigma^2$ at nominal 95\% level. Weibull model.
Success proportion= 0.25. 100 clusters. Sample sizes: 300 (black), 1000 (red) and 10000 (blue).
True values: $\beta_{success-scale}$ = -0.5 , $\beta_{success-shape}$= -0.05 , $\beta_{score-scale}$= -1 , $\beta_{score-shape}$= -0.1}
	\label{CoveragePL100clustersWeibull4}
\end{figure}

\begin{figure}[ht]
	\centering
	\includegraphics[scale=1]{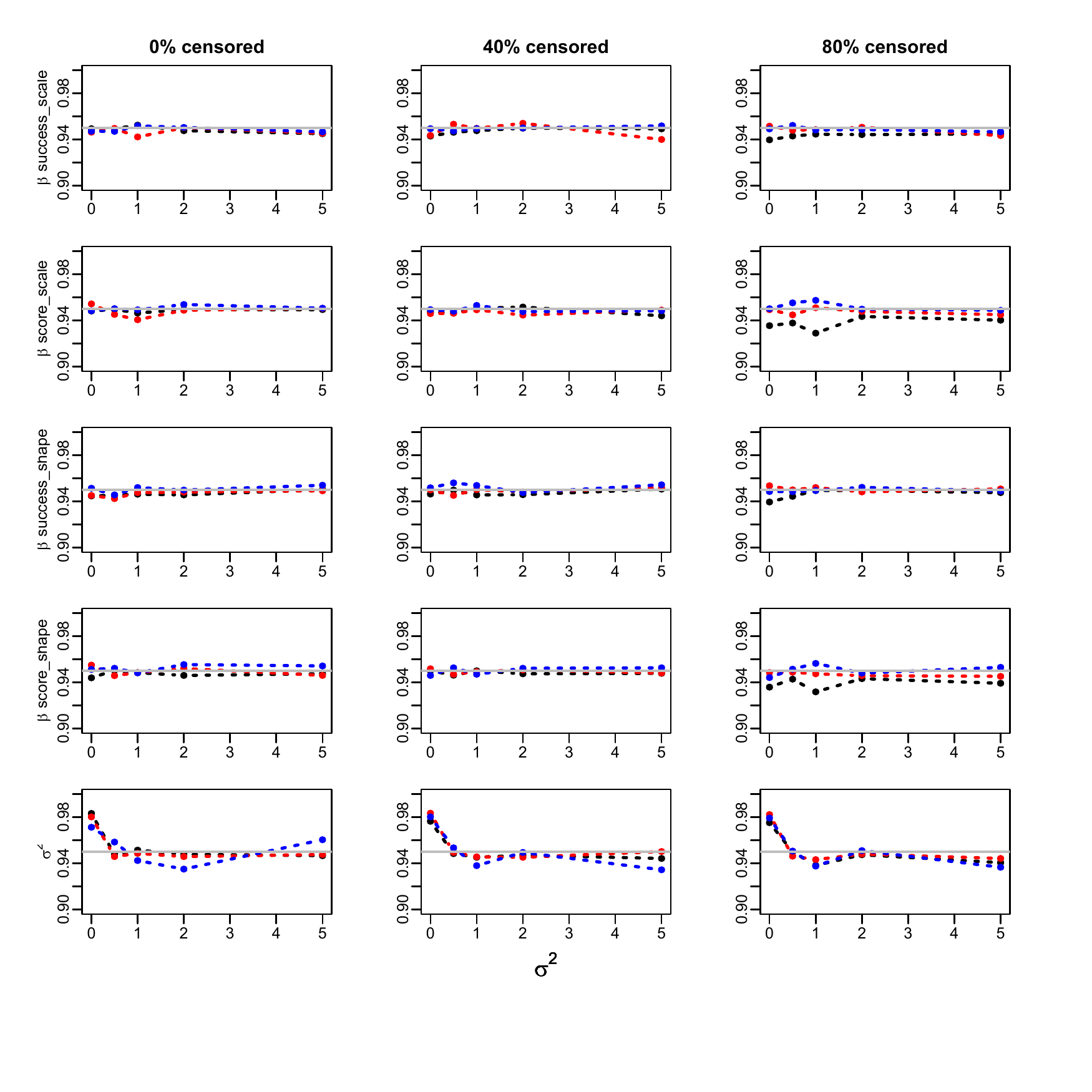}
	\caption{Coverage of the profile likelihood based confidence intervals for the Cox regression parameters and $\sigma^2$ at nominal 95\% level. Weibull model.
Success proportion= 0.5. 100 clusters. Sample sizes: 300 (black), 1000 (red) and 10000 (blue).
True values: $\beta_{success-scale}$ = 0.5 , $\beta_{success-shape}$= 0.05 , $\beta_{score-scale}$= 1 , $\beta_{score-shape}$= 0.1}
	\label{CoveragePL100clustersWeibull5}
\end{figure}

\begin{figure}[ht]
	\centering
	\includegraphics[scale=1]{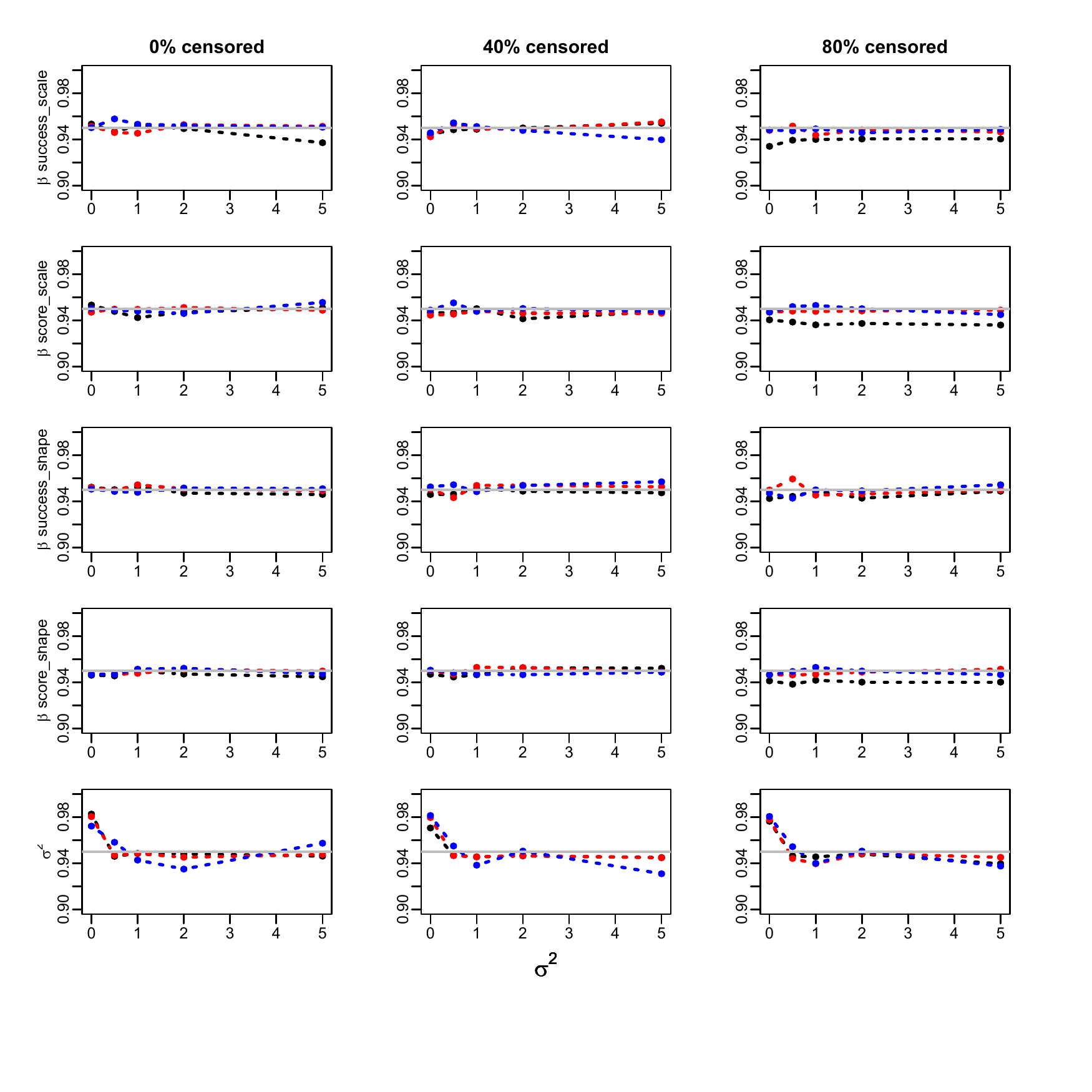}
	\caption{Coverage of the profile likelihood based confidence intervals for the Cox regression parameters and $\sigma^2$ at nominal 95\% level. Weibull model.
Success proportion= 0.5. 100 clusters. Sample sizes: 300 (black), 1000 (red) and 10000 (blue).
True values: $\beta_{success-scale}$ = 0.5 , $\beta_{success-shape}$= -0.05 , $\beta_{score-scale}$= 1 , $\beta_{score-shape}$= - 0.1}
	\label{CoveragePL100clustersWeibull6}
\end{figure}

\begin{figure}[ht]
	\centering
	\includegraphics[scale=1]{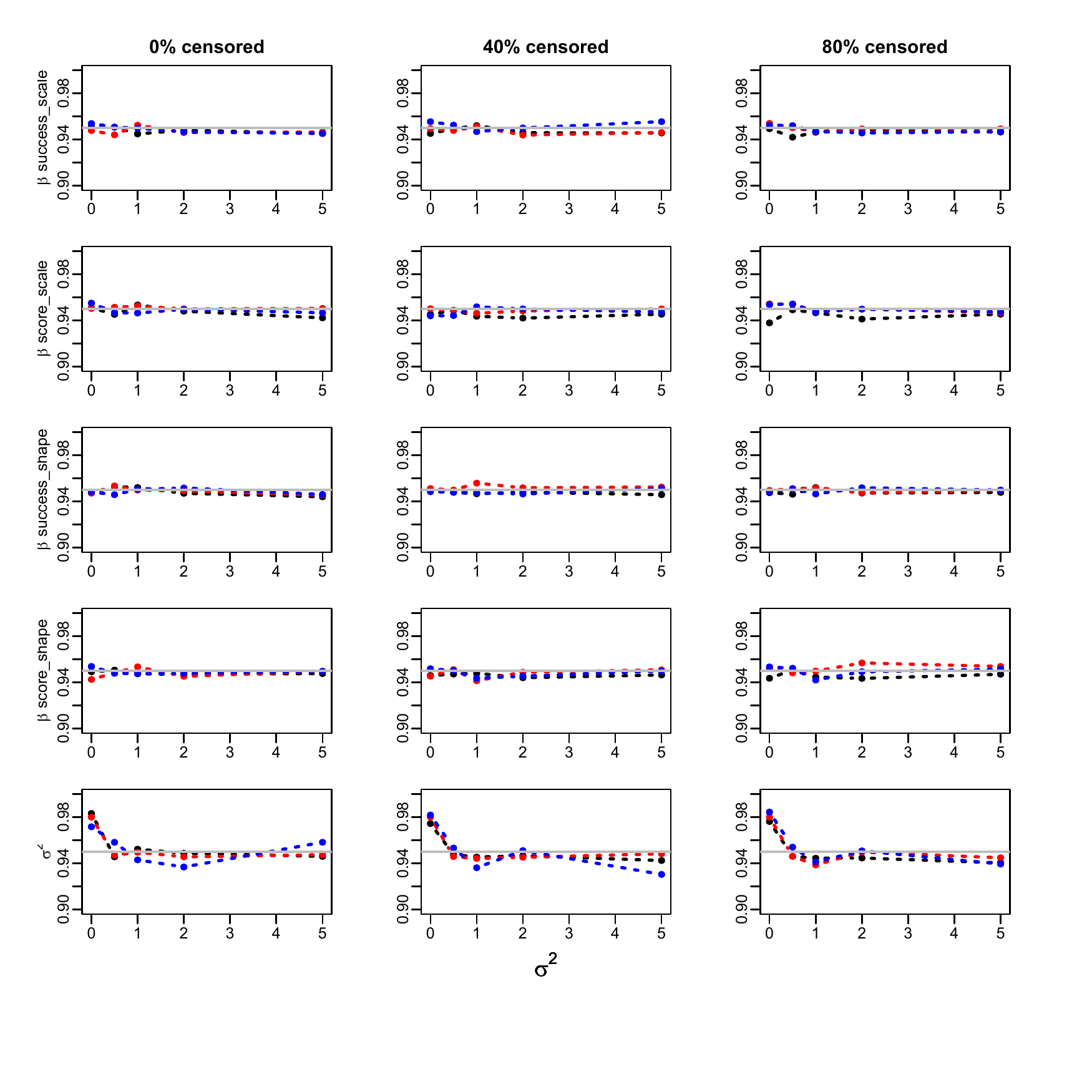}
	\caption{Coverage of the profile likelihood based confidence intervals for the Cox regression parameters and $\sigma^2$ at nominal 95\% level. Weibull model.
Success proportion= 0.5. 100 clusters. Sample sizes: 300 (black), 1000 (red) and 10000 (blue).
True values: $\beta_{success-scale}$ = -0.5 , $\beta_{success-shape}$= 0.05 , $\beta_{score-scale}$= -1 , $\beta_{score-shape}$= 0.1}
	\label{CoveragePL100clustersWeibull7}
\end{figure}

\begin{figure}[ht]
	\centering
	\includegraphics[scale=1]{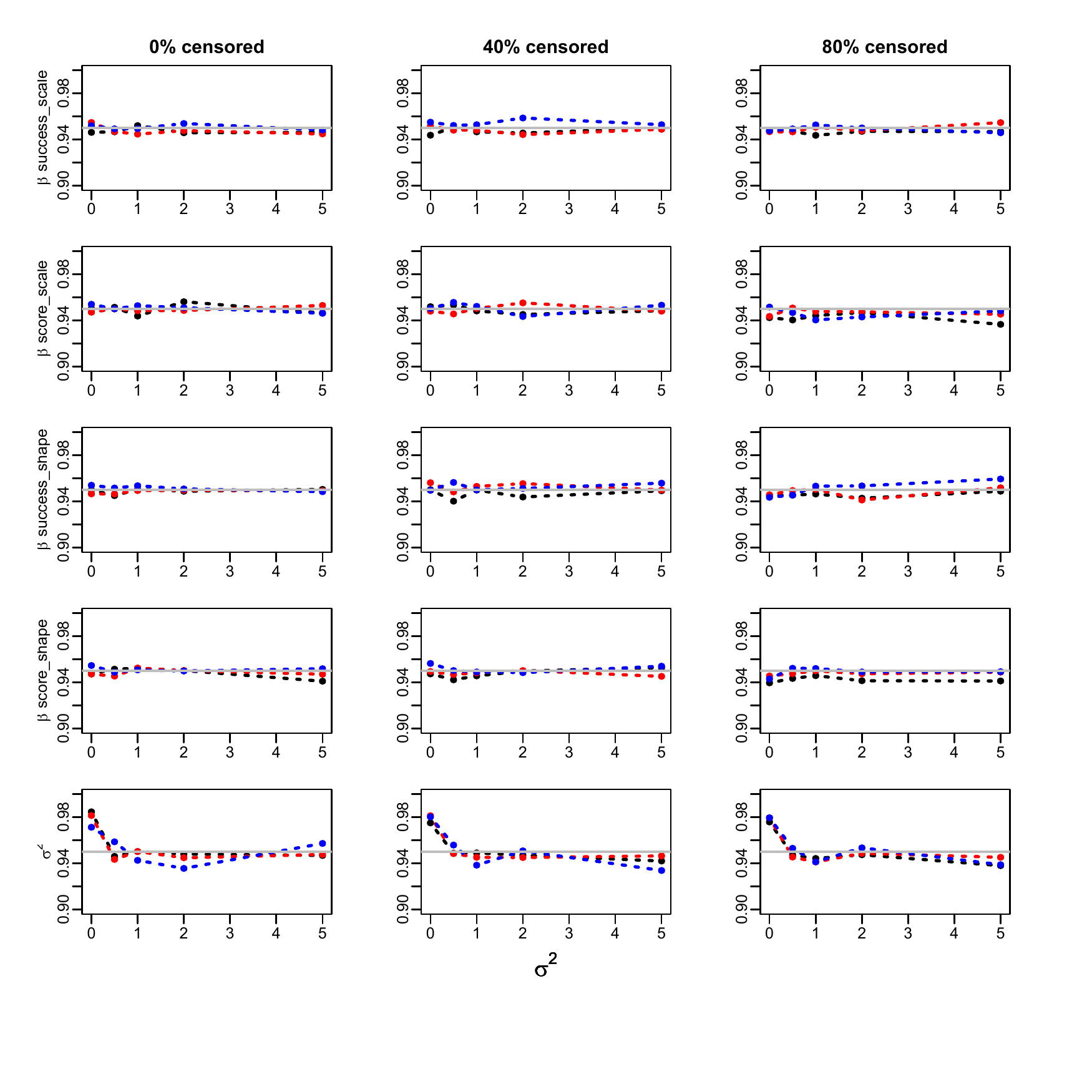}
	\caption{Coverage of the profile likelihood based confidence intervals for the Cox regression parameters and $\sigma^2$ at nominal 95\% level. Weibull model.
Success proportion= 0.5. 100 clusters. Sample sizes: 300 (black), 1000 (red) and 10000 (blue).
True values: $\beta_{success-scale}$ = -0.5 , $\beta_{success-shape}$= -0.05 , $\beta_{score-scale}$= -1 , $\beta_{score-shape}$= -0.1}
	\label{CoveragePL100clustersWeibull8}
\end{figure}


 \begin{figure}[ht]
	\centering
	\includegraphics[scale=1]{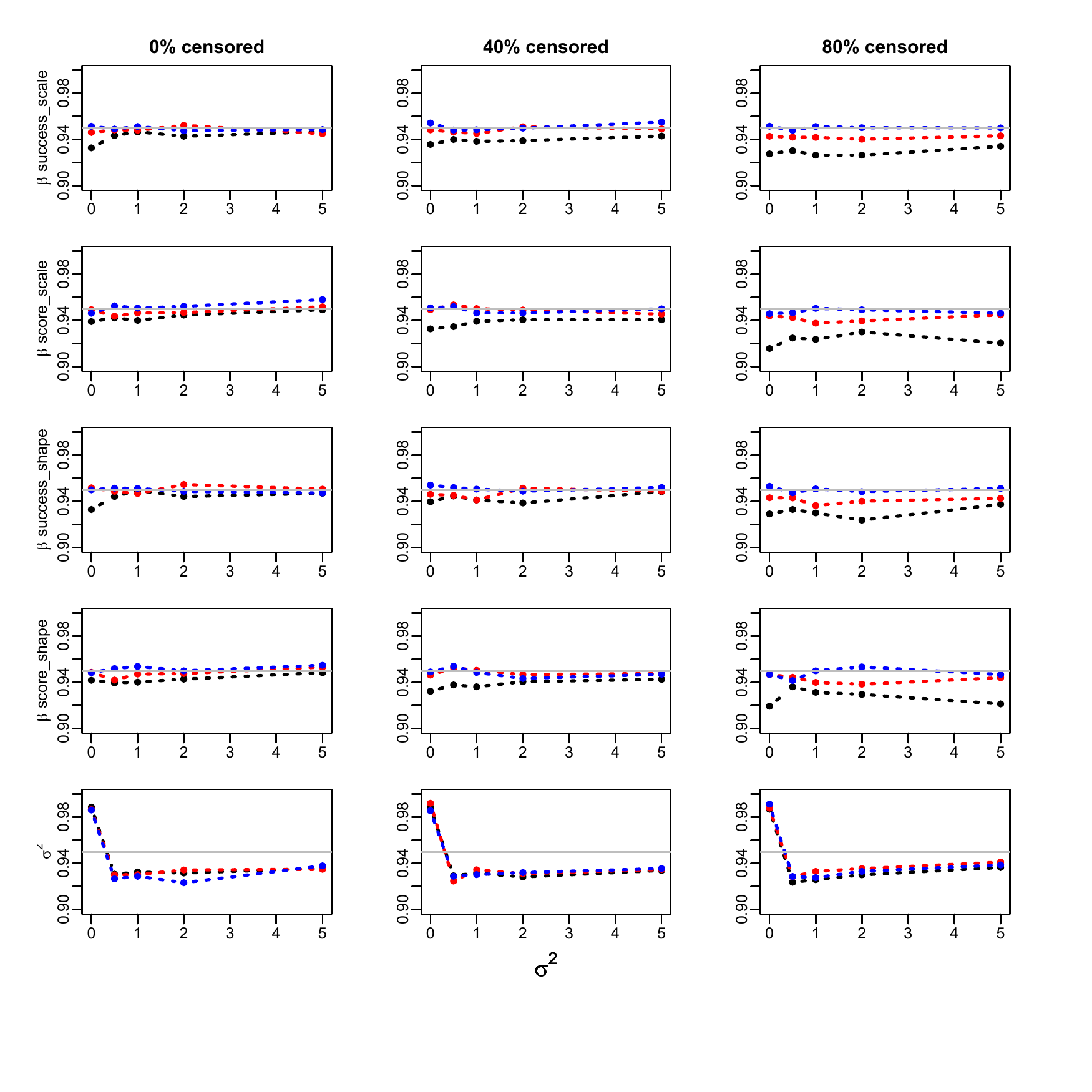}
	\caption{Coverage of the profile likelihood based confidence intervals for the Cox regression parameters and $\sigma^2$ at nominal 95\% level. Gompertz model.
Success proportion= 0.25. 10 clusters. Sample sizes: 300 (black), 1000 (red) and 10000 (blue).
True values: $\beta_{success-scale}$ = 0.5 , $\beta_{success-shape}$= 0.05 , $\beta_{score-scale}$= 1 , $\beta_{score-shape}$= 0.1}
	\label{CoveragePL10clustersGompertz1}
\end{figure}

\begin{figure}[ht]
	\centering
	\includegraphics[scale=1]{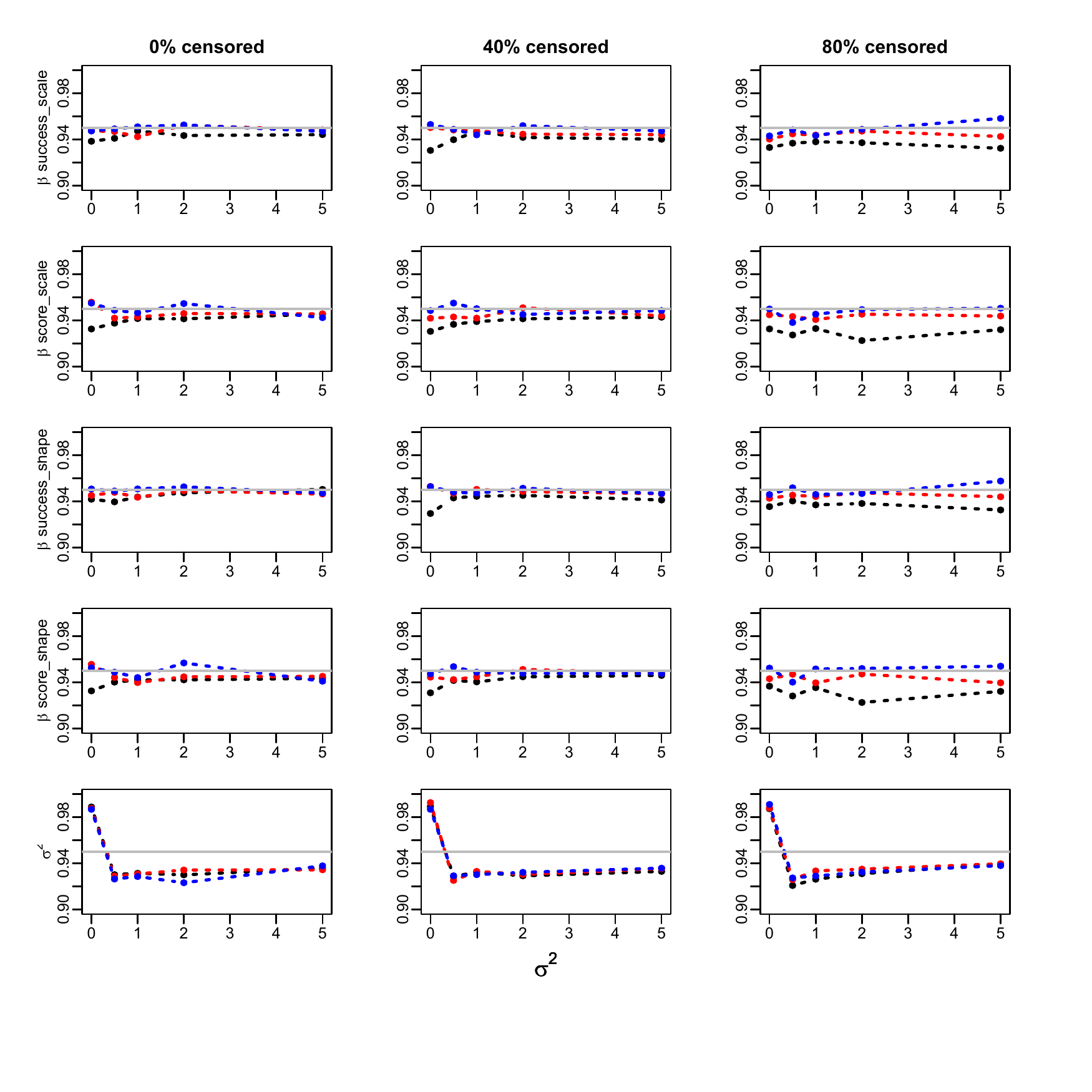}
	\caption{Coverage of the profile likelihood based confidence intervals for the Cox regression parameters and $\sigma^2$ at nominal 95\% level. Gompertz model.
Success proportion= 0.25. 10 clusters. Sample sizes: 300 (black), 1000 (red) and 10000 (blue).
True values: $\beta_{success-scale}$ = 0.5 , $\beta_{success-shape}$= -0.05 , $\beta_{score-scale}$= 1 , $\beta_{score-shape}$= - 0.1}
	\label{CoveragePL10clustersGompertz2}
\end{figure}

\begin{figure}[ht]
	\centering
	\includegraphics[scale=1]{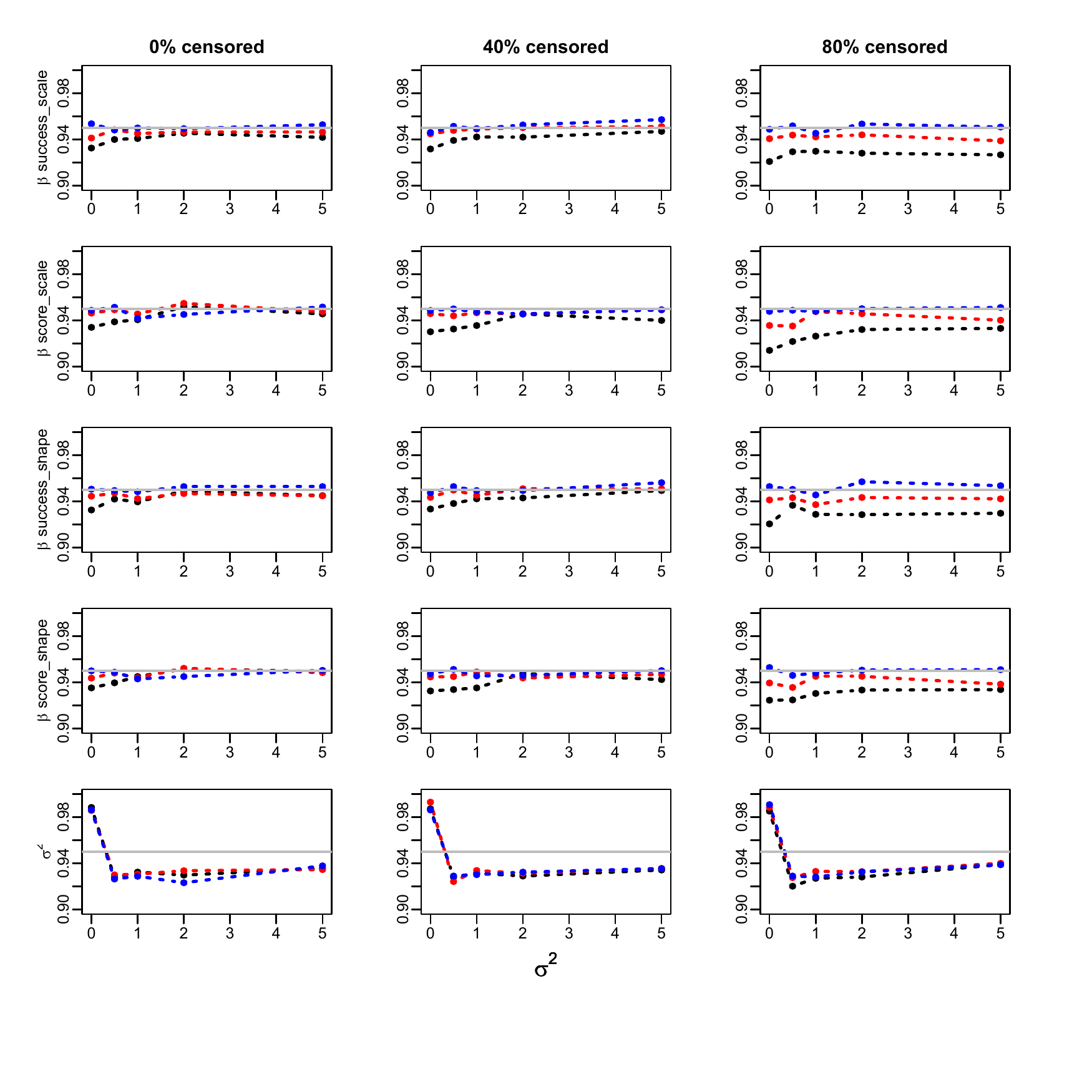}
	\caption{Coverage of the profile likelihood based confidence intervals for the Cox regression parameters and $\sigma^2$ at nominal 95\% level. the Gompertz model.
Success proportion= 0.25. 10 clusters. Sample sizes: 300 (black), 1000 (red) and 10000 (blue).
True values: $\beta_{success-scale}$ = -0.5 , $\beta_{success-shape}$= 0.05 , $\beta_{score-scale}$= -1 , $\beta_{score-shape}$= 0.1}
	\label{CoveragePL10clustersGompertz3}
\end{figure}

\begin{figure}[ht]
	\centering
	\includegraphics[scale=1]{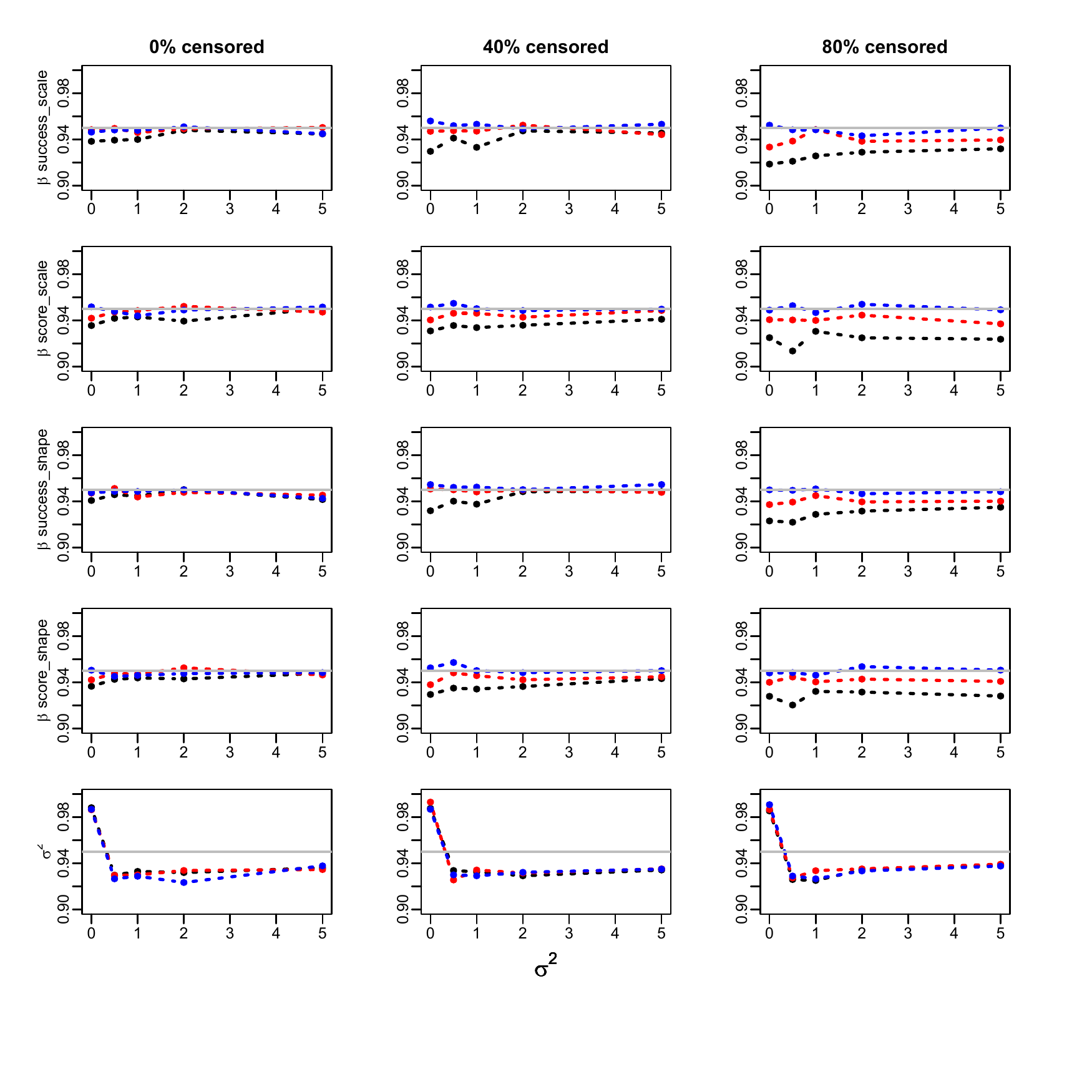}
	\caption{Coverage of the profile likelihood based confidence intervals for the Cox regression parameters and $\sigma^2$ at nominal 95\% level. the Gompertz model.
Success proportion= 0.25. 10 clusters. Sample sizes: 300 (black), 1000 (red) and 10000 (blue).
True values: $\beta_{success-scale}$ = -0.5 , $\beta_{success-shape}$= -0.05 , $\beta_{score-scale}$= -1 , $\beta_{score-shape}$= -0.1}
	\label{CoveragePL10clustersGompertz4}
\end{figure}

\begin{figure}[ht]
	\centering
	\includegraphics[scale=1]{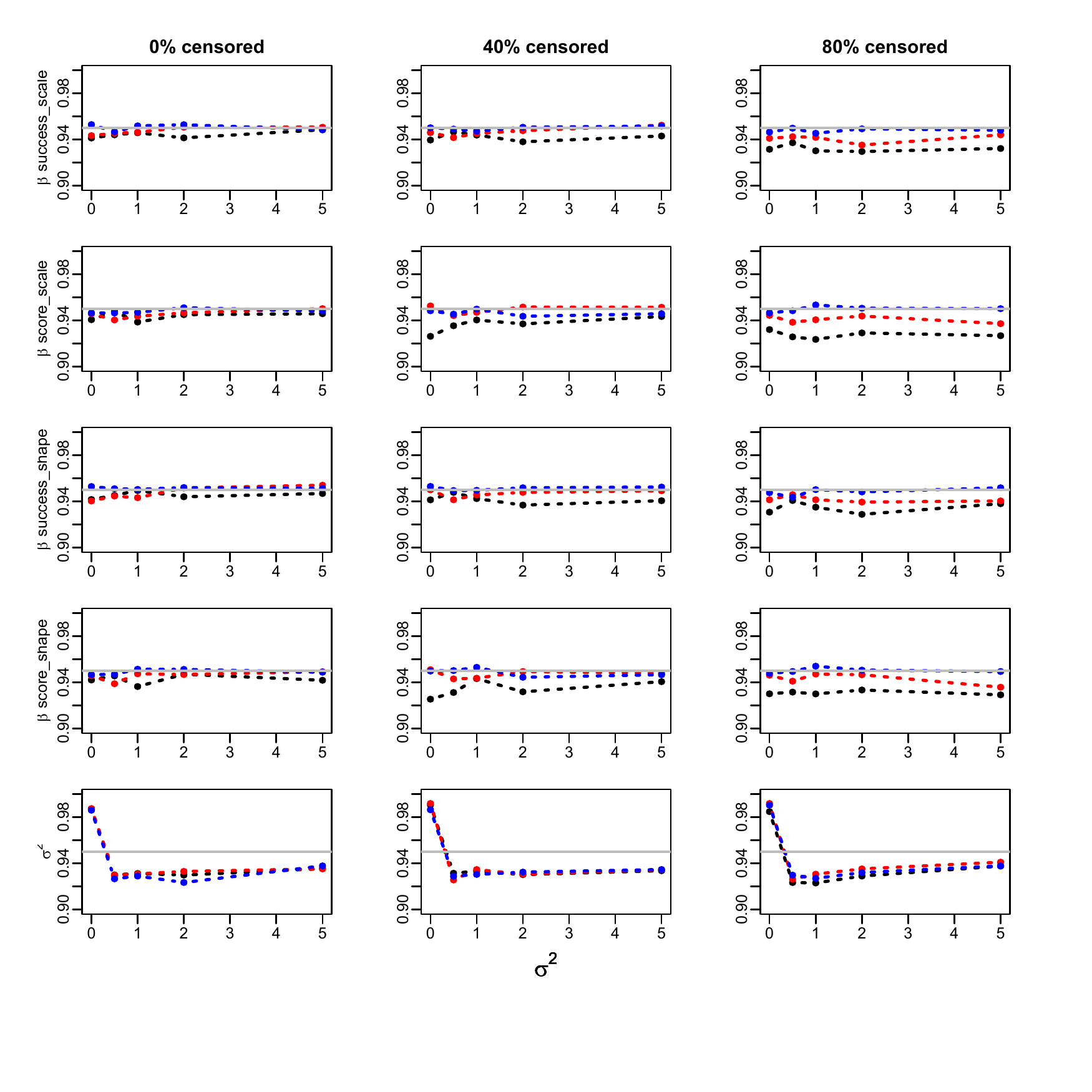}
	\caption{Coverage of the profile likelihood based confidence intervals for the Cox regression parameters and $\sigma^2$ at nominal 95\% level. Gompertz model.
Success proportion= 0.5. 10 clusters. Sample sizes: 300 (black), 1000 (red) and 10000 (blue).
True values: $\beta_{success-scale}$ = 0.5 , $\beta_{success-shape}$= 0.05 , $\beta_{score-scale}$= 1 , $\beta_{score-shape}$= 0.1}
	\label{CoveragePL10clustersGompertz5}
\end{figure}

\begin{figure}[ht]
	\centering
	\includegraphics[scale=1]{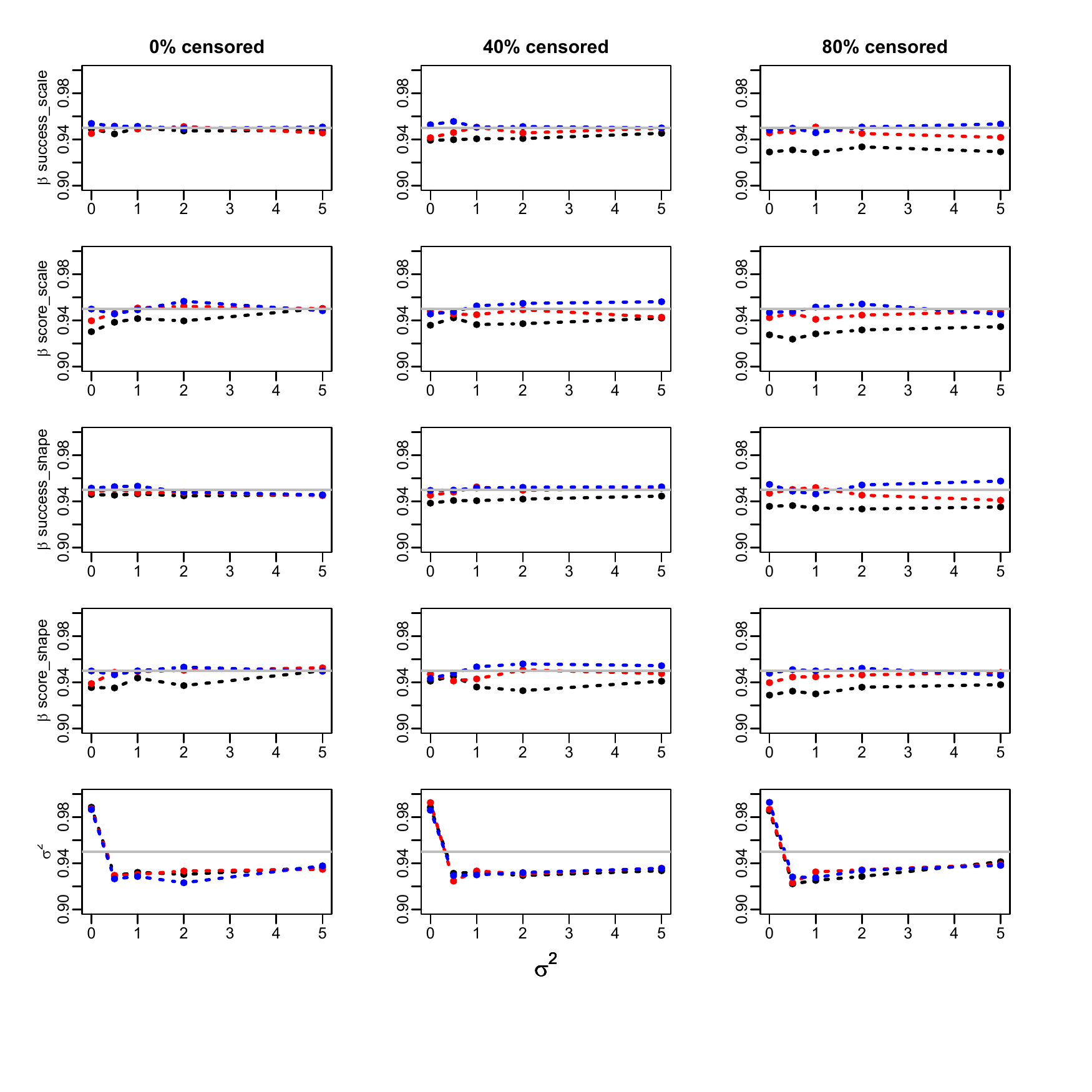}
	\caption{Coverage of the profile likelihood based confidence intervals for the Cox regression parameters and $\sigma^2$ at nominal 95\% level. Weibull model.
Success proportion= 0.5. 10 clusters. Sample sizes: 300 (black), 1000 (red) and 10000 (blue).
True values: $\beta_{success-scale}$ = 0.5 , $\beta_{success-shape}$= -0.05 , $\beta_{score-scale}$= 1 , $\beta_{score-shape}$= - 0.1}
	\label{CoveragePL10clustersGompertz6}
\end{figure}

\begin{figure}[ht]
	\centering
	\includegraphics[scale=1]{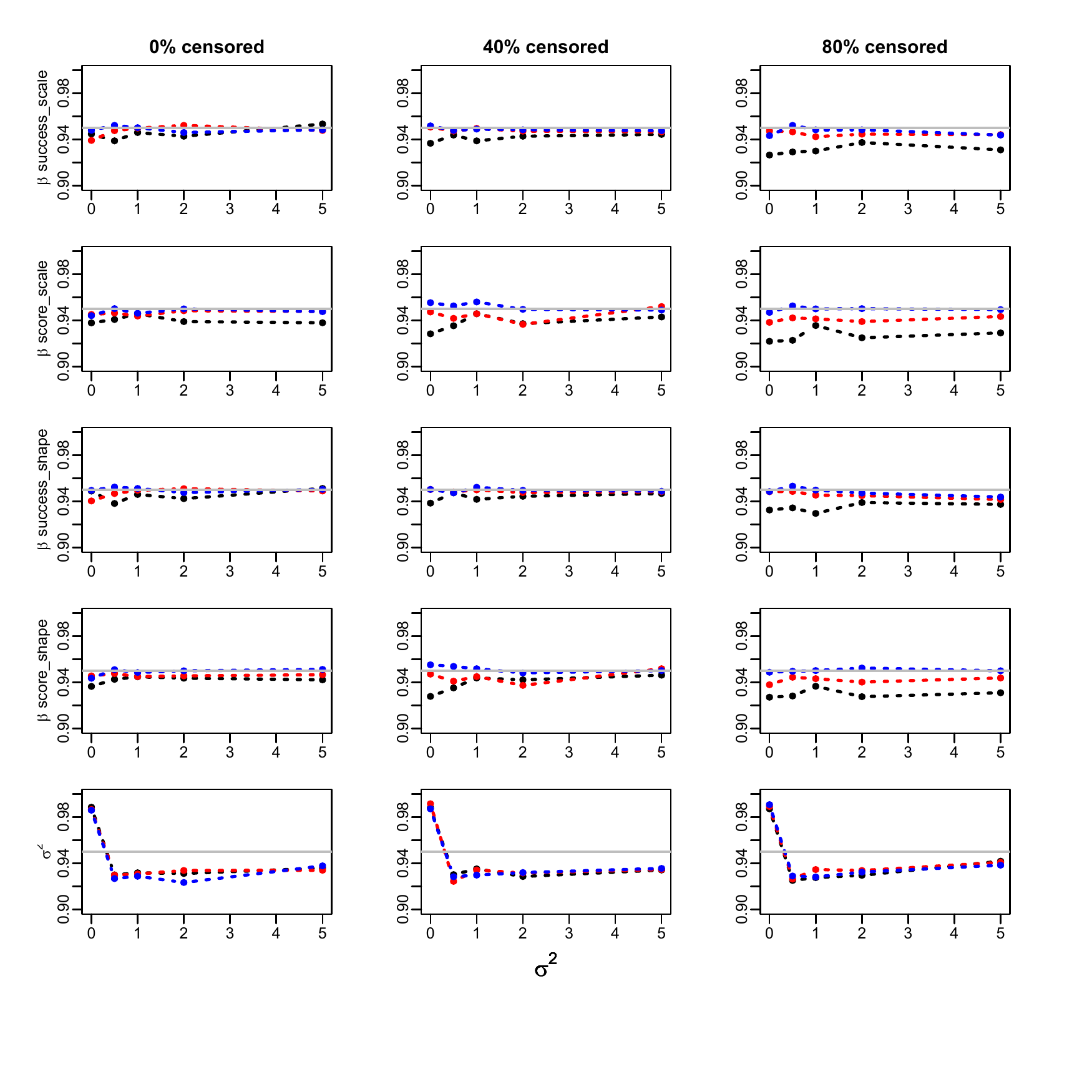}
	\caption{Coverage of the profile likelihood based confidence intervals for the Cox regression parameters and $\sigma^2$ at nominal 95\% level. Gompertz model.
Success proportion= 0.5. 10 clusters. Sample sizes: 300 (black), 1000 (red) and 10000 (blue).
True values: $\beta_{success-scale}$ = -0.5 , $\beta_{success-shape}$= 0.05 , $\beta_{score-scale}$= -1 , $\beta_{score-shape}$= 0.1}
	\label{CoveragePL10clustersGompertz7}
\end{figure}

\begin{figure}[ht]
	\centering
	\includegraphics[scale=1]{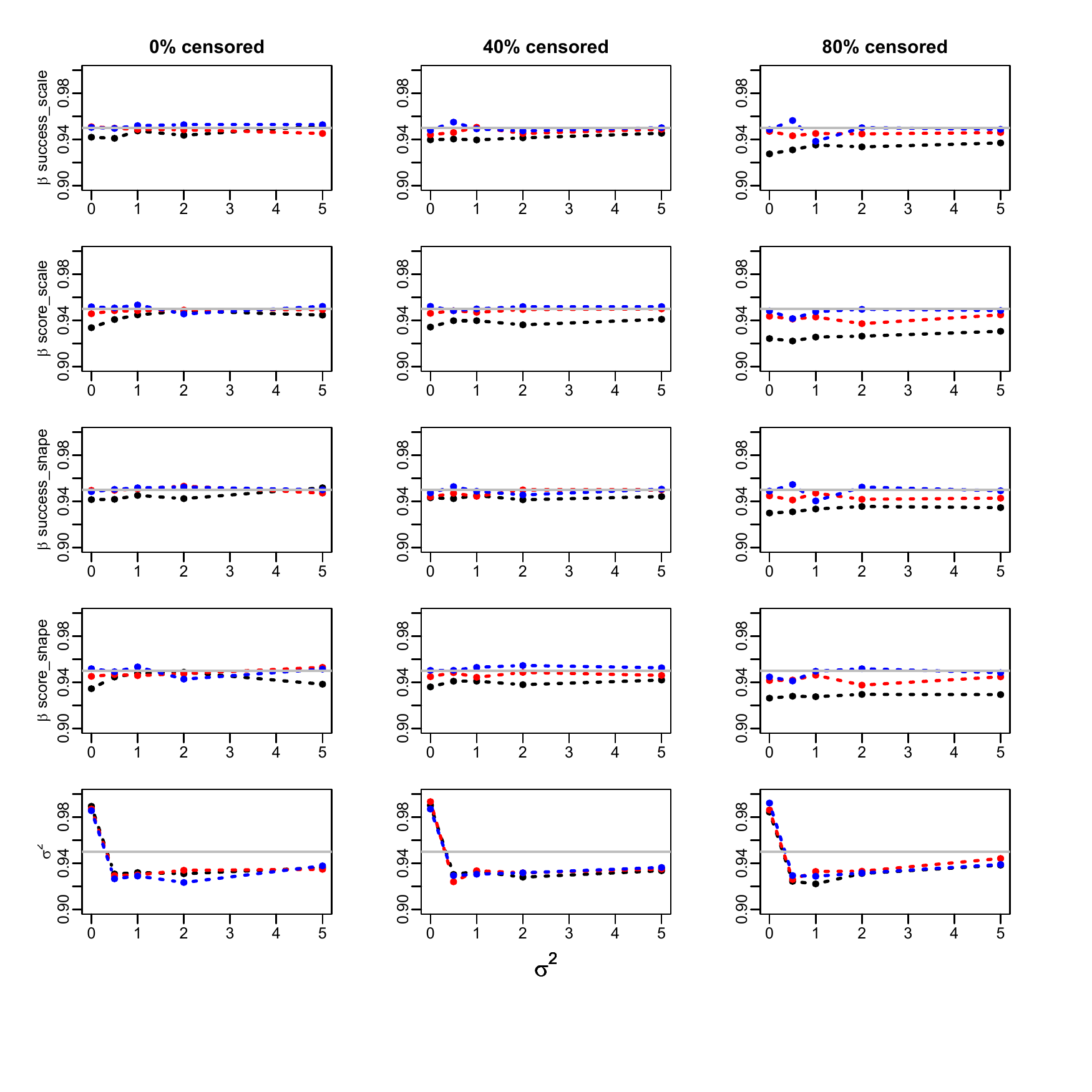}
	\caption{Coverage of the profile likelihood based confidence intervals for the Cox regression parameters and $\sigma^2$ at nominal 95\% level. Gompertz model.
Success proportion= 0.5. 10 clusters. Sample sizes: 300 (black), 1000 (red) and 10000 (blue).
True values: $\beta_{success-scale}$ = -0.5 , $\beta_{success-shape}$= -0.05 , $\beta_{score-scale}$= -1 , $\beta_{score-shape}$= -0.1}
	\label{CoveragePL10clustersGompertz8}
\end{figure}

 \begin{figure}[ht]
	\centering
	\includegraphics[scale=1]{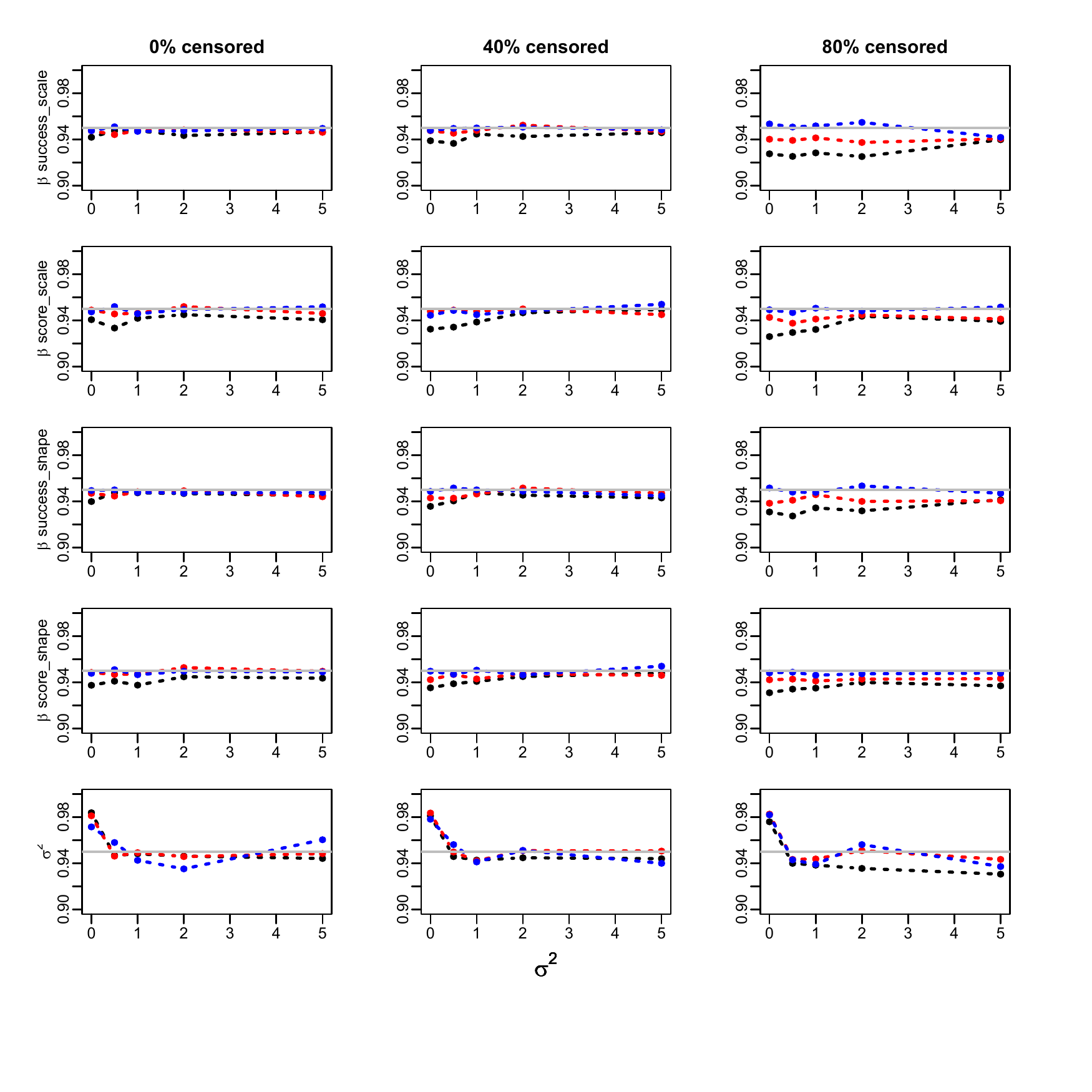}
	\caption{Coverage of the profile likelihood based confidence intervals for the Cox regression parameters and $\sigma^2$ at nominal 95\% level. Gompertz model.
Success proportion= 0.25. 100 clusters. Sample sizes: 300 (black), 1000 (red) and 10000 (blue).
True values: $\beta_{success-scale}$ = 0.5 , $\beta_{success-shape}$= 0.05 , $\beta_{score-scale}$= 1 , $\beta_{score-shape}$= 0.1}
	\label{CoveragePL100clustersGompertz1}
\end{figure}

\begin{figure}[ht]
	\centering
	\includegraphics[scale=1]{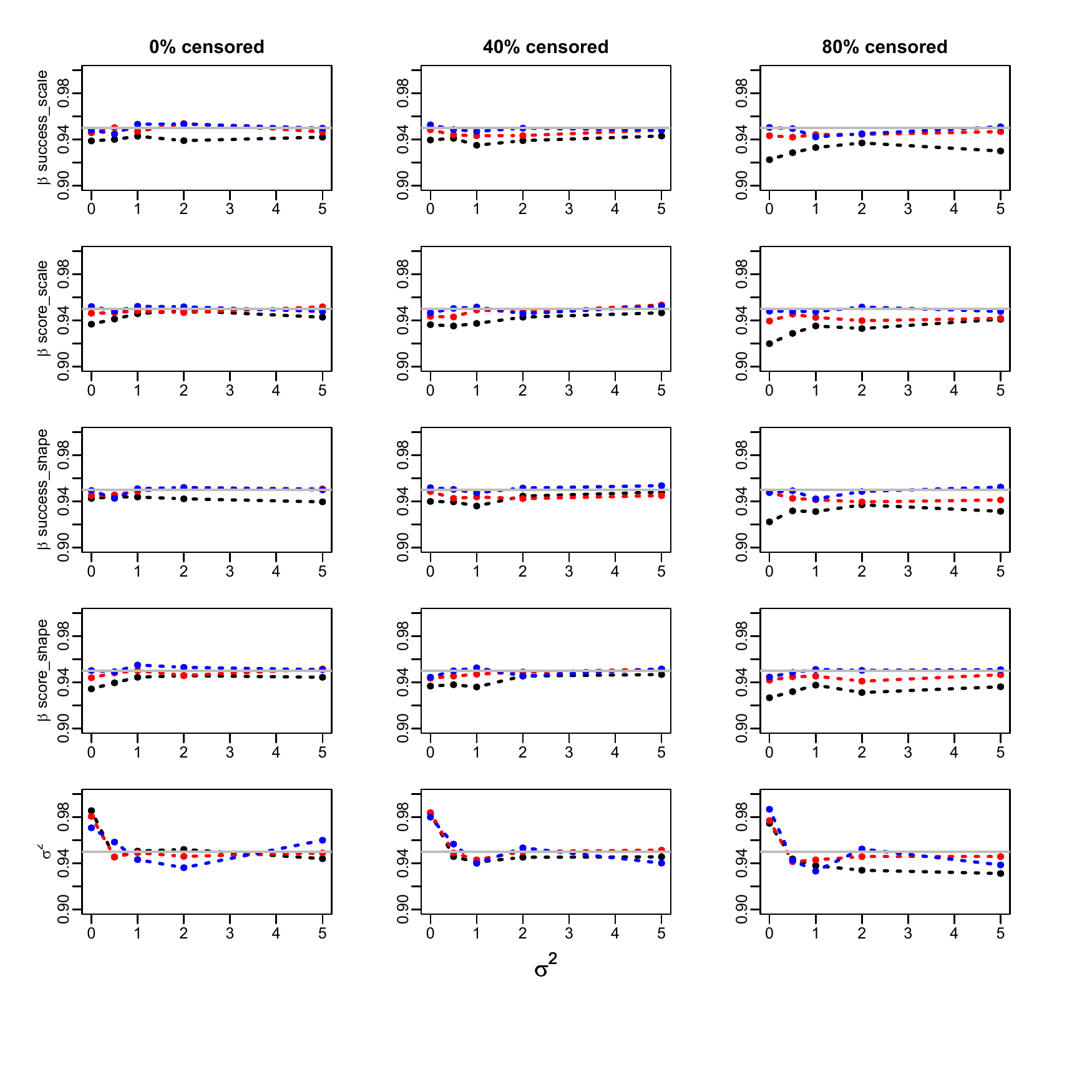}
	\caption{Coverage of the profile likelihood based confidence intervals for the Cox regression parameters and $\sigma^2$ at nominal 95\% level. Gompertz model.
Success proportion= 0.25. 100 clusters. Sample sizes: 300 (black), 1000 (red) and 10000 (blue).
True values: $\beta_{success-scale}$ = 0.5 , $\beta_{success-shape}$= -0.05 , $\beta_{score-scale}$= 1 , $\beta_{score-shape}$= - 0.1}
	\label{CoveragePL100clustersGompertz2}
\end{figure}

\begin{figure}[ht]
	\centering
	\includegraphics[scale=1]{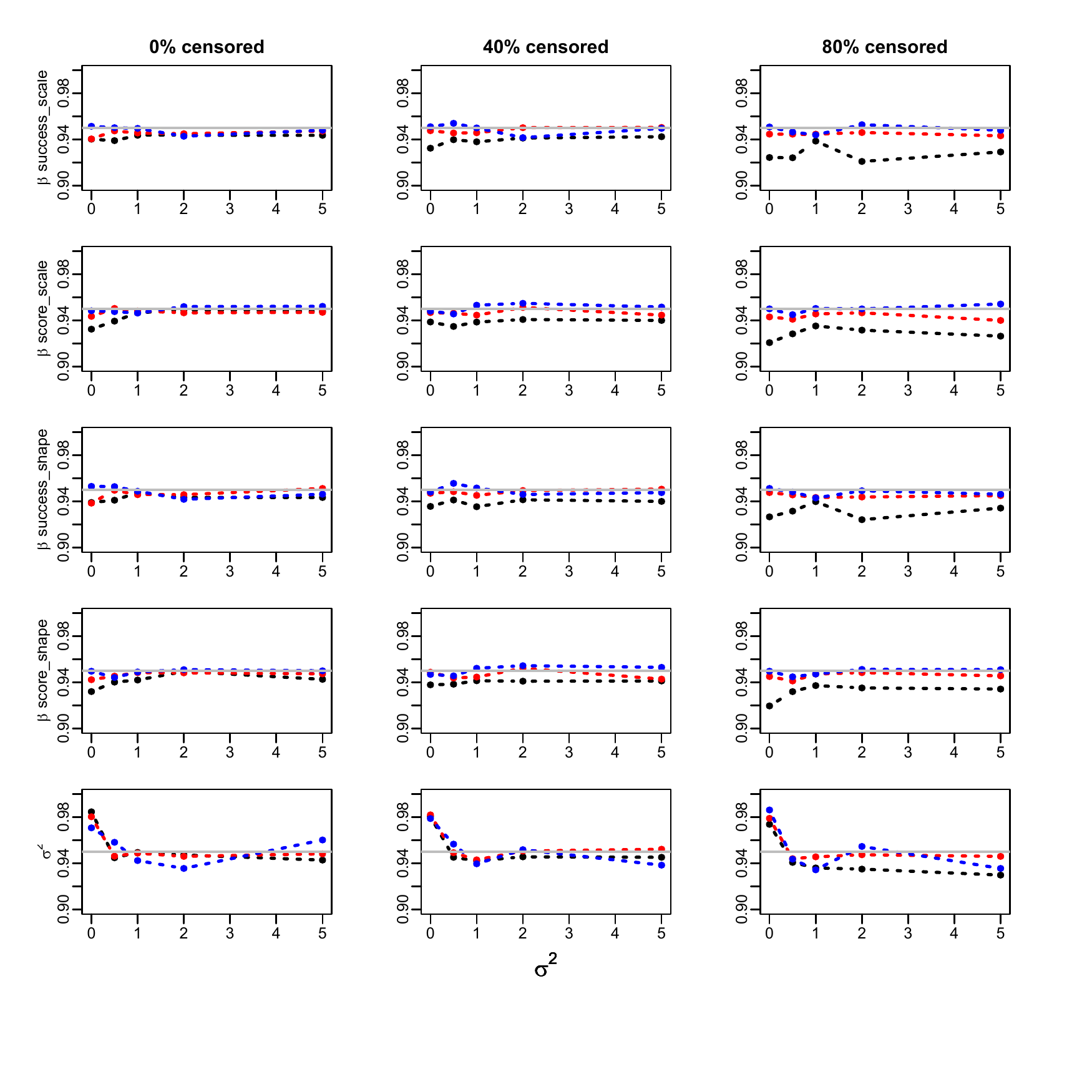}
	\caption{Coverage of the profile likelihood based confidence intervals for the Cox regression parameters and $\sigma^2$ at nominal 95\% level. Gompertz model.
Success proportion= 0.25. 100 clusters. Sample sizes: 300 (black), 1000 (red) and 10000 (blue).
True values: $\beta_{success-scale}$ = -0.5 , $\beta_{success-shape}$= 0.05 , $\beta_{score-scale}$= -1 , $\beta_{score-shape}$= 0.1}
	\label{CoveragePL100clustersGompertz3}
\end{figure}

\begin{figure}[ht]
	\centering
	\includegraphics[scale=1]{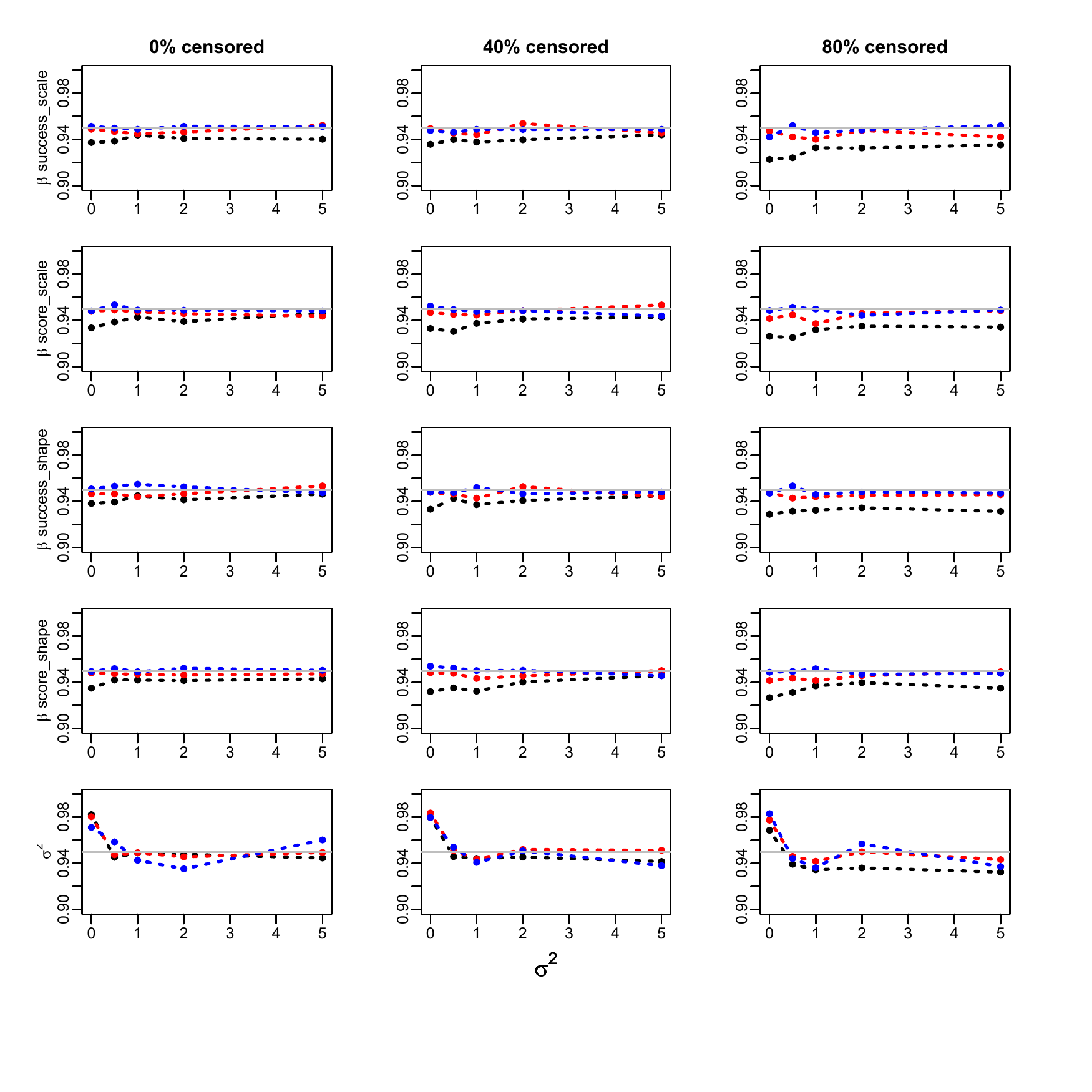}
	\caption{Coverage of the profile likelihood based confidence intervals for the Cox regression parameters and $\sigma^2$ at nominal 95\% level. Gompertz model.
Success proportion= 0.25. 100 clusters. Sample sizes: 300 (black), 1000 (red) and 10000 (blue).
True values: $\beta_{success-scale}$ = -0.5 , $\beta_{success-shape}$= -0.05 , $\beta_{score-scale}$= -1 , $\beta_{score-shape}$= -0.1}
	\label{CoveragePL100clustersGompertz4}
\end{figure}

\begin{figure}[ht]
	\centering
	\includegraphics[scale=1]{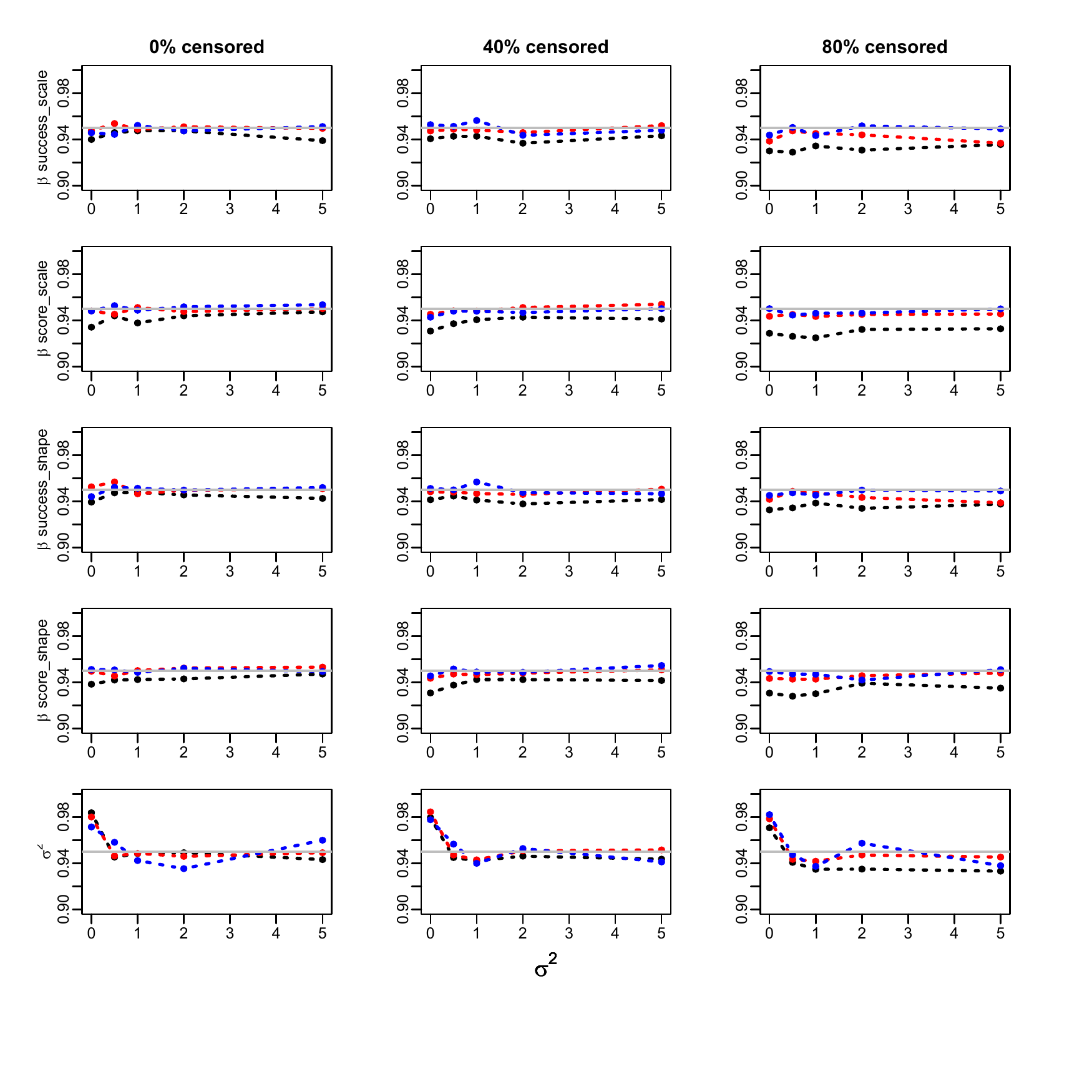}
	\caption{Coverage of the profile likelihood based confidence intervals for the Cox regression parameters and $\sigma^2$ at nominal 95\% level. Gompertz model.
Success proportion= 0.5. 100 clusters. Sample sizes: 300 (black), 1000 (red) and 10000 (blue).
True values: $\beta_{success-scale}$ = 0.5 , $\beta_{success-shape}$= 0.05 , $\beta_{score-scale}$= 1 , $\beta_{score-shape}$= 0.1}
	\label{CoveragePL100clustersGompertz5}
\end{figure}

\begin{figure}[ht]
	\centering
	\includegraphics[scale=1]{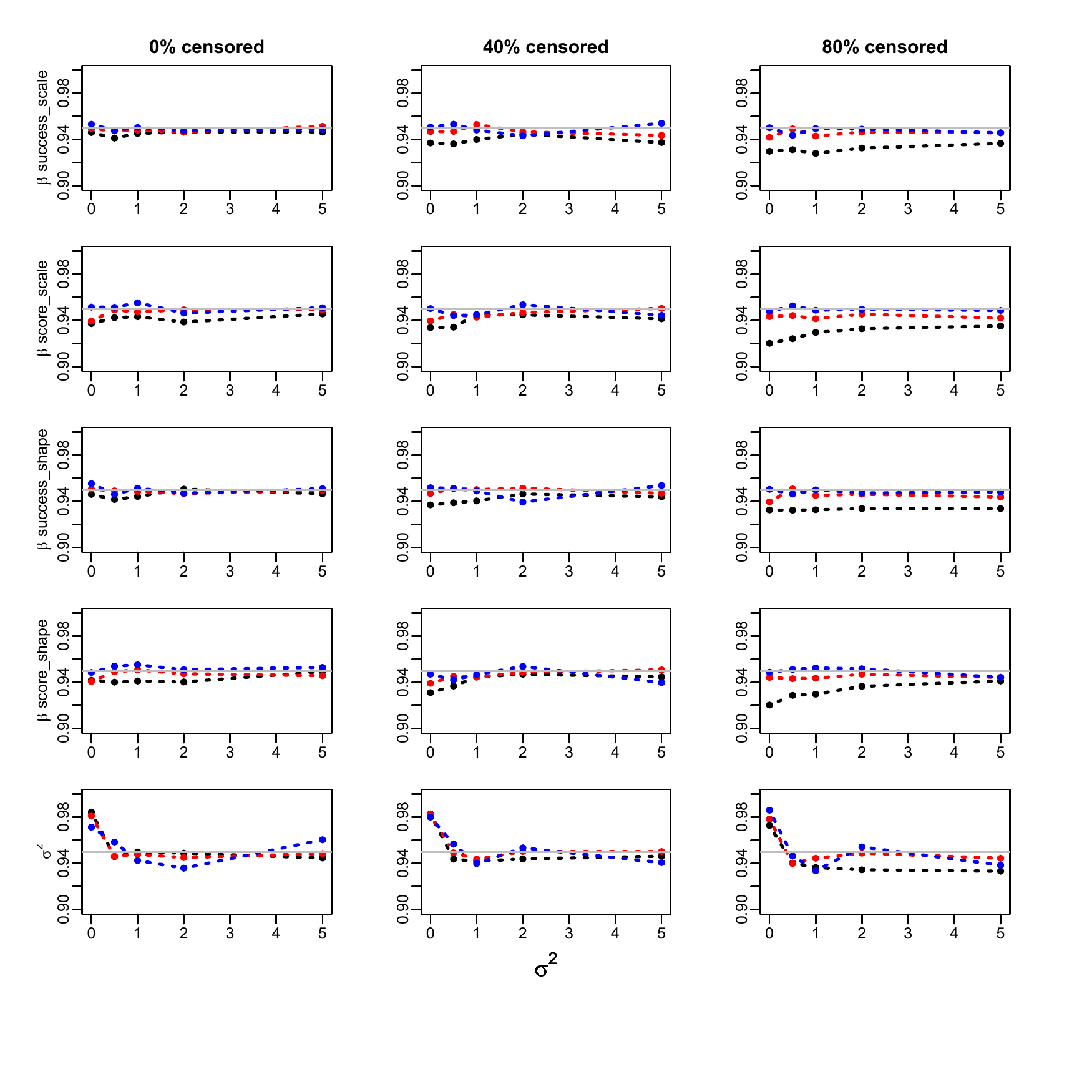}
	\caption{Coverage of the profile likelihood based confidence intervals for the Cox regression parameters and $\sigma^2$ at nominal 95\% level. Gompertz model.
Success proportion= 0.5. 100 clusters. Sample sizes: 300 (black), 1000 (red) and 10000 (blue).
True values: $\beta_{success-scale}$ = 0.5 , $\beta_{success-shape}$= -0.05 , $\beta_{score-scale}$= 1 , $\beta_{score-shape}$= - 0.1}
	\label{CoveragePL100clustersGompertz6}
\end{figure}

\begin{figure}[ht]
	\centering
	\includegraphics[scale=1]{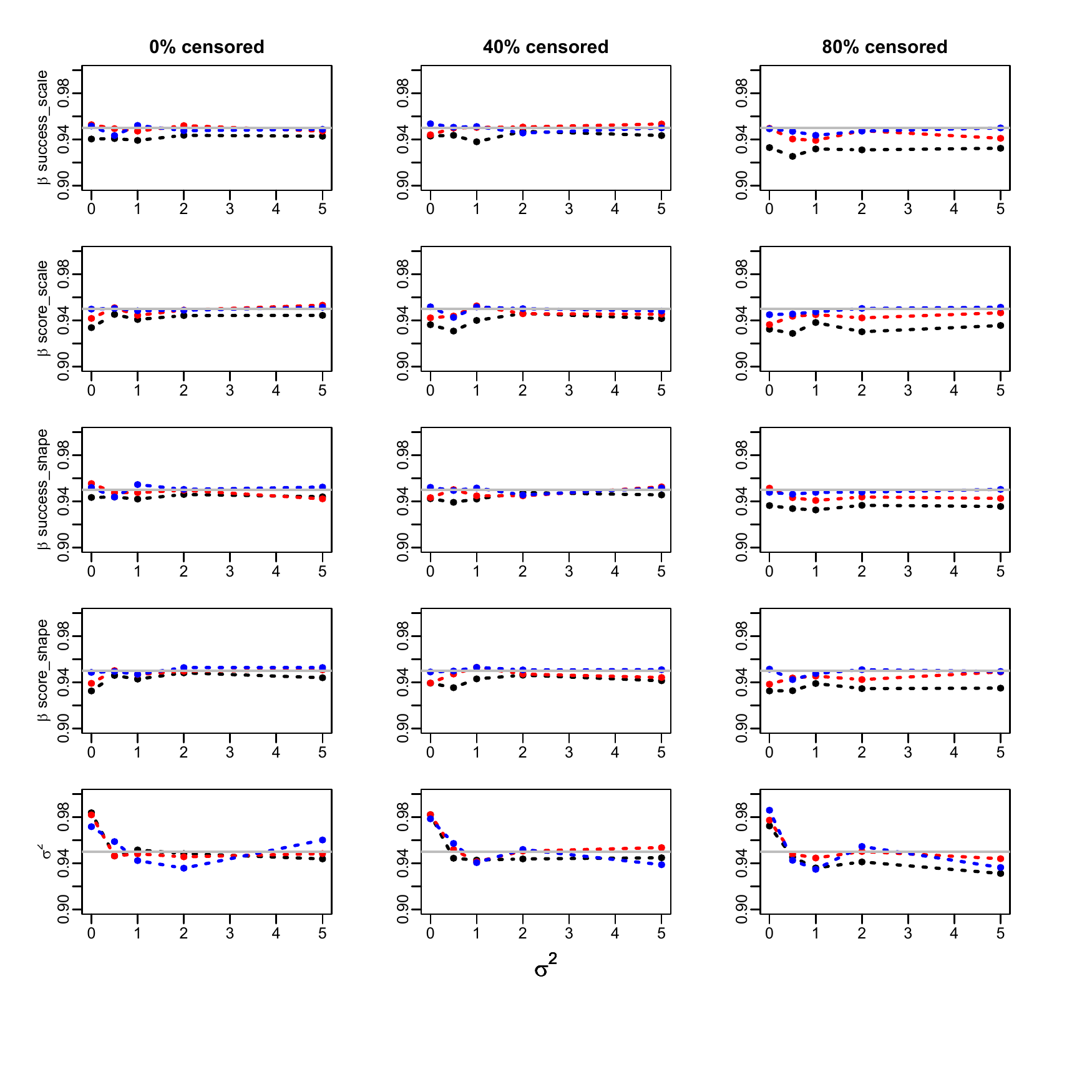}
	\caption{Coverage of the profile likelihood based confidence intervals for the Cox regression parameters and $\sigma^2$ at nominal 95\% level. Gompertz model.
Success proportion= 0.5. 100 clusters. Sample sizes: 300 (black), 1000 (red) and 10000 (blue).
True values: $\beta_{success-scale}$ = -0.5 , $\beta_{success-shape}$= 0.05 , $\beta_{score-scale}$= -1 , $\beta_{score-shape}$= 0.1}
	\label{CoveragePL100clustersGompertz7}
\end{figure}

\begin{figure}[ht]
	\centering
	\includegraphics[scale=1]{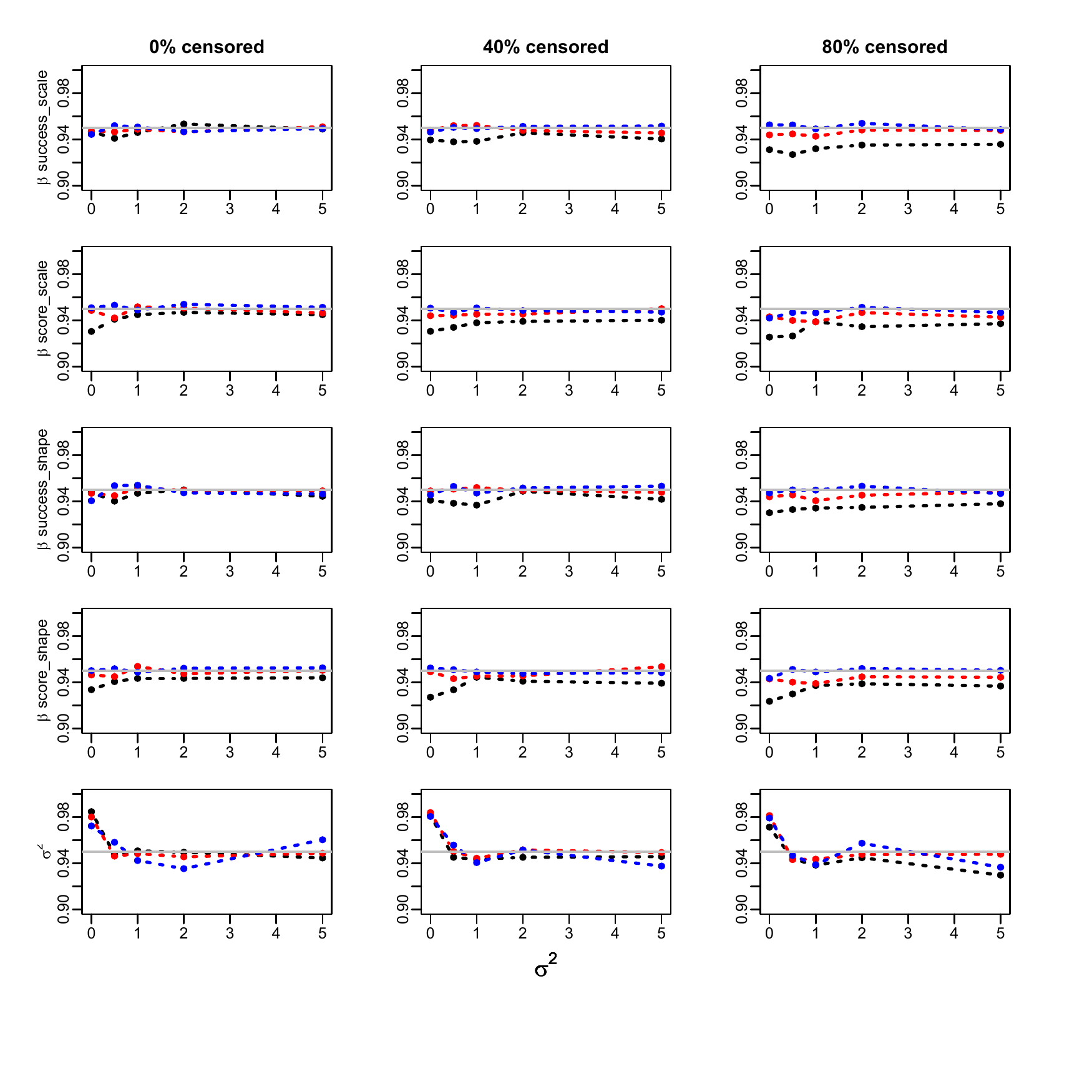}
	\caption{Coverage of the profile likelihood based confidence intervals for the Cox regression parameters and $\sigma^2$ at nominal 95\% level. Gompertz model.
Success proportion= 0.5. 100 clusters. Sample sizes: 300 (black), 1000 (red) and 10000 (blue).
True values: $\beta_{success-scale}$ = -0.5 , $\beta_{success-shape}$= -0.05 , $\beta_{score-scale}$= -1 , $\beta_{score-shape}$= -0.1}
	\label{CoveragePL100clustersGompertz8}
\end{figure}

\clearpage
\setcounter{figure}{0}
\setcounter{section}{0}
\renewcommand{\thefigure}{E.\arabic{figure}}

\section*{E: Coverage of the standard error based confidence intervals for  the scale   and shape  parameters of the  baseline distribution}

Each figure corresponds to a particular  baseline distribution  (Weibull or Gompertz), a value of the probability of success $p_{success}$ (= 0.25 or 0.5), a value for the number of clusters $N_{cl}$ (=10, 100) and a particular choice of the signs of the Cox regression parameters (+ + + +, + - + -, - + - + and - - - -).\\

The absolute values of the Cox regression parameters are held constant at $\beta_{success-scale}$ = 0.5 , $\beta_{success-shape}$= 0.05 , $\beta_{score-scale}$= 1 , $\beta_{score-shape}$= 0.1.

For each combination of a censoring proportion  (= 0, 40\%, 80\%), a panel plots, versus the frailty variance $\sigma^2$ (= 0, 0.5, 1, 2, 3, 4, 5), the empirical coverage of the true value of the  scale parameter $a$  or the true value of  the shape parameter $b$  by a SE-based confidence interval at 95\% nominal level, for three sample sizes (300, 1000 and 10000) . \\

\clearpage

 \begin{figure}[ht]
	\centering
	\includegraphics[scale=1]{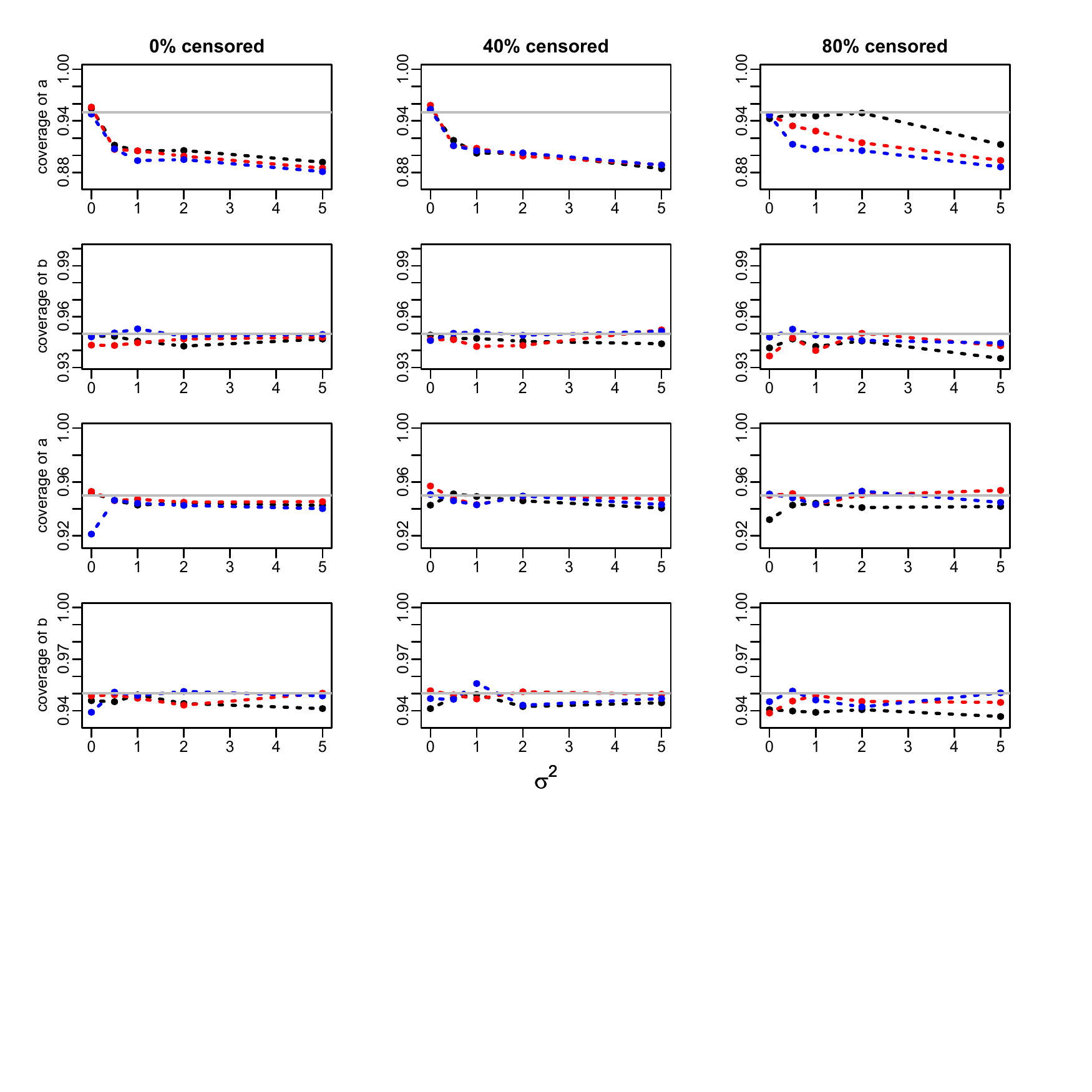}
	\caption{Coverage of the standard error based confidence intervals for the $a$ and $b$ parameters in the Weibull model.
Success proportion= 0.25.  Sample sizes: 300 (black), 1000 (red) and 10000 (blue).
True values: $\beta_{success-scale}$ = 0.5 , $\beta_{success-shape}$= 0.05 , $\beta_{score-scale}$= 1 , $\beta_{score-shape}$= 0.1.
Top two rows: 10 clusters; bottom two rows: 100 clusters.}
	\label{CoverageSE10_100clustersWeibull1}
\end{figure}

\begin{figure}[ht]
	\centering
	\includegraphics[scale=1]{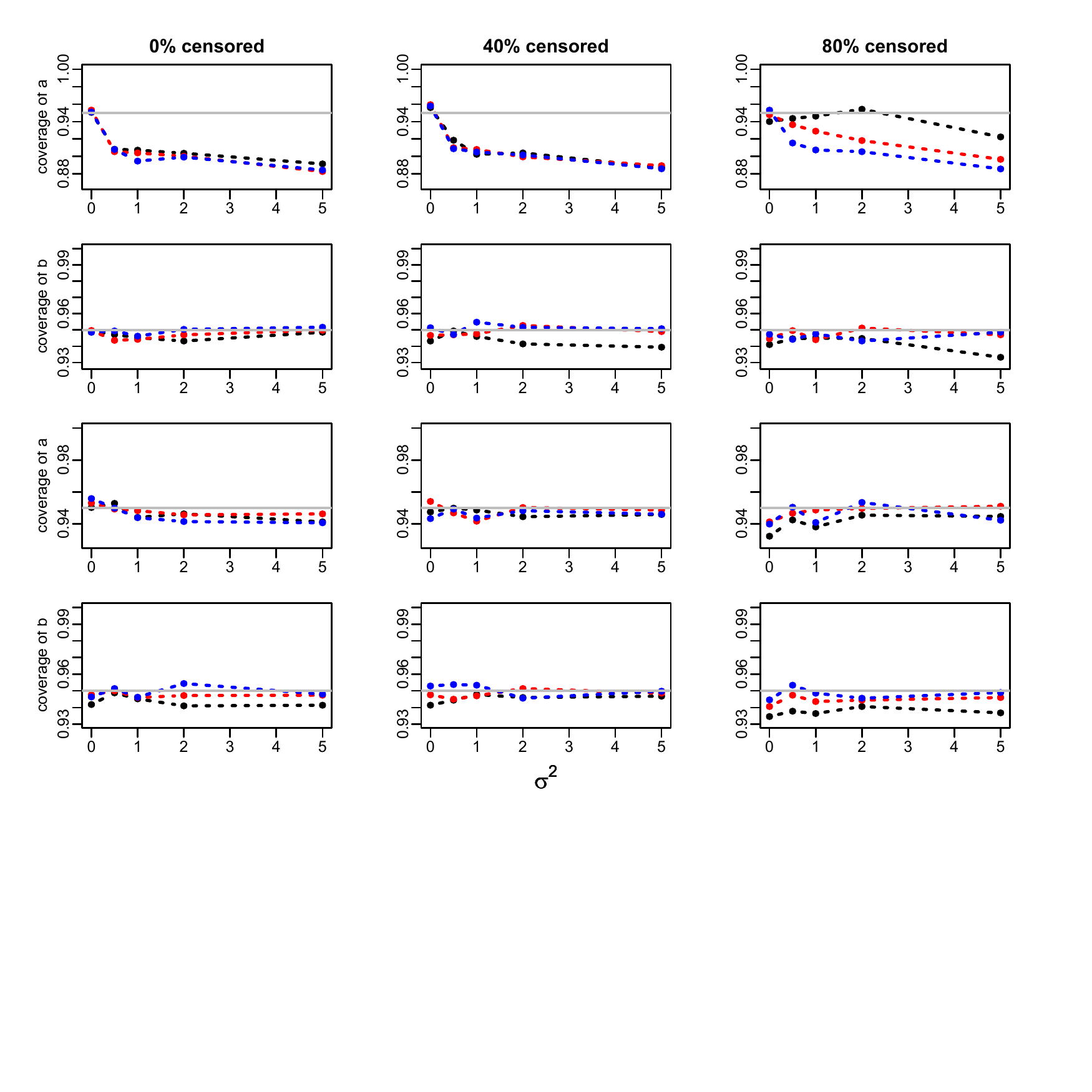}
	\caption{Coverage of the standard error based confidence intervals for the $a$ and $b$ parameters  in the Weibull model.
Success proportion= 0.25.  Sample sizes: 300 (black), 1000 (red) and 10000 (blue).
True values: $\beta_{success-scale}$ = 0.5 , $\beta_{success-shape}$= -0.05 , $\beta_{score-scale}$= 1 , $\beta_{score-shape}$= - 0.1.
Top two rows: 10 clusters; bottom two rows: 100 clusters.}
	\label{CoverageSE10_100clustersWeibull2}
\end{figure}

\begin{figure}[ht]
	\centering
	\includegraphics[scale=1]{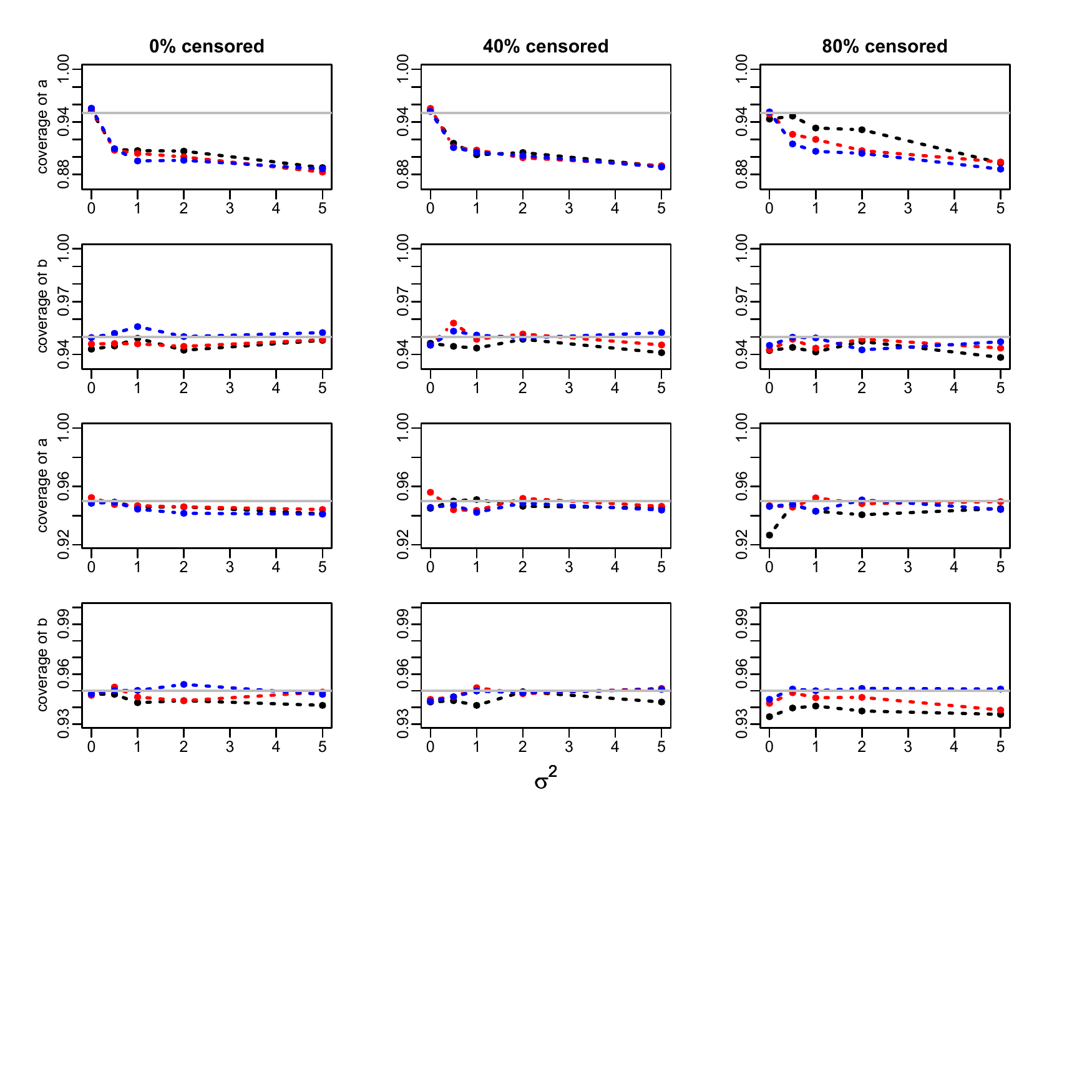}
	\caption{Coverage of the standard error based confidence intervals for the $a$ and $b$ parameters  in the Weibull model.
Success proportion= 0.25.  Sample sizes: 300 (black), 1000 (red) and 10000 (blue).
True values: $\beta_{success-scale}$ = -0.5 , $\beta_{success-shape}$= 0.05 , $\beta_{score-scale}$= -1 , $\beta_{score-shape}$= 0.1.
Top two rows: 10 clusters; bottom two rows: 100 clusters.}
	\label{CoverageSE10_100clustersWeibull3}
\end{figure}

\begin{figure}[ht]
	\centering
	\includegraphics[scale=1]{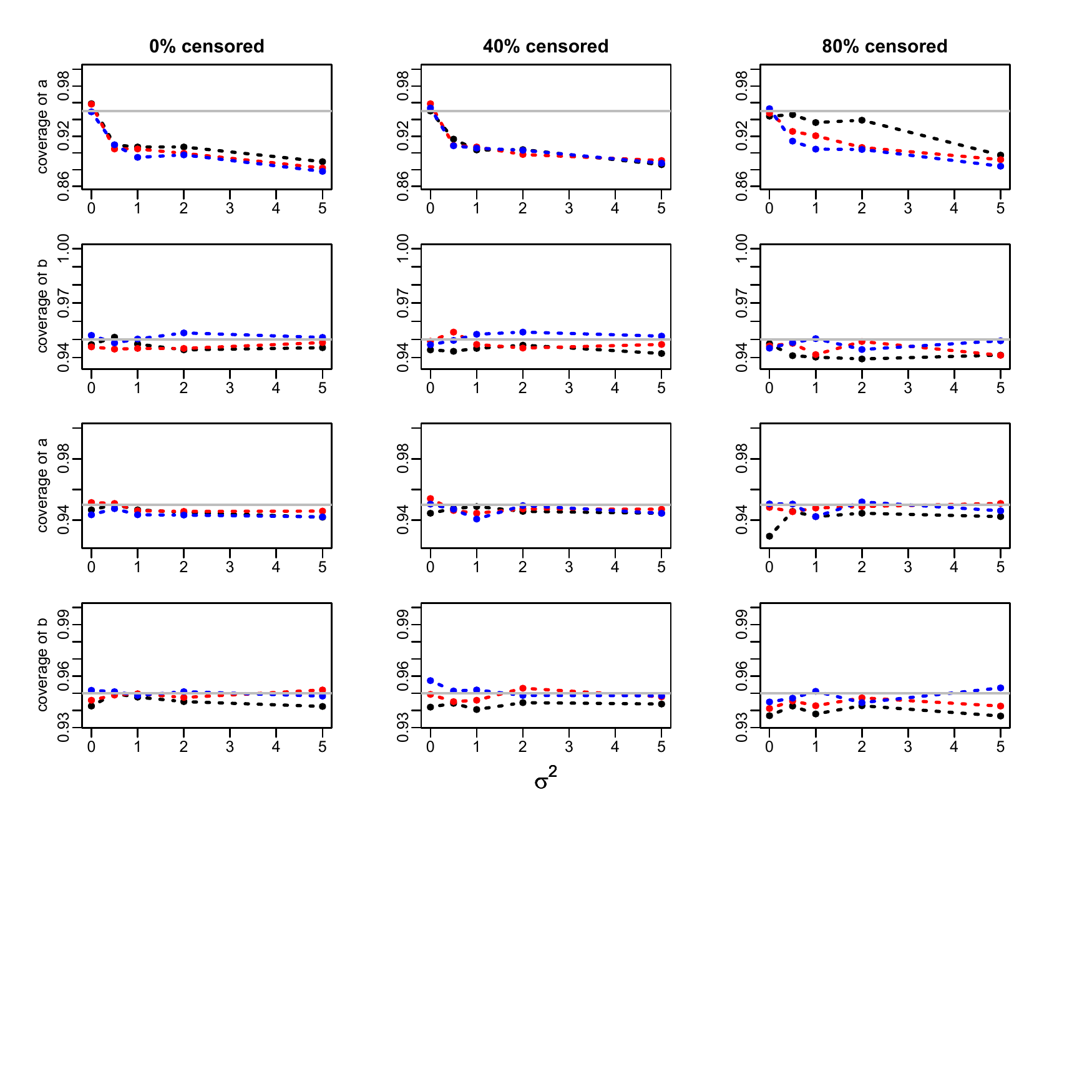}
	\caption{Coverage of the standard error based confidence intervals for the $a$ and $b$ parameters  in the Weibull model.
Success proportion= 0.25. Sample sizes: 300 (black), 1000 (red) and 10000 (blue).
True values: $\beta_{success-scale}$ = -0.5 , $\beta_{success-shape}$= -0.05 , $\beta_{score-scale}$= -1 , $\beta_{score-shape}$= -0.1.
Top two rows: 10 clusters; bottom two rows: 100 clusters.}
	\label{CoverageSE10_100clustersWeibull4}
\end{figure}

\begin{figure}[ht]
	\centering
	\includegraphics[scale=1]{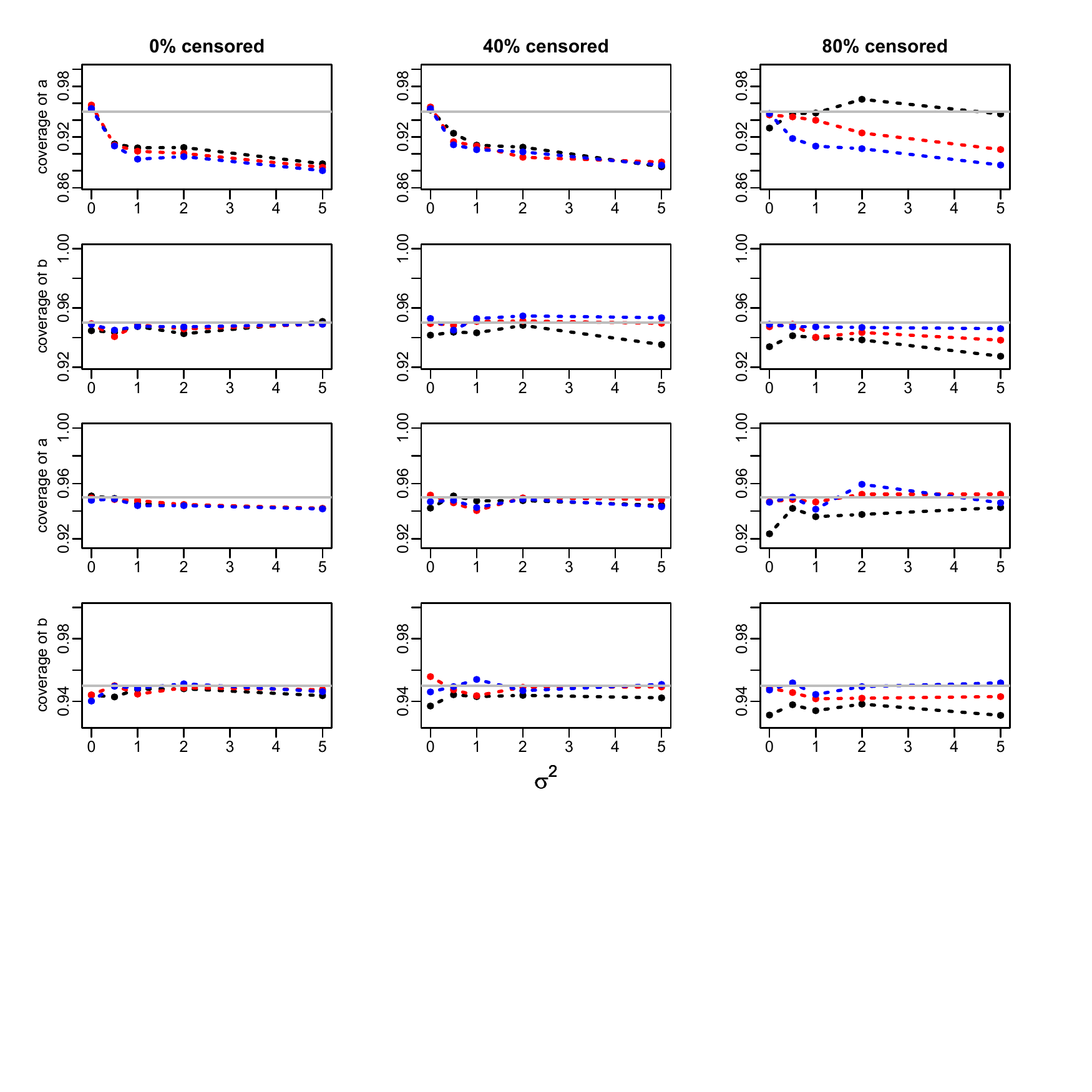}
	\caption{Coverage of the standard error based confidence intervals for the $a$ and $b$ parameters  in the Weibull model.
Success proportion= 0.5.  Sample sizes: 300 (black), 1000 (red) and 10000 (blue).
True values: $\beta_{success-scale}$ = 0.5 , $\beta_{success-shape}$= 0.05 , $\beta_{score-scale}$= 1 , $\beta_{score-shape}$= 0.1.
Top two rows: 10 clusters; bottom two rows: 100 clusters.}
	\label{CoverageSE10_100clustersWeibull5}
\end{figure}

\begin{figure}[ht]
	\centering
	\includegraphics[scale=1]{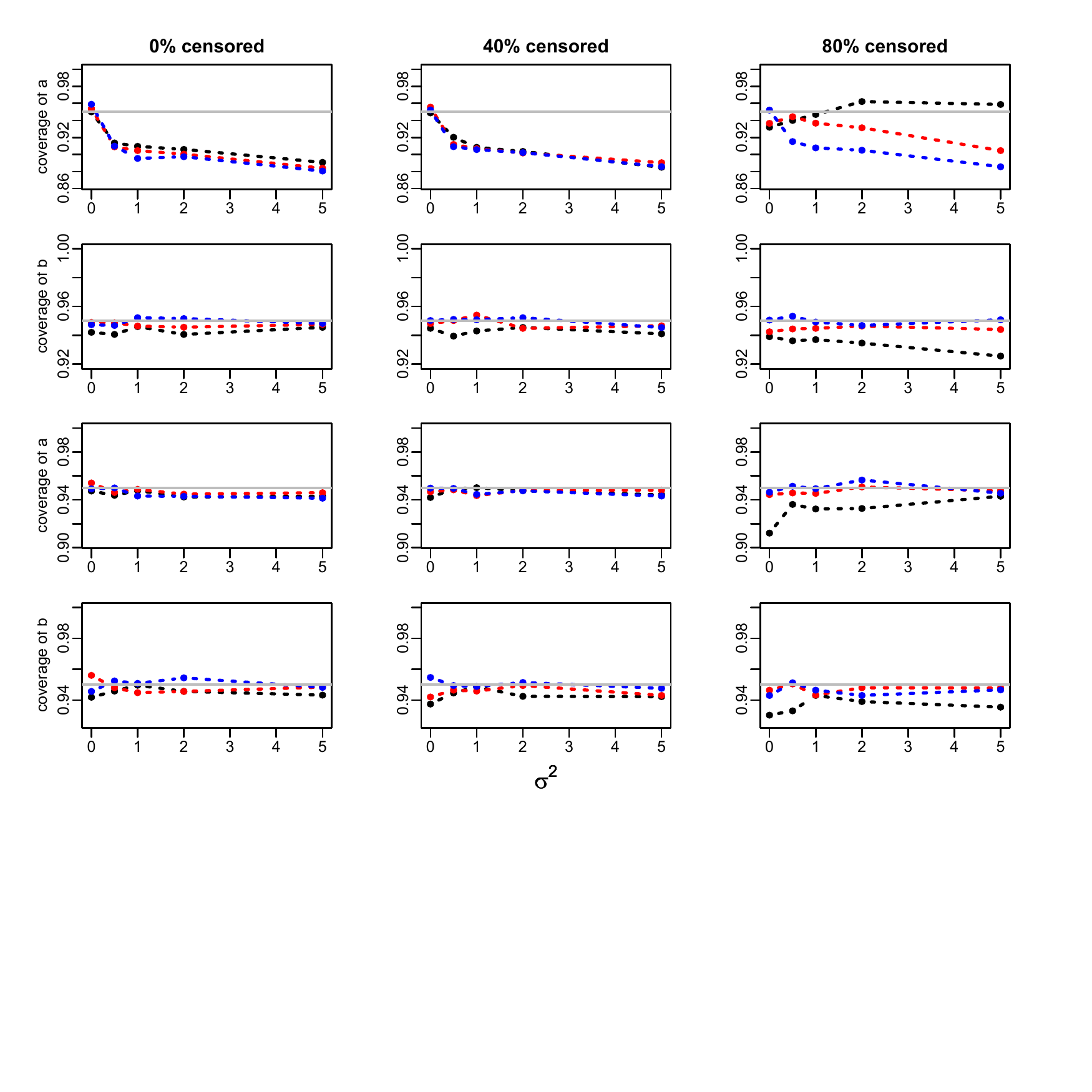}
	\caption{Coverage of the standard error based confidence intervals for the $a$ and $b$ parameters  in the Weibull model.
Success proportion= 0.5.  Sample sizes: 300 (black), 1000 (red) and 10000 (blue).
True values: $\beta_{success-scale}$ = 0.5 , $\beta_{success-shape}$= -0.05 , $\beta_{score-scale}$= 1 , $\beta_{score-shape}$= - 0.1.
Top two rows: 10 clusters; bottom two rows: 100 clusters.}
	\label{CoverageSE10_100clustersWeibull6}
\end{figure}

\begin{figure}[ht]
	\centering
	\includegraphics[scale=1]{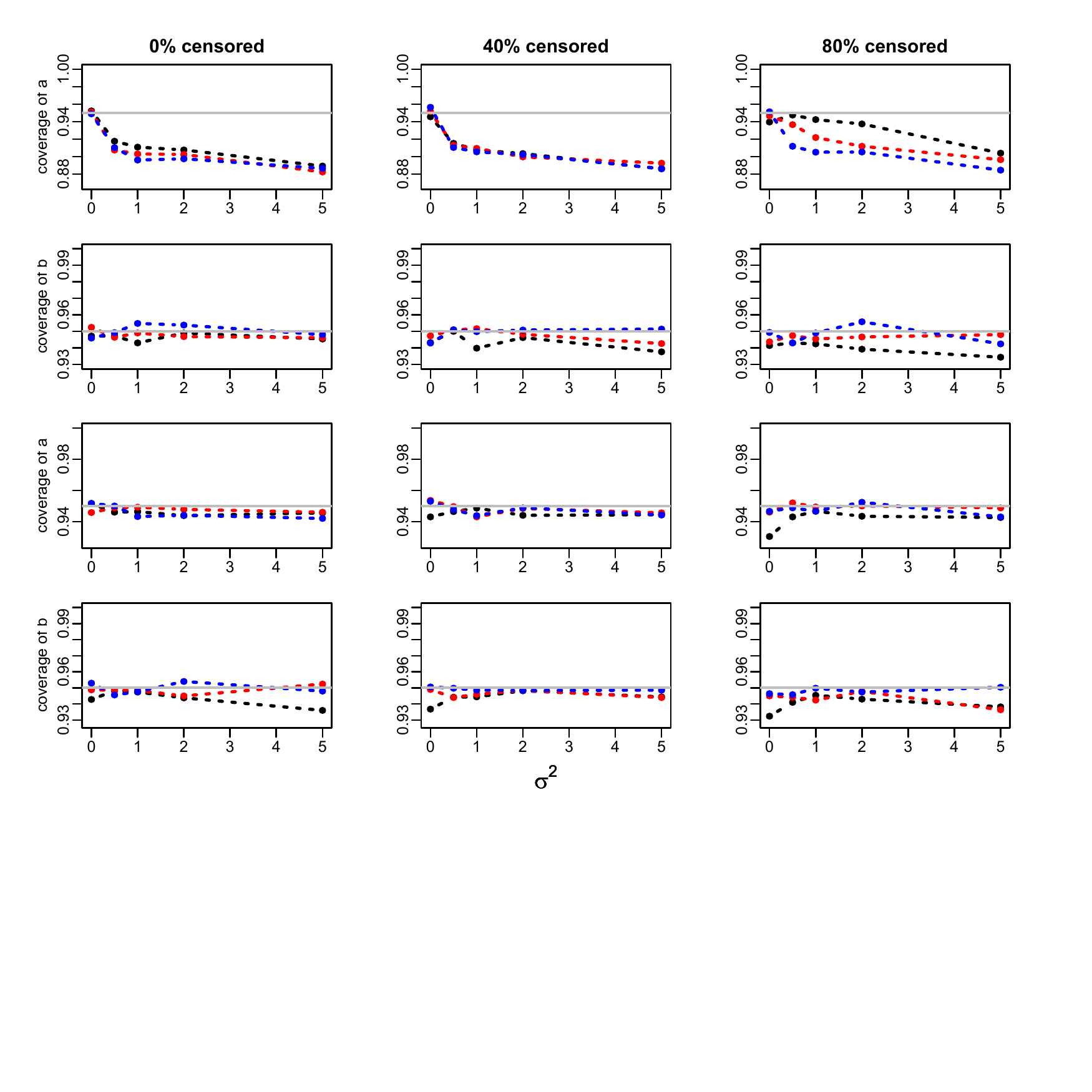}
	\caption{Coverage of the standard error based confidence intervals for the $a$ and $b$ parameters in the Weibull model.
Success proportion= 0.5.  Sample sizes: 300 (black), 1000 (red) and 10000 (blue).
True values: $\beta_{success-scale}$ = -0.5 , $\beta_{success-shape}$= 0.05 , $\beta_{score-scale}$= -1 , $\beta_{score-shape}$= 0.1.
Top two rows: 10 clusters; bottom two rows: 100 clusters.}
	\label{CoverageSE10_100clustersWeibull7}
\end{figure}

\begin{figure}[ht]
	\centering
	\includegraphics[scale=1]{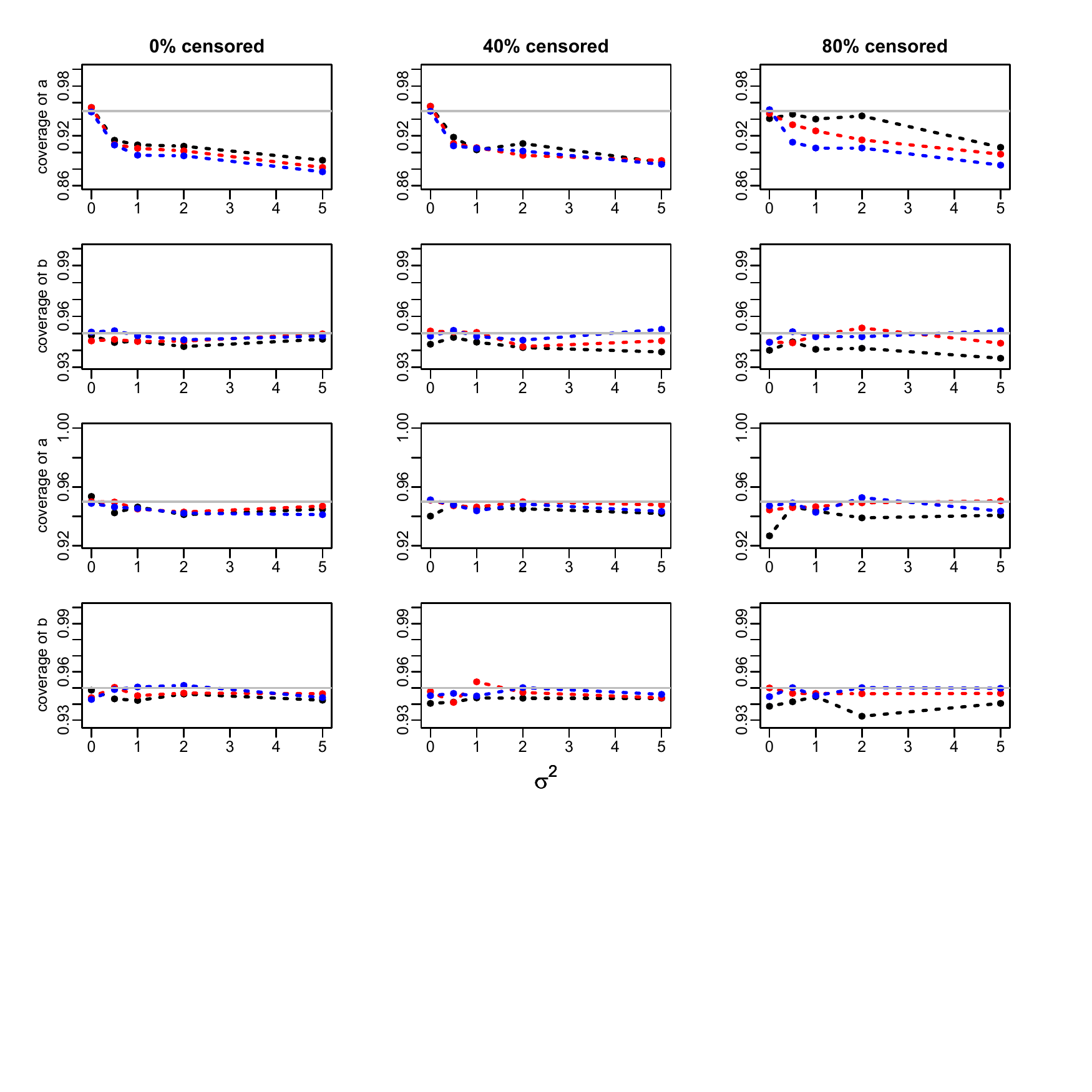}
	\caption{Coverage of the standard error based confidence intervals for the $a$ and $b$ parameters  in the Weibull model.
Success proportion= 0.5.  Sample sizes: 300 (black), 1000 (red) and 10000 (blue).
True values: $\beta_{success-scale}$ = -0.5 , $\beta_{success-shape}$= -0.05 , $\beta_{score-scale}$= -1 , $\beta_{score-shape}$= -0.1.
Top two rows: 10 clusters; bottom two rows: 100 clusters.}
    \label{CoverageSE10_100clustersWeibull8}
\end{figure}


\begin{figure}[ht]
	\centering
	\includegraphics[scale=1]{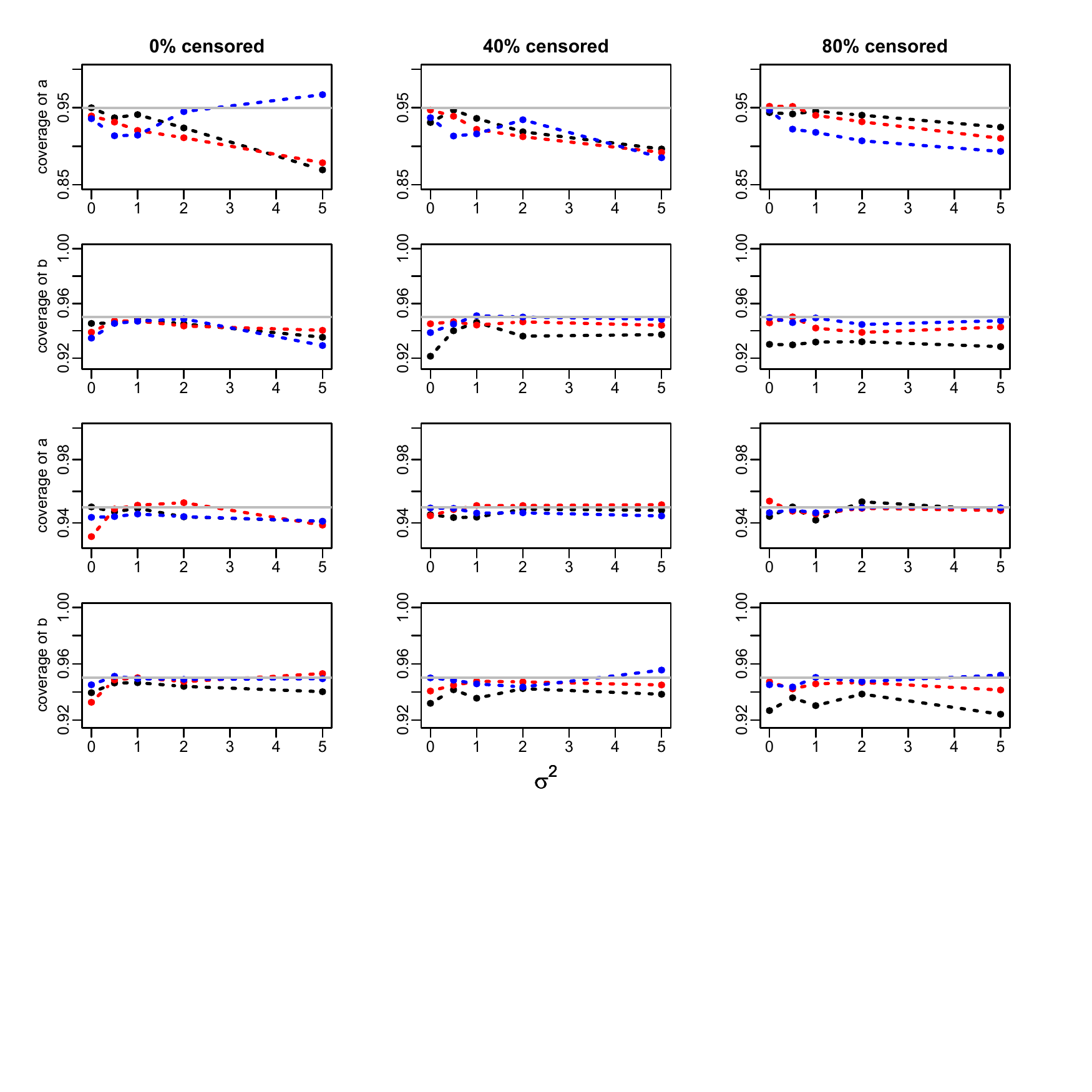}
	\caption{Coverage of the standard error based confidence intervals for the $a$ and $b$ parameters in the Gompertz model.
Success proportion= 0.25.  Sample sizes: 300 (black), 1000 (red) and 10000 (blue).
True values: $\beta_{success-scale}$ = 0.5 , $\beta_{success-shape}$= 0.05 , $\beta_{score-scale}$= 1 , $\beta_{score-shape}$= 0.1.
Top two rows: 10 clusters; bottom two rows: 100 clusters.}
	\label{CoverageSE10_100clustersGompertz1}
\end{figure}

\begin{figure}[ht]
	\centering
	\includegraphics[scale=1]{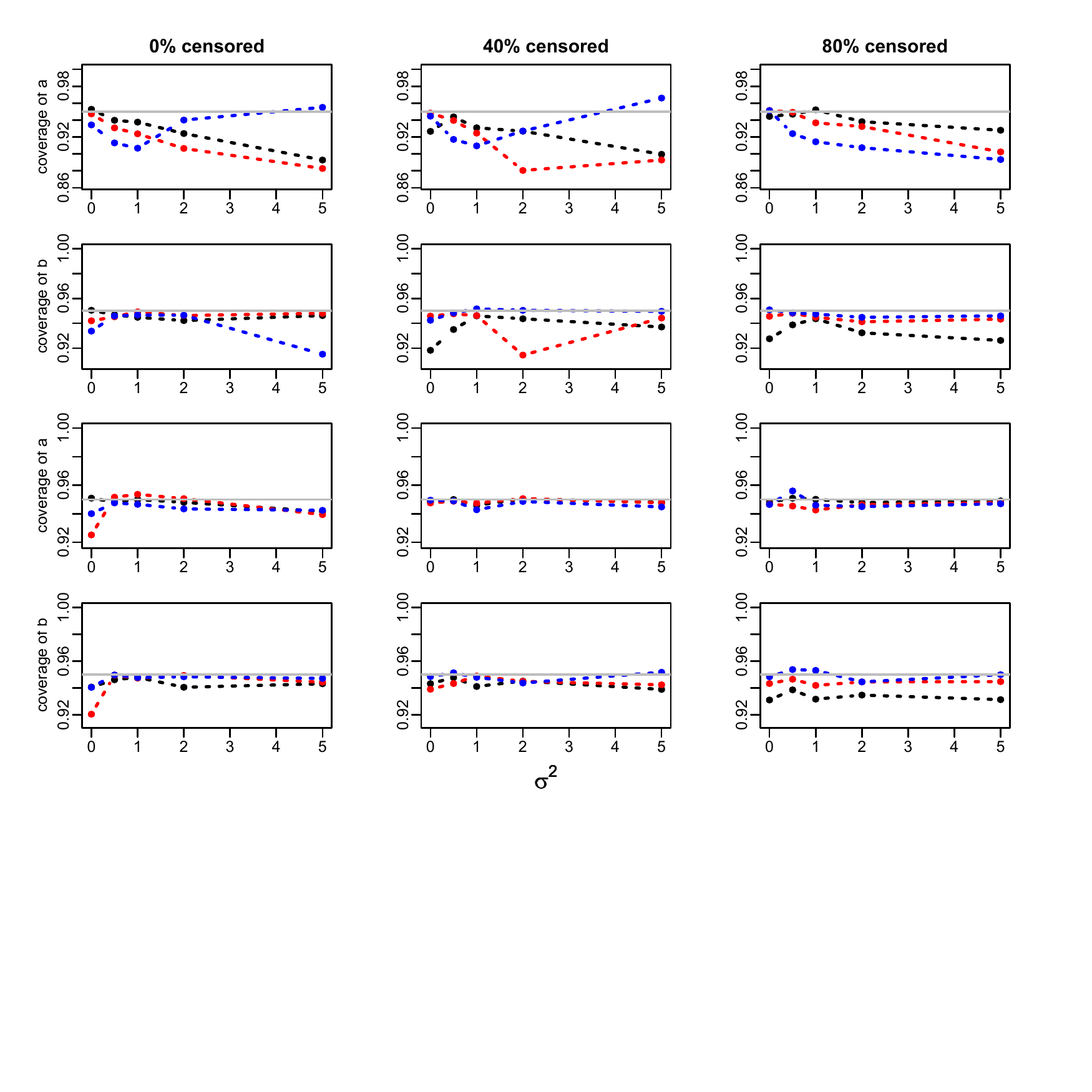}
	\caption{Coverage of the standard error based confidence intervals for the $a$ and $b$ parameters  in the Gompertz model.
Success proportion= 0.25.  Sample sizes: 300 (black), 1000 (red) and 10000 (blue).
True values: $\beta_{success-scale}$ = 0.5 , $\beta_{success-shape}$= -0.05 , $\beta_{score-scale}$= 1 , $\beta_{score-shape}$= - 0.1.
Top two rows: 10 clusters; bottom two rows: 100 clusters.}
	\label{CoverageSE10_100clustersGompertz2}
\end{figure}

\begin{figure}[ht]
	\centering
	\includegraphics[scale=1]{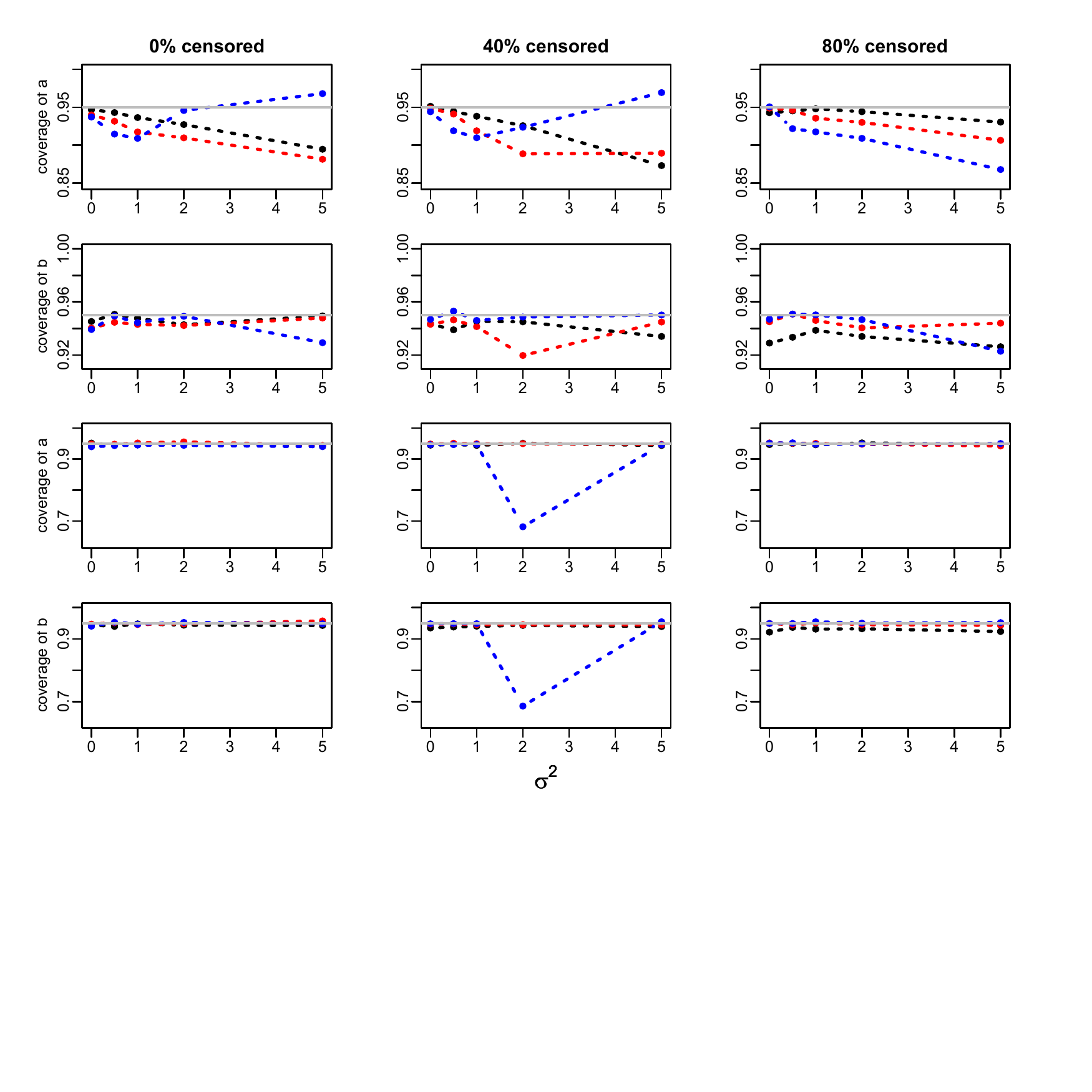}
	\caption{Coverage of the standard error based confidence intervals for the $a$ and $b$ parameters  in the Gompertz model.
Success proportion= 0.25.  Sample sizes: 300 (black), 1000 (red) and 10000 (blue).
True values: $\beta_{success-scale}$ = -0.5 , $\beta_{success-shape}$= 0.05 , $\beta_{score-scale}$= -1 , $\beta_{score-shape}$= 0.1.
Top two rows: 10 clusters; bottom two rows: 100 clusters.}
	\label{CoverageSE10_100clustersGompertz3}
\end{figure}

\begin{figure}[ht]
	\centering
	\includegraphics[scale=1]{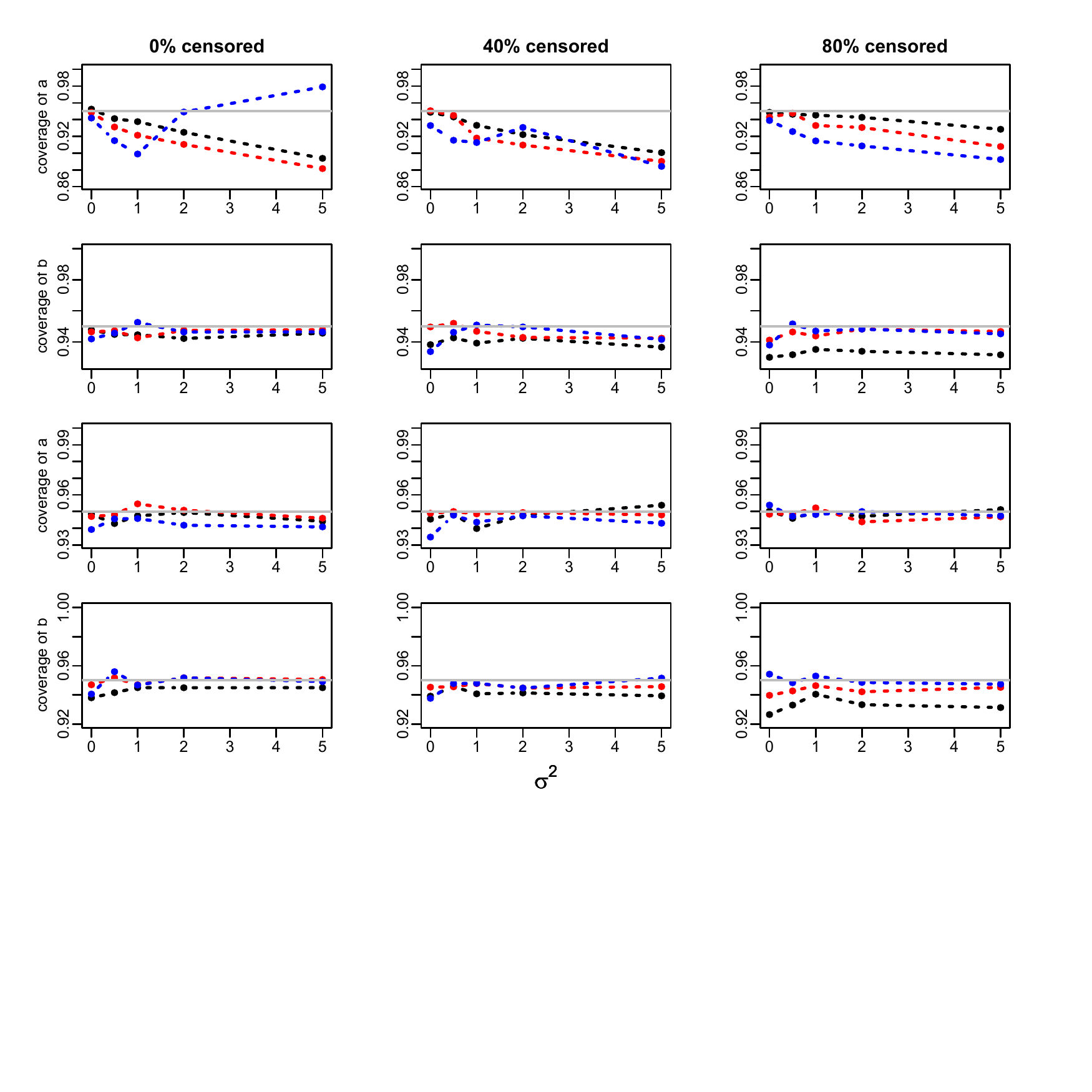}
	\caption{Coverage of the standard error based confidence intervals for the $a$ and $b$ parameters  in the Gompertz model.
Success proportion= 0.25.  Sample sizes: 300 (black), 1000 (red) and 10000 (blue).
True values: $\beta_{success-scale}$ = -0.5 , $\beta_{success-shape}$= -0.05 , $\beta_{score-scale}$= -1 , $\beta_{score-shape}$= -0.1.
Top two rows: 10 clusters; bottom two rows: 100 clusters.}
	\label{CoverageSE10_100clustersGompertz4}
\end{figure}

\begin{figure}[ht]
	\centering
	\includegraphics[scale=1]{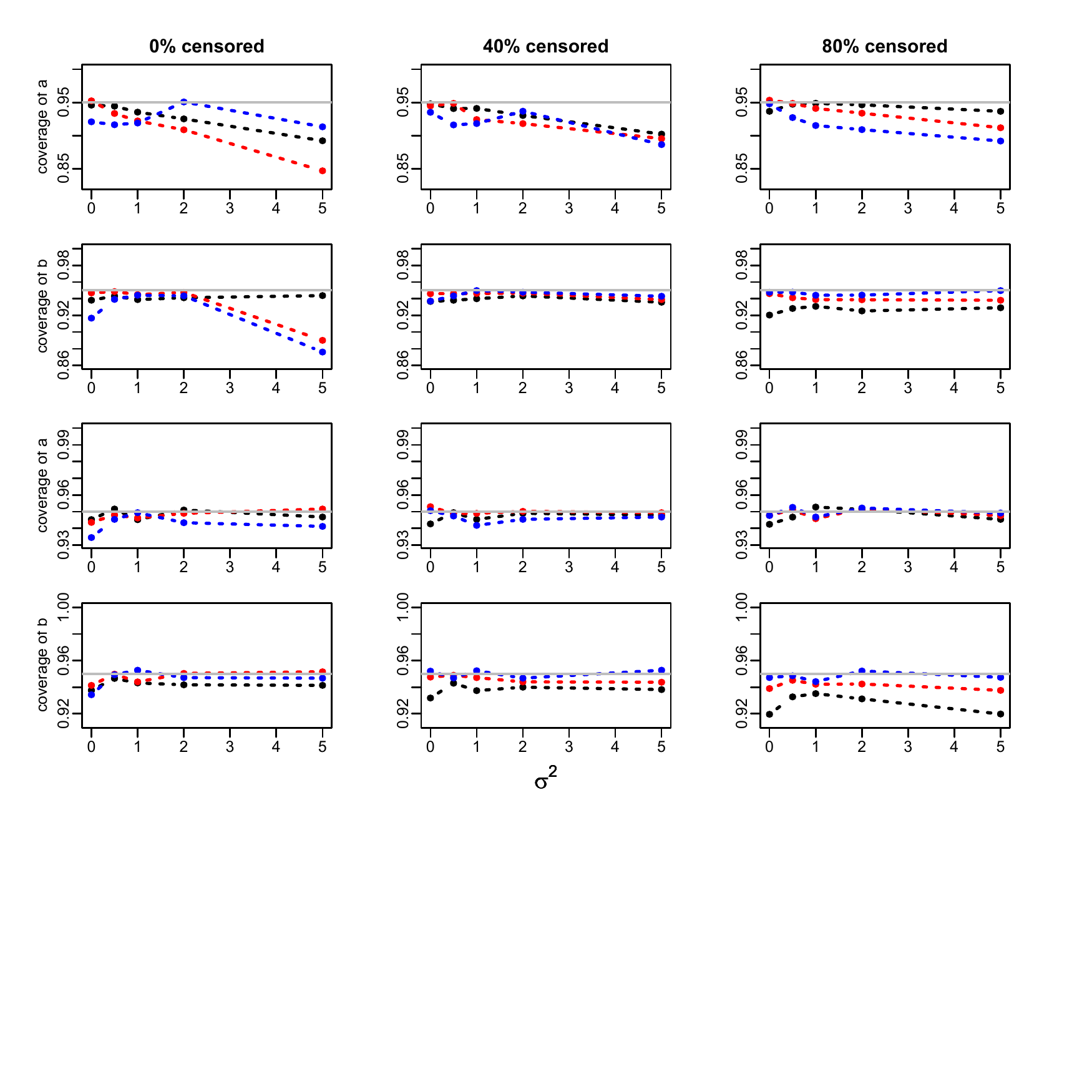}
	\caption{Coverage of the standard error based confidence intervals for the $a$ and $b$ parameters  in the Gompertz model.
Success proportion= 0.5.  Sample sizes: 300 (black), 1000 (red) and 10000 (blue).
True values: $\beta_{success-scale}$ = 0.5 , $\beta_{success-shape}$= 0.05 , $\beta_{score-scale}$= 1 , $\beta_{score-shape}$= 0.1.
Top two rows: 10 clusters; bottom two rows: 100 clusters.}
	\label{CoverageSE10_100clustersGompertz5}
\end{figure}

\begin{figure}[ht]
	\centering
	\includegraphics[scale=1]{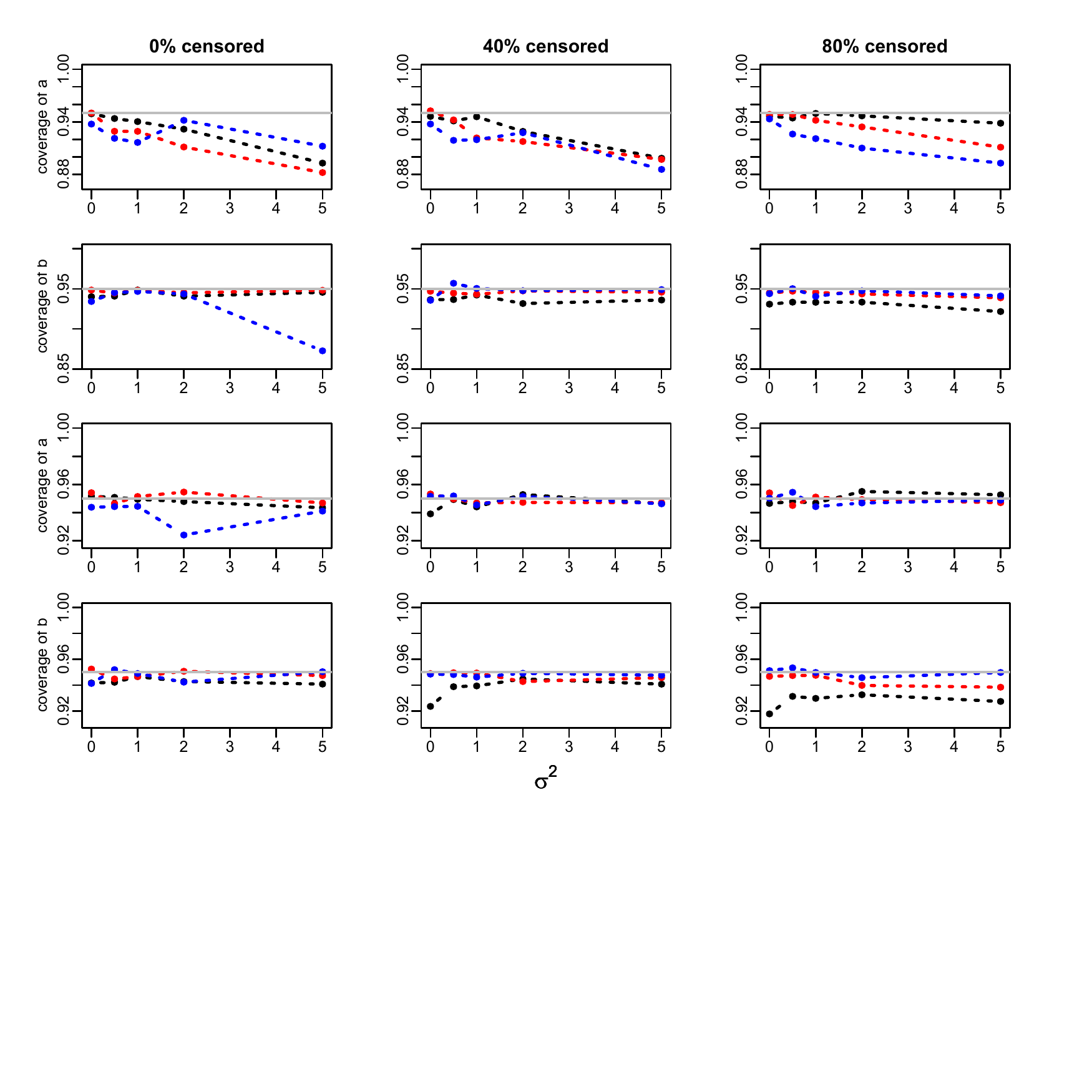}
	\caption{Coverage of the standard error based confidence intervals for the $a$ and $b$ parameters  in the Gompertz model.
Success proportion= 0.5.  Sample sizes: 300 (black), 1000 (red) and 10000 (blue).
True values: $\beta_{success-scale}$ = 0.5 , $\beta_{success-shape}$= -0.05 , $\beta_{score-scale}$= 1 , $\beta_{score-shape}$= - 0.1.
Top two rows: 10 clusters; bottom two rows: 100 clusters.}
	\label{CoverageSE10_100clustersGompertz6}
\end{figure}

\begin{figure}[ht]
	\centering
	\includegraphics[scale=1]{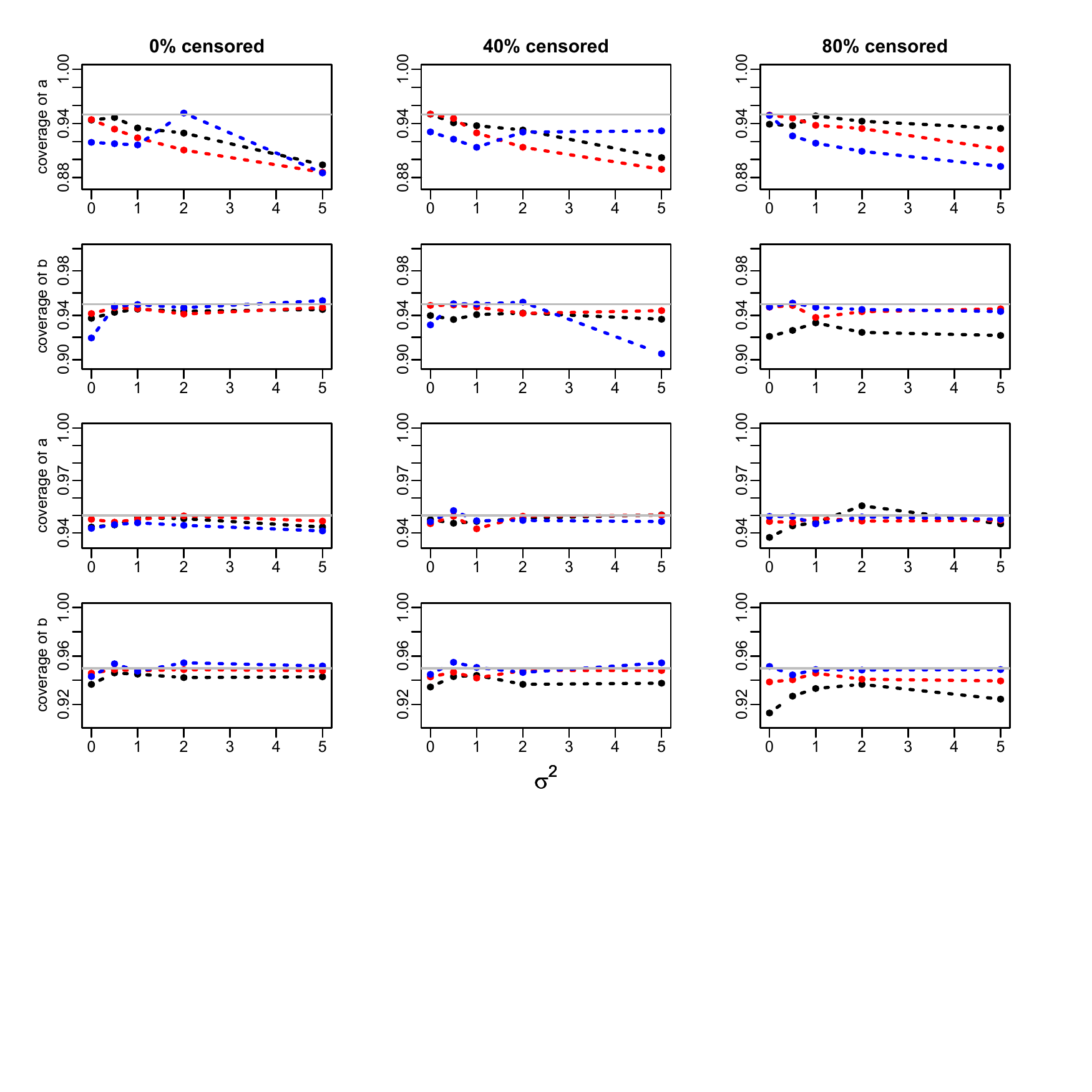}
	\caption{Coverage of the standard error based confidence intervals for the $a$ and $b$ parameters in the Gompertz model.
Success proportion= 0.5.  Sample sizes: 300 (black), 1000 (red) and 10000 (blue).
True values: $\beta_{success-scale}$ = -0.5 , $\beta_{success-shape}$= 0.05 , $\beta_{score-scale}$= -1 , $\beta_{score-shape}$= 0.1.
Top two rows: 10 clusters; bottom two rows: 100 clusters.}
	\label{CoverageSE10_100clustersGompertz7}
\end{figure}

\begin{figure}[ht]
	\centering
	\includegraphics[scale=1]{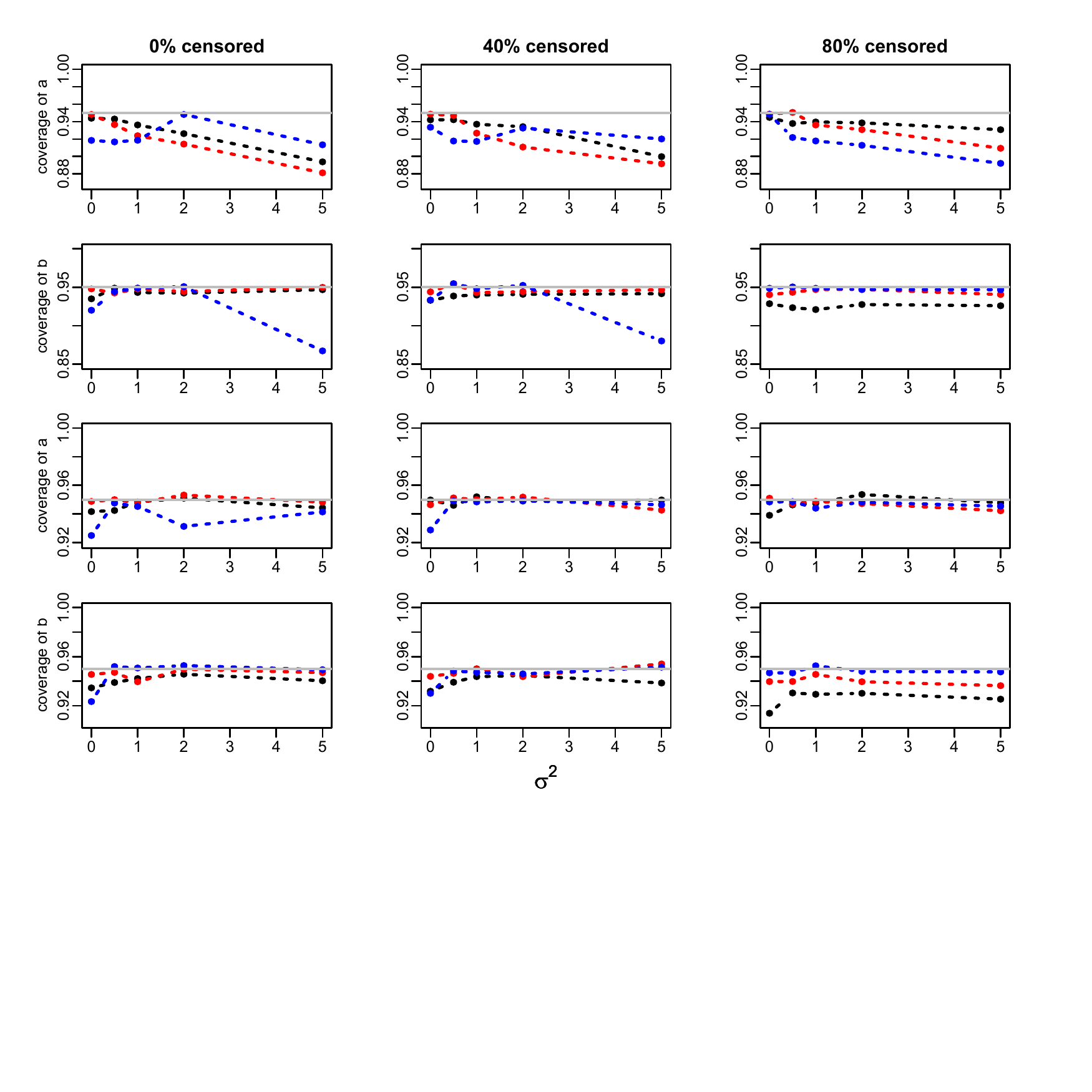}
	\caption{Coverage of the standard error based confidence intervals for the $a$ and $b$ parameters  in the Gompertz model.
Success proportion= 0.5.  Sample sizes: 300 (black), 1000 (red) and 10000 (blue).
True values: $\beta_{success-scale}$ = -0.5 , $\beta_{success-shape}$= -0.05 , $\beta_{score-scale}$= -1 , $\beta_{score-shape}$= -0.1.
Top two rows: 10 clusters; bottom two rows: 100 clusters.}
	\label{CoverageSE10_100clustersGompertz8}
\end{figure}

\clearpage

\setcounter{figure}{0}
\setcounter{section}{0}
\renewcommand{\thefigure}{F.\arabic{figure}}

\section*{F: Coverage of the standard-error-based confidence intervals for the Cox regression parameters and  the frailty variance}

Each figure corresponds to a particular  baseline distribution  (Weibull or Gompertz), a value of the probability of success $p_{success}$ (= 0.25 or 0.5), a value for the number of clusters $N_{cl}$ (=10, 100) and a particular choice of the signs of the Cox regression parameters (+ + + +, + - + -, - + - + and - - - -).\\

The absolute values of the Cox regression parameters are held constant at $\beta_{success-scale}$ = 0.5 , $\beta_{success-shape}$= 0.05 , $\beta_{score-scale}$= 1 , $\beta_{score-shape}$= 0.1. 

For each combination of a censoring proportion  (= 0, 40\%, 80\%), a panel plots, versus the frailty variance $\sigma^2$ (= 0, 0.5, 1, 2, 3, 4, 5), the empirical coverage of the true value of a Cox regression parameter or the true value of $\sigma^2$ by a SE-based confidence interval at 95\% nominal level, for three sample sizes (300, 1000 and 10000) . \\

\clearpage


\begin{figure}[ht]
	\centering
	\includegraphics[scale=1]{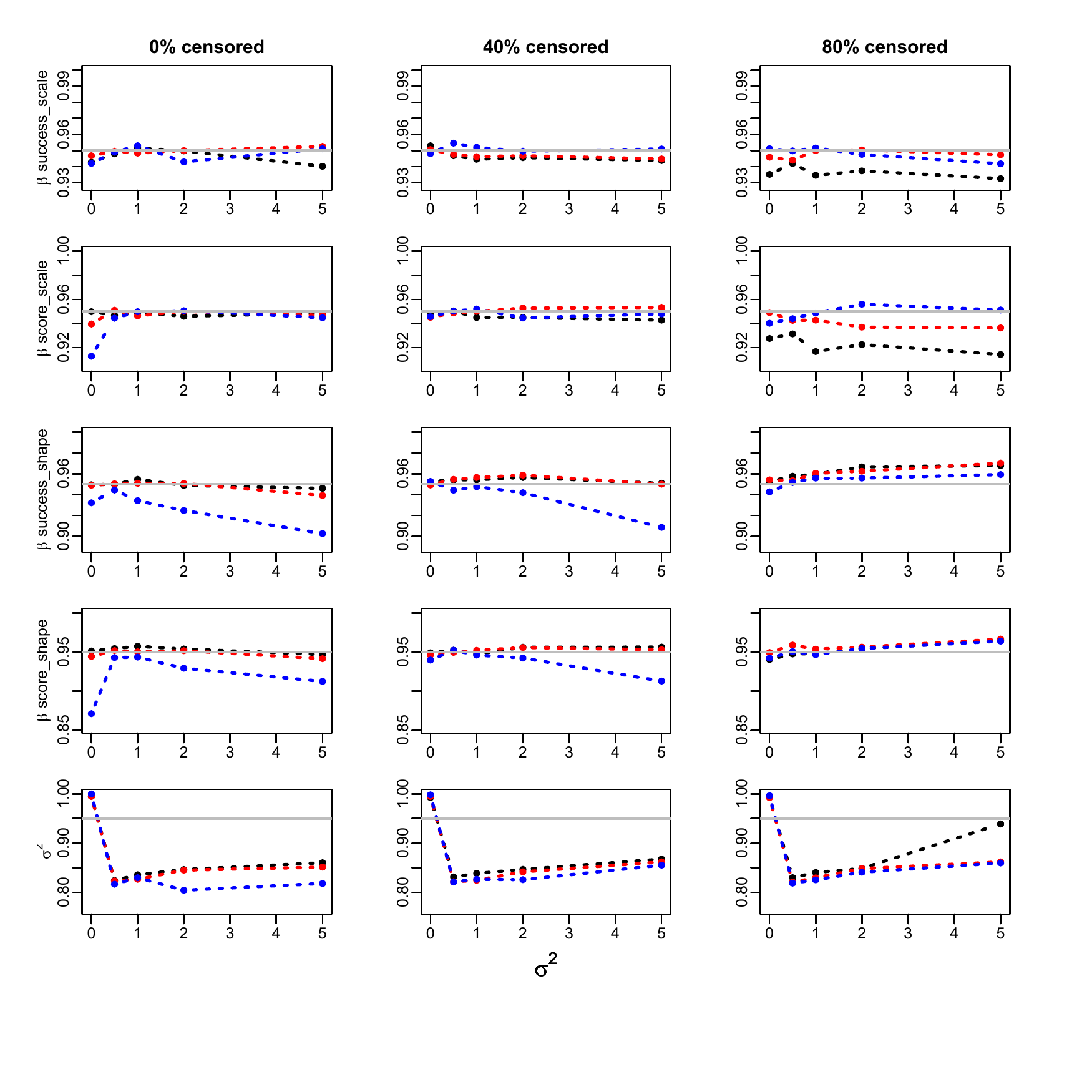}
	\caption{Coverage of the standard error based confidence intervals for the Cox regression parameters and $\sigma^2$ at nominal 95\% level. Weibull model.
Success proportion= 0.25. 10 clusters. Sample sizes: 300 (black), 1000 (red) and 10000 (blue).
True values: $\beta_{success-scale}$ = 0.5 , $\beta_{success-shape}$= 0.05 , $\beta_{score-scale}$= 1 , $\beta_{score-shape}$= 0.1}
	\label{CoverageSE10clustersWeibull1}
\end{figure}

\begin{figure}[ht]
	\centering
	\includegraphics[scale=1]{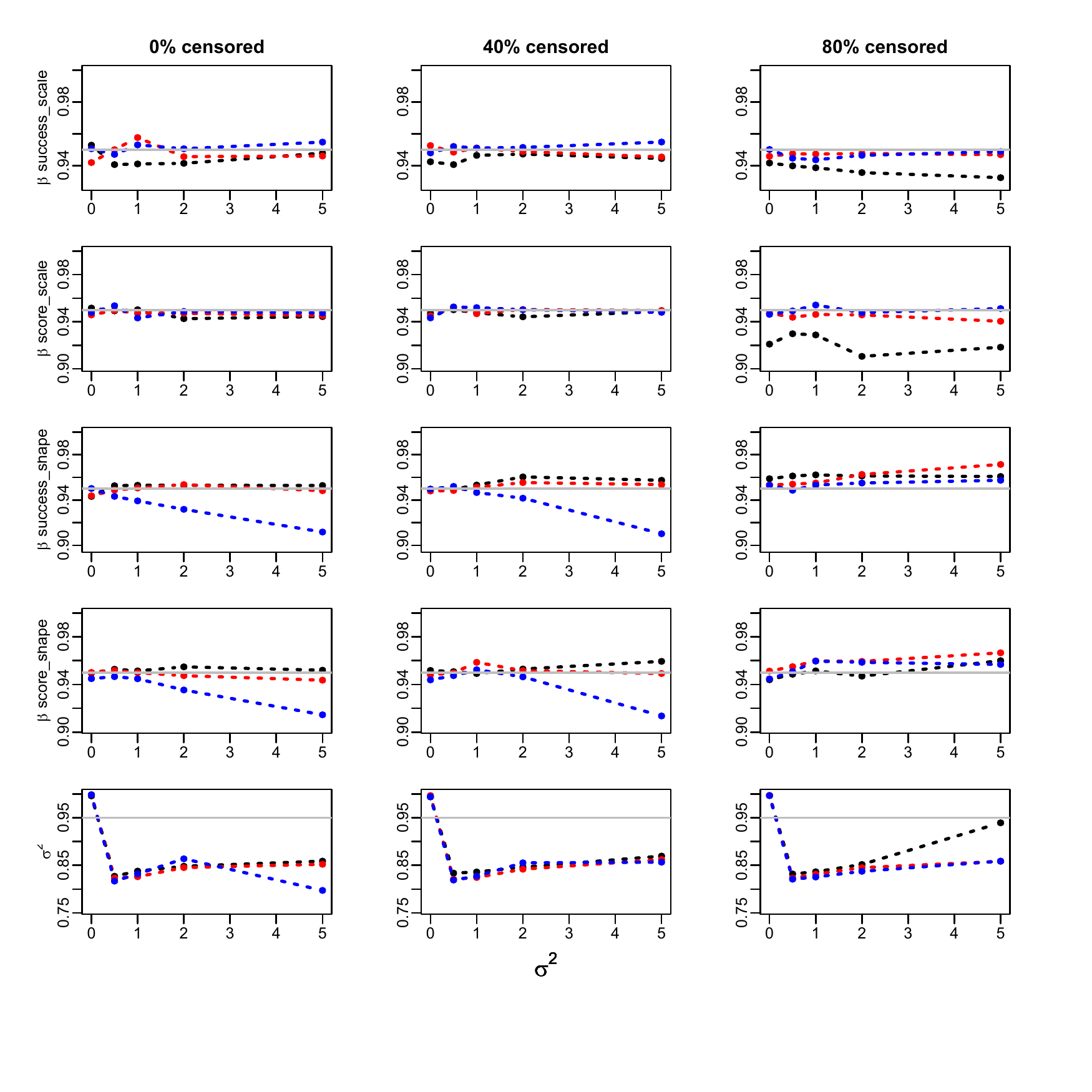}
	\caption{Coverage of the standard error based confidence intervals for the Cox regression parameters and $\sigma^2$ at nominal 95\% level.  Weibull model.
Success proportion= 0.25. 10 clusters. Sample sizes: 300 (black), 1000 (red) and 10000 (blue).
True values: $\beta_{success-scale}$ = 0.5 , $\beta_{success-shape}$= -0.05 , $\beta_{score-scale}$= 1 , $\beta_{score-shape}$= - 0.1}
	\label{CoverageSE10clustersWeibull2}
\end{figure}

\begin{figure}[ht]
	\centering
	\includegraphics[scale=1]{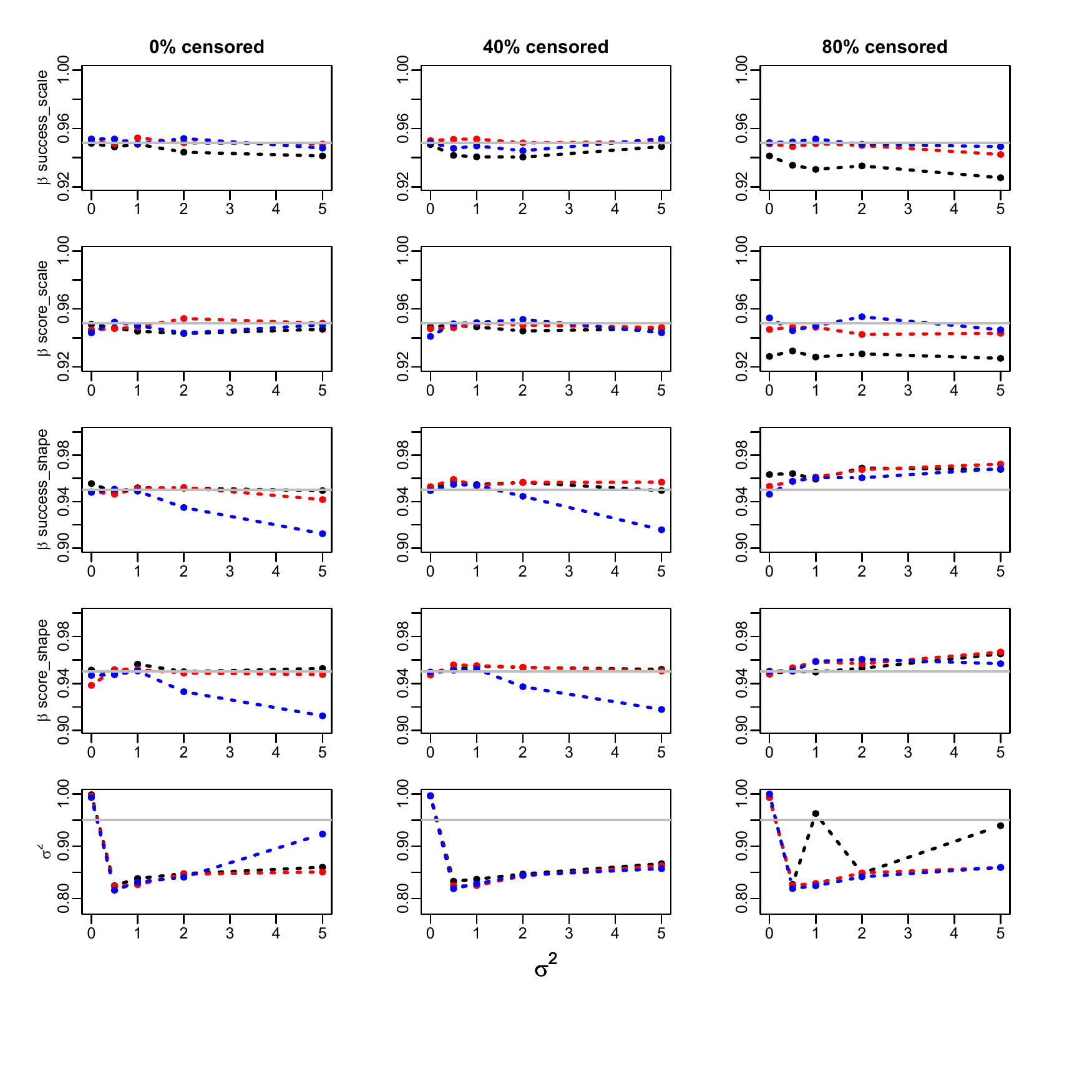}
	\caption{Coverage of the standard error based confidence intervals for the Cox regression parameters and $\sigma^2$ at nominal 95\% level. Weibull model.
Success proportion= 0.25. 10 clusters. Sample sizes: 300 (black), 1000 (red) and 10000 (blue).
True values: $\beta_{success-scale}$ = -0.5 , $\beta_{success-shape}$= 0.05 , $\beta_{score-scale}$= -1 , $\beta_{score-shape}$= 0.1}
	\label{CoverageSE10clustersWeibull3}
\end{figure}

\begin{figure}[ht]
	\centering
	\includegraphics[scale=1]{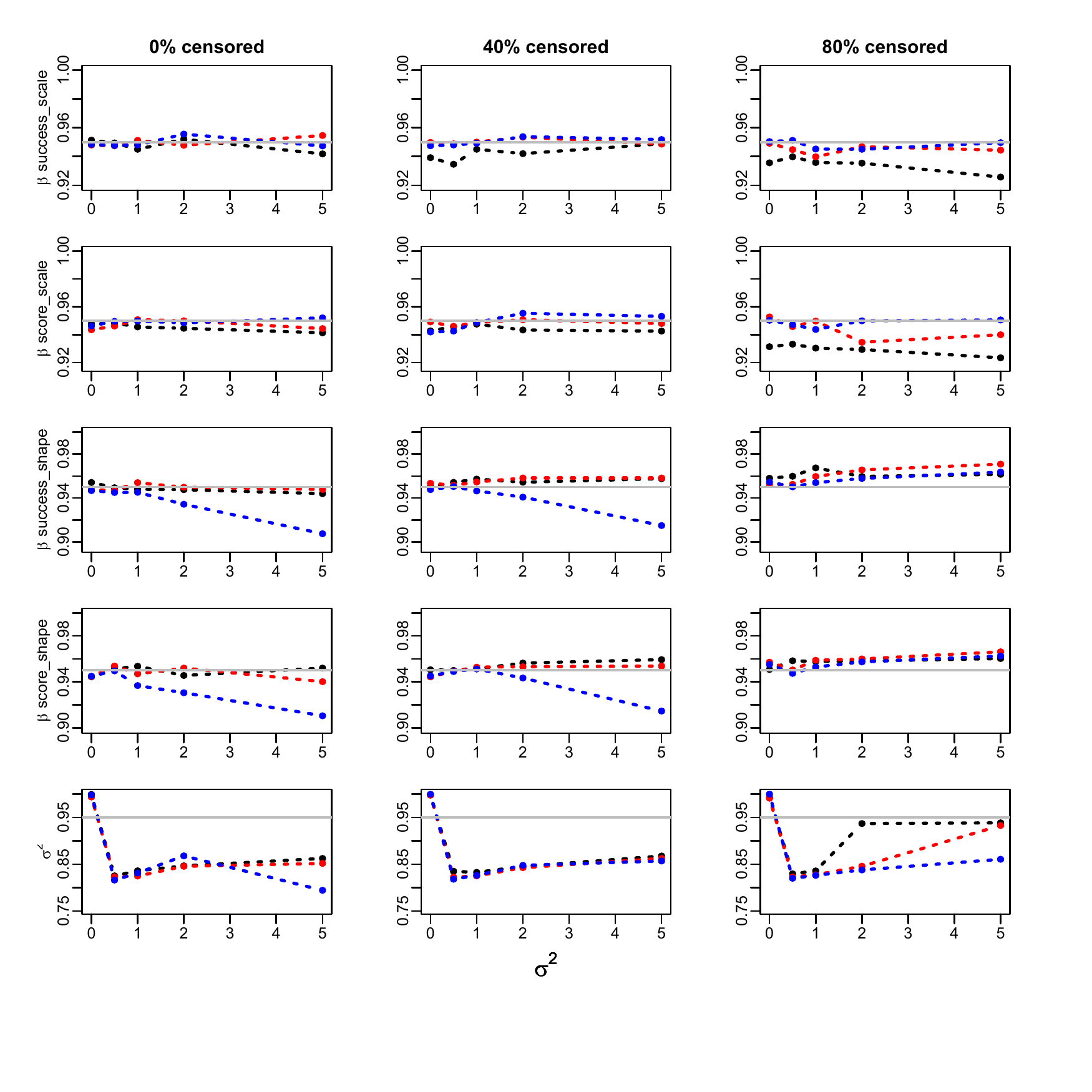}
	\caption{Coverage of the standard error based confidence intervals for the Cox regression parameters and $\sigma^2$ at nominal 95\% level. Weibull model.
Success proportion= 0.25. 10 clusters. Sample sizes: 300 (black), 1000 (red) and 10000 (blue).
True values: $\beta_{success-scale}$ = -0.5 , $\beta_{success-shape}$= -0.05 , $\beta_{score-scale}$= -1 , $\beta_{score-shape}$= -0.1}
	\label{CoverageSE10clustersWeibull4}
\end{figure}

\begin{figure}[ht]
	\centering
	\includegraphics[scale=1]{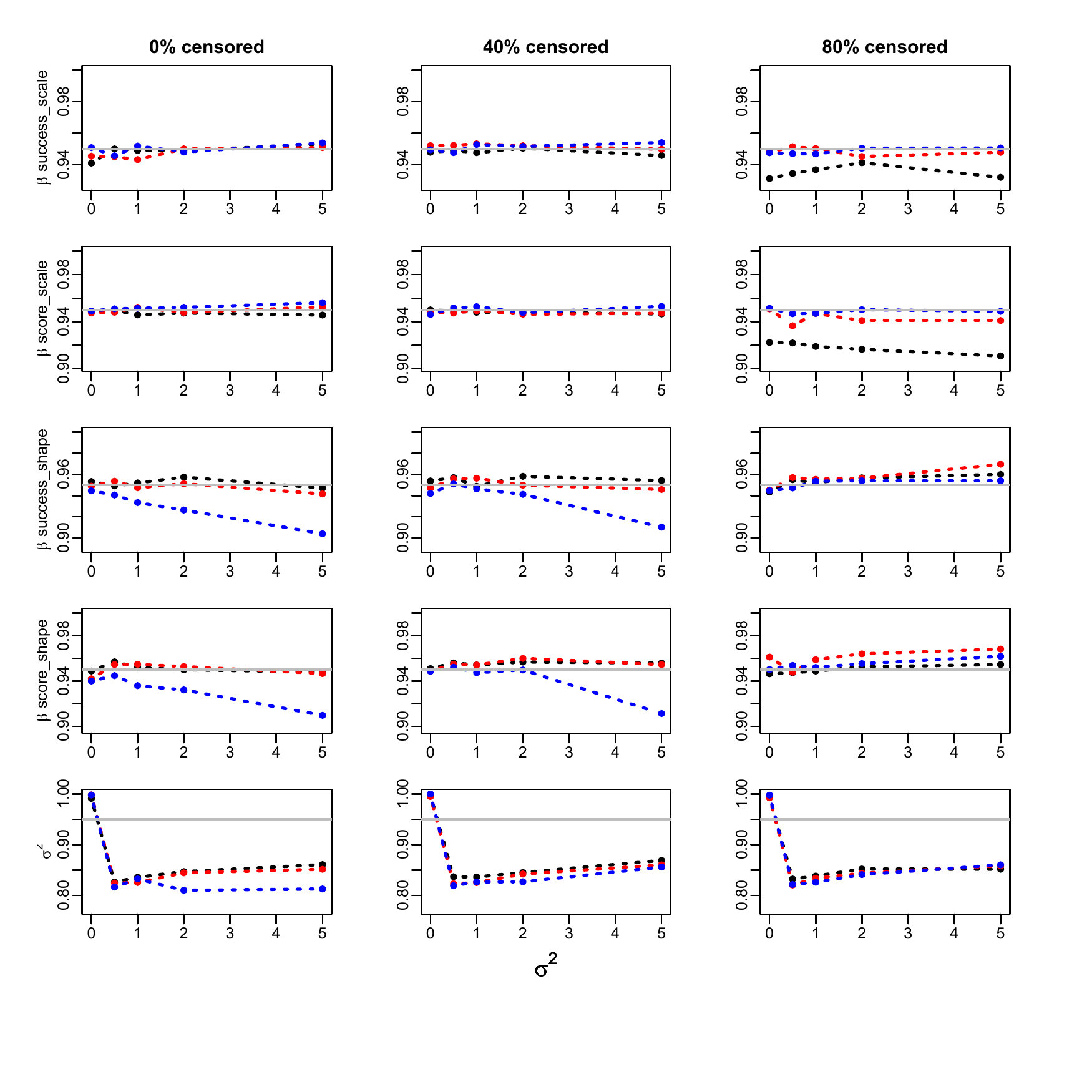}
	\caption{Coverage of the standard error based confidence intervals for the Cox regression parameters and $\sigma^2$ at nominal 95\% level. Weibull model.
Success proportion= 0.5. 10 clusters. Sample sizes: 300 (black), 1000 (red) and 10000 (blue).
True values: $\beta_{success-scale}$ = 0.5 , $\beta_{success-shape}$= 0.05 , $\beta_{score-scale}$= 1 , $\beta_{score-shape}$= 0.1}
	\label{CoverageSE10clustersWeibull5}
\end{figure}

\begin{figure}[ht]
	\centering
	\includegraphics[scale=1]{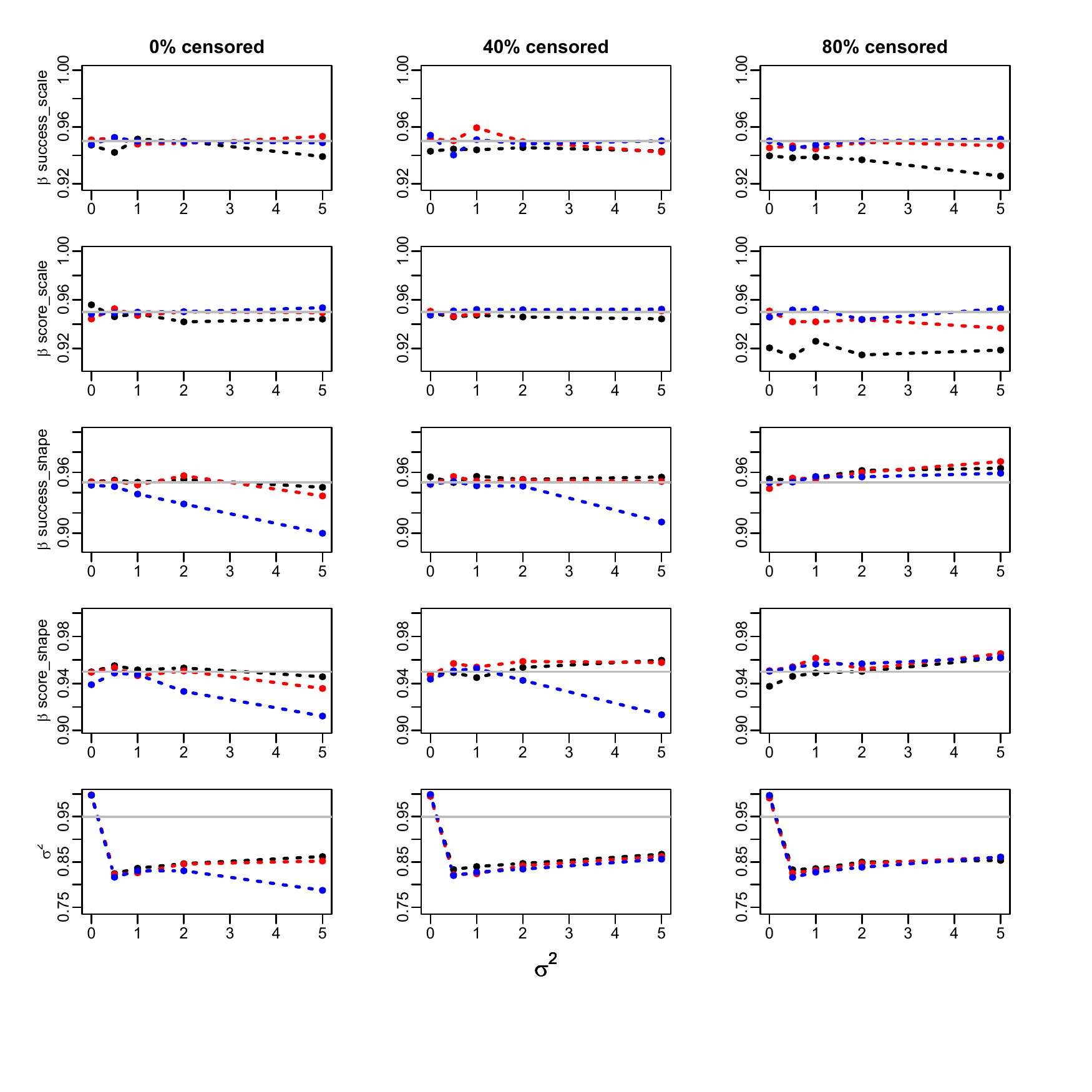}
	\caption{Coverage of the standard error based confidence intervals for the Cox regression parameters and $\sigma^2$ at nominal 95\% level. Weibull model.
Success proportion= 0.5. 10 clusters. Sample sizes: 300 (black), 1000 (red) and 10000 (blue).
True values: $\beta_{success-scale}$ = 0.5 , $\beta_{success-shape}$= -0.05 , $\beta_{score-scale}$= 1 , $\beta_{score-shape}$= - 0.1}
	\label{CoverageSE10clustersWeibull6}
\end{figure}

\begin{figure}[ht]
	\centering
	\includegraphics[scale=1]{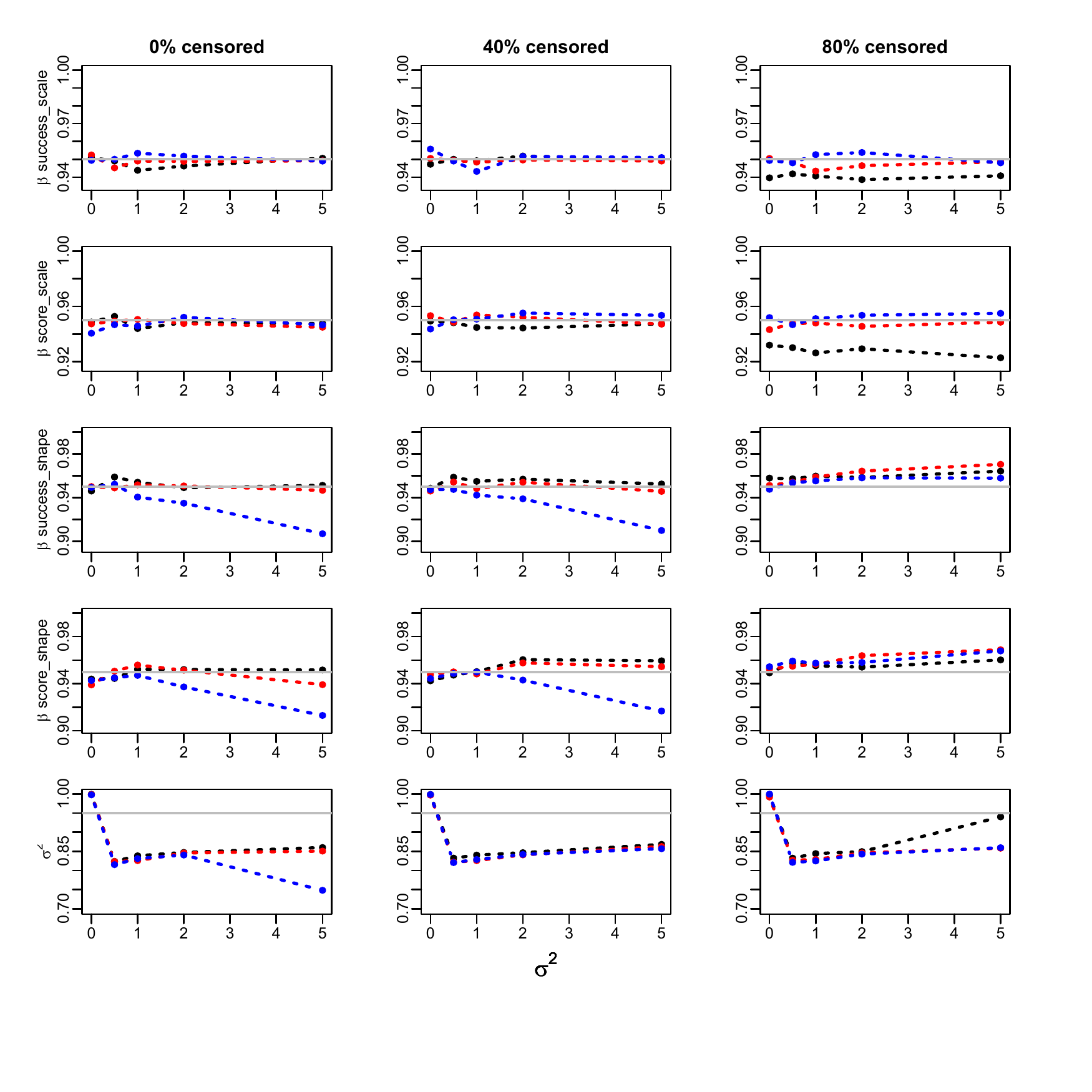}
	\caption{Coverage of the standard error based confidence intervals for the Cox regression parameters and $\sigma^2$ at nominal 95\% level. Weibull model.
Success proportion= 0.5. 10 clusters. Sample sizes: 300 (black), 1000 (red) and 10000 (blue).
True values: $\beta_{success-scale}$ = -0.5 , $\beta_{success-shape}$= 0.05 , $\beta_{score-scale}$= -1 , $\beta_{score-shape}$= 0.1}
	\label{CoverageSE10clustersWeibull7}
\end{figure}

\begin{figure}[ht]
	\centering
	\includegraphics[scale=1]{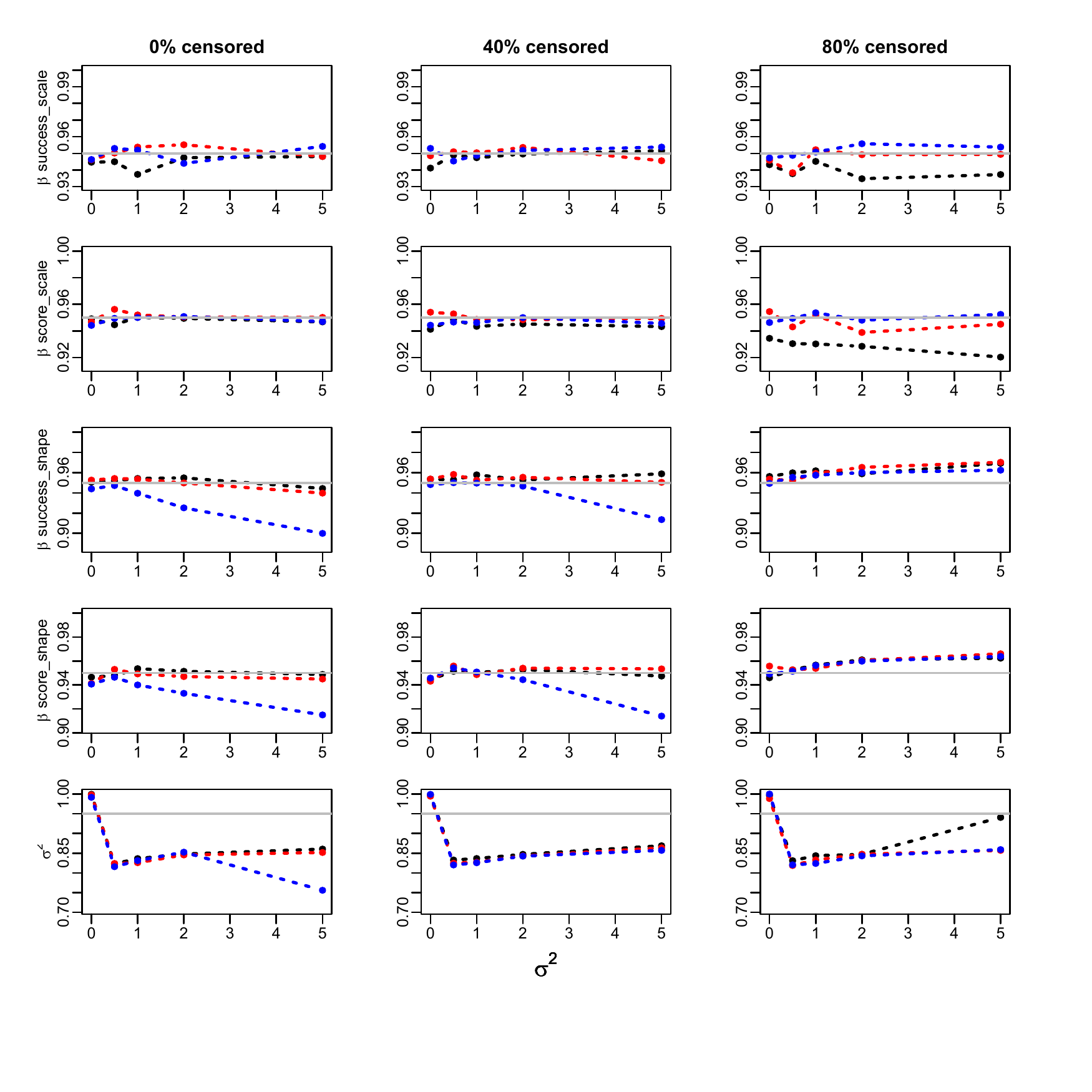}
	\caption{Coverage of the standard error based confidence intervals for the Cox regression parameters and $\sigma^2$ at nominal 95\% level. Weibull model.
Success proportion= 0.5. 10 clusters. Sample sizes: 300 (black), 1000 (red) and 10000 (blue).
True values: $\beta_{success-scale}$ = -0.5 , $\beta_{success-shape}$= -0.05 , $\beta_{score-scale}$= -1 , $\beta_{score-shape}$= -0.1}
	\label{CoverageSE10clustersWeibull8}
\end{figure}

 \begin{figure}[ht]
	\centering
	\includegraphics[scale=1]{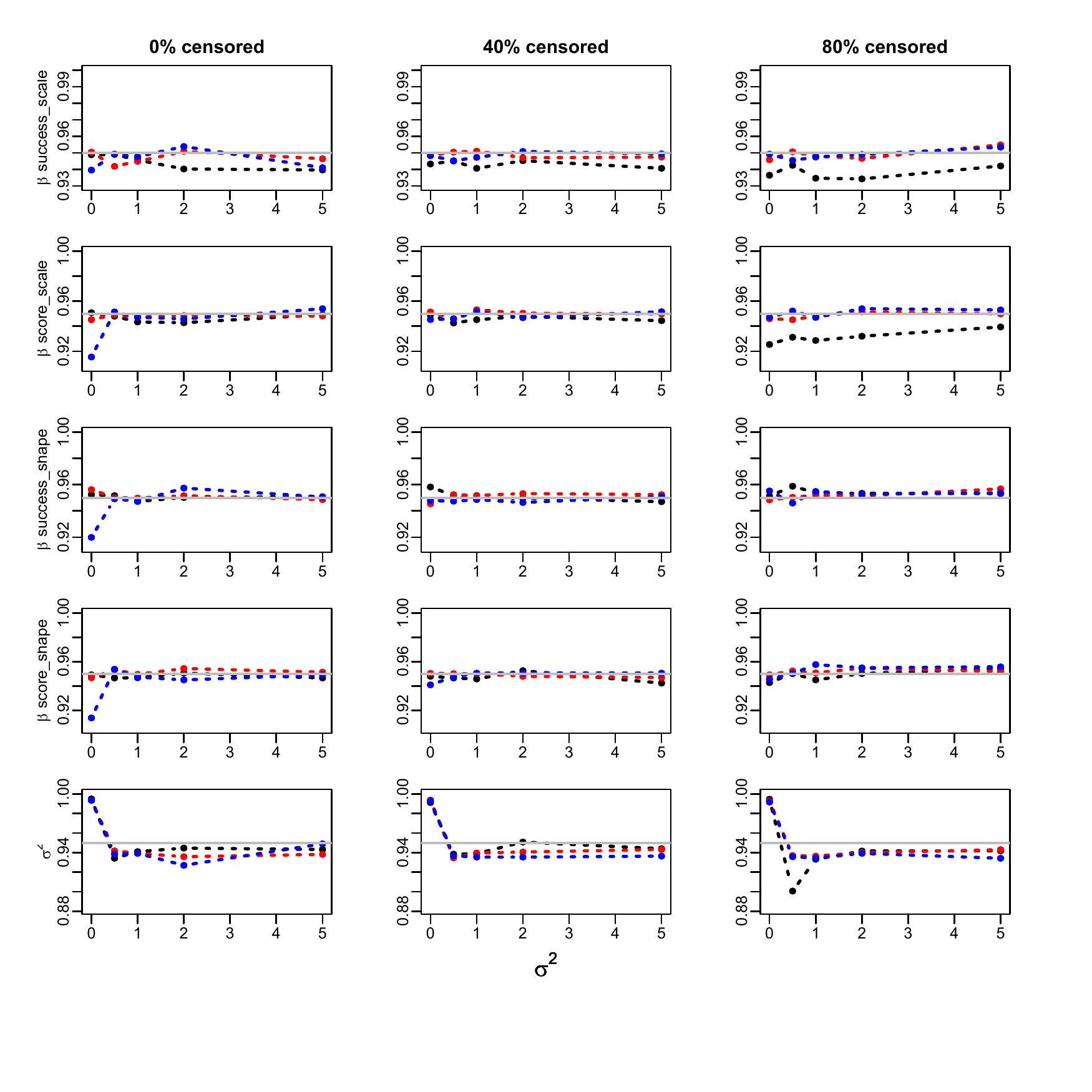}
	\caption{Coverage of the standard error based confidence intervals for the Cox regression parameters and $\sigma^2$ at nominal 95\% level. Weibull model.
Success proportion= 0.25. 100 clusters. Sample sizes: 300 (black), 1000 (red) and 10000 (blue).
True values: $\beta_{success-scale}$ = 0.5 , $\beta_{success-shape}$= 0.05 , $\beta_{score-scale}$= 1 , $\beta_{score-shape}$= 0.1}
	\label{CoverageSE100clustersWeibull1}
\end{figure}

\begin{figure}[ht]
	\centering
	\includegraphics[scale=1]{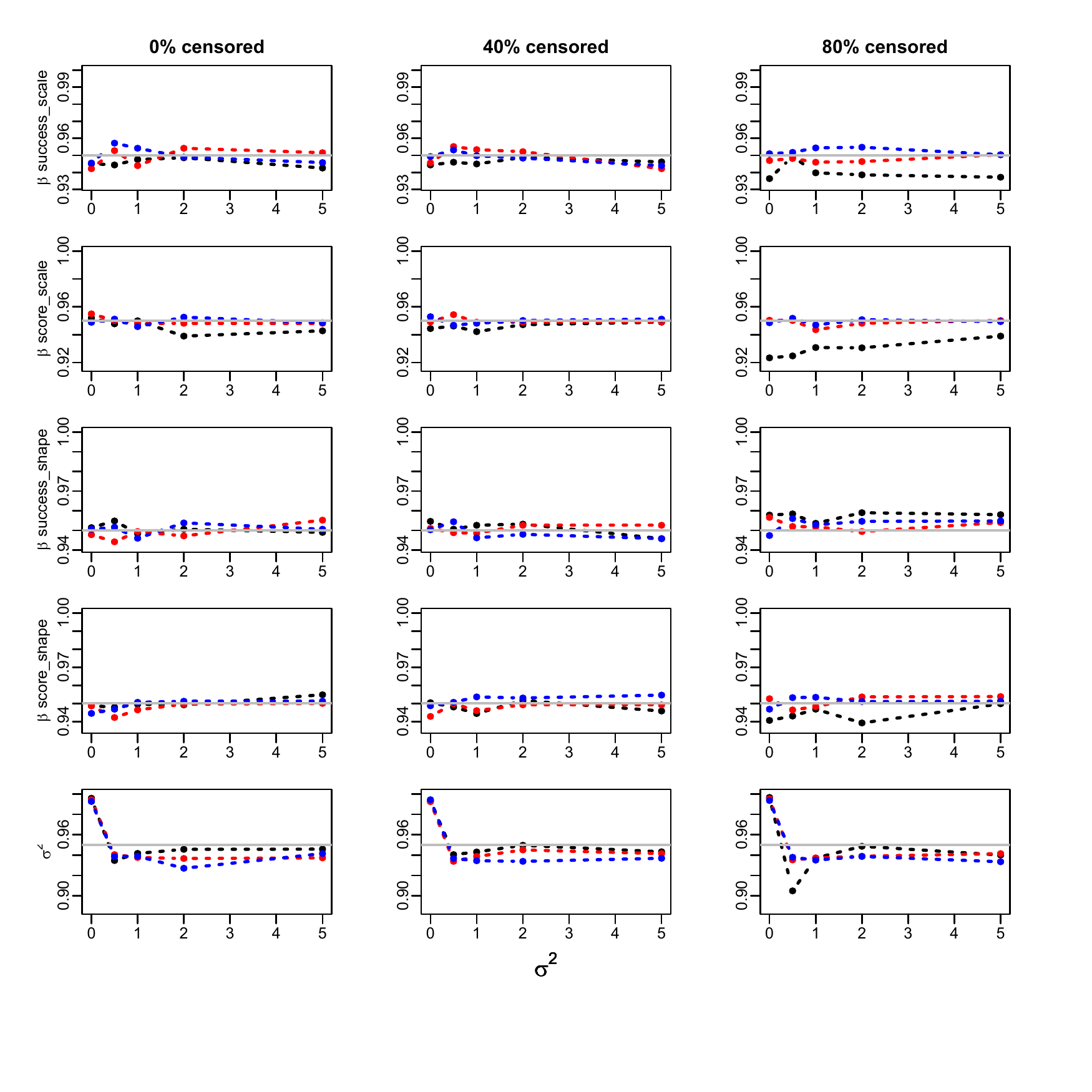}
	\caption{Coverage of the standard error based confidence intervals for the Cox regression parameters and $\sigma^2$ at nominal 95\% level. Weibull model.
Success proportion= 0.25. 100 clusters. Sample sizes: 300 (black), 1000 (red) and 10000 (blue).
True values: $\beta_{success-scale}$ = 0.5 , $\beta_{success-shape}$= -0.05 , $\beta_{score-scale}$= 1 , $\beta_{score-shape}$= - 0.1}
	\label{CoverageSE100clustersWeibull2}
\end{figure}

\begin{figure}[ht]
	\centering
	\includegraphics[scale=1]{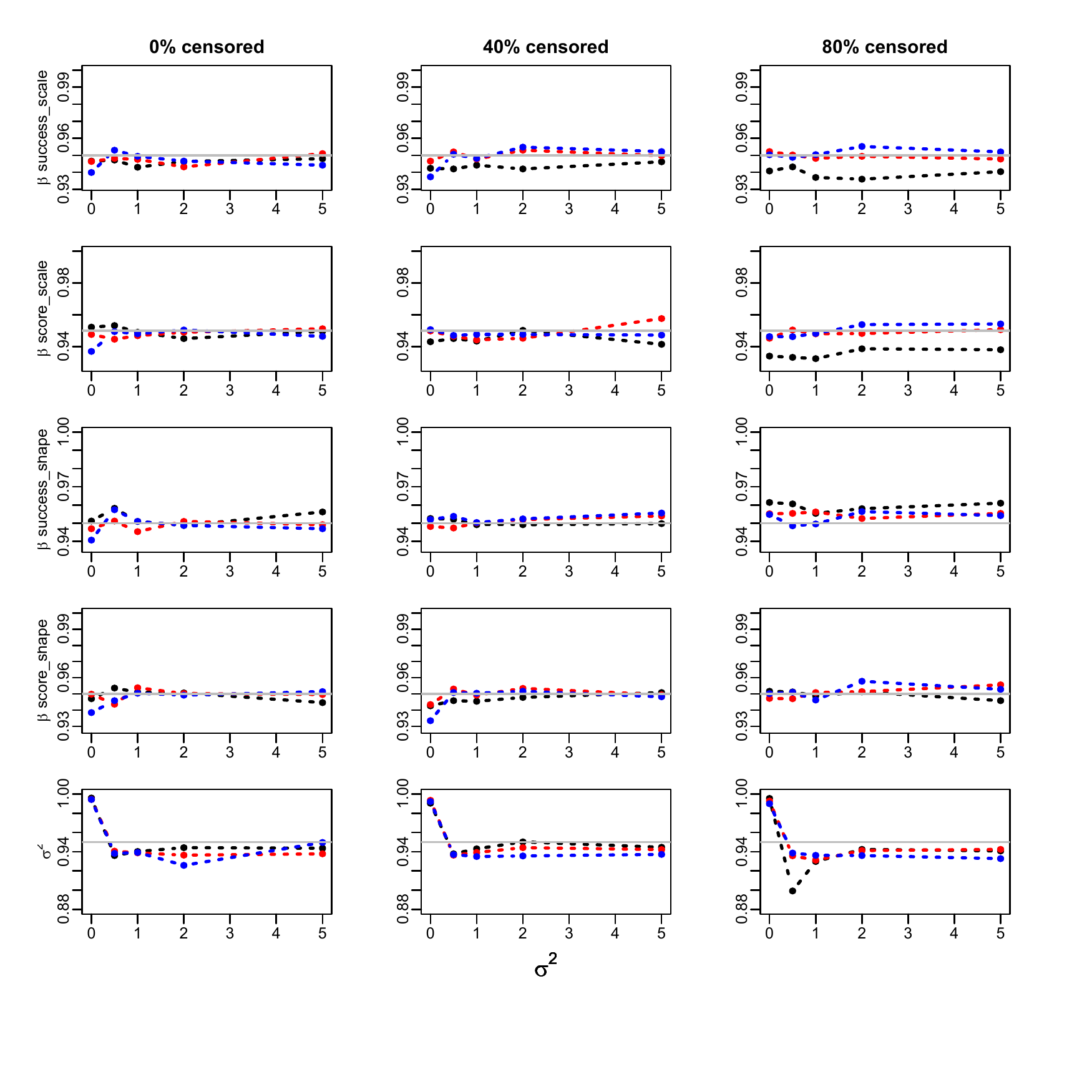}
	\caption{Coverage of the standard error based confidence intervals for the Cox regression parameters and $\sigma^2$ at nominal 95\% level. Weibull model.
Success proportion= 0.25. 100 clusters. Sample sizes: 300 (black), 1000 (red) and 10000 (blue).
True values: $\beta_{success-scale}$ = -0.5 , $\beta_{success-shape}$= 0.05 , $\beta_{score-scale}$= -1 , $\beta_{score-shape}$= 0.1}
	\label{CoverageSE100clustersWeibull3}
\end{figure}

\begin{figure}[ht]
	\centering
	\includegraphics[scale=1]{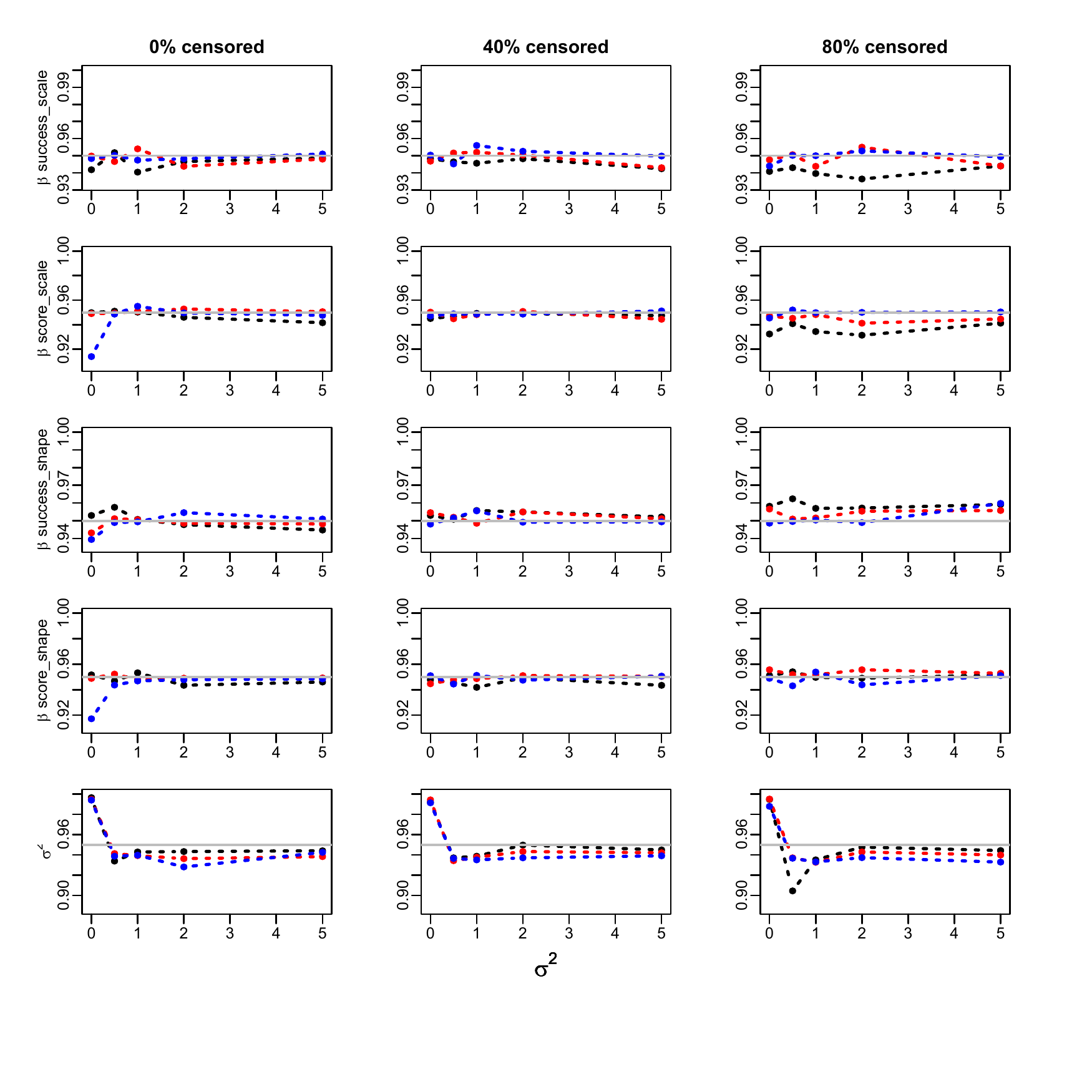}
	\caption{Coverage of the standard error based confidence intervals for the Cox regression parameters and $\sigma^2$ at nominal 95\% level. Weibull model.
Success proportion= 0.25. 100 clusters. Sample sizes: 300 (black), 1000 (red) and 10000 (blue).
True values: $\beta_{success-scale}$ = -0.5 , $\beta_{success-shape}$= -0.05 , $\beta_{score-scale}$= -1 , $\beta_{score-shape}$= -0.1}
	\label{CoverageSE100clustersWeibull4}
\end{figure}

\begin{figure}[ht]
	\centering
	\includegraphics[scale=1]{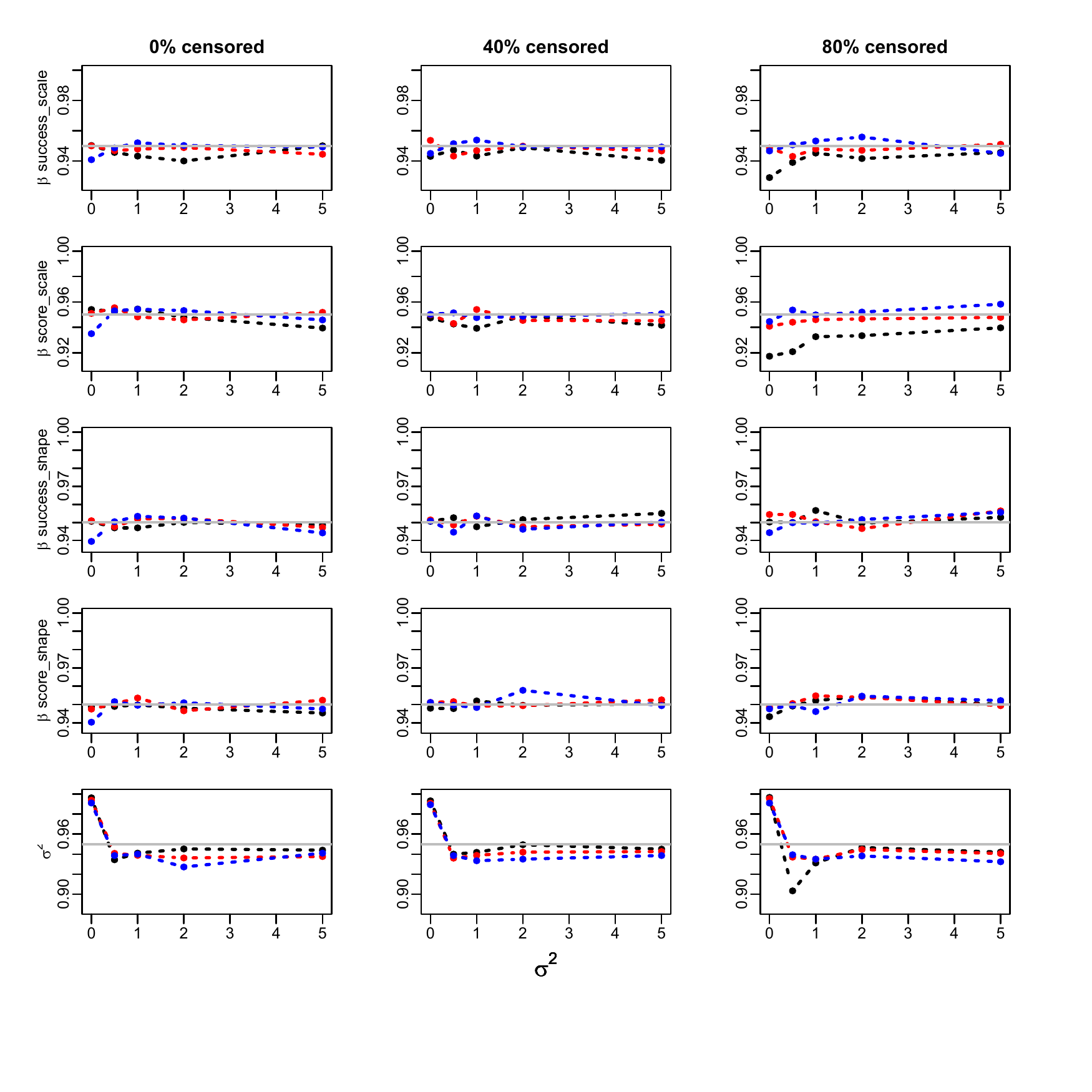}
	\caption{Coverage of the standard error based confidence intervals for the Cox regression parameters and $\sigma^2$ at nominal 95\% level. Weibull model.
Success proportion= 0.5. 100 clusters. Sample sizes: 300 (black), 1000 (red) and 10000 (blue).
True values: $\beta_{success-scale}$ = 0.5 , $\beta_{success-shape}$= 0.05 , $\beta_{score-scale}$= 1 , $\beta_{score-shape}$= 0.1}
	\label{CoverageSE100clustersWeibull5}
\end{figure}

\begin{figure}[ht]
	\centering
	\includegraphics[scale=1]{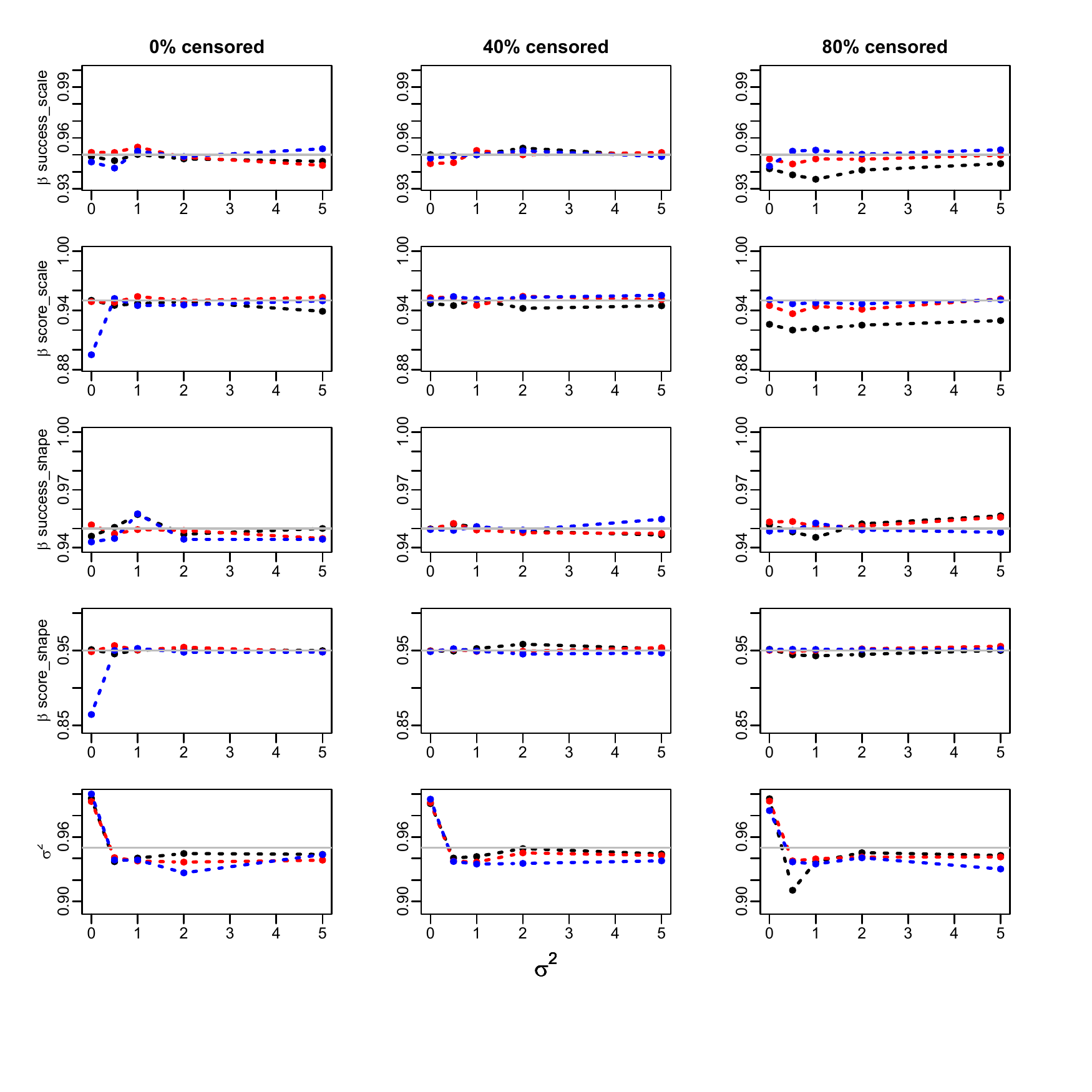}
	\caption{Coverage of the standard error based confidence intervals for the Cox regression parameters and $\sigma^2$ at nominal 95\% level. Weibull model.
Success proportion= 0.5. 100 clusters. Sample sizes: 300 (black), 1000 (red) and 10000 (blue).
True values: $\beta_{success-scale}$ = 0.5 , $\beta_{success-shape}$= -0.05 , $\beta_{score-scale}$= 1 , $\beta_{score-shape}$= - 0.1}
	\label{CoverageSE100clustersWeibull6}
\end{figure}

\begin{figure}[ht]
	\centering
	\includegraphics[scale=1]{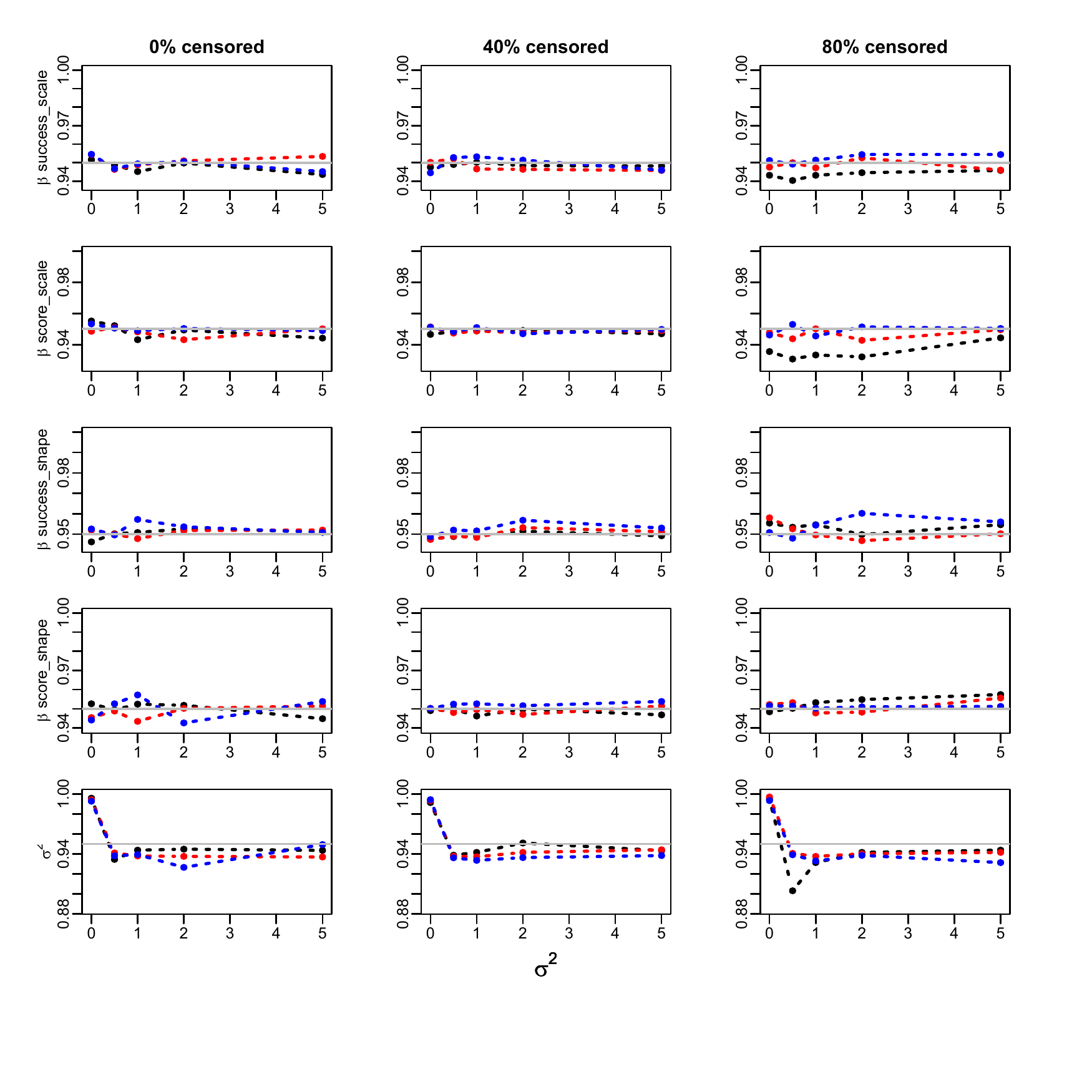}
	\caption{Coverage of the standard error based confidence intervals for the Cox regression parameters and $\sigma^2$ at nominal 95\% level. Weibull model.
Success proportion= 0.5. 100 clusters. Sample sizes: 300 (black), 1000 (red) and 10000 (blue).
True values: $\beta_{success-scale}$ = -0.5 , $\beta_{success-shape}$= 0.05 , $\beta_{score-scale}$= -1 , $\beta_{score-shape}$= 0.1}
	\label{CoverageSE100clustersWeibull7}
\end{figure}

\begin{figure}[ht]
	\centering
	\includegraphics[scale=1]{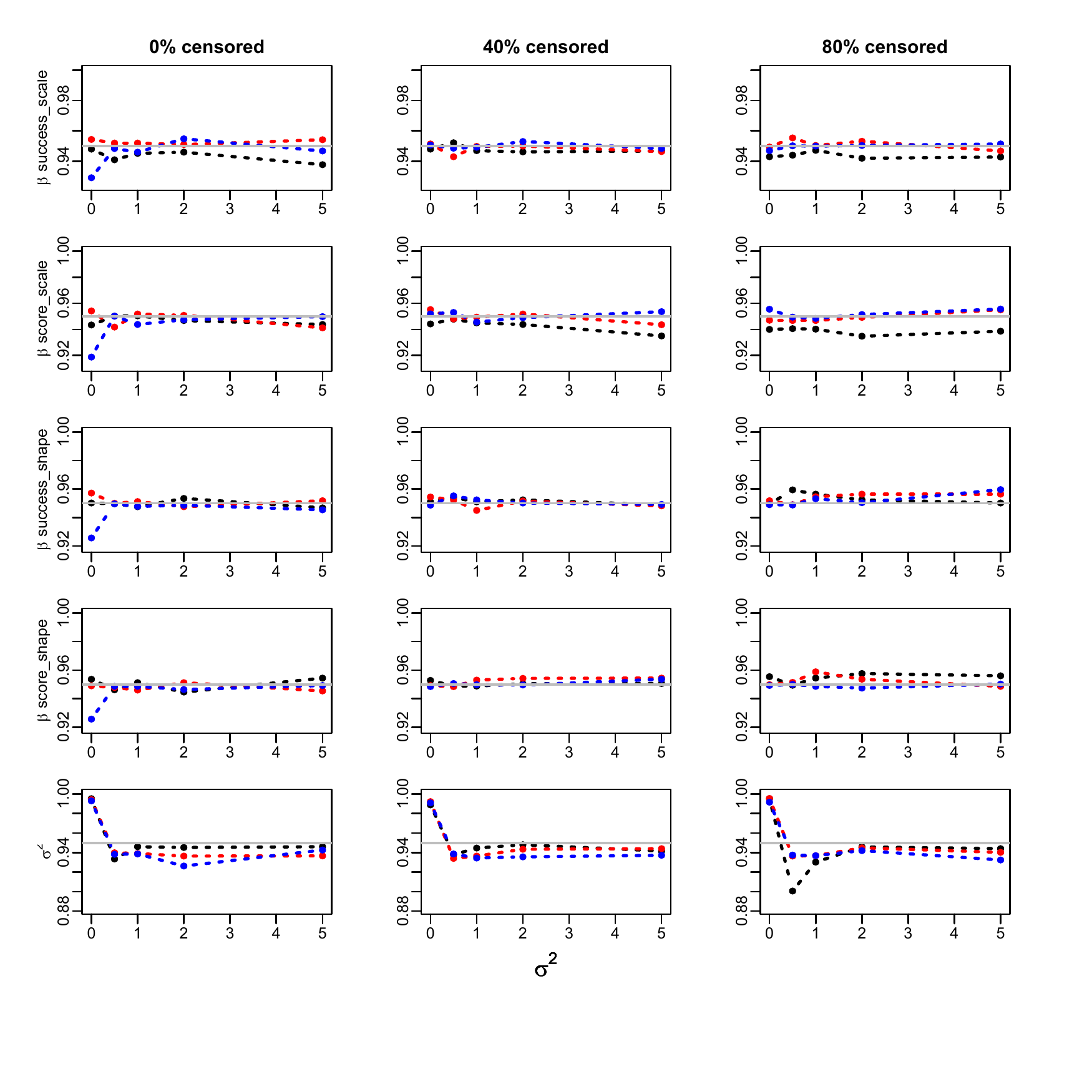}
	\caption{Coverage of the standard error based confidence intervals for the Cox regression parameters and $\sigma^2$ at nominal 95\% level. Weibull model.
Success proportion= 0.5. 100 clusters. Sample sizes: 300 (black), 1000 (red) and 10000 (blue).
True values: $\beta_{success-scale}$ = -0.5 , $\beta_{success-shape}$= -0.05 , $\beta_{score-scale}$= -1 , $\beta_{score-shape}$= -0.1}
	\label{CoverageSE100clustersWeibull8}
\end{figure}


 \begin{figure}[ht]
	\centering
	\includegraphics[scale=1]{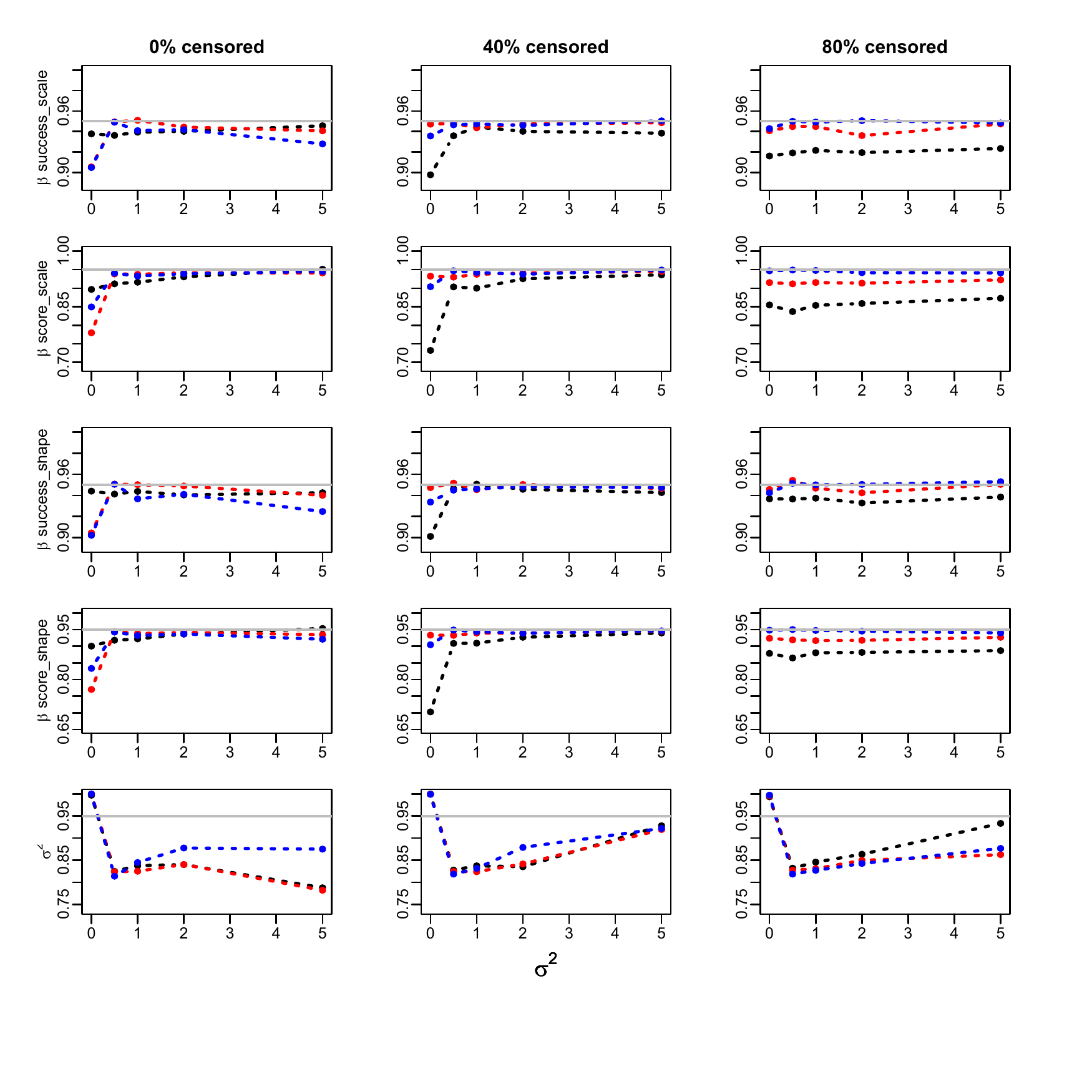}
	\caption{Coverage of the standard error based confidence intervals for the Cox regression parameters and $\sigma^2$ at nominal 95\% level. Gompertz model.
Success proportion= 0.25. 10 clusters. Sample sizes: 300 (black), 1000 (red) and 10000 (blue).
True values: $\beta_{success-scale}$ = 0.5 , $\beta_{success-shape}$= 0.05 , $\beta_{score-scale}$= 1 , $\beta_{score-shape}$= 0.1}
	\label{CoverageSE10clustersGompertz1}
\end{figure}

\begin{figure}[ht]
	\centering
	\includegraphics[scale=1]{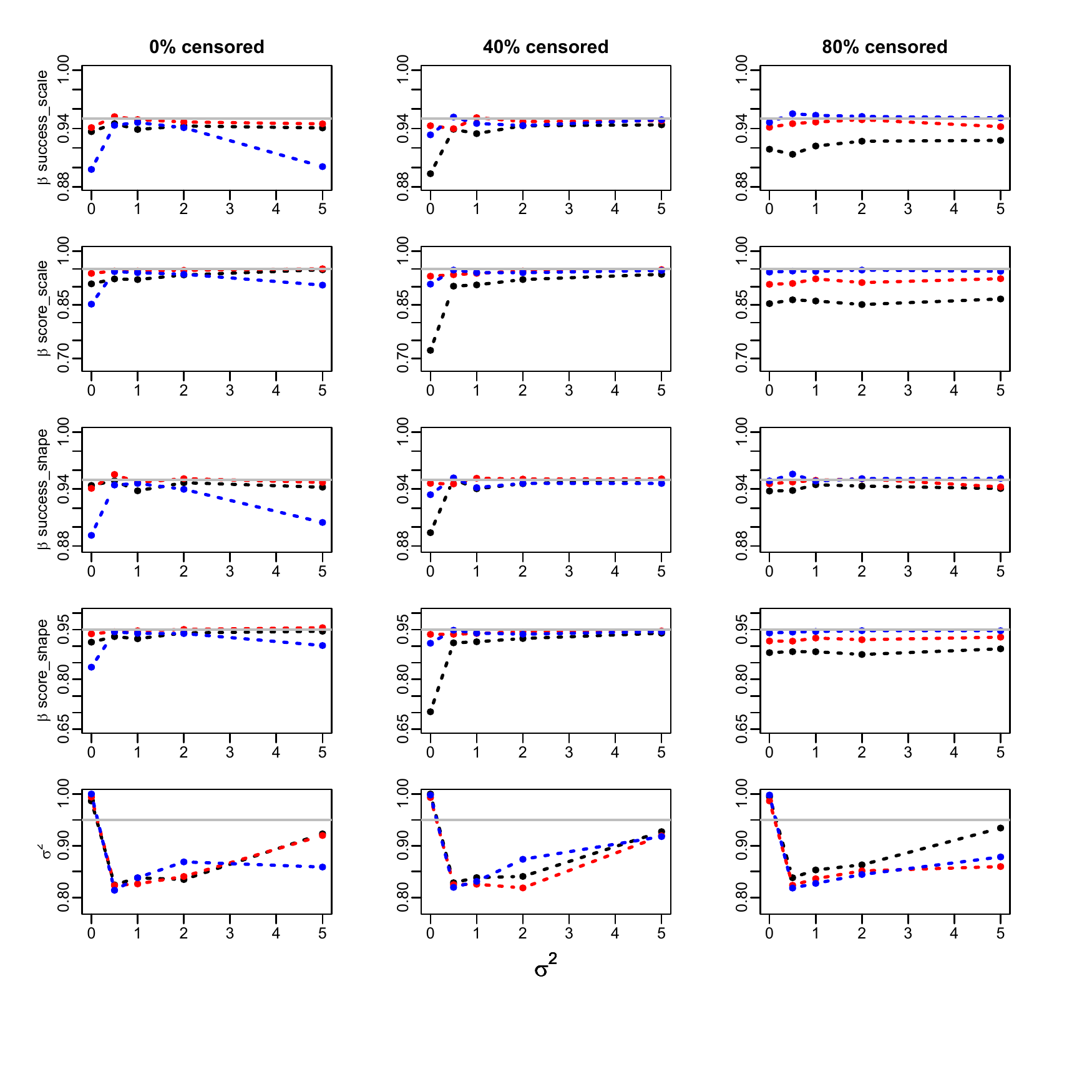}
	\caption{Coverage of the standard error based confidence intervals for the Cox regression parameters and $\sigma^2$ at nominal 95\% level. Gompertz model.
Success proportion= 0.25. 10 clusters. Sample sizes: 300 (black), 1000 (red) and 10000 (blue).
True values: $\beta_{success-scale}$ = 0.5 , $\beta_{success-shape}$= -0.05 , $\beta_{score-scale}$= 1 , $\beta_{score-shape}$= - 0.1}
	\label{CoverageSE10clustersGompertz2}
\end{figure}

\begin{figure}[ht]
	\centering
	\includegraphics[scale=1]{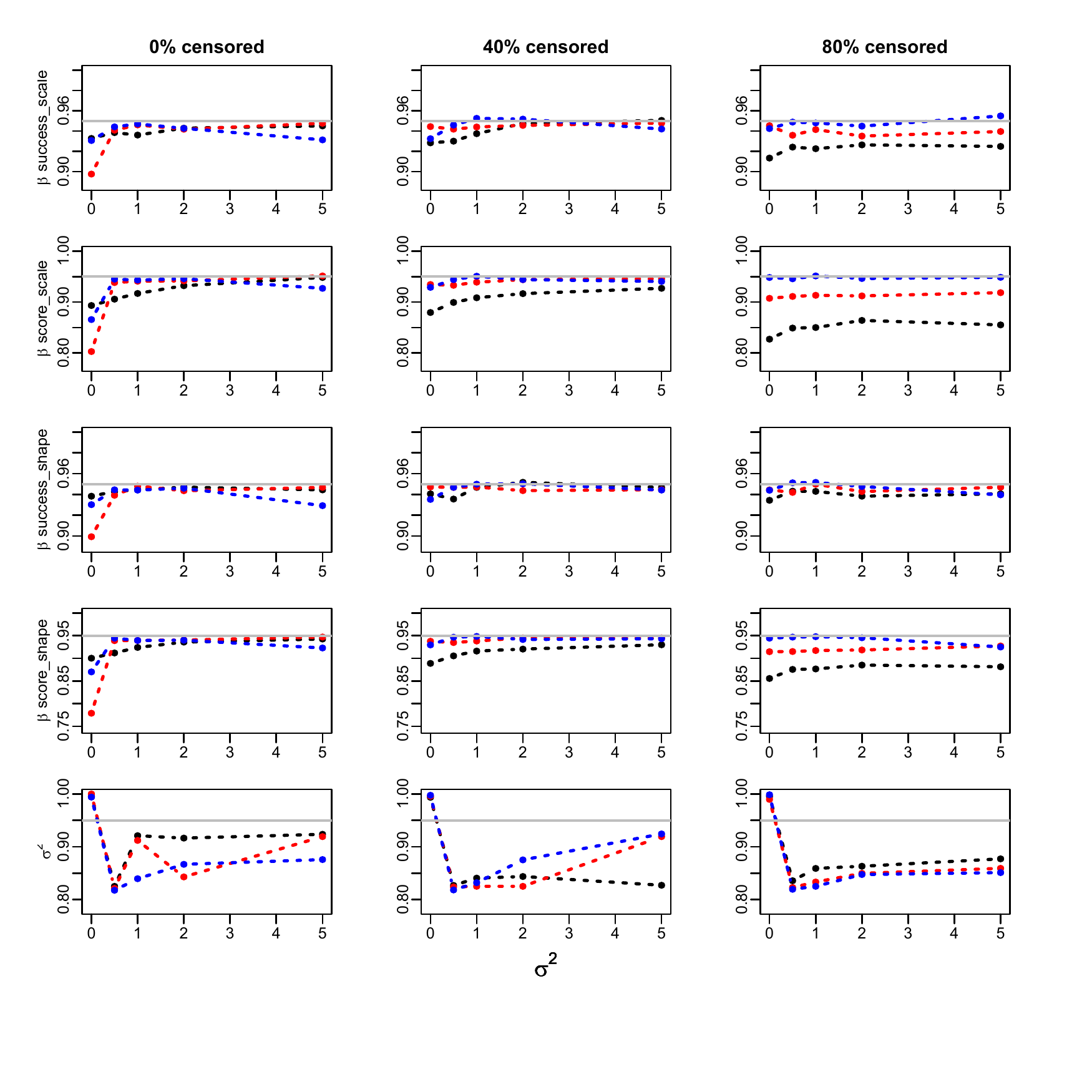}
	\caption{Coverage of the standard error based confidence intervals for the Cox regression parameters and $\sigma^2$ at nominal 95\% level. the Gompertz model.
Success proportion= 0.25. 10 clusters. Sample sizes: 300 (black), 1000 (red) and 10000 (blue).
True values: $\beta_{success-scale}$ = -0.5 , $\beta_{success-shape}$= 0.05 , $\beta_{score-scale}$= -1 , $\beta_{score-shape}$= 0.1}
	\label{CoverageSE10clustersGompertz3}
\end{figure}

\begin{figure}[ht]
	\centering
	\includegraphics[scale=1]{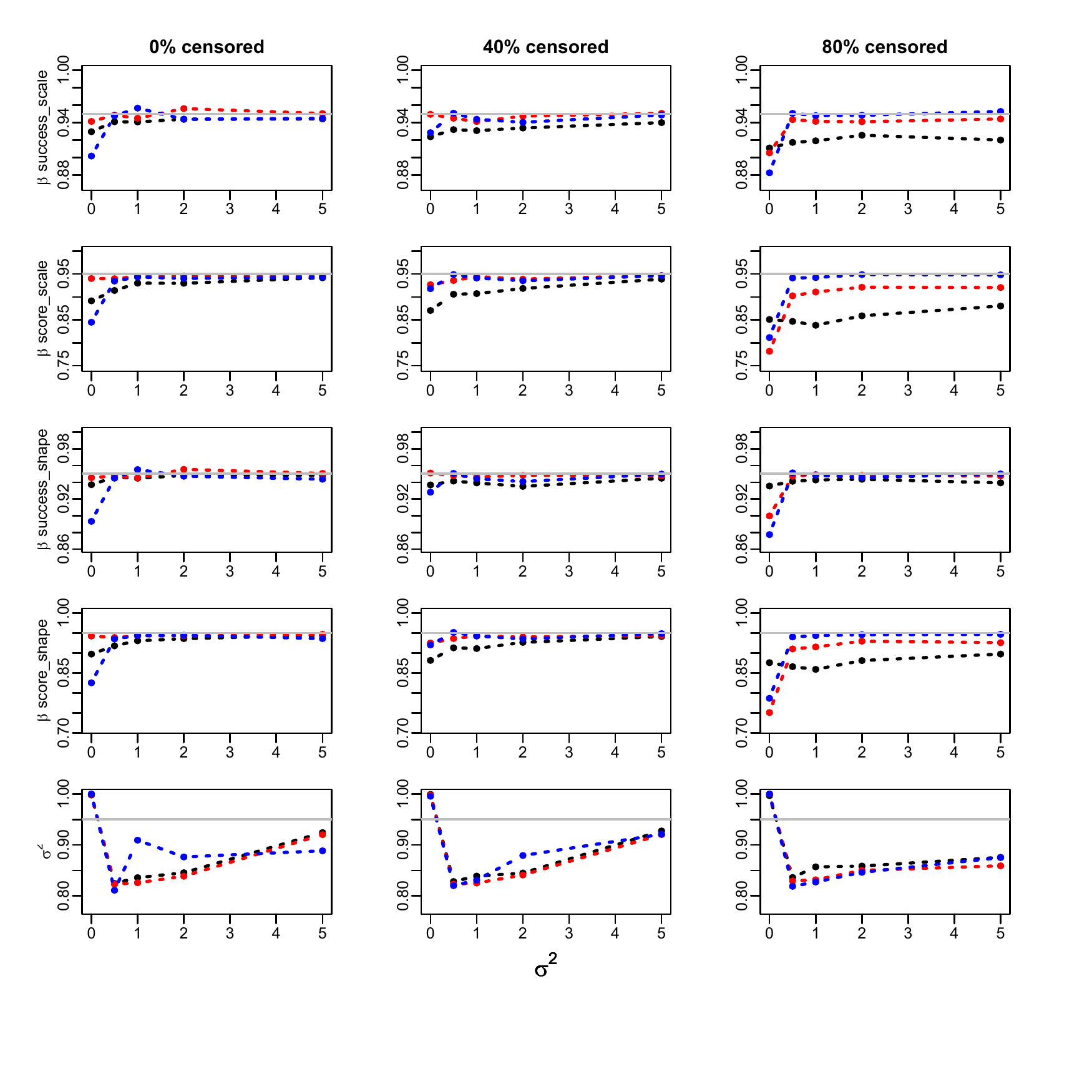}
	\caption{Coverage of the standard error based confidence intervals for the Cox regression parameters and $\sigma^2$ at nominal 95\% level. the Gompertz model.
Success proportion= 0.25. 10 clusters. Sample sizes: 300 (black), 1000 (red) and 10000 (blue).
True values: $\beta_{success-scale}$ = -0.5 , $\beta_{success-shape}$= -0.05 , $\beta_{score-scale}$= -1 , $\beta_{score-shape}$= -0.1}
	\label{CoverageSE10clustersGompertz4}
\end{figure}

\begin{figure}[ht]
	\centering
	\includegraphics[scale=1]{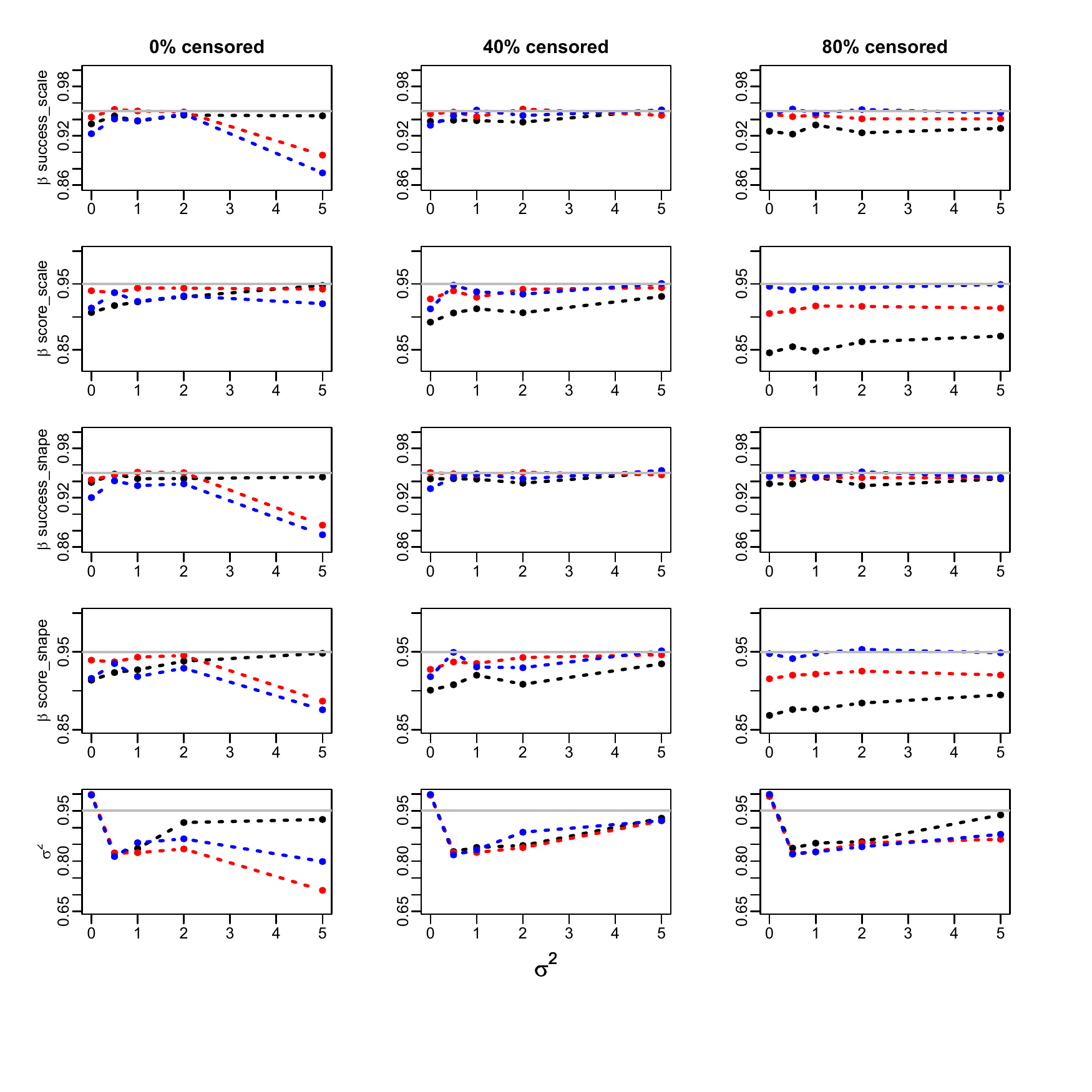}
	\caption{Coverage of the standard error based confidence intervals for the Cox regression parameters and $\sigma^2$ at nominal 95\% level. Gompertz model.
Success proportion= 0.5. 10 clusters. Sample sizes: 300 (black), 1000 (red) and 10000 (blue).
True values: $\beta_{success-scale}$ = 0.5 , $\beta_{success-shape}$= 0.05 , $\beta_{score-scale}$= 1 , $\beta_{score-shape}$= 0.1}
	\label{CoverageSe10clustersGompertz5}
\end{figure}

\begin{figure}[ht]
	\centering
	\includegraphics[scale=1]{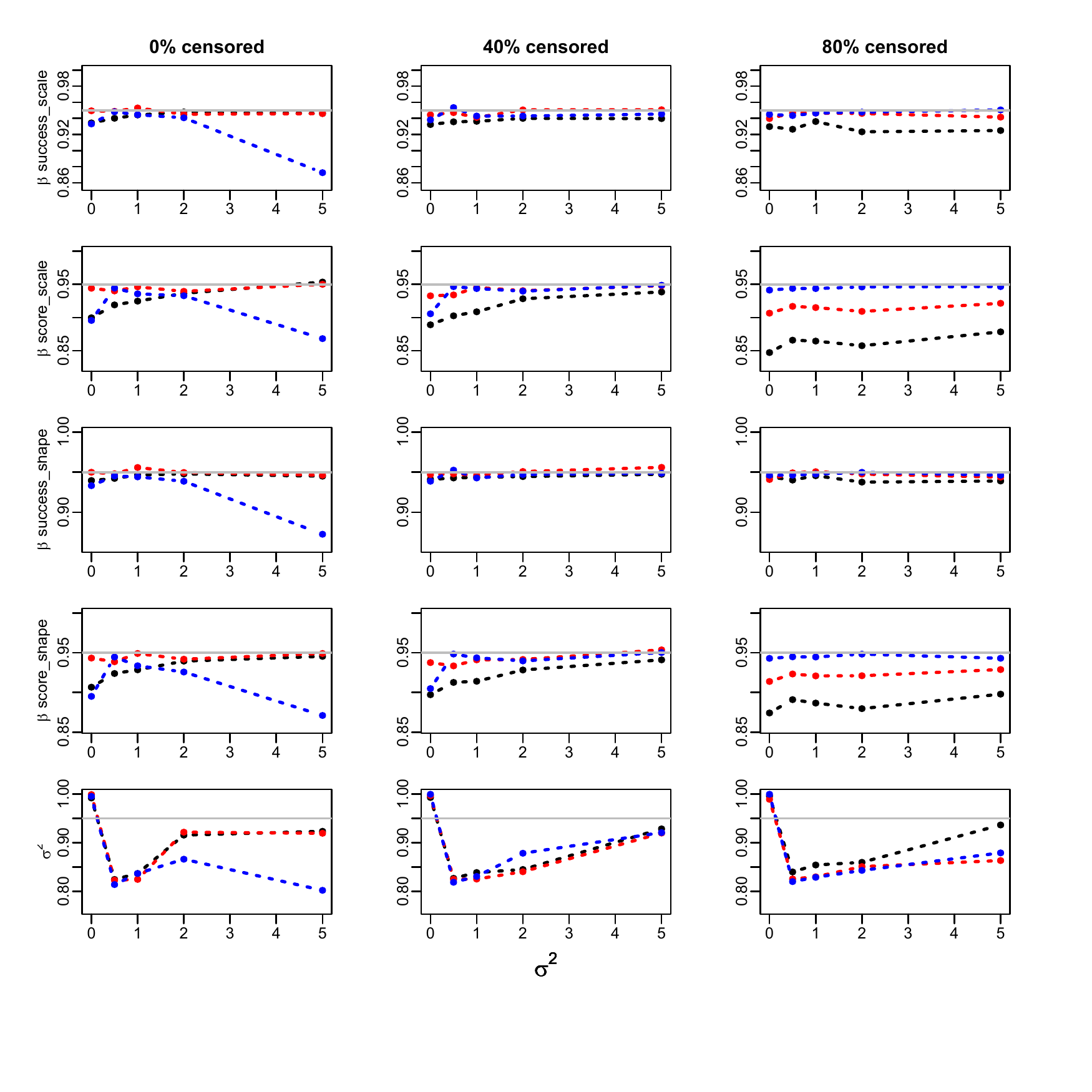}
	\caption{Coverage of the standard error based confidence intervals for the Cox regression parameters and $\sigma^2$ at nominal 95\% level. Weibull model.
Success proportion= 0.5. 10 clusters. Sample sizes: 300 (black), 1000 (red) and 10000 (blue).
True values: $\beta_{success-scale}$ = 0.5 , $\beta_{success-shape}$= -0.05 , $\beta_{score-scale}$= 1 , $\beta_{score-shape}$= - 0.1}
	\label{CoverageSE10clustersGompertz6}
\end{figure}

\begin{figure}[ht]
	\centering
	\includegraphics[scale=1]{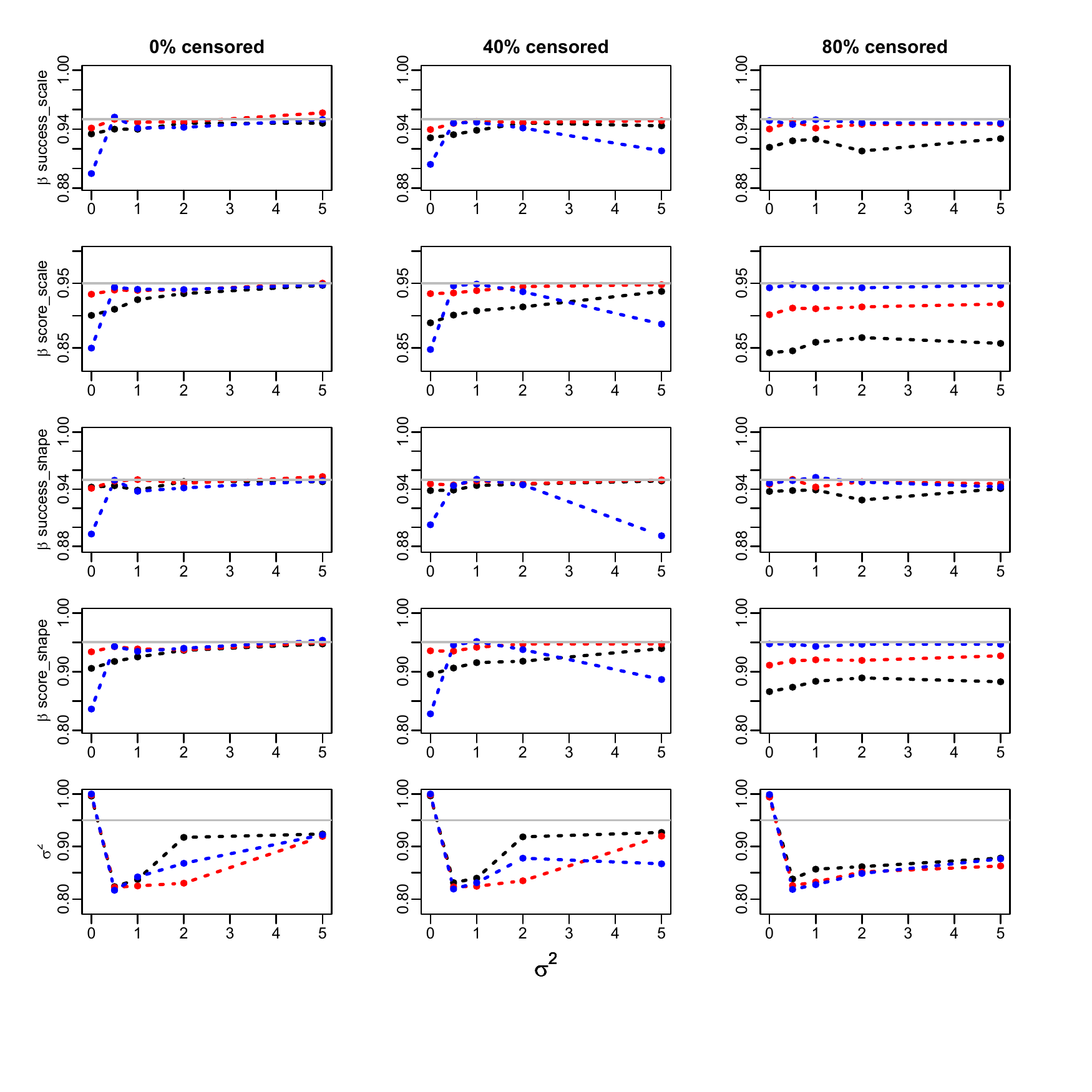}
	\caption{Coverage of the standard error based confidence intervals for the Cox regression parameters and $\sigma^2$ at nominal 95\% level. Gompertz model.
Success proportion= 0.5. 10 clusters. Sample sizes: 300 (black), 1000 (red) and 10000 (blue).
True values: $\beta_{success-scale}$ = -0.5 , $\beta_{success-shape}$= 0.05 , $\beta_{score-scale}$= -1 , $\beta_{score-shape}$= 0.1}
	\label{CoverageSE10clustersGompertz7}
\end{figure}

\begin{figure}[ht]
	\centering
	\includegraphics[scale=1]{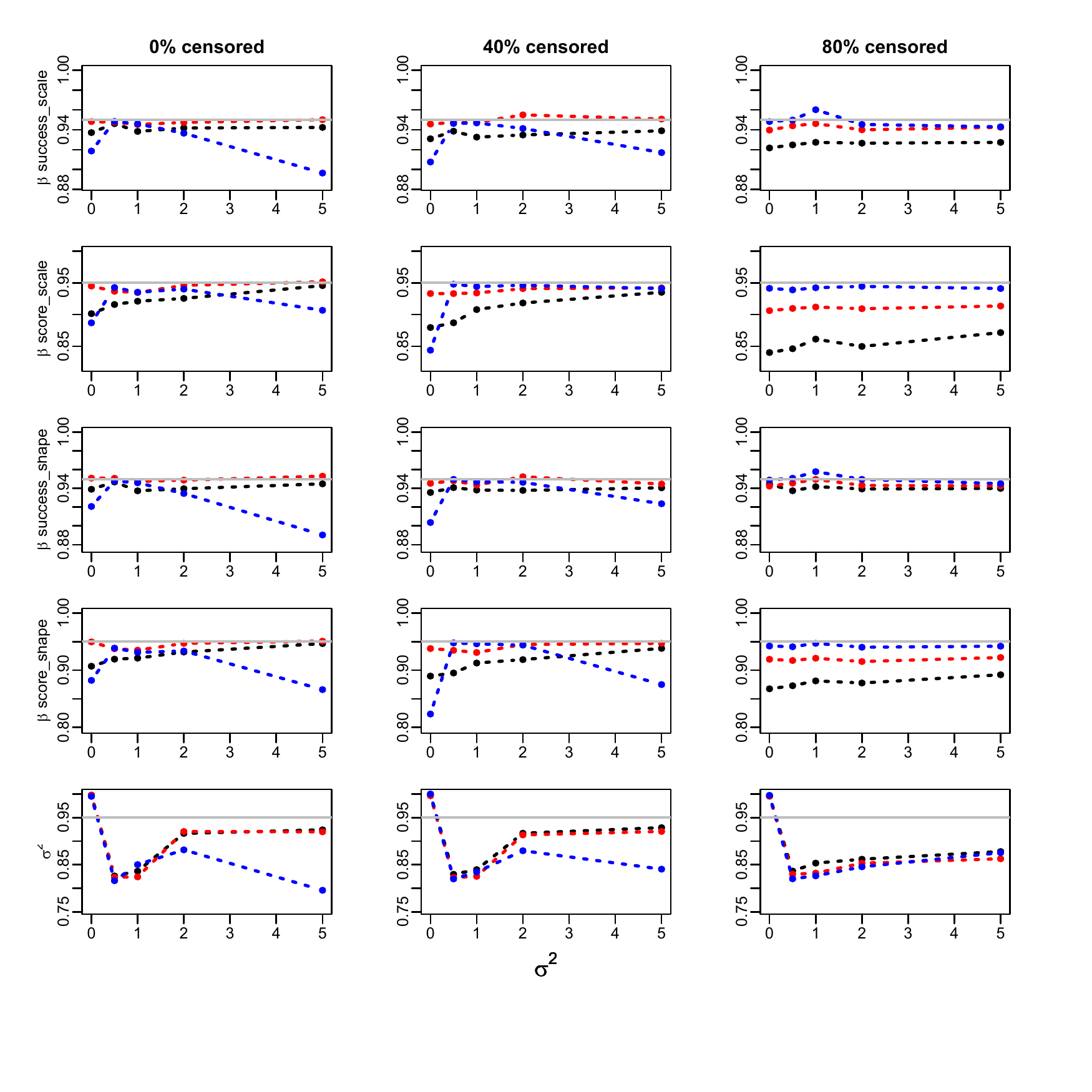}
	\caption{Coverage of the standard error based confidence intervals for the Cox regression parameters and $\sigma^2$ at nominal 95\% level. Gompertz model.
Success proportion= 0.5. 10 clusters. Sample sizes: 300 (black), 1000 (red) and 10000 (blue).
True values: $\beta_{success-scale}$ = -0.5 , $\beta_{success-shape}$= -0.05 , $\beta_{score-scale}$= -1 , $\beta_{score-shape}$= -0.1}
	\label{CoverageSE10clustersGompertz8}
\end{figure}

 \begin{figure}[ht]
	\centering
	\includegraphics[scale=1]{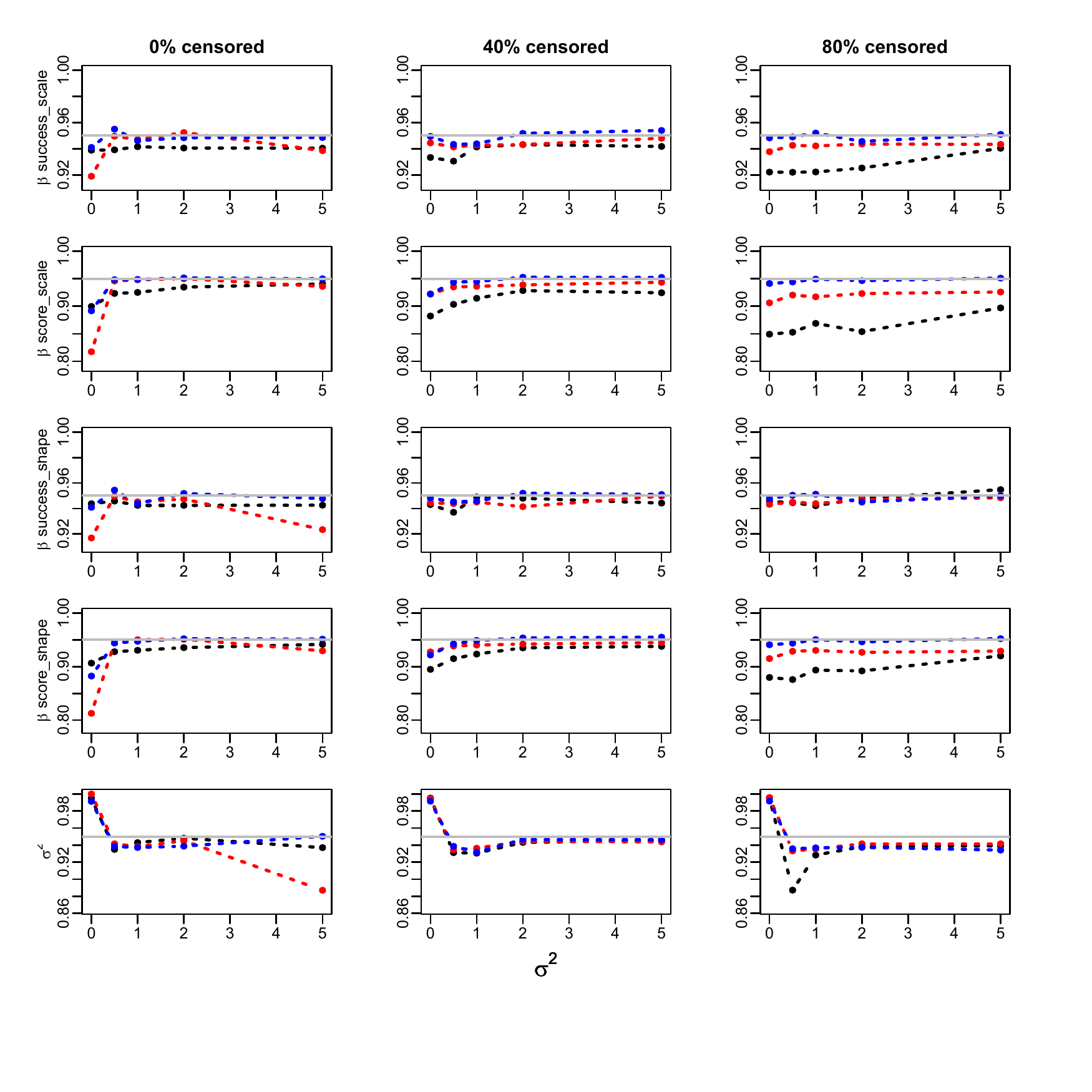}
	\caption{Coverage of the standard error based confidence intervals for the Cox regression parameters and $\sigma^2$ at nominal 95\% level. Gompertz model.
Success proportion= 0.25. 100 clusters. Sample sizes: 300 (black), 1000 (red) and 10000 (blue).
True values: $\beta_{success-scale}$ = 0.5 , $\beta_{success-shape}$= 0.05 , $\beta_{score-scale}$= 1 , $\beta_{score-shape}$= 0.1}
	\label{CoverageSE100clustersGompertz1}
\end{figure}

\begin{figure}[ht]
	\centering
	\includegraphics[scale=1]{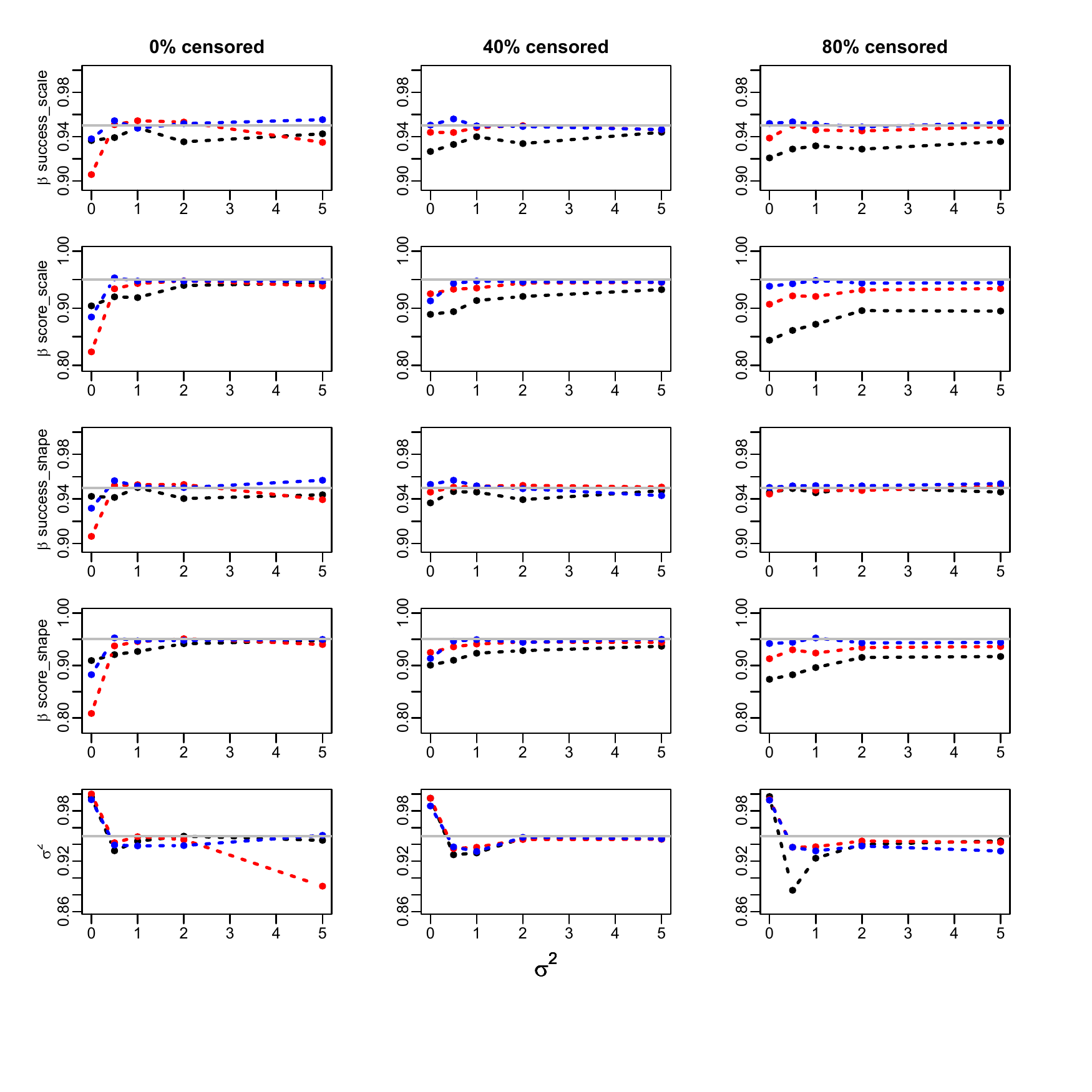}
	\caption{Coverage of the standard error based confidence intervals for the Cox regression parameters and $\sigma^2$ at nominal 95\% level. Gompertz model.
Success proportion= 0.25. 100 clusters. Sample sizes: 300 (black), 1000 (red) and 10000 (blue).
True values: $\beta_{success-scale}$ = 0.5 , $\beta_{success-shape}$= -0.05 , $\beta_{score-scale}$= 1 , $\beta_{score-shape}$= - 0.1}
	\label{CoverageSE100clustersGompertz2}
\end{figure}

\begin{figure}[ht]
	\centering
	\includegraphics[scale=1]{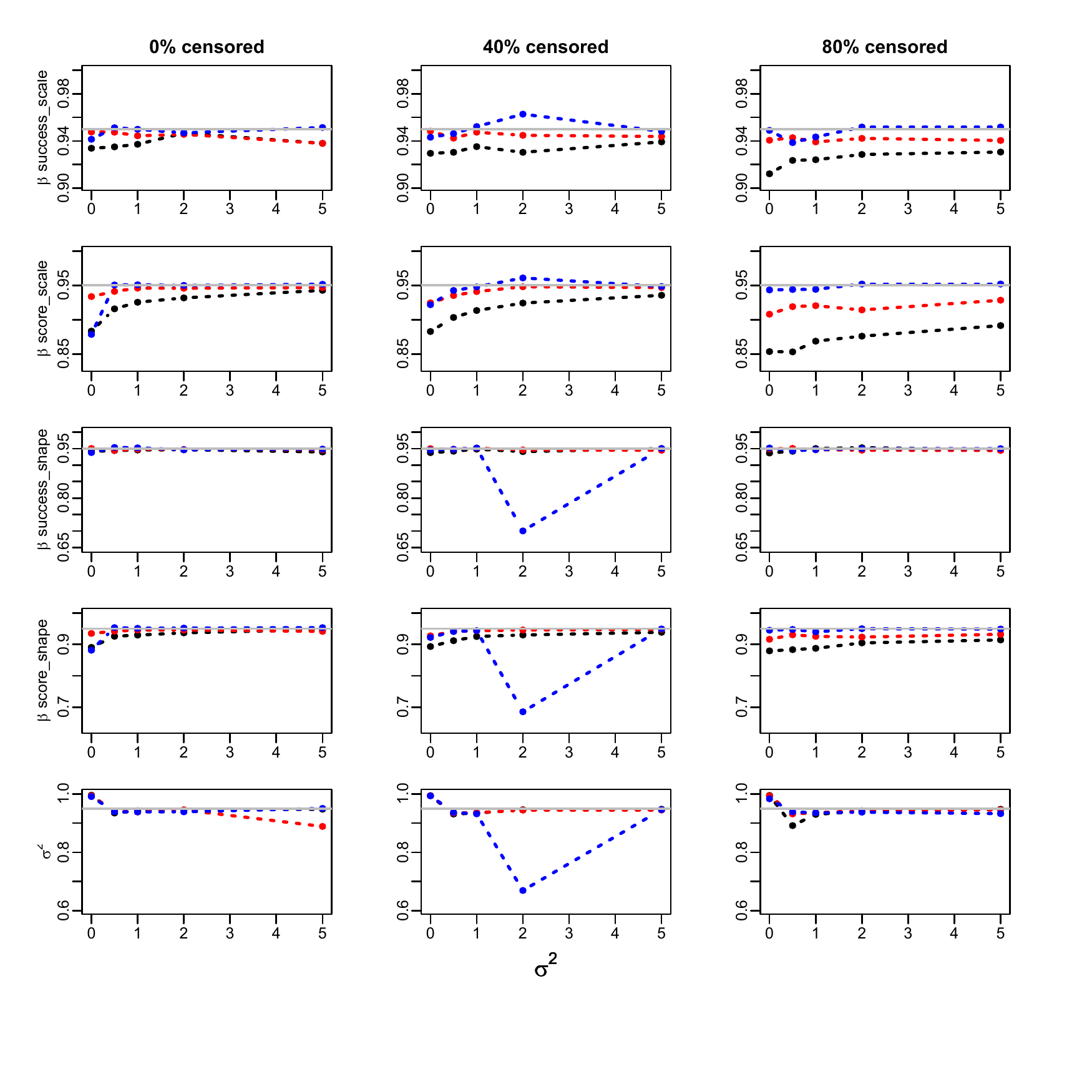}
	\caption{Coverage of the standard error based confidence intervals for the Cox regression parameters and $\sigma^2$ at nominal 95\% level. Gompertz model.
Success proportion= 0.25. 100 clusters. Sample sizes: 300 (black), 1000 (red) and 10000 (blue).
True values: $\beta_{success-scale}$ = -0.5 , $\beta_{success-shape}$= 0.05 , $\beta_{score-scale}$= -1 , $\beta_{score-shape}$= 0.1}
	\label{CoverageSE100clustersGompertz3}
\end{figure}

\begin{figure}[ht]
	\centering
	\includegraphics[scale=1]{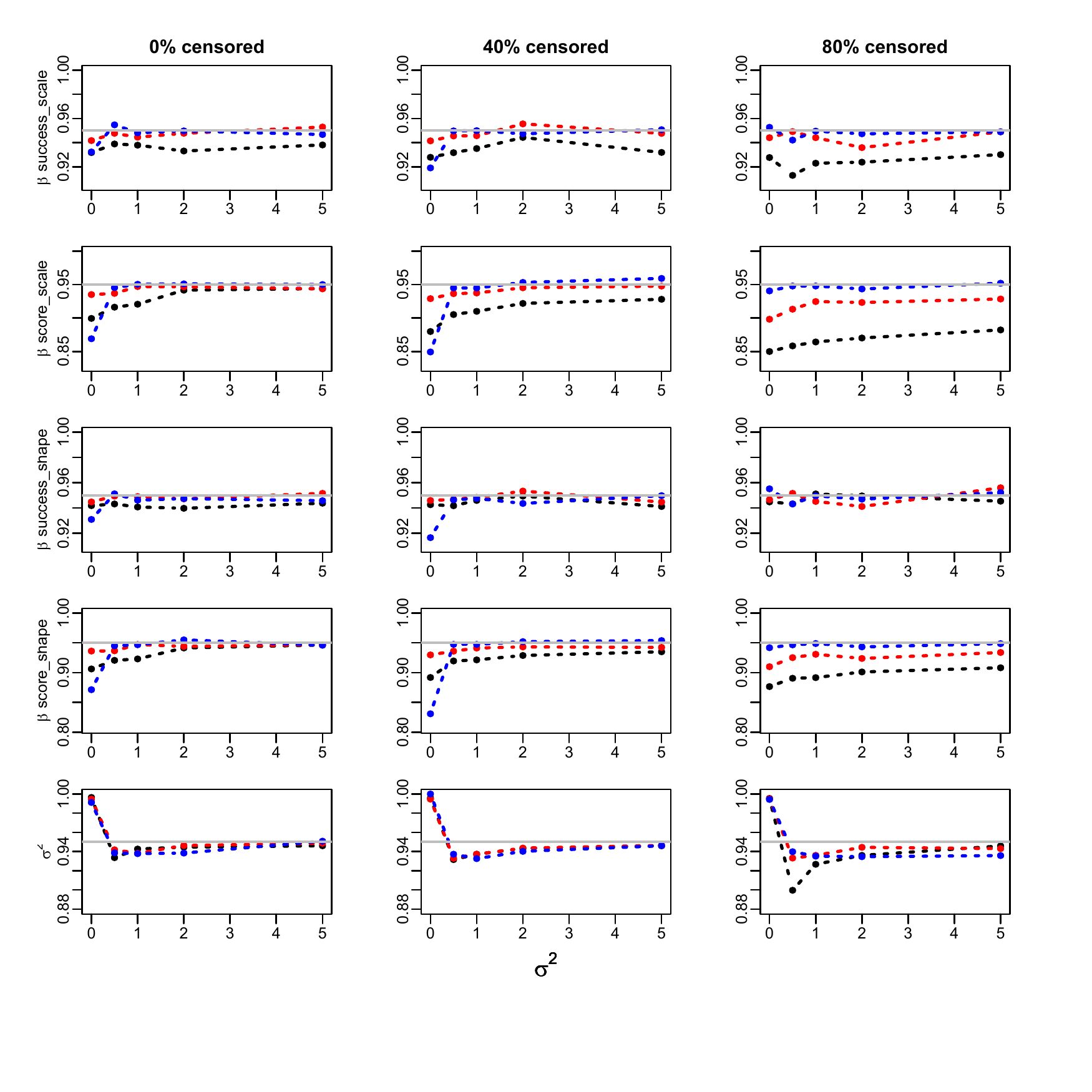}
	\caption{Coverage of the standard error based confidence intervals for the Cox regression parameters and $\sigma^2$ at nominal 95\% level. Gompertz model.
Success proportion= 0.25. 100 clusters. Sample sizes: 300 (black), 1000 (red) and 10000 (blue).
True values: $\beta_{success-scale}$ = -0.5 , $\beta_{success-shape}$= -0.05 , $\beta_{score-scale}$= -1 , $\beta_{score-shape}$= -0.1}
	\label{CoverageSE100clustersGompertz4}
\end{figure}

\begin{figure}[ht]
	\centering
	\includegraphics[scale=1]{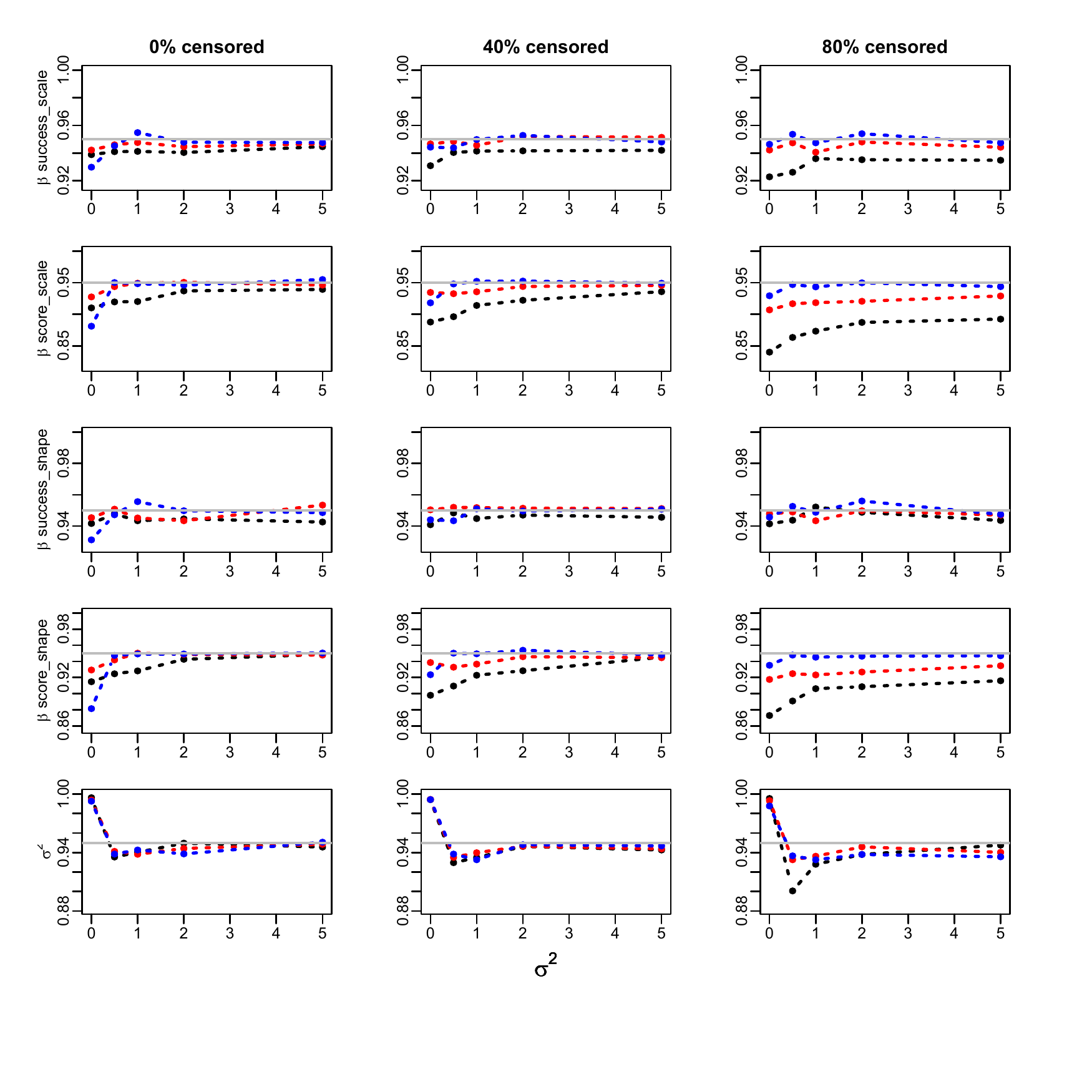}
	\caption{Coverage of the standard error based confidence intervals for the Cox regression parameters and $\sigma^2$ at nominal 95\% level. Gompertz model.
Success proportion= 0.5. 100 clusters. Sample sizes: 300 (black), 1000 (red) and 10000 (blue).
True values: $\beta_{success-scale}$ = 0.5 , $\beta_{success-shape}$= 0.05 , $\beta_{score-scale}$= 1 , $\beta_{score-shape}$= 0.1}
	\label{CoverageSE100clustersGompertz5}
\end{figure}

\begin{figure}[ht]
	\centering
	\includegraphics[scale=1]{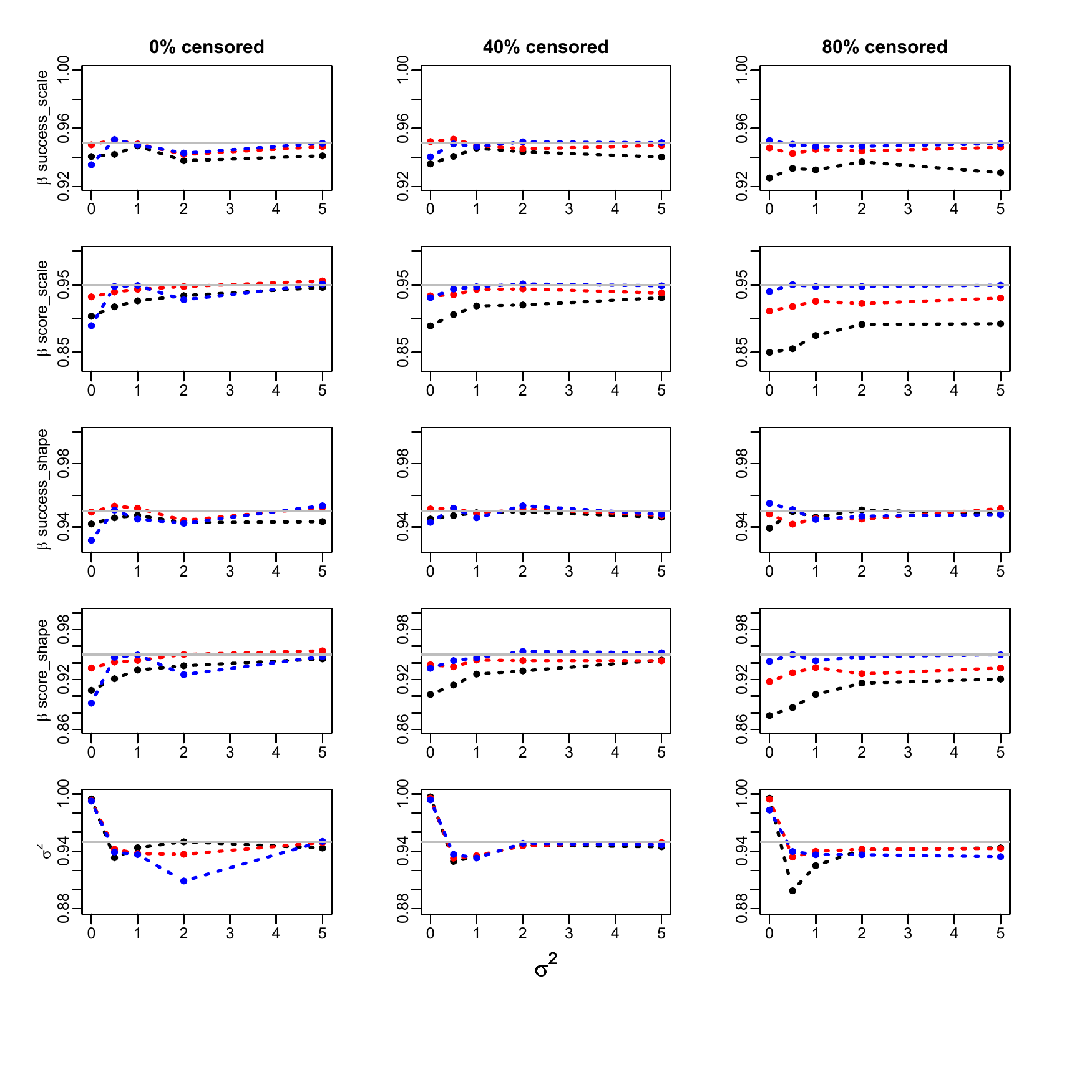}
	\caption{Coverage of the standard error based confidence intervals for the Cox regression parameters and $\sigma^2$ at nominal 95\% level. Gompertz model.
Success proportion= 0.5. 100 clusters. Sample sizes: 300 (black), 1000 (red) and 10000 (blue).
True values: $\beta_{success-scale}$ = 0.5 , $\beta_{success-shape}$= -0.05 , $\beta_{score-scale}$= 1 , $\beta_{score-shape}$= - 0.1}
	\label{CoverageSE100clustersGompertz6}
\end{figure}

\begin{figure}[ht]
	\centering
	\includegraphics[scale=1]{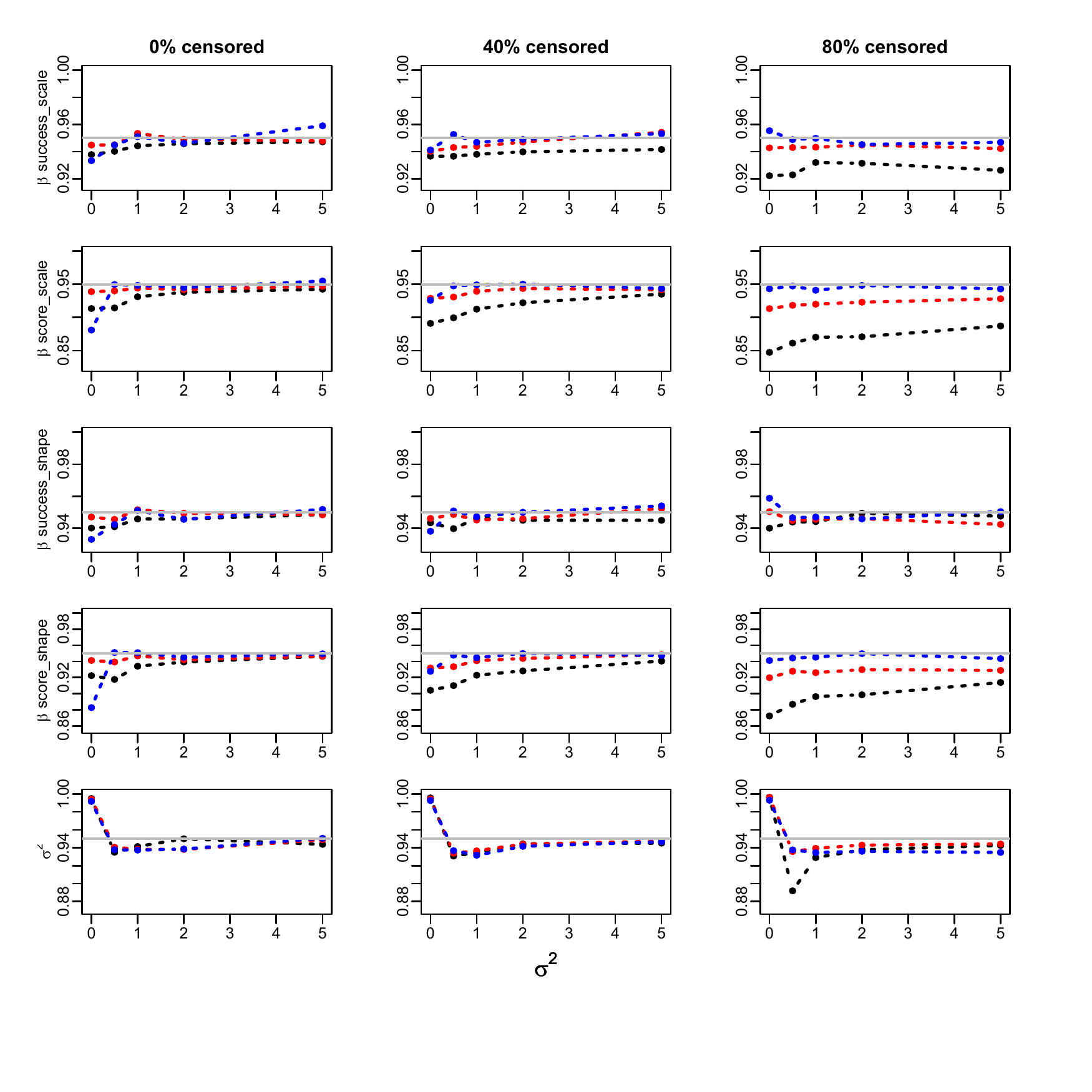}
	\caption{Coverage of the standard error based confidence intervals for the Cox regression parameters and $\sigma^2$ at nominal 95\% level. Gompertz model.
Success proportion= 0.5. 100 clusters. Sample sizes: 300 (black), 1000 (red) and 10000 (blue).
True values: $\beta_{success-scale}$ = -0.5 , $\beta_{success-shape}$= 0.05 , $\beta_{score-scale}$= -1 , $\beta_{score-shape}$= 0.1}
	\label{CoverageSE100clustersGompertz7}
\end{figure}

\begin{figure}[ht]
	\centering
	\includegraphics[scale=1]{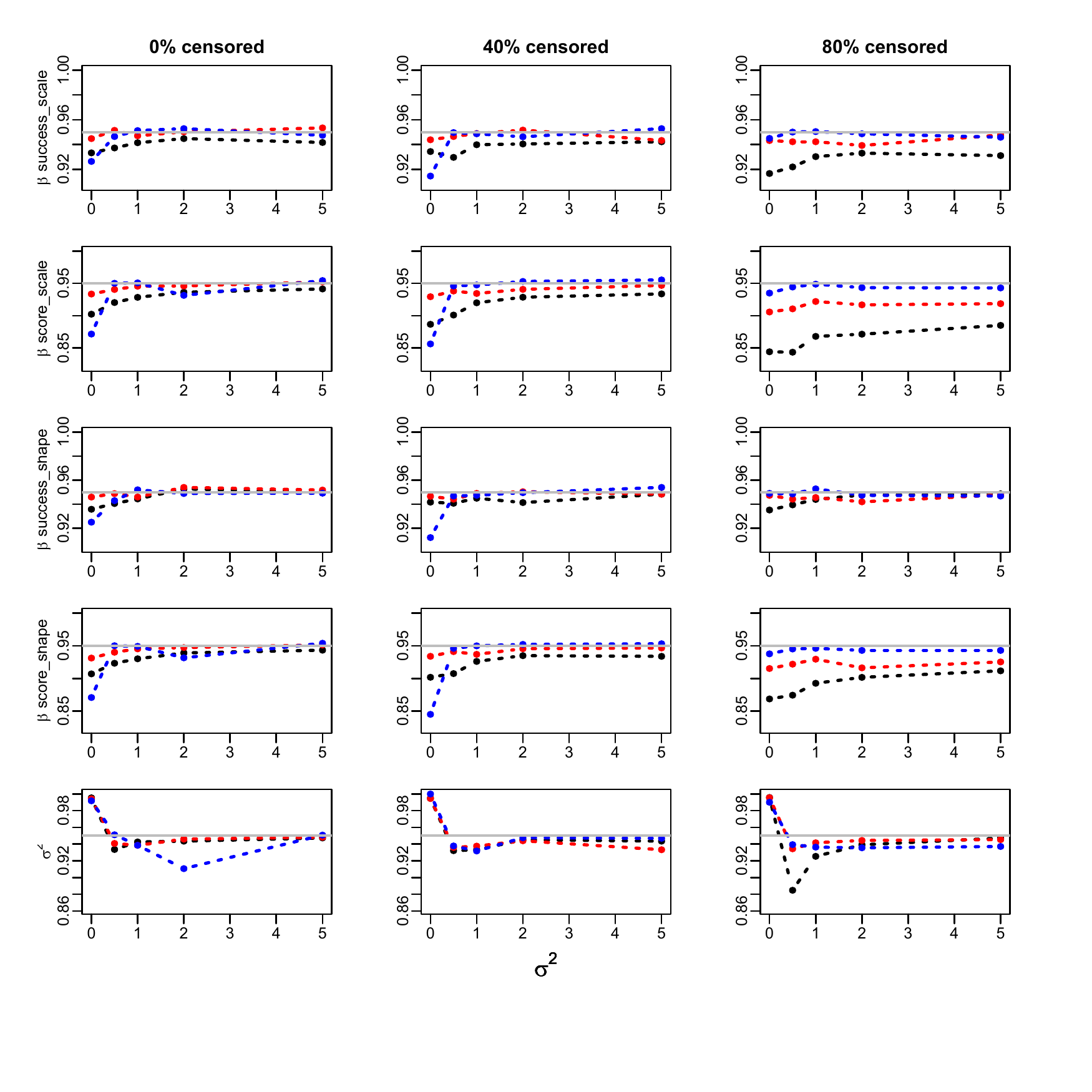}
	\caption{Coverage of the standard error based confidence intervals for the Cox regression parameters and $\sigma^2$ at nominal 95\% level. Gompertz model.
Success proportion= 0.5. 100 clusters. Sample sizes: 300 (black), 1000 (red) and 10000 (blue).
True values: $\beta_{success-scale}$ = -0.5 , $\beta_{success-shape}$= -0.05 , $\beta_{score-scale}$= -1 , $\beta_{score-shape}$= -0.1}
	\label{CoverageSE100clustersGompertz8}
\end{figure}

\end{document}